# Euclid Imaging Consortium

Imaging the dark Universe with Euclid

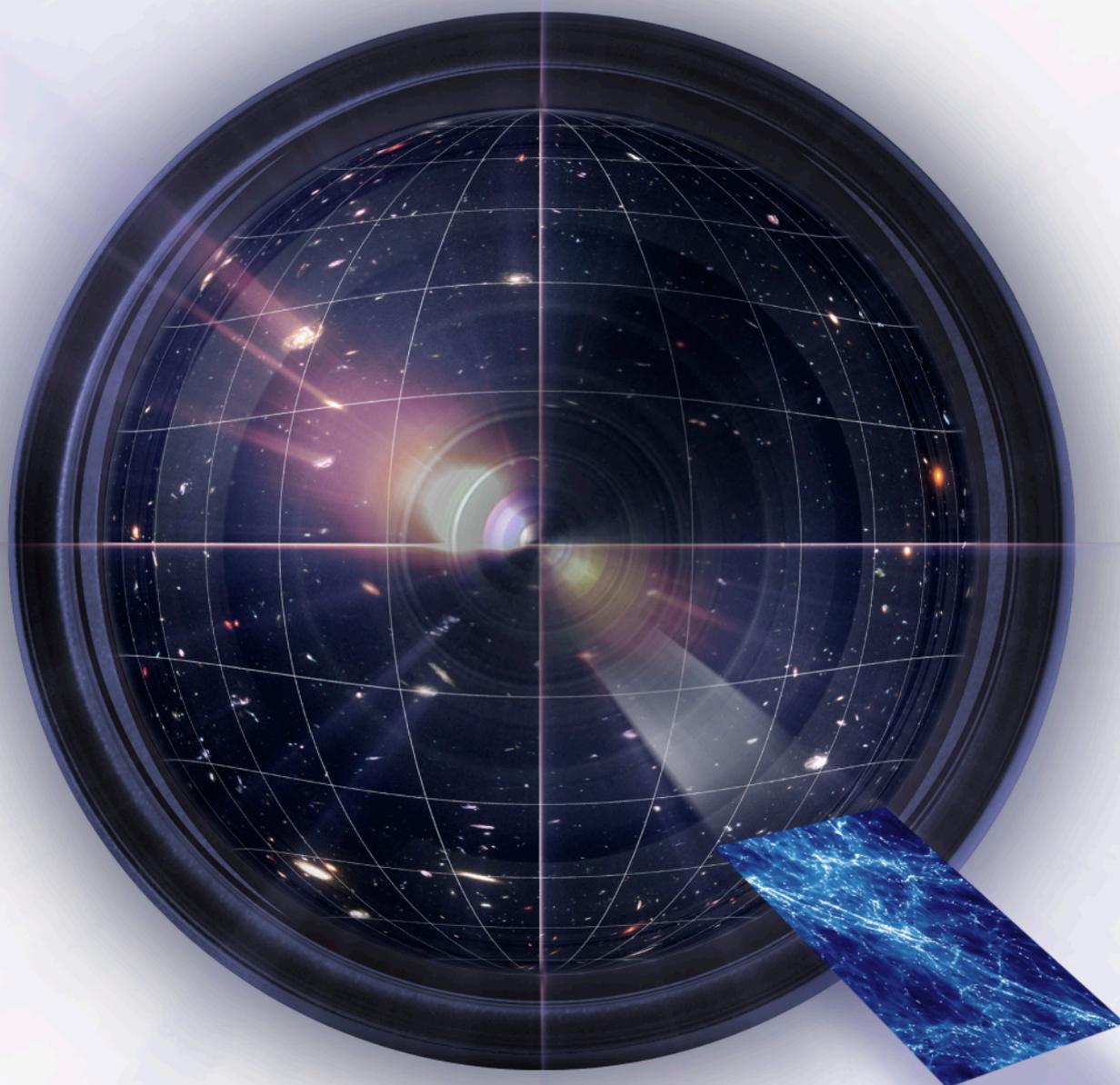

# Science Book

Euclid Assessment Phase

# EIC Science Book

by the Euclid Imaging Consortium

EDITED BY:

Alexandre Réfrégier

and

Adam Amara
Thomas Kitching
Anaïs Rassat
Roberto Scaramella
Jochen Weller

1st January 2010

Euclid Imaging Consortium

# Euclid Imaging Consortium Members

**PI**: Alexandre Refregier (CEA Saclay)
**Co-PIs**: Ralf Bender (MPE Germany), Mark Cropper (MSSL, UK), Roberto Scaramella (INAF-Oss. Roma, Italy), Simon Lilly (ETH Zurich, Switzerland), Nabila Aghanim (IAS Orsay, France), Francisco Castander (IEEC Barcelona, Spain), Jason Rhodes (JPL, USA)

**Study Manager**: Jean-Louis Augueres (CEA Saclay)
**System engineer**: Jerome Amiaux (CEA Saclay)
**Instrument scientists**: Mario Schweitzer (MPE), Mark Cropper (MSSL) Olivier Boulade (CEA Saclay), Adam Amara (ETH Zurich), Jeff Booth (JPL), Anna Maria di Giorgio (INAF-IFSI)

**Co-Investigators: France**: Marian Douspis (IAS Orsay), Yannick Mellier (IAP Paris), Olivier Boulade (CEA Saclay), **Germany**: Peter Schneider (U. Bonn), Oliver Krause (MPIA Heidelberg), Frank Eisenhauer (MPE Garching), **Italy**: Lauro Moscardini (U. Bologna), Luca Amendola (INAF-Oss. Roma and University of Heidelberg), Fabio Pasian (INAF-OATS), **Spain**: Ramon Miquel (IFAE Barcelona), Eusebio Sanchez (CIEMAT Madrid), **Switzerland**: George Meylan (EPFL-UniGE), Marcella Carollo (ETH Zurich), Francois Widi (EPFL-UniGE), **UK**: John Peacock (IfA Edinburgh), Sarah Bridle (UCL, London), Ian Bryson (IfA Edinburgh), **USA**: Jeff Booth (JPL), Steven Kahn (Stanford U.)

**Science working group coordinators**:
**Weak Lensing**: Adam Amara (ETH Zurich), Andy Taylor (IfA Edinburgh)
**Clusters/CMB**: Jochen Weller (U. Munich/MPE), Nabila Aghanim (IAS Orsay)
**BAO (photometric)**: Francisco Castander (IEEC Barcelona), Anais Rassat (CEA Saclay)
**Supernovae**: Isobel Hook (Oxford, and Obs. Rome), Massimo Della Valle (Oss. Capodimonte/ICRA)
**Theory**: Luca Amendola (INAF-Oss. Roma), Martin Kunz (U. Sussex/U. Geneva)
**Strong Lensing**: Matthias Bartelmann (U. Heidelberg), Leonidas Moustakas (JPL)
**Galaxy Evolution**: Marcella Carollo (ETH Zurich), Hans-Walter Rix (MPIA)
**Galactic studies and planets**: Eva Grebel (U. Heidelberg), Jean-Philippe Beaulieu (IAP Paris),
**Photometric Redshifts**: Ofer Lahav (UCL), Adriano Fontana (INAF Oss. Roma)
**Image Simulations**: Jason Rodes (JPL), Lauro Moscardini (U. Bologna)
**Communication**: Marian Douspis (IAS Orsay)

**Science**:
**France**: Karim Benabed (IAP), Emmanuel Bertin (IAP), Pascal Bordé (IAS Orsay), Frederic Bournaud (CEA Saclay), Marian Douspis (IAS Orsay), Herve Dole (IAS Orsay), Raphael Gavazzi (IAP), Guilaine Lagache (IAS Orsay), M. Langer (IAS Orsay), Henry McCracken (IAP), Christophe, Magneville (CEA Saclay), M. Ollivier (IAS Orsay), Nathalie Palanque-Delabrouille (CEA Saclay), Stephane Paulin-Henriksson (CEA Saclay), Sandrine Pires (CEA Saclay), Jean-Luc Starck (CEA Saclay), Romain Teyssier (CEA Saclay/U. Zurich)
**Germany**: Wolfgang Brandner (MPIA), Hans Böhringer (MPE), Frank Eisenhauer (MPE), Natascha Foerster-Schreiber (MPE), Bertrand Goldman (MPIA), Oliver Krause (MPIA), Peter Melchior (ITA - Heidelberg), Steffi Phleps (MPE), Roberto Saglia (MPE), Stella Seitz (MPE), Peter Schneider (Univ. Bonn), Benjamin Joachimi (Univ. Bonn), Joachim Wambsgans (Univ. Heidelberg), David Wilman (MPE)
**Italy**: Vicenzo Antonuccio (INAF-Oss. Catania), Carlo Baccigalupi (SISSA), Fabio Bellagamba (U. di Bologna), Silvio Bonometto (U. Milano Bicocca), Andrea Grazian (INAF- Oss. Roma),

Marco Lombardi (U. di Milano), Roberto Mainini (INAF Oss. Roma), Filippo Mannucci (INAF - IRA), Roberto Maoli (U. Roma La Sapienza), Sabino Matarrese (U. di Padova), Alessandro Melchiorri (U. Roma La Sapienza), Massimo Meneghetti (INAF- Oss. Bologna), Julian Merten (INAF Oss. Bologna), Lauro Moscardini (U. di Bologna), Claudia Quercellini (U. Roma Tor Vergata), Mario Radovich (INAF Oss. Capodimonte), Anna Romano (U. Roma La Sapienza), Roberto Scaramella (INAF Oss. Roma), Andrea Zacchei (INAF - Oss. Trieste)

**Spain**: Ricard Casas (IEEC Barcelona), Martin Crocce (IEEC Barcelona), Pablo Fosalba (IEEC Barcelona), Enrique Gaztanaga (IEEC Barcelona), Ignacio Sevilla (CIEMAT Madrid), Eusebio Sanchez (CIEMAT Madrid), Juan Garcia-Bellido (UAM Madrid)

**Switzerland**: Rongmon Bordoloi (ETH Zurich), Julien Carron (ETH Zurich), Frederic Courbin (EPFL Lausanne), Stephane Paltani (ISDC/U. Geneva), Justin Read (U. Zurich/U. Leicester), Robert Smith (U. Zurich)

**UK**: Filipe Abdalla (UCL), Manda Banerji (UCL), David Bacon (U. Portsmouth), Kirk Donnacha (UCL), Ignacio Ferreras (MSSL-UCL), Gert Hutsi (UCL), Alan Heavens (IfA Edinburgh), Tom Kitching (IfA Edinburgh), Lindsay King (IoA Cambridge), Ofer Lahav (UCL), Katarina Markovic (UCL/Ludwigs-Maximilians U.), Richard Massey (IfA Edinburgh), Michael Schneider (U. Durham), Fabrizio Sidoli (UCL), Jiayu Tang (UCL/IPMU), Shaun Thomas (UCL), Lisa Voigt (UCL)

**USA**: Joel Berge (JPL), Benjamin Dobke (JPL), Richard Ellis (Caltech), David Johnston (Northwest. U.), Michael Seiffert (JPL), Ali Vanderveld (JPL)

**Other**: Eduardo Cypriano (U. Sao Paulo), Hakon Dahle (U. Oslo), Konrad Kuijken (Leiden)

**Instrument**:

**France**: Christophe Cara (CEA Saclay), Sandrine Cazaux (CEA Saclay), Arnaud Claret (CEA Saclay), Philippe Daniel-Thomas (CEA Saclay), Eric Doumayrou (CEA Saclay), Cydalise Dusmesnil (IAS Orsay), Jean-Jacques Fourmond (IAS Orsay), Jean-Claude Leclech (IAS Orsay), G. Morinaud (IAS Orsay), Samuel Ronayette (CEA Saclay), Zihong Sun (CEA Saclay), Tristan VanDenBerghe (IAS Orsay)

**Germany**: Reiner Hofmann (MPE), Rory Holmes (MPIA), Reinhard Katterloher (MPE), Oliver Krause (MPIA)

**Italy**: Marco Frailis (INAF Oss. Trieste), Pasquale Cerulli Irelli (INAF IFSI), Stefano Gallozzi (INAF Oss. Roma), Michele Maris (INAF Oss. Trieste), Renato Orfei (INAF IFSI), Fabio Pasian (INAF Oss. Trieste), Andrea Zacchei (INAF - Oss. Trieste)

**Spain**: Laia Cardiel (IFAE Barcelona), Francesc Madrid (IEEC Barcelona), Santiago Serrano (IEEC Barcelona)

**Switzerland**: Adrian Glauser (ETH Zurich), Francois Wildi (ISDC/U. Geneva)

**UK**: Eli Atad-Ettedgui (UK-ATC), Ian Bryson (UK-ATC), Richard Cole (MSSL), Jason Gow (Open U), Phil Guttridge (MSSL) , Andrew Holland (Open U), Neil Murray (Open U), Kerrin Rees (MSSL), Phil Thomas (MSSL), Dave Walton (MSSL)

**USA**: Parker Fagrelius (JPL), Kirk Gilmore (Stanford), Steven Kahn (Stanford), Andrew Rasmussen (Stanford), Suresh Seshadri (JPL), Roger Smith (Caltech), Harry Teplitz (IPAC/Caltech)

# ACKNOWLEDGEMENTS

---

We thank the national agencies and institutes in the EIC member countries for their support of the consortium during the assessment phase. We are particularly grateful to CNES, CEA, CNRS, CNAP in France, DLR and the Max Planck Society in Germany, STFC in the UK, ASI (contract "Euclid-DUNE" No. I/064/08/0) in Italy, ETH Zurich, EPF Lausanne and U. Geneva in Switzerland, the Spanish Ministerio de Educacion y Ciencia (grant ESP2007-29984-E) and the Spanish Ministerio de Ciencia e Innovacion (grant AYA2008-03347-E/ESP) in Spain, and the Jet Propulsion Laboratory, run under a contract for NASA by the California Institute of Technology in the US.

We also thank ESA for their coordination of the Euclid Assessment phase, and the members of the Euclid Science Study team and of the ENIS consortium for scientific interactions.

# CONTENTS

# Part I

# INTRODUCTION

# EIC Science Book Overview

*Authors: Alexandre Refregier(CEA Saclay) for the EIC collaboration*

**Abstract**

The energy density of the universe is dominated by dark energy and dark matter, two mysterious components which pose some of the most important questions in fundamental science today. Euclid is a high-precision survey mission designed to answer these questions by mapping the geometry of the dark universe. Euclid's Visible-NIR imaging and spectroscopy of the entire extragalactic sky will further produce extensive legacy science for various fields of astronomy. Over the 2008-2009 period, Euclid has been the object of an ESA Assessment Phase in which the study of the Euclid Imaging instrument was under the responsibility of the Euclid Imaging Consortium (EIC). In this chapter, we provide an overview of the EIC Science Book which presents the studies done by the EIC science working groups in the context of this study phase. We first give a brief description of the Euclid mission and of the imaging instrument and surveys. We then summarise the primary and legacy science which will be achieved with the Euclid imaging surveys, along with the simulations and data handling scheme which have been developed to optimise the instrument and ensure its science performance.

## 1.1 Introduction

The energy density of the universe is dominated by dark energy and dark matter, two mysterious components which pose some of the most important questions in fundamental science today. Euclid is a high-precision survey mission designed to answer these questions by mapping the geometry of the dark universe. Euclid's Visible-NIR imaging and spectroscopy of the entire extragalactic sky will further produce extensive legacy science for various fields of astronomy. The Euclid payload consist of a wide-field imager in the visible and NIR and a NIR spectrometer. During the 2008-2009 period, Euclid has been the object of an ESA Assessment Phase during which the study of the imaging instrument was under the responsibility of the Euclid Imaging Consortium (EIC; see members and organisation on page 3). This EIC Science book presents the work done by the EIC science working groups in the context of this Assessment Phase. It is a supporting document within the EIC data pack delivered at the end of the assessment phase. Some of this work has appeared in scientific publications which are listed in Chapter 2. A description of the overall Euclid mission is presented in the Euclid Assessment Study Report (Euclid Yellow Book).

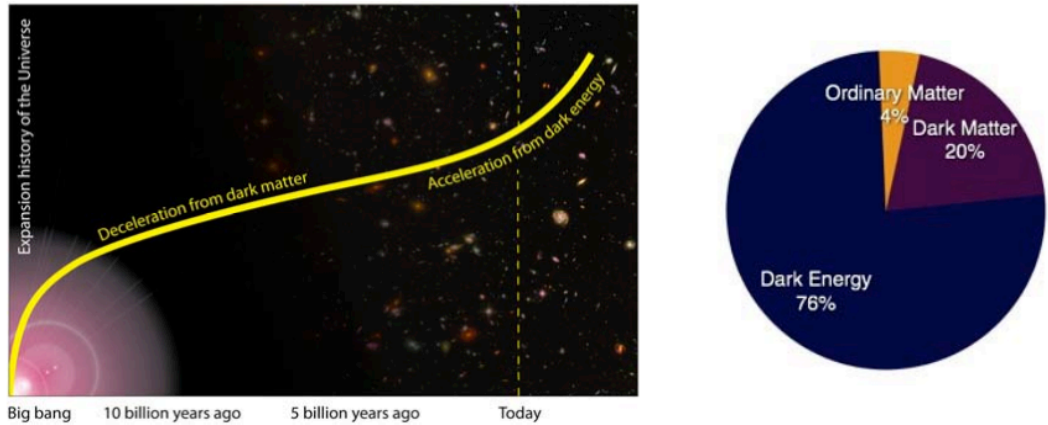

Figure 1.1: The Dark Universe. Left: Evolution of the universe from a homogeneous state after the big bang through cooling and expansion. Right: Mass-energy budget at our cosmological epoch showing that the universe is dominated by two mysterious components, dark energy and dark matter, whose nature poses some of the most important questions in fundamental physics.

In this overview of the EIC Science Book, we first give a brief description of the Euclid mission and then describe the imaging instrument and surveys. We then summarise the primary and legacy science which will be achieved with the imaging instruments, as well as the simulations and data handling scheme which have been developed to optimise the instrument and ensure its scientific performance.

## 1.2   The Euclid Mission

Euclid is a candidate ESA mission to map the geometry and evolution of the dark universe with unprecedented precision (see Figure 1.1). Its primary goal is to place high accuracy constraints on Dark Energy, Dark Matter, Gravity and cosmic initial conditions using two independent cosmological probes: weak gravitational lensing (WL) and baryonic acoustic oscillations (BAO). For this purpose, Euclid will measure the shape and spectra of galaxies over the entire extragalactic sky in the visible and NIR, out to redshift 2, thus covering the period over which dark energy accelerated the universe expansion (<10 Billion years). Galaxy clusters and the Integrated Sachs-Wolfe effect will be used as secondary cosmological probes. The Euclid datasets will also provide unique legacy surveys for the study of galaxy evolution, large-scale structure, the search for high redshift objects and for various other fields of astronomy.

The baseline mission is based on a telescope with a primary mirror of 1.2 m diameter (see Figure 1.2). The payload baseline comprises wide field instruments (0.5 deg$^2$): an imaging instrument comprising a visible and a NIR channel, and a NIR spectroscopic instrument. The visible channel is used to measure the shapes of galaxies for weak lensing, with a resolution of 0.18 arcsec in a wide visible red band (R+I+Z, $0.55 - 0.92\mu$). The NIR photometric channel provides three NIR bands (Y, J, H, spanning $1.0 - 1.6\mu$) with a resolution of 0.3 arcsec. The baseline for the NIR spectroscopic channel operates in the wavelength range 1.0–2.0 micron in slitless mode at a spectral resolution $R \sim 500$, employing 0.5" pixels. The optional spectroscopic implementation is slit spectroscopy using digital micro-mirror devices (DMDs).

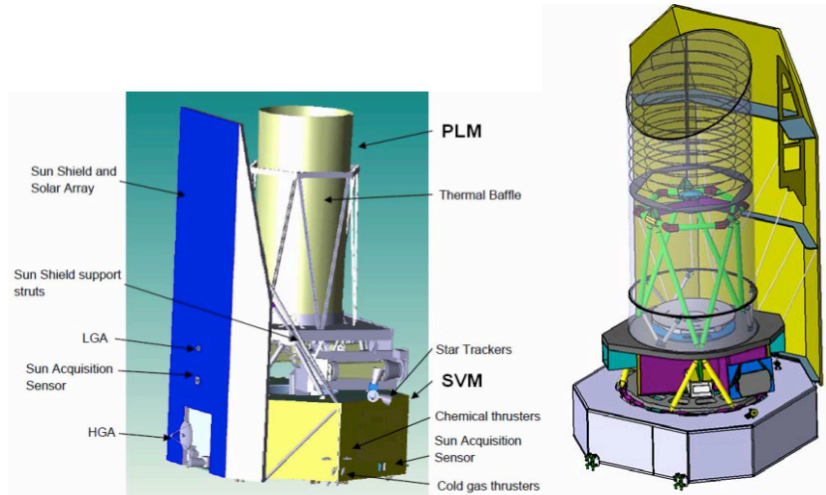

Figure 1.2: Euclid Spacecraft concept from EADS (left) and Thales-Alenia (right).

The mission will perform a wide survey of the entire extragalactic sky (20,000 deg$^2$) down to 24.5 AB magnitude in the visible, thus providing 30–40 resolved galaxies per amin$^2$. For all galaxies, photometric redshifts are obtained from the broad-band visible and near-IR measurements and complementary ground-based observations in other visible bands. For 40–60 million galaxies with $H\alpha$ line flux level $> 3 - 4 \times 10^{-16}$ erg s$^{-1}$ cm$^{-2}$ the slitless spectrometer will directly measure the redshift with a success rate in excess of 35%. A deep survey will also be performed to monitor the stability of the spacecraft and payload and for legacy science.

The Euclid mission concept followed from two dark energy related missions which were proposed to ESAs Cosmic Vision programme and were jointly selected: the Dark UNiverse Explorer (DUNE) and the SPectroscopic All-sky Cosmology Explorer (SPACE). Initial studies in 2008 led to the Euclid merged concept. Euclid has been the object of an Assessment Phase in 2008-2009 in the context of ESA's Cosmic Vision programme and is a candidate for the first medium mission launch slot in late 2017. A description of the overall Euclid mission is presented in the Euclid Assessment Study Report (Euclid Yellow Book).

## 1.3 The Euclid Imaging Instrument and Surveys

The Euclid baseline payload consist of a Korsch telescope with a primary mirror of 1.2 m diameter. The design of the telescope optics is such that it provides a large field of view and that it can feed the imaging channels and a spectroscopic channel. The imaging instrument (see Figure 1.3) are optimised for weak lensing and consists of a CCD based optical imaging channel (VIS), a NIR imaging photometry channel (NIP), a Common Mechanical Assembly (COMA) and a Payload Data Handling Unit (PDHU).

VIS will measure the shapes of galaxies with a resolution of 0.18 arcsec (PSF FWHM) with 0.1 arcsec pixels in one wide visible band (R+I+Z). NIP contains three NIR bands (Y, J, H), employing HgCdTe NIR detector with 0.3 arcsec pixels. To accomplish the wide and deep surveys within the nominal mission lifetime of 5 years, each channel has a large field of view of about 0.5

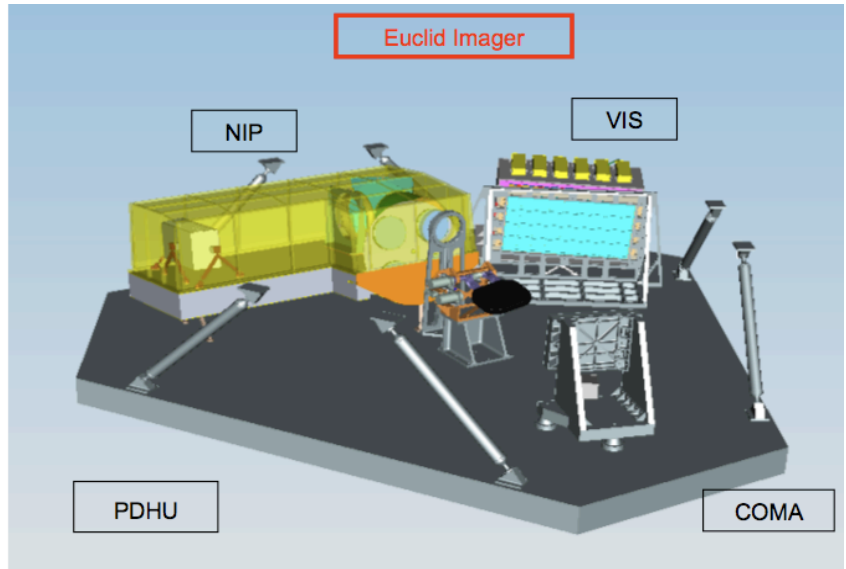

Figure 1.3: The Euclid Imaging instrument showing the visible imaging camera (VIS), the NIR imaging camera (NIP), the common mechanical assembly (COMA) and the Payload data Handling Unit (PDHU).

deg$^2$, and the system design is optimised for a sky survey with fast attitude slews to support a step-and-stare tiling mode. To meet the survey depth and sensitivity, the telescope has a well baffled design and is cooled to minimise background noise. For the NIR detectors, on-board on-the-ramp processing will be performed, i.e. combining image frames to lower the noise. The NIR related optics and detectors are cooled down to $\sim 100$ K.

Euclid's primary wide survey will cover 20,000 deg$^2$, i.e. the entire extragalactic sky, thus measuring shapes and redshifts of galaxies to redshift 2 as required for weak lensing and BAO. For weak lensing, Euclid will measure the shape of over 2 billion galaxies with a density of 30–40 resolved galaxies per amin$^2$ in one broad visible R+I+Z band (550-920 nm) down to AB mag 24.5 ($10\sigma$ extended). The photometric redshifts for these galaxies will reach a precision of $\sigma(z)/(1+z) = 0.03 - 0.05$. They will be derived from three additional Euclid NIR band (Y,J,H in the range 0.92–2.0 micron) reaching AB mag 24 ($5\sigma$ point sources) in each, complemented by ground based photometry in visible bands derived through engaged discussions and collaborations with the Pan-STARRS, DES and LSST ground based projects (see support letters from the DES and Pan-STARRS projects in Part VI). To measure the shear from the galaxy ellipticities a tight control is imposed on possible instrumental effects and will lead to the variance of the shear systematic errors to be less than $10^{-7}$. Euclid will also perform a deep survey, about 2 mag deeper than the wide survey and which will cover an area of about 40 deg$^2$. Although unique as a self standing survey, the deep survey will also be used to monitor the stability of the spacecraft and payload through repeated visits of the same regions.

## 1.4 Primary Science

Over the last decades, a combination of observations has led to the emergence and confirmation of the concordance cosmological model (see Figure 1.1). Remarkably, the energy density of the resulting universe is dominated by two mysterious components. First, 74% of its energy density is in the form of Dark Energy, which is causing the Universe expansion to accelerate. Another 22% of the energy in the Universe is in the form of dark matter, which exerts a gravitational attraction as normal matter, but does not emit light. One possibility to explain one or both of these puzzles is that Einstein's General Relativity, and thus our understanding of gravity, needs to be revised on cosmological scales. Together, dark energy and dark matter pose some of the most important questions in fundamental physics today.

The Euclid Imaging instrument is optimised for the Weak gravitational Lensing cosmological probe. Weak lensing is a method to map the dark matter and measure dark energy through the distortions of galaxy images by mass inhomogeneities along the line-of-sight. The Euclid imaging surveys will also make use of several secondary cosmological probes such as the Integrated Sachs Wolfe Effect (ISW), galaxy clusters to provide additional measurements of the cosmic geometry and structure growth. Weak lensing requires a high image quality for the shear measurements, near-infrared imaging capabilities to measure photometric redshifts for galaxies at redshifts $z > 1$, a very high degree of system stability to minimize systematic effects, and the ability to survey the entire extra-galactic sky. Such a combination of requirements cannot be met from the ground, and demands a wide-field Visible/NIR space mission. A central design driver for Euclid is the ability to provide tight control of systematic effects in space-based conditions.

With this capability, the Euclid imaging instrument will contribute to the four Euclid primary science objectives in fundamental cosmology: (1) Euclid will measure the dark energy equation of state parameters $w_0$ and $w_a$ to a precision of 2% and 10% from the geometry and structure growth of the Universe. Euclid will thus achieve a Dark Energy Figure of Merit of 500 (1500) without (with) Planck Priors, thus improving by a factor of 50 (150) upon current knowledge. (2) Euclid will test the validity of General Relativity against modified gravity theories, and measure the growth factor exponent $\gamma$ to an accuracy of 2%. (3) Euclid will study the properties of dark matter by mapping its distribution, testing the Cold Dark Matter paradigm and measuring the sum of the neutrino masses to a few 0.01eV in combination with Planck. (4) Euclid will improve the constraints on the initial condition parameters by a factor of 2–30 compared to Planck alone. Euclid is therefore poised to uncover new physics by challenging all sectors of the cosmological model. The Euclid survey can thus be thought as the low-redshift, 3-dimensional analogue and complement to the map of the high-redshift universe provided by CMB experiments. The primary science which should be achieved by the Euclid Imaging Surveys is described in Parts II, III and IV

## 1.5 Legacy Science

Beyond these breakthroughs in fundamental cosmology, the Euclid imaging surveys will provide unique legacy science in various fields of astrophysics. In the area of galaxy evolution and formation, Euclid will deliver high quality morphologies and masses for billions of galaxies out to $z \sim 2$, over the entire extra-galactic sky, with a resolution 4 times better and 3 NIR magnitudes deeper than ground based surveys. The Euclid deep survey will probe the "dark ages" of galaxy formation as it is predicted to find thousands of galaxies at $z > 6$, of which about 100 could be at $z > 10$ i.e. probing the era of reionisation of the Universe. These high redshift galaxies and quasars will be critical targets for JWST and E-ELT. Euclid will also augment the Gaia survey of our Milky Way, taking it several magnitudes deeper. Below $V = 20$, Euclid will provide complementary information to Gaia, adding infrared colours for every Gaia star it observes; hence breaking the age-metallicity degeneracy, which is critical for the chemical enrichment history of our Galaxy. Also, the all-sky coverage of Euclid will detect nearby extremely low surface brightness tidal streams of stars thus allowing us to probe the formation and evolution of our own Galaxy. The Euclid Imaging surveys will also provide key measurements of the mass function of galaxy clusters (esp. in combination with eROSITA, Planck and SZ telescopes), and of over $10^5$ strong lensing systems, and should find thousands of intermediate-redshift supernovae in the near-IR. Euclid could also undertake a programme to detect Earth-mass planets in the habitable zone through the microlensing technique. The legacy science which will be achieved by the Euclid Imaging surveys is described in Part IV.

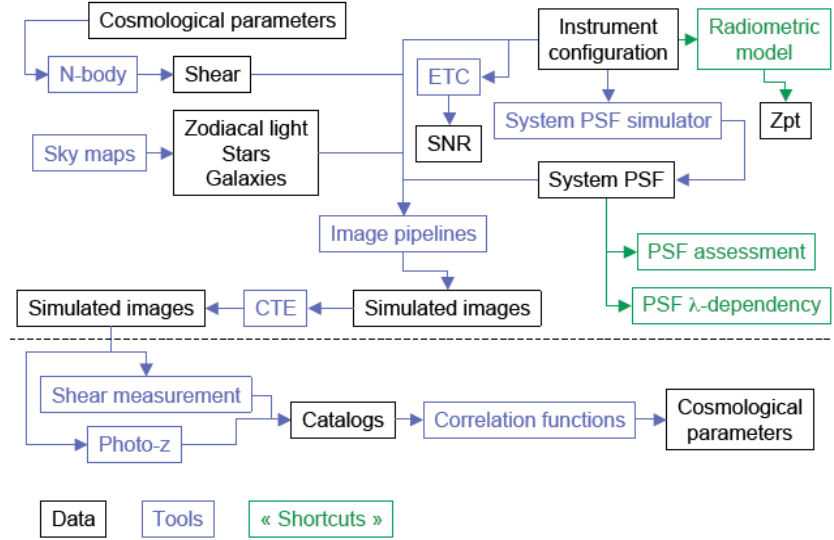

Figure 1.4: Overview of the EIC simulation tools.

## 1.6 Simulations

The Euclid Imaging Consortium has performed simulations at each stage of the performance monitoring and prediction chain. A comparison of the key results with our requirements is provided. In this book, we set out additional details of our simulations. An overview of the entire simulations process is provided in Figure 1.4 and we give details of individual components in Part V.

## 1.7 Data handling

The processing of the imaging data will be handled through the Euclid data handling architecture which is based on existing space-based and ground-based projects (see Figure 1.5 for an overview). Dedicated teams from the EIC and the spectroscopic consortium (ENIS) will process the data during all phases of the mission through and Instrument Operations Centres (IOCs) and several Science Data Centres (SDCs). The IOCs will be responsible for the first level standard data processing (calibration, removal of the instrumental effects, etc) and for requesting to the SOC corrective action in operations. The SDCs will be in charge of second and third level data products and the development of simulation pipelines. The data handling system includes a common archive, the Euclid Mission Archive (EMA) which will support the sharing of data within the project, the reporting of quality controls and a built in redundancy in the key processing tasks. A subset of the qualified data of the EMA will form the Euclid Legacy Archive (ELA) which will be delivered to the astronomical community at large. The high-precision requirements for the weak lensing analysis will be achieved within this architecture using redundant cross-check, built-in simulations, quality control, and software development from existing weak lensing pipelines.

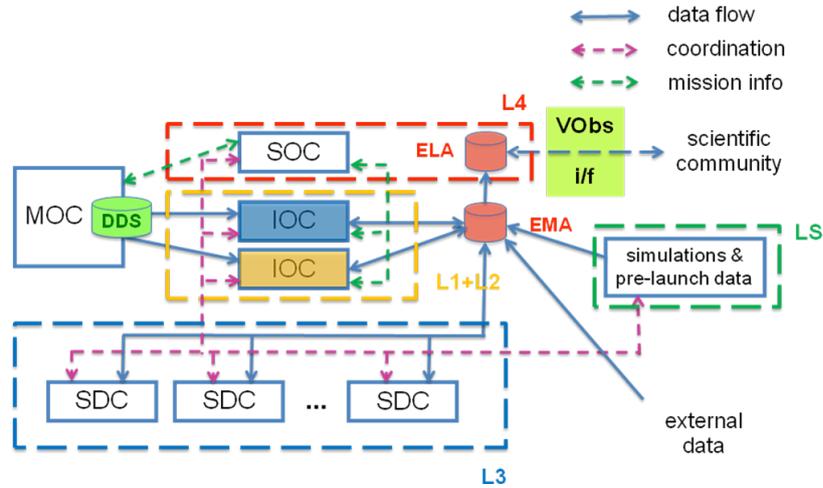

Figure 1.5: Overview of the Euclid Ground segment as proposed in the EIC-ENIS Ground Segment document.

# Publications

## 2.1 Collaboration Papers

The following are publications from the DUNE/EIC Collaboration:

- *The Dark Universe Explorer (DUNE): Proposal to ESA's Cosmic Vision*; DUNE collaboration, Experimental Astronomy, 2008, arxiv:0802.2522

- *Summary of the DUNE Mission Concept ; DUNE collaboration*, To appear in Proc. SPIE, 2008, arxiv:0807.4036

- *The focal plane instrumentation for the DUNE mission* ; DUNE collaboration, To appear in Proc. SPIE, 2008, arxiv:0807.4037

- *Description of the DUNE concept as defined during the CNES Phase 0 study*; Refregier et al., Proc. of the SPIE Symposium, Orlando, 6265,58 (arXiv:0610062)

- *NODI optical design for the DUNE CNES phase 0 concept*; Grange et al., Proc. of the SPIE Symposium, Orlando, 6265,129

## 2.2 Science working groups and related papers

The following are publications by EIC Science Working Group members done in the context of the Euclid Assessment phase:

- *Optimal Surveys for Weak Lensing Tomography*; Amara, A.; Refregier, A., MNRAS, 381, 1018, (2007)

- *Systematic Bias in Cosmic Shear: Beyond the Fisher Matrix*; Amara, A.; Refregier, R., MNRAS, 391, 228, (2008)

- *Requirements on Systematics: The Interplay Between Software and Hardware*; Amara A.; Refregier A.; Paulin-Henriksson S., submitted to MNRAS, arXiv:0905.3176

- *Photo-z for weak lensing tomography from space: the role of optical and near-IR photometry*; Abdalla, F., Amara, A.; Capak, P.; Cypriano, E.; Lahav, O.; Rhodes, J., MNRAS, 387, 969, (2008)

- *Measuring the dark side (with weak lensing)*; Amendola, L.; Kunz M.; Sapone, D.; JCAP, 4, 013, (2008)

- *Measuring Dark Matter Substructure with Galaxy-Galaxy Flexion Statistics*; Bacon, D.; Amara, A.; Read, J., submitted to MNRAS , arXiv:0909.5133

- *Optimal capture of non-Gaussianity in weak lensing surveys: power spectrum, bispectrum and halo counts*; Berge, J.; Amara, A.; Refregier, A., submitted to ApJ, arXiv:0909.0529

- *Dark energy constraints from cosmic shear power spectra: impact of intrinsic alignments on photometric redshift requirements*; Bridle, S.; King, L.; NJP, 9, 44, (2007)

- *Results of the GREAT08 Challenge: An image analysis competition for cosmological lensing*; Bridle, S.; Balan, S.; Bethge, M.; Gentile, M.; Harmeling, S.; Heymans, C.; Hirsch, M.; Hosseini, R.; Jarvis, M.; Kirk, D.; Kitching, T.; Kuijken, K.; Lewis, A.; Paulin-Henriksson, S.; Scholkopf, B.; Velander, M.; Voigt, L.; Witherick, D.; Amara, A.; Bernstein, G.; Courbin, F.; Gill, M.; Heavens, A.; Mandelbaum, R.; Massey, R.; Moghaddam, B; Rassat, A.; Refregier, A.; Rhodes, J.; Schrabback, T.; Shawe-Taylor, J.; Shmakova, M.; van-Waerbeke, L.; Wittman, D.; submitted to MNRAS, arXiv0908.0945

- *Optimising large galaxy surveys for ISW detection*; Douspis, M.; Castro, P.; Caprini, C.; Aghanim, N., A&A, 485, 395, (2008)

- *Gravitational Flexion by Elliptical Dark Matter Haloes*; Hawken, A.; Bridle, S.; submitted to PRD, arXiv0903.3870

- *On model selection forecasting, Dark Energy and modified gravity* Heavens A.; Kitching, T.; Verde, L.; MNRAS, 380, 1029, (2007)

- *Fisher matrix decomposition for dark energy prediction*; Kitching, T.; Amara, A., MNRAS, 398, 213, (2009)

- *Cosmological Systematics Beyond Nuisance Parameters : Form Filling Functions*; Kitching, T.; Amara, A.; Abdalla, F.; Joachimi, B.; Refregier, A., MNRAS, 399, 2107, (2009)

- *Finding Evidence for Massive Neutrinos using 3D Weak Lensing*; Kitching, T.; Heavens, A.; Verde, L.; Serra, P.; Melchiorri, A.; PhRvD, 77, 3008, (2008)

- *Systematic effects on dark energy from 3D weak shear*; Kitching, T.; Taylor, A.; Heavens, A., MNRAS, 389, 173, (2008)

- *Pixel-based correction for Charge Transfer Inefficiency in the Hubble Space Telescope Advanced Camera for Surveys*; Massey, R.; Stoughton, C.; Leauthaud, A.; Rhodes, J.; Koekemoer, A.; Ellis, R.; Shaghoulian, E., MNRAS, 1564, (2009)

- *Realistic simulations of gravitational lensing by galaxy clusters: extracting arc parameters from mock DUNE images*; Meneghetti, M.; et al., A&A, 462, 403, (2008)

- *Requirements on PSF Calibration for Dark Energy from Cosmic Shear*; Paulin-Henriksson, S.; Amara, A.; Voigt, L.; Refregier, A.; Bridle S., A&A, 484, 67. (2009)

- *Optimal PSF modeling for weak lensing : complexity and sparsity*; Paulin-Henriksson S.; Refregier A.; Amara A., A&A, 500, 647, (2009)

- *Deconstructing Baryon Acoustic Oscillations: A Comparison of Methods*; Rassat, A.; Amara, A.; Amendola, L.; Castander, F.; Kitching, T.; Kunz, M.; Refregier, A.; Wang, Y.; Weller, J.; submitted to MNRAS, arXiv0810.0003

- *A halo model for intrinsic alignments of galaxy ellipticities*; Schneider, M.; Bridle, S., submitted to PRD, arXiv0903.3870

- *Constraining Modified Gravity and Growth with Weak Lensing*; Thomas, S.; Abdalla, F.; Weller, J., MNRAS, 395, 197, (2009)

- *Full-Sky Weak Lensing Simulation with 70 Billion Particles*; Teyssier, R.; Pires, S.; Prunet, S.; Aubert, D.; Pichon, C.; Amara, A.; Benabed, K.; Colombi, S.; Refregier, A.;, Starck, J.-L., A&A, 497, 335, (2009)

- *Limitations of model fitting methods for lensing shear estimation*; Voigt, L.; Bridle, S.; MNRAS, 1560, (2009)

# Part II

# THEORY

# Fundamental Cosmology with Euclid: A Theoretical Perspective

*Authors: Martin Kunz (University of Geneva), Luca Amendola (University of Heidelberg & INAF Rome), Thomas Kitching (University of Edinburgh), Adam Amara (ETH Zurich), David Bacon (University of Portsmouth), Alan Heavens (University of Edinburgh), Jochen Weller (University of Munich).*

### Abstract

This Chapter provides an overview of the four primary science objectives of Euclid, with a special emphasis on the dark energy / modified gravity aspects. We discuss the cosmological phenomenology of dark energy and modified gravity at the level of first order perturbation theory. We show that in addition to the equation of state parameter $w(z)$ we require two additional functions of scale and time to characterise the scalar sector for cosmological observations. We argue that measuring these two functions will be an important goal for observational cosmology in the next decades, and discuss how Euclid will be able to measure them. We also discuss how Euclid can constrain particle properties with a particular emphasis on dark matter and neutrino constraints. We highlight the synergy of Euclid with CMB experiments that will measure the initial conditions of the perturbations. We finally highlight the synergy between Euclid and future large European projects.

## 3.1 Introduction

In the early nineties, the model of the Universe most widely accepted by theoretical cosmologists was flat and filled with matter. But slowly doubts increased whether this simple model was really an adequate reflection of reality. Early measurements of the large scale structure and of velocity fields indicated that there was not enough matter to make the Universe flat. In 1998, just over a decade ago, two teams using supernovae to measure luminosity distances made an amazing discovery: the expansion rate of the universe was accelerating. Combined with the other data and new measurements of the anisotropies in the cosmic microwave background (CMB), a coherent picture started to emerge. Even though the Universe appears to be spatially flat, matter only makes up 25% of the critical energy today, and the rest is "something else".

The best-known candidate for the 75% of "something else" is the cosmological constant $\Lambda$, originally introduced and then abandoned by Einstein (see e.g. (Copeland *et al.* 2006) for a review on dark energy models). According to particle physics there should be a cosmological constant from the zero point energy of quantum fields. But how big is the zero-point contribution to $\Lambda$? We know that general relativity (GR), being a classical theory, breaks down at some high energy scale. Sensible choices for a high-energy cut-off would be the Planck scale, or if SUSY is realised in nature, the supersymmetry breaking scale. SUSY also may lead to an exact cancellation of the zero-point energy, this cancellation makes SUSY an interesting concept in dark energy research, but from experimental particle physics we know that it is definitely broken below 100 GeV. The classical world we observe around us is not supersymmetric, however dark matter may be a relic population of the lightest supersymmetric particle.

Comparing the measured values required to make up the missing 75% of the critical energy density today with the theoretical value from quantum field zero-point energy, there is a discrepancy of 120 orders of magnitude (when cutting at the Planck scale) or still over 40 orders of magnitude (when cutting at the lowest possible SUSY breaking scale). Even worse, the quantum contributions from gravity coupling to well-understood physics, like electrons, should already be much larger than the observed size of $\Lambda$! It is always possible to subtract a "bare" value, but there is no guidance from theory. This cancellation may then be a window to physics taking place high in the energy scale, where quantum gravity is important, and which we cannot probe in any other way. This is one of the reasons why the "something else" in the Universe is so exciting. It may change our understanding of the Universe as completely as the development of quantum mechanics did in the last century. To highlight just one possibility: maybe our world has 10 spatial dimensions, but we only experience 3 of them, and the dark energy is nature's way of telling us about all the others! If so, then nature may have left a few more clues lying around in between the stars and galaxies, and one of the main science goals for the Euclid mission is to go and look for those clues.

But it is not only the 75% dark energy that is puzzling. Both big-bang nucleosynthesis (BBN) and the CMB indicate that baryons (the "normal" matter) only make up about 4% of the energy density today. The remaining 21% are an unknown substance called dark matter because it is apparently invisible but clusters at least on large scales like pressureless matter. We do not know yet what the dark matter is, but candidates exist in extensions of the standard model of particle physics, for example in SUSY. Some aspects of the nature of dark matter is expected to show up in the precise way that it clusters to form the halos of galaxies, clusters and generally the large scale structure in the Universe. Such matter concentrations bend light and act like gravitational lenses, and the Euclid Imaging Instrument is uniquely qualified to detect the weak distortions in the shapes of background galaxies due to dark matter concentrations in between those galaxies and us.

Yet another cosmic riddle concerns the origin of the tiny perturbations which have grown over time under the influence of gravity to finally become galaxies. Most cosmologists accept the inflationary model in which the Universe underwent a phase of extremely rapid expansion early in its evolution, similar to what seems to be starting again now. During this expansion, tiny quantum perturbations were enlarged and froze once they became larger than the horizon at the time. Much later they re-entered the horizon and started to collapse under the influence of gravity and to form the observed structure. The large-scale structure therefore retains the imprint of what happened in the very early Universe.

The cosmological standard model is very successful at describing the current measurements, however it lacks a physical description for over 95% of the "stuff" in the Universe, and we do not know with any certainty how the seeds for the observed structures like galaxies were generated. In this chapter we will describe how Euclid in general and the EIC in particular can make major advances in our understanding of all those questions.

## 3.2 Phenomenology vs fundamental theory of the dark energy

Although a plethora of models have been constructed to explain observed late-time accelerated expansion, none of them is really appealing on theoretical grounds. An alternative approach in this situation, especially when designing a future experiment, is to ask what can actually be measured by cosmological observations, and how we can exploit the data to learn something about the nature of the dark energy phenomenon, and generally about cosmology.

A successful example for such a phenomenological parameterisation in the dark energy context is the equation of state parameter of the dark energy component, $w \equiv p/\rho$. If we can consider the Universe as evolving like a homogeneous and isotropic Friedmann-Lemaître-Robertson-Walker (FLRW) universe, then the only observationally accessible quantity is the expansion rate of the universe $H$, given by the Friedmann equation,

$$H(a)^2 = \left(\frac{\dot{a}}{a}\right)^2 = \frac{8\pi G}{3} \left(\rho_m(a) + \rho_{\rm DE}(a)\right). \tag{3.1}$$

For simplicity we neglect here radiation and assume that space is flat; we also take the value of the speed of light to be $c = 1$. This equation governs the expansion law of the Universe as whole and can be studied with geometrical tests: luminosity and angular diameter distances are determined by integrals of $1/H$; and $H$ can be directly measured by a number of methods, among which are radial baryonic acoustic oscillations (BAO) (Blake & Glazebrook 2003; Seo & Eisenstein 2003) or the dipole of the supernova distribution (Bonvin *et al.* 2006).

For any dark energy or modified gravity model, it is possible to compute $H(a)$ and compare it to the observational data. However, this is not necessary: *any* expansion history can be created through adding a (possibly effective) dark energy component which is parameterised through a choice of $w(a)$: The dark energy (or anything else) is described by its homogeneous energy density $\rho_{\rm DE}$ and the isotropic pressure $p_{\rm DE}$, corresponding to the $T^0_0$ and $T^i_i$ elements respectively of the energy momentum tensor (in the global FLRW frame). Any other non-zero component $T^j_i$ would require us to go beyond the FLRW description of the Universe. The evolution of $\rho$ is then governed by the "covariant conservation" equation $T^\nu_{\mu;\nu} = 0$ which is just

$$\dot{\rho}_{\rm DE} = -3H(\rho_{\rm DE} + p_{\rm DE}) = -3H(1+w)\rho_{\rm DE}. \tag{3.2}$$

Instead of working forward from a specific model, this allows us to work backwards from the observations. For any model one can predict its own effective $w(a)$ and compare it to observational limits. Also, even though this $w(a)$ is a phenomenological quantity, we can immediately draw conclusions: if the observed $w$ ever deviates significantly from $-1$ then a cosmological constant is ruled out, and if $w < -1$ then canonical scalar field models of the dark energy are in trouble.

## 3.3 Beyond the background

Once $H(a)$ has been measured with the needed accuracy, then $w(a)$ can be extracted. This then allows us for example to reconstruct a scalar field potential that would give rise to this expansion history. However, there is an ambiguity present since it is also possible to construct a function $f$ so that a modified gravity model leads to the same expansion history. In other words, it is possible to get the same global expansion law by changing either side of Einstein's equations, the metric side or the matter side. The expansion history alone does therefore not allow us to distinguish between large classes of models. Can we do better?

Arguably this is not the right question: we *have to* do better! A well-known example is afforded by the growth-rate of the matter perturbations: once the expansion history $H(a)$ is

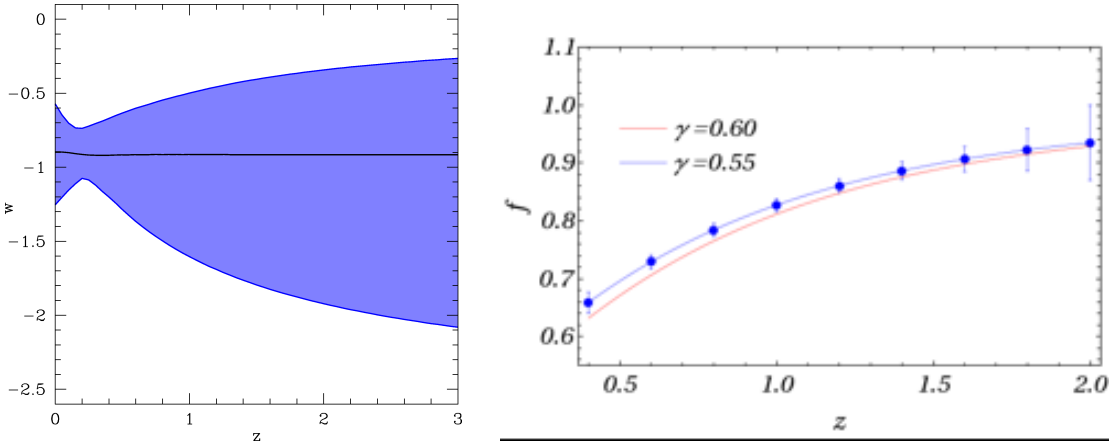

Figure 3.1: Left: Current 95% constraints on $w(z)$ from CMB data (WMAP3), SN-Ia data (Union sample) and BAO (SDSS) using the $w_0, w_a$ parametrisation for a quintessence-like model. Right: Examples of different growth rates with predicted error bars for Euclid. The two curves can be clearly distinguished. DGP with $\gamma \approx 0.68$ would be even more discrepant.

measured, the dark matter perturbations evolve according to

$$\ddot{\delta}_m + 2H\dot{\delta}_m = 4\pi G \rho_m \delta_m. \tag{3.3}$$

We can therefore predict how the density contrast $\delta_m(a)$ grows. A common parametrisation is in terms of a parameter $\gamma$ (Wang & Steinhardt 1998; Lue *et al.* 2004),

$$\frac{d \log \delta_m}{d \log a} = \Omega_m(a)^\gamma. \tag{3.4}$$

In this case, $\gamma$ is uniquely fixed by the expansion history, which in turn depends on $w(a)$. A good fit to the full numerical result is $\gamma \approx 0.55 + 0.05(1 + w)$ (Linder 2005). However, in deriving Eq. (3.3) we have used the Poisson equation, $\Delta\phi = 4\pi G \rho \delta$, *and* we have assumed both that it is valid, and that only matter contributes to the total density perturbation $\rho\delta$ on the right hand side. In general this is not the case, so that we need to go beyond the background description of the Universe and understand dark energy and modified gravity at the level of perturbation theory. Then we will find that they make different predictions for $\gamma$, even if they lead to the same $H(a)$! The same is true for weak lensing, the CMB, redshift space distortions, and so on. But how many new parameters are necessary to describe all possible measurements?

The metric is in principle composed of ten arbitrary functions $g_{\mu\nu}$. But coordinate invariance allows us to eliminate four of those, and at the level of first order perturbation theory, the remaining six can be split into two scalar, two vector and two tensor degrees of freedom. If we work in the Newtonian gauge, and limit ourselves to scalar (density) perturbations, we thus have to add two functions $\phi$ and $\psi$ to the metric, which play a role very similar to gravitational potentials. The line element then becomes

$$ds^2 = a(\tau)^2 \left[ -(1 + 2\psi)d\tau^2 + (1 - 2\phi)dx^2 \right]. \tag{3.5}$$

These potentials can be considered as being similar to $H$ in that they enter the description of the space-time geometry (but they are functions of scale/position as well as time). Also the energy momentum tensor becomes more general, and $\rho$ is complemented by perturbations $\delta\rho$ as well as a velocity $v_i$. The pressure $p$ now can also have perturbations $\delta p$ and there can further be an anisotropic stress $\pi$.

The reason why we grouped the new parameters in this way is to emphasise their role: at the background level, the evolution of the universe is described by $H$, which is linked to $\rho$ by the Einstein equations, and $p$ controls the evolution of $\rho$ but is a priori a free quantity describing the physical properties of the fluid. Now in addition there are $\phi$ and $\psi$ describing the Universe, and they are linked to $\delta\rho$ and $v$ of the fluids through the Einstein equations. $\delta p$ and $\pi$ in turn describe the fluids. Actually, there is a simplification: the total anisotropic stress $\pi$ directly controls the difference between the potentials, $\phi - \psi$.

This means that a general dark energy component can be described by phenomenological parameters similar to $w$, even at the level of first order perturbation theory. This description adds two new parameters $\delta p$ and $\pi$, which are both functions of scale as well as time. These parameters fully describe the dark energy fluid, and they can in principle be measured.

However, recently much interest has arisen in modifying GR itself to explain the accelerated expansion without a dark energy fluid. What happens if we try to reconstruct our parameters in this case? Is it possible at all?

Let us assume that the (dark) matter is three-dimensional and conserved, and that it does not have any direct interactions beyond gravity. We assume further that it and the photons move on geodesics of the same (possibly effective) $3 + 1$ dimensional space-time metric. In this case we can write the modified Einstein equations as

$$X_{\mu\nu} = -8\pi G T_{\mu\nu} \tag{3.6}$$

where the matter energy momentum tensor still obeys $T^\nu_{\mu;\nu} = 0$. While in GR this is a consequence of the Bianchi identities, this is now no longer the case and so this is an additional condition on the behaviour of the matter[1].

In this case, we can construct $Y_{\mu\nu} = X_{\mu\nu} - G_{\mu\nu}$, so that $G_{\mu\nu}$ is the Einstein tensor of the 3+1 dimensional space-time metric and we have that

$$G_{\mu\nu} = -8\pi G T_{\mu\nu} - Y_{\mu\nu}. \tag{3.7}$$

Up to the prefactor we can consider $Y$ to be the energy momentum tensor of a dark energy component. This component is also covariantly conserved since $T$ is and since $G$ obeys the Bianchi identities. The equations governing the matter are going to be exactly the same, by construction, so that the effective dark energy described by $Y$ mimics the modified gravity model (Hu & Sawicki 2007; Kunz *et al.* 2008).

By looking at $Y$ we can then for example extract an effective anisotropic stress and an effective pressure perturbation and build a dark energy model which mimics the modified gravity model and leads to *exactly* the same observational properties (Kunz & Sapone 2007). This provides a clear target for future experiments: their job is to measure the two additional functions describing $Y$ as precisely as possible. These functions can then provide clear hints about the nature of the dark energy phenomenon. For example, scalar field models have generically a sound horizon that could be detected in the data as it suppresses the dark energy perturbations on smaller scales (Weller & Lewis 2003; Bean & Doré 2004; Sapone & Kunz 2009). Modified gravity models on the other hand have generically a non-zero effective anisotropic stress, while scalar field models usually have $\pi = 0$ (Mukhanov *et al.* 1992; Boisseau *et al.* 2000; Kunz & Sapone 2007). Since the parameters of $Y$ are just effective quantities for a modified gravity model, they will probably look very unnatural, so that Bayesian model comparison methods may provide an avenue for distinguishing between different competing possibilities.

---

[1]This condition could be relaxed due to the dark degeneracy, since all visible components are conserved to the best of our current knowledge.

We can choose more or less freely which two additional functions we want to introduce to describe the dark energy perturbations or the modifications of gravity at the linear perturbation level. However, it is important to understand that we need *two*. A single extra parameter, for example only $\gamma$, does not suffice. In this case, we make effectively an additional choice, which may not be the one that we wanted. In the following section we take a closer look at the two gravitational potentials $\phi(k, a)$ and $\psi(k, a)$ since they are very close to the measurements and have a direct geometric interpretation. An alternative parametrisation used in Chapter 6 stays close to the potentials, but replaces them with dimensionless quantities $Q$ and $\eta$ that allow for simpler parameterisations, very similar to $w$ compared to $H$. They are defined through

$$k^2 \phi = -4\pi G a^2 Q \rho_m \delta_m, \tag{3.8}$$

$$\psi = (1+\eta)\phi. \tag{3.9}$$

Since weak lensing probes the combination $\phi + \psi$ it is often useful to replace one of the two parameters by the combination $\Sigma \equiv Q(1+\eta/2)$ (Amendola *et al.* 2008). There is also a relation between $\gamma$ and $Q(1+\eta)$. In the limit of small deviations from the canonical GR values, we have (Amendola *et al.* 2008)

$$\gamma = \gamma_s \left( 1 + \frac{Q(1+\eta) - 1}{(w-1)(1-\Omega_m)} \right) \tag{3.10}$$

where $\gamma_s$ is the standard value (i.e. for uncoupled, unclustered dark energy in Einstein gravity). A significant deviation of $\gamma$ from $\gamma_s$ would amount to the discovery that either dark energy clusters or that gravity is modified.

## 3.4 Extracting the potentials with Euclid

Here we provide a high-level overview on how to extract the potentials from a survey like the one planned for the Euclid satellite project. A much more detailed look at how lensing measures these quantities is presented in Chapter 6.

Since we actually observe galaxies, not the dark matter directly, we would really like to reconstruct not only the potentials but also the dark matter perturbation variables and the bias $b$. All together, we therefore would like to extract the set $\{\delta_m, v_m, \phi, \psi, b\}$. Here $v_m$ is the divergence of the dark matter velocity field.

Since the matter behaves in the usual way, we can use the conservation equations,

$$\delta'_m = 3\phi' - \frac{v_m}{Ha}, \tag{3.11}$$

$$v'_m = -v_m + \frac{k^2}{Ha}\psi, \tag{3.12}$$

where $' = d/d\ln a$. These equations determine effectively two of the variables. We therefore need three independent measurements. For example, if we could measure $v_m(k, a)$ then we would immediately know $\psi$ as well. This may not be impossible to do: one of our assumptions is that galaxies are test-particles that are accelerated by gravity in the same way as the dark matter, at least on scales where we can neglect baryonic physics like gas pressure. One of the three measurements should therefore be a probe of the velocity field. Apart from direct measurements, a promising approach is to extract the redshift space distortions of the galaxy power spectrum in redshift space,

$$P_z = (1 + \beta\mu^2)^2 P_r \tag{3.13}$$

where $P_r = b^2 \delta_m^2$ is the real-space galaxy power spectrum, which is affected by bias and does not directly probe the matter power spectrum.

Using this direction-dependence of the power spectrum, it is possible to separate out the velocity power spectrum, for example by comparing the power spectrum for $\mu = 1$ and $\mu = 0$. Integrating the conservation equation for $\delta_m$ and neglecting the $\phi'$ term, it is then further possible to recover separately the matter power spectrum $\delta_m^2$ and the bias $b$. Since light bending is described by $\phi + \psi$, we can finally use the lensing signal to recover $\phi$.

How this is best done in practice is an open and difficult question. Maybe it is preferable to parameterise all quantities of interest and to constrain all the parameters simultaneously by the data. The main point remains that it is possible in principle for Euclid alone to measure the potentials (which encode information about dark energy) and the dark matter perturbations as well as the bias, although at least CMB data would be additionally used in any realistic scenario to improve the constraints. Further data sets allow to cross-check the results and to test for the presence of systematic problems within the Euclid data set.

## 3.5 Dark Matter

The weak lensing probes of Euclid will measure $\phi + \psi$ with great accuracy. So far we have discussed how to extract the minute changes in the gravitational potentials due to different dark energy or modified gravity models. However, the largest contribution to the gravitational potential is expected to be due to the clustering of dark matter. For this reason, EIC is really an indirect dark matter detection and characterisation machine of tremendous power.

We have already discussed how the EIC in combination with the Euclid spectrograph will measure the large-scale properties of the dark matter, $\delta_m(k, t)$ and $v_m(k, t)$, and disentangle them from effects due to the dark energy or modified gravity. This allows immediately strong bounds on physical properties of the dark matter, for example to put limits on its velocity dispersion which would lead to a cut-off in the perturbation spectrum.

The weak lensing survey allows to go well beyond this global view of the dark matter. It will enable us to directly construct a 3D map of the dark matter distribution in the Universe. It will also allow us to characterise the dark matter sub-structure of cluster halos, and constrain galactic dark matter halo properties.

Many extensions of the standard model of particle physics predict the existence of new long-lived or stable particles that can play the role of the dark matter (or at least of part of it). Two of the best-motivated such particles are the axion and the lightest supersymmetric particle (LSP). The axion appears in models where the Peccei-Quinn symmetry is introduced to solve the so-called strong CP problem of QCD. When this extra symmetry is spontaneously broken, a (pseudo-) Nambu-Goldstone boson appears, the axion. This particle would have extremely weak interactions and would be very difficult to find, except through its possible conversion to photons or with cosmological measurements.

The LSP is a dark matter candidate in supersymmetric extensions of the standard model which also obeys R parity. R-parity prevents any SUSY reactions that could violate lepton/baryon number conservation hence SUSY particles can decay into lighter SUSY particles but not into standard model particles, so that the lightest SUSY particle (the LSP) is stable under these assumptions. Depending on the model parameters, there are various candidates for the LSP that include the neutralino, gravitino and axino. The expected mass and cross sectional ranges of these particles cover many orders of magnitude.

If the majority of the dark matter consists of such new fundamental particles then the EIC will be able to constrain many of its properties like the mass and the interaction cross section, independently of particle accelerators. Indeed cosmological observations from experiments such as Euclid are the only way to verify any candidate as being dark matter.

The LSP is an example of so-called "cold" dark matter (CDM). These are particles that are non-relativistic when they decouple from the photon background. The opposite is called "hot" dark matter (HDM). An example of hot dark matter that is known to be present is due to the neutrinos. The observed oscillations between different neutrino flavours (for recent reviews see (King 2007; Lesgourgues & Pastor 2006; Amsler *et al.* 2008)) implies that neutrinos are massive since the mass eigenstates $|\nu_i\rangle$ and the interaction eigenstates $|\nu_\alpha\rangle$ can be rotated with respect to each other,

$$|\nu_\alpha\rangle = U_{\alpha i}|\nu_i\rangle. \tag{3.14}$$

In this case, a neutrino flavour which is created in interactions can turn into a different flavour as it propagates through space. Neutrino oscillation experiments measure only the (absolute) mass differences between the mass eigenstates but are not sensitive to the absolute mass scale. Current constraints on the neutrino mass differences are $\delta m_{23} \sim 0.05$ eV and $\delta m_{12} \sim 0.007$ eV. Hence the hierarchy (order of $m_1$, $m_2$, $m_3$) and total mass scale of the mass eigenstates is unknown. Proposed $\beta$ decay experiments (e.g. KATRIN (KATRIN collaboration 2001; Kristiansen & Elgarøy 2008)) are expected to reach a precision of about $\sim 0.35$ eV. But if the neutrino masses are of the order of the mass differences (as for the other particle types) then a better accuracy is required, for which cosmological probes will be needed. These can detect the cumulative effect of this small mass on cosmological scales since relativistic particles diffuse out of overdensities and so suppress structure formation on small scales. The EIC will provide strong limits on the sum of neutrino masses, and may even be able to determine individual neutrino masses, and as a result distinguish between the normal and inverted hierarchies.

For these reasons there is a natural complementarity between Euclid and the LHC as well as dark matter and neutrino direct detection experiments, as we also discuss below. We will discuss the EIC dark matter science with a focus on weak lensing in much more detail in Chapter 7.

## 3.6 Initial Conditions

In measuring the properties of the dark matter and the dark energy or modified gravity, Euclid studies the evolution of cosmological perturbations. According to our current understanding, these perturbations were generated at a very early stage in the history of the Universe, called inflation. The very precise measurements of Euclid will not only allow us to characterise the evolution of the perturbations, but it will also be possible to study their initial properties. In this way, Euclid will probe inflation itself.

The power spectrum of the initial perturbations is conventionally parametrised through a (weakly) scale dependent scalar index $n_s(k)$,

$$P(k) \equiv \left\langle |\delta_{\rm in}(k)|^2 \right\rangle \propto k^{n_s(k)}. \tag{3.15}$$

If the Universe had been expanding exactly exponentially as the perturbations on observable scales were generated, then $n_s = 1$ which is also called a Harrison-Zel'dovich (HZ) or scale-invariant spectrum. Inflation models predict small deviations from exact scale invariance, so that a precise characterisation of $n_s(k)$ will permit to constrain the space of allowed inflation models. Most inflation models also lead to a small deviation from a Gaussian distribution of the perturbations, and Euclid will be able to put strong limits on such non-Gaussianities. In addition, features in $n_s(k)$ may be correlated with a deviation from Gaussianity in the initial perturbations.

CMB probes like Planck provide an excellent measurement of the initial power spectrum and very sensitive tests of non-Gaussianity on large scales. Euclid/EIC will rival the precision of Planck and provide independent measurements, allowing a cross-validation with the results of Planck and tighter combined limits. It will also extend the measurements to much smaller scales, allowing a longer lever arm to determine deviations from scale invariance with greater precision. Euclid is therefore an excellent complement to CMB probes.

## 3.7 Synergy of Euclid with LHC, EELT and Gaia

It is a hallmark of cosmology to require the input from many branches of physics and astronomy. The same applies to an observational project like Euclid where its scientific objectives will be strengthened and complemented by other planned surveys and experiments of the next decade and beyond. We mentioned already in passing the Planck mission, that will provide the best picture of the early universe, and will allow a firm initial point in the reconstruction of the cosmic dynamics. We expect at least three other large projects to interact constructively with Euclid's goals: the Large Hadron Collider (LHC) at CERN, the European Extremely Large Telescope (EELT), and the Gaia astrometric mission. This impressive array of European-led projects hold the promise to maintain Europe's decisive role in physics and astrophysics for the next decades. Let us discuss briefly the area of overlap with Euclid.

The primary scientific target for the LHC accelerator is the search for the last missing brick of the particle standard' model, the Higgs boson. Although its discovery could be classified among the biggest triumphs of technology and human inventiveness, its existence is taken almost for granted by particle physicists and many eyes are already pointed towards the next step, the search for new physics at high energy scales, i.e. for features not predicted within the standard model. One possibility would be the discovery of supersymmetry. This would have far-reaching implications for cosmology, since the lightest supersymmetric particle is seen as one of the main candidates for the particles making up the dark matter. In this case, the comparison between the collider data of LHC and the cosmological measurements of Euclid would allow to check whether this model is consistent and allow for a much better determination of particle properties than is possible with either experiment alone. An even more exciting discovery would be the existence of extra spatial dimensions, as predicted by e.g. string theory. The energy of the LHC beam could for the first time scrape the surface of our three-dimensional world and reveal a new extra-dimensional Universe. This would completely change the foundations of particle physics and cosmology, providing some scenarios of dark energy (and inflation) based upon extra-dimensional physics with a sound experimental basis.

The EELT, if realised, will be the largest optical telescope on Earth. Its 42 meter primary mirror will collect a huge quantity of light from distant sources and provide the astrophysics community with many new opportunities in all fields, from extrasolar planet search to high-resolution spectroscopy. One of the scientific goals of EELT is the detection of the so-called Sandage effect or redshift drift. That is, the detection of slight changes in redshift due to cosmological expansion between observations several years apart. Several authors anticipated the use of the redshift drift as the ultimate probe to reconstruct the cosmic expansion in real time, without assuming standard candles and with very few sources of possible systematics. Moreover, observations along two light-cones, instead of one, allows us to distinguish at a fundamental level a genuine accelerated expansion in a homogeneous universe from an apparent expansion induced by a large-scale inhomogeneity as, for instance, in void models.

Finally, the Gaia satellite, to be launched in 2011, will provide astrometric positions of up to a billion stars and a million quasars with unprecedented accuracy, thereby fixing to exceptional precision both the first rung of the distance ladder, the star parallax, and the cosmic reference frame. Although Gaia is not primarily a cosmological probe, precise quasar astrometry could be employed to rule out the possibility that dark energy causes an anisotropic expansion or that we live off-center in a huge void. These possibilities, exotic as they might appear, are still relatively unconstrained and the very fact that we can still speculate about them shows how little we know about dark energy and how important is to integrate all sources of information.

The joint efforts of Euclid, Gaia and EELT hold the potential to map the full three-dimensional cosmic kinematics and dynamics of our Universe for a large fraction of its history and of its volume.

## 3.8 Conclusions

We have discussed how both dark energy and modified gravity cosmologies can be described at the level of first-order perturbation theory by adding two functions to the equation of state parameter $w(z)$. This allows the construction of a phenomenological parametrisation for the analysis of e.g. CMB data, weak lensing surveys or galaxy surveys, which depend in essential ways on the behaviour of the perturbations.

Different choices are possible for these two functions. For example, one can directly use the gravitational potentials $\phi(k, t)$ and $\psi(k, t)$ and so try to measure the metric. Alternatively one can use the parameters which describe the dark energy: the pressure perturbation $\delta p(k, t)$ and the anisotropic stress $\pi(k, t)$. The pressure perturbation can be replaced by a sound speed $c_s^2$, which then has to be allowed to depend on scale and time. As argued above, these parameters can *also* describe modified gravity models, in which case they do of course not have a direct physical counterpart. Also other choices are possible, like $Q$ and $\Sigma$ which provide a parametrisation similar to $w$ compared to $H$. The important message is that *only two new functions* are required (although they are functions of scale $k$ and time $t$). Together with $w(z)$, they span the complete model space for both modified gravity and dark energy models in the cosmological context (i.e. without direct couplings, and for $3 + 1$ dimensional matter and radiation moving on geodesics of a single metric). By measuring them, we extract the *full information* from cosmological data sets to first order.

These phenomenological functions are useful in different contexts. Firstly they can be used to analyse data sets and to look for *general* departures from e.g. a scalar field dark energy model. If measured, they can also give clues to the physical nature of whatever makes the expansion of the Universe accelerate. Finally, they are useful to forecast the performance of future experiments in e.g. allowing one to rule out scalar field dark energy, since for this explicit alternatives are needed.

In order to measure the two functions, we will require at least two independent data sets spanning the desired range in scale and redshift. If we use galaxy clustering, we will also need to determine the bias of the galaxies, adding a third function. The proposed Euclid satellite mission, with its combination of weak lensing *and* galaxy clustering (including growth and redshift space distortions) is uniquely qualified to measure all of these functions. Especially the weak lensing measurements will play a crucial role in distinguishing the two gravitational potentials $\phi$ and $\psi$ and in providing limits on the difference between the two.

Weak lensing is a powerful tool which can probe many more physical phenomena, beyond the dark energy question. We have briefly commented on measurements of the properties of the dark matter and of the initial perturbations. Other examples include searches for topological defects (Thomas *et al.* 2009) and extrasolar planets based on lensing (see Chapter 17). The combination of all Euclid data sets can also measure the curvature at different redshifts (Clarkson *et al.* 2008) and so test the Copernican principle and constrain large voids and backreaction effects (Larena *et al.* 2009). Some of these applications are discussed elsewhere in this book. Finally we should not forget that there will be many synergies between EIC, Euclid and other current and future projects like Planck, LHC, EELT and Gaia.

# Part III

# WEAK GRAVITATIONAL LENSING

# Weak Lensing with Euclid Imaging

*Authors: Adam Amara (ETH Zurich), Andrew Taylor (U. Edinburgh), Filipe Abdalla (UCL), Nabila Aghanim (IAS), Luca Amendola (INAF - OSS.Roma/U. Heidelberg), Vincenzo Antonuccio (INAF - Cantania Astrophys. Obs.), David Bacon (U. Portsmouth), Manda Banerji (UCL/IoA Cambridge), Joel Bergé (JPL), Rongmon Bordoloi (ETH Zurich), Julien Carron (ETH Zurich), Frederic Courbin (EPF Lausanne), Eduardo Cypriano (UCL/U. Sao Paulo), Hakon Dahle (U. Oslo), Benjamin Dobke (JPL), Adam Hawken (UCL), Alan Heavens (U. Edinburgh), Benjamin Joachimi (U. Bonn/UCL), David Johnston (Nothwestern U.), Lindsay King (IoA Cambridge), Donnacha Kirk (UCL), Thomas Kitching (U. Edinburgh), Konrad Kuijken (U. Leiden), Ofer Lahav (UCL), Marco Lombardi (U. Milano), Roberto Maoli (U. Roma La Sapienza), Katarina Markovic (UCL/Ludwigs-Maximilians U.), Richard Massey (U. Edinburgh), Massimo Meneghetti (INAF - Oss. Bologna), Stéphane Paulin-Henriksson (CEA Saclay), Mario Radovich (INAF - Oss. Capodimonte), Anaïs Rassat (CEA Saclay), Justin Read (U. Zurich/U. Leicester), Alexandre Réfrégier (CEA Saclay), Jason Rhodes (JPL), Anna Romano (U. Roma La Sapienza), Roberto Scaramella (INAF - Oss. Roma), Michael Schneider (U. Durham), Michael Seiffert (JPL), Fabrizio Sidoli (UCL), Robert Smith (U. Zurich), Jiayu Tang (UCL/IPMU), Romain Teyssier (CEA Saclay/U. Zurich), Shaun Thomas (UCL), Lisa Voigt (UCL).*

**Abstract**

During the Euclid assessment phase, the EIC Weak Lensing Working Group has focused on two aspects of the Euclid mission: (1) developing the weak lensing related science case; and (2) support the technical studies of the instrument and mission by setting and maintaining the weak lensing science requirements. In this section we give a brief overview of this work, some of which is discussed in more detail in later chapters.

## 4.1 Introduction

During the Euclid definition phase, the EIC Weak Lensing Working Group (WLWG) has been actively and consistently working on the concept of weak lensing from space, with the overarching aim of creating the Euclid weak lensing surveys that will be of unprecedented quality and size. This will enable advances to be made in all aspects of cosmology. The WLWG has focused on providing a comprehensive science case for the Euclid imaging surveys and supporting the technical studies of the instrument. The science case will tackle all of the fundamental cosmological questions, including dark energy, dark matter, probing the initial conditions of the Universe and testing gravity itself. To enable these science goals, the WLWG has focused on the determination and mitigation of weak lensing related systematic effects, which will be of paramount importance for Euclid. To achieve and refine these goals, the WLWG has held weekly teleconferences and regular face-to-face meetings.

## 4.2 Basics of Weak Lensing

The light from galaxies is slightly deflected by the intervening matter distribution, causing a distortion in the shapes of the galaxies that we measure. Weak lensing is the subtlest of the lensing effects (see Figure 4.1), where an intrinsically circular galaxy would appear to us as an ellipse. However, since galaxies themselves are intrinsically elliptical, the lensing effect cannot be measured on individual galaxies. Instead, weak lensing must be measured statistically by measuring the shapes of several millions of galaxies (Blandford *et al.* 1991; Bartelmann & Schneider 2001; Refregier 2003). The weak lensing effect can be most simply captured in a linear distortion of a galaxies image, this can be expressed in a matrix transformation as follows:

$$\begin{pmatrix} x_2 \\ y_2 \end{pmatrix} = \begin{pmatrix} 1 - \gamma_1 & -\gamma_2 \\ -\gamma_2 & 1 + \gamma_1 \end{pmatrix} \begin{pmatrix} x_1 \\ y_1 \end{pmatrix}. \tag{4.1}$$

The "shear" has two components: a positive $\gamma_1$ stretches an image along the x-axis and compresses along the y-axis; and a positive $\gamma_2$ stretches an image along the diagonal $y = x$ and compresses along $y = -x$. The coordinates $(x_1, y_1)$ denote a point on the original galaxy image (in the absence of lensing) and $(x_2, y_2)$ denotes the new position of this point on the distorted (lensed) image (there is also an isotropic scaling that we do not focus here, but it can also be used to probe cosmology; see e.g. Barber & Taylor (2003)). The strength of the shear (lensing) signal depends on both the amount of matter and the distances between the observer, lens and source. This allows lensing to measure the geometry of the Universe and the growth of structure.

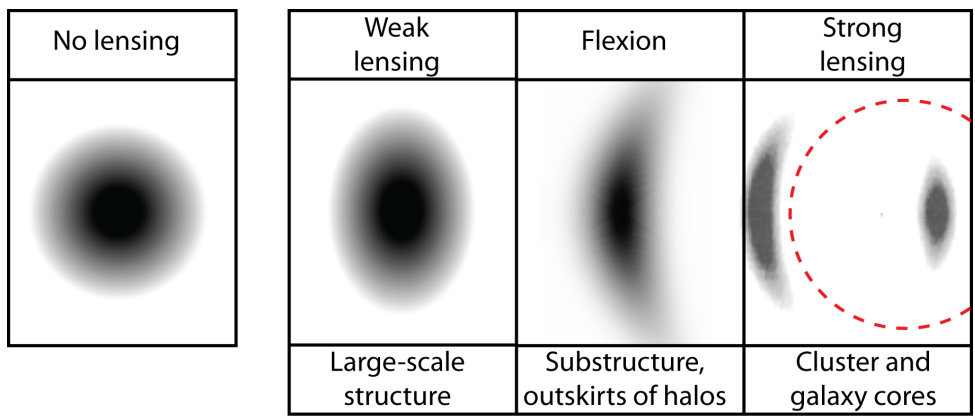

Figure 4.1: Examples of lensed images. On the far left we see a pre-lensing image. The first order lensing effect (weak lensing/cosmic shear) is to stretch the image, thus turning it from a circle to an ellipse. As the gravitational lensing signal becomes stronger, higher order shape distortions are induced, changing the circle into a banana-shaped arc. This effect is known as flexion (Goldberg & Natarajan 2002). On the far right we see that the strong lensing regime, gravitational lensing can be so extreme that multiple images of the same object can be seen.

## 4.3 Developing the Science Case

The WLWG has been investigating the wide array of sectors that can be measured through gravitational lensing. In Chapter 5, we show how an imaging survey can be designed to maximise our constraints on the dark energy and the broader cosmology parameters. We show that for all the parameters in the standard concordance model, measurement errors are minimised for an ultra wide survey that is able to reach a median redshift of roughly $z = 1$. In Chapter 7, we review in more detail how it will be able to study dark matter. We show that the 3D matter power spectrum can be constrained to a high precision, which in conjunction with CMB measurements can be used to measure the make up of the early Universe, such as the slope and roll of the initial (post inflation) power spectrum (see for example Heavens *et al.* 2006). We also show how, on the sub-atomic scales, Euclid will constrain the temperature and mass of dark matter particles and neutrinos to a high precision.

Euclid will also be able to test possible explanations for accelerated expansion beyond dark energy. These include a need to modify our model gravity. In Chapter 6, we show how Euclid will be able to detect deviations from Einstein gravity (the implications of which have already been discussed in Chapter 3). Finally, the large area survey envisioned for Euclid will allow unique cross-correlations between Euclid and other other cosmological experiments, such as Planck and eROSITA.

### 4.3.1 Summary of the Predictions

Table 4.1 shows the projected errors that Euclid will be able to achieve using the techniques enabled by the visible and near-IR imaging surveys. For these calculations, we have assumed an eight-parameter cosmological model, where curvature has been allowed, with the fiducial values given below:

- Density of matter density: $\Omega_m = 0.25$

- Density of dark energy: $\Omega_\Lambda = 0.75$

- Density of baryons: $\Omega_b = 0.0445$

- Power spectrum normalisation: $\sigma_8 = 0.8$

- Hubble parameter: $h = 0.7$

- Spectral index: $n_s = 1$

- Dark energy Equation of state pivot point parameter: $w_p = -0.95$

- Dark energy Equation of state parameter: $w_a = 0.0$

For the dark energy component, we use a simple parameterisation where the equation of state ($w$) is modelled as a linear expansion with scale factor ($a$) (Linder 2003; Chevallier & Polarski 2001),

$$w(a) = w_p + (a_p - a)w_a, \tag{4.2}$$

with $a_p$ being the scale (redshift) at which the errors on $w_0$ and $w_p$ are uncorrelated.

For the lensing survey, we have assumed a 20,000 square degree survey with 30 galaxies per square arcminutes and median redshift of 0.8. We have also assumed that the redshift of the galaxies can be measured to a precision of $\sigma(z) = 0.05(1 + z)$. This configuration corresponds to the minimal survey requirements for Euclid (see Table 4.2). If Euclid is able to meet its goals (see EIC Requirements Document) we can expect stronger constraints. From Table 4.1, we see that the weak lensing survey alone is able to make significant improvements over today in most sectors of the cosmological model. This is especially true of the dark energy sector, where we see more than an order of magnitude improvement on the dark energy FoM from the stand alone lensing survey (with no prior) as compared to today's data (which is a combination of CBM, SNe, BAO). When all of the imaging probes (galaxy clustering through photoz - Chapter 20, cluster counts - Chapter 11 and ISW - Chapter 12) are combined we see over a factor of two improvement in the FoM. Finally, we see that when the Euclid imaging probes are combined with Planck we are able to reach sub-percent level precision on almost all cosmological parameters and a factor of five improvement in DE FoM compared to Euclid alone. This shows that although the CMB is not able to directly measure dark energy, it provides a data set that works extremely well Euclid by boosting the Euclid dark energy measurements due to degeneracy breaking in other sectors. We also show in the left panel of Figure 4.2 the error ellipses in the equation of state parameters space for these configurations. These are shown in relation to distinct regions for the dark energy theories. Thus far, these error predictions only consider the statistical errors. Below we discuss how we can deal with systematic errors.

|  | $\Delta w_p$ | $\Delta w_a$ | $\Delta\Omega_m$ | $\Delta\Omega_\Lambda$ | $\Delta\Omega_b$ | $\Delta\sigma_8$ | $\Delta n_s$ | $\Delta h$ | DE FoM |
|---|---|---|---|---|---|---|---|---|---|
| Current + WMAP | 0.13 | - | 0.01 | 0.015 | 0.0015 | 0.026 | 0.013 | 0.013 | $\sim 10$ |
| Planck | - | - | 0.008 | - | 0.0007 | 0.05 | 0.005 | 0.007 | - |
| Weak Lensing | 0.03 | 0.17 | 0.006 | 0.04 | 0.012 | 0.013 | 0.02 | 0.1 | 180 |
| EIC probes | 0.018 | 0.15 | 0.004 | 0.02 | 0.007 | 0.009 | 0.014 | 0.07 | 400 |
| EIC + Planck | 0.013 | 0.08 | 0.001 | 0.004 | 0.0005 | 0.0016 | 0.003 | 0.002 | 1000 |

Table 4.1: The error predictions for an eight-parameter cosmological model achievable with the probes enabled by Euclid's imaging. These are compared against current measurements (Komatsu *et al.* 2009). We also show the results when Euclid constrains are combined with those coming from Planck.

## 4.4 Code comparison

Within the WLWG, a number of independent routines have been developed for calculating the expected errors for future weak lensing missions. These have largely focused on constraints coming from weak lensing tomography. In this approach, galaxies are first divided into redshift bins (typically 10 bins are used) and the two point correlation function of shear, both within the bin and between different bins, is used to measure cosmological parameters. We have conducted a detailed code comparison program using four tomographic routines. These calculate the expected errors using the Fisher matrix formalism (e.g. Tegmark *et al.* 1997), and they have been used in a number of published works (these include, among others, Amara & Réfrégier 2007; Bridle & King 2007; Thomas *et al.* 2009; Joachimi & Schneider 2009). For these calculations, we have found sub-percent level agreement on fixed error prediction and a better than 10% agreement on marginalised errors, which corresponds to the level of accuracy expected from a Fisher matrix approach. One of these codes (the package `iCosmo`) has been made public (Refregier *et al.* 2008), along with a user-friendly website (Kitching *et al.* 2009a). Another approach is to measure the full 3D correlation function of shear without first binning the galaxies (Kitching *et al.* (2007)

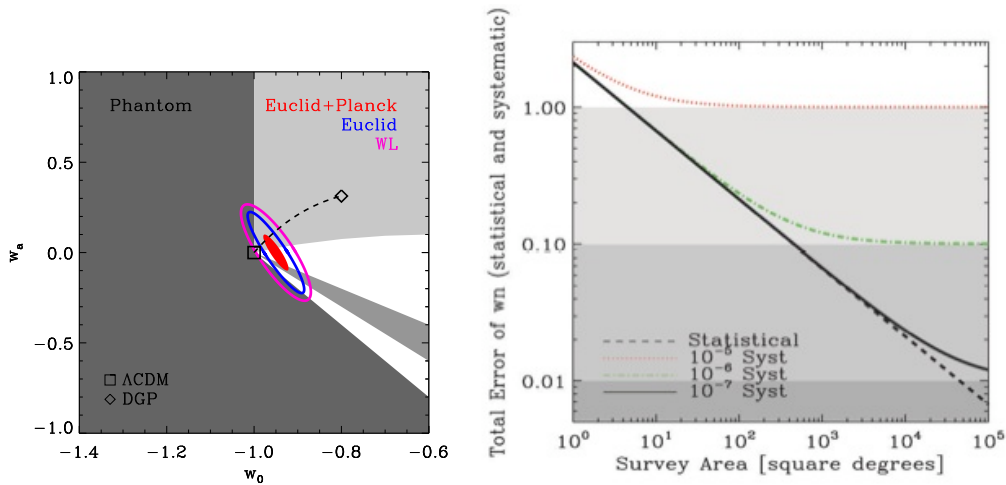

Figure 4.2: Left Panel: Predictions for constraints on the equation of state parameters $w_0$ and $w_a$ coming from weak lensing and the other imaging probes. We also show the expected errors when combined with Planck. The background grey regions show distinct theoretical regions. These are: dark grey - phantom models; middle grey - Freeze models; and light grey - Thaw models (see Chapter 3 for more details). Right Panel: The total error on the equation of state parameter, which includes both statistical and systematic, as a function of area and the level of systematic floor.

and Heavens *et al.* (2006)). Expected errors from this method have also been compared against those coming from weak lensing tomography, and we find consistent error predictions with the 3D lensing method performing slightly better, as expected. Unless stated elsewhere, our fiducial cosmology model used the transfer function by Eisenstein & Hu (1998) and a non-linear correction calculated using halofit (Smith *et al.* 2003). We have found that our predictions are not very sensitive to the choice of transfer function and that the choice of non-linear prescription can have a 50% effect on the projected errors. This highlights the importance of accurate predictions of the non-linear growth of structure, which is discussed in Chapter 22. Code comparison tests have also been performed for predictions using higher order correlations, such as the bispectrum (see Chapter 5) and modified gravity models (see Chapter 6 for more details).

## 4.5   Mitigation of Systematic effects

The potential of weak gravitational lensing as a uniquely powerful way of making precision cosmology measurements is now well established (Peacock *et al.* 2006; Albrecht *et al.* 2006). The key issue rests on our ability to control sources of systematic errors to the point where they remain sub-dominant to the statistical ones. There are a number of potential sources of systematic error. From the theory and the modelling side, care is needed to understand: (i) the non-linear growth of structure; (ii) the impact of baryons on small scale structure growth; and (iii) the intrinsic correlated alignment of galaxy ellipticities even in the absence of lensing. Tackling these problems will require a combination of high resolution computer simulations to better understand the physical processes involved, as well as the development of statistical techniques to minimise the impact of any theoretical uncertainty. In Chapter 22, we discuss the numerical simulations that have been performed by EIC members to tackle these issues; and in Chapter 10 we give a detailed discussion of intrinsic alignment, along with mitigation strategies.

Given the simplicity of the gravitational lensing effect, these sources of theoretical uncertainty will likely be overcome with continued work along the lines highlighted in this document. The main remaining challenge for weak lensing studies will then be the control of measurement errors to the precision that will be needed to meet the ambitious Euclid targets.

Weak lensing measurement rely on the accurate measurements of galaxy shapes and redshifts. For practical considerations coming from the large number of galaxies that are needed ($\sim 2$ billion), the redshifts of galaxies will be measured photometrically. In Chapter 20, we highlight the challenges to this approach and ways that our targets can be met. Galaxy shapes are most easily measured in the visible, and the process is outlined below in section 4.5.2.

### 4.5.1 Setting Systematic Limits

To set limits on the level of systematics that we can tolerate, we have developed a number of methods for propagating these errors. The first approach is to add nuisance parameters to the Fisher matrix that is used to perform the error predictions for cosmology parameters. We would then marginalise over these nuisance parameters to determine their effect. A good example of this comes from setting the requirement on the knowledge of the mean redshift of galaxies in each redshift bin. By leaving the mean redshift of the galaxy distributions as a free parameters, we effectively try to simultaneously measure these along with the cosmology parameters (Kitching *et al.* 2008b; Amara & Réfrégier 2007; Abdalla *et al.* 2008; Zhang *et al.* 2009; Schrabback *et al.* 2009). Using only shear data, there is a serious degradation of cosmology errors if no prior information is known about the mean redshift of the galaxies. To remove this negative effect, we have found that the mean redshift of the galaxies in each bin needs to be known to a precision of $\langle z \rangle = 0.002(1 + z)$. In Chapter 20 we show how this goal can be achieved.

The difficulty in adding nuisance parameters to the Fisher matrix comes from the fact that we have to decide how to parameterise the unknown. Another approach that has been developed by members of the EIC is to study the biases that an unknown systematic will induce on cosmological parameters and to use this to set tolerance limits on possible uncertainties. Using this approach Amara & Réfrégier (2008), we were able to quantify the systematic uncertainty in term of the quantity $\sigma_{sys}^2$, which is the induced variance coming from the difference between the measured and true correlation function. For a Euclid-like survey, the systematics need to be controlled to a level better than $\sigma_{sys}^2 = 10^{-7}$. This result was also confirmed by Kitching *et al.* (2009b), who developed the method further. In Figure 4.2 we show how the measurement error of a survey depends on the $\sigma_{sys}^2$ as a function of survey area. We see that as the area of the survey is increased, the statistical error on the equation of state parameter decrease. However, a given mission will be limited by the Root Mean Squared Error (RMSE) of the measurement, which comes from the quadratic sum of statistical error and biases, $RMSE(p) = \sqrt{\sigma^2(p) + b^2(p)}$. We see that depending on the level of the systematics of a survey, the RMSE will plateau at a given survey area at which point the mission becomes systematics limited. Given the fact that current lensing data is limited to areas $< 100$ square degrees, we have estimated the current systematic level to be $\sigma_{sys}^2 \sim 4 \times 10^{-6}$ (Paulin-Henriksson & et al. 2009). The Euclid targets are close to two orders of magnitude more demanding in $\sigma_{sys}^2$, and this motivates our need to have space-based high quality imaging that does not suffer from adverse effects of the atmosphere.

### 4.5.2 Shape Measurement

The process of measuring galaxy shapes has a central significance in securing the weak lensing science goals of Euclid. For example, if the ellipticity (or flexion) signal measured is biased or offset relative to the true underlying galaxy ellipticity, then this can propagate through the weak lensing statistics into a bias in the cosmological parameter estimation, as discussed above. The process that galaxy images go through is encapsulated in Figure 4.3. A galaxy is sheared by the gravitational potential along the line of sight. This sheared galaxy is then further convolved with a PSF, pixelated and is observed in the presence of noise. The shape measurement problem is to disentangle the steps subsequent to the shearing process and to measure this shear to a high accuracy. Members of the EIC WLWG are at the forefront in finding solutions to this problem, with EIC members having developed a number of shape measurement approaches. These includes *lens*fit (Miller *et al.* 2007; Kitching *et al.* 2008a), shapelets (Refregier & Bacon 2003; Massey & Refregier 2005), im2shape (Bridle & et al. 2004). Shape measurement methods have always met the demands of contemporary weak lensing data. However, this is an area in which further improvements must be made.

To improve on the current methods, a roadmap of simulations and testing has been developed. In Bridle *et al.* (2008, 2009), the first GRavitational lEnsing Accuracy Testing (GREAT) challenge was launched. These are a set of simulations in which the shear is introduced in a controlled manner. Shape measurement algorithms can then be used on these simulations and their performance measured against the input. GREAT08 ran for 6 months during 2008 as a blind challenge. During this short time, a factor of 2 improvement was gained over existing methods. In some simulated conditions, the most successful methods met and surpassed the requirements set by Euclid, but further work is ongoing to broaden this success. The next suite of simulations in this challenge is GREAT10 (Kitching & et al. 2010), which will increase the complexity over that of the GREAT08 challenge by introducing variable shear and PSF.

### The Forward Process.

**Galaxies:** Intrinsic galaxy shapes to measured image:

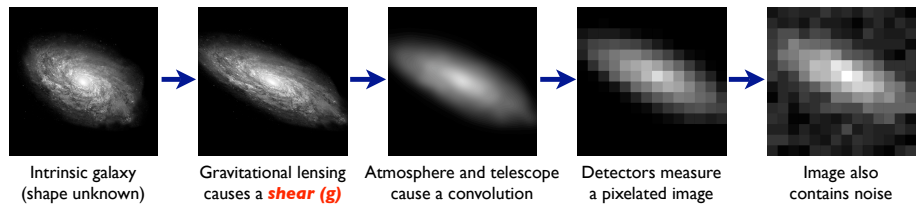

**Stars:** Point sources to star images:

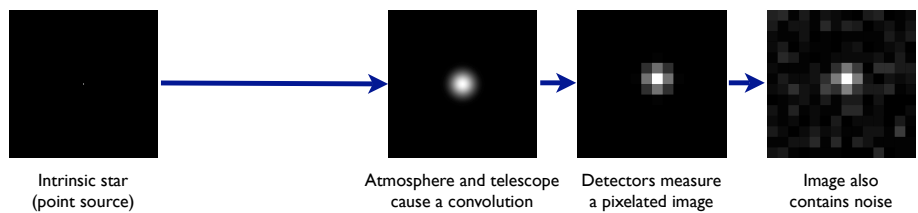

Figure 4.3: Illustration of the processes that affect galaxy and star images. The intrinsic shape of a galaxy is gravitationally lensed by intervening matter causing the cosmic shear effect that we plan to measure. After this, the galaxy image becomes blurred due to the PSF (in space this would come only from the instrument), pixelated by the detectors. The final image will also have noise. Star images suffer from many of these effects but crucially their images are not gravitationally lensed. We are therefore able to use star images to correct galaxy images to recover the shear signal. This is discussed in more detail in Chapter 8 (Figure taken from Bridle *et al.* 2008)

## 4.6 Summary of Top Level Requirements

In Table 4.2 we summarise the top level requirements coming from weak lensing on the survey geometry and the level to which the systematics need to be controlled. These top level requirements form the inputs to the EIC requirements document. These then flow down into lower level requirements, such as PSF profile, etc. The requirements are also used as targets for the work described in the sections that follow. In Part V of this science document we describe the set of simulations that we have developed to verify that the Euclid survey will fulfil these requirements. Specifically we describe image simulations (Chapter 19), photometric redshift simulations (Chapter 20), radiometric simulations (Chapter 21) and cosmological simulations (Chapter 22).

| Description | Quantity | Requirement |
|---|---|---|
| Survey Geometry: Area: Errors on dark energy parameters depend on the area of the survey | $A_s$ | $> 20000$ deg$^2$ |
| Survey Geometry: density of galaxies: And the effective number density of galaxies useful for gravitational lensing (Neff) | $N_{eff}$ | $> 30$ gals/amin$^2$ [Goal: $> 40$] |
| Survey Geometry: galaxy redshift: Redshift distribution of the lensing galaxies | $z_m$ | $> 0.8$ |
| Shape Measurement: To reach the above cosmological objectives, systematic effects shall be controlled to a level where they do not dominate over the statistical errors. This is done by controlling the variance of the residual shear systematics.( $\sigma^2_{sys}$) | $\sigma^2_{sys}$ | $< 10^{-7}$ |
| Photometric Redshifts statistical: The statistical rms error $\sigma(\overline{z})$ in the photo-zs in the range $0.2 < z < 2.0$ | $\sigma(z)/(1+z)$ | $< 0.05$ |
| Photometric Redshifts error in the mean: The mean of the redshift distribution n(z) of each bin must be known to high precision | $\sigma(\overline{z_i})/(1+\overline{z_i})$ | $< 0.002$ |

Table 4.2: Summary of the top level requirement for precision measurements with Euclid. The area, number density, median redshift and photometric uncertainty set the statistical potential of a weak lensing survey (see Chapter 5). To meet this potential, the shape measurement errors need to be controlled, as well as having accurate measurements of the mean redshift of the galaxies.

# Weak Lensing Survey Optimisation for Dark Energy Measurement

*Authors: Joel Bergé (JPL/Caltech), Adam Amara (ETH Zurich), Alan Heavens (ROE Edinburgh), Benjamin Joachimi (Bonn University), Tom Kitching (ROE Edinburgh), Alexandre Réfrégier (CEA Saclay), Andy Taylor (ROE Edinburgh), + EIC Weak Lensing Working Group.*

**Abstract**

We describe the work done by the EIC Weak Lensing working group members to optimise the survey's configuration and the extraction of cosmological parameters. We present the weak lensing techniques that we use, shear power spectrum, shear tomography, 3D shear, shear-ratio geometric test, and how we design a weak lensing to optimise the dark energy measurement using those techniques. We then present non-Gaussian measurements, like the weak lensing bispectrum and weak-lensing-selected-cluster counts, and show how we use them to optimise the extraction of cosmological information.

## 5.1 Introduction

We present how we optimise both the survey geometry and the weak lensing measurements that will allow the optimal extraction of cosmological information. Besides the usual shear power spectrum tomography, we will use more recent techniques such the full 3D weak lensing and the shear-ratio geometric test. Non-Gaussian-sensitive measurements, like the weak lensing bispectrum and weak-lensing-selected cluster counts, are complementary to those, and allow us to optimise cosmological parameters measurements. Section 5.2 presents the survey geometry optimisation : we show that to optimise dark energy measurement, for a given observation time, we need a large survey with median redshit $z_m \approx 1$. Section 5.3 shows how to optimise dark energy measurement by combining complementary weak lensing statistics, such as combining the weak lensing power spectrum with the weak lensing bispectrum or weak-lensing-selected cluster counts. We parameterise the varying dark energy equation of state $w(a) = w_0 + (1 - a)w_a$, where $a$ is the scale factor.

## 5.2 Optimising the survey geometry

The shear power spectrum, the Fourier transform of the real-space shear two-point correlation function, shows a strong dependence on the dark matter density parameter $\Omega_m$ and on the dark matter power spectrum amplitude $\sigma_8$. It has already shown its efficiency in constraining cosmological parameters on current data (e.g. Massey *et al.* 2007; Fu *et al.* 2008). The source galaxies' redshift information is usually taken into account when measuring the shear power spectrum by binning the galaxies' population into redshift slices. It has been demonstrated, both theoretically (e.g. Hu 1999) and observationally (Massey *et al.* 2007; Schrabback *et al.* 2009) that this tomographic method helps to tighten constraints on cosmological parameters. Moreover, tomography helps control systematics such as galaxy shapes intrinsic alignments. Heavens (2003) introduced a 3D cosmic shear analysis (see also Taylor 2001), a technique that takes into account the full 3D information, and avoids the loss of information suffered by tomography when binning galaxies into redshift. Jain & Taylor (2003) introduced the shear-ratio geometric test. By taking the ratio of the galaxy-shear correlation functions at different redshift, one is left with a purely geometric quantity that can be efficiently used to measure dark energy.

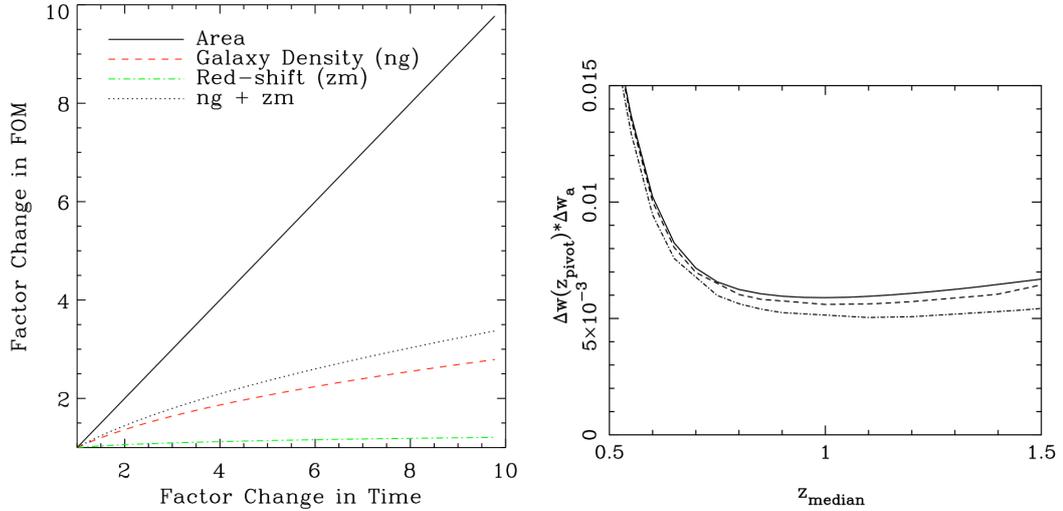

Figure 5.1: Survey geometry optimisation. Left : Gain in FoM, as measured with shear tomography, when observation time is dedicated to increasing the survey's area, galaxy density $n_g$ or median redshift $z_m$. Increasing the area is synonymous with a wide survey, whereas increasing $n_g$ and $z_m$ is synonymous with a deep survey (from Amara & Réfrégier 2007). Right : Inverse FoM's dependence on the median redshift of the survey, as measured with 3D shear, for a given survey area and for 5-band (solid), 9-band (dashed) and 17-band (dot-dashed) photometric redshift survey (from Heavens *et al.* 2006).

Amara & Réfrégier (2007) have investigated the dependence of the precision on the cosmological parameters, as measured with power spectrum tomography, on the survey's characteristics. The left panel of Fig. 5.1 shows the gain in the dark energy Figure of Merit (FoM) when changing the survey's characteristics. Increasing the area provides a much more pronounced gain in the FoM compared to increasing the number density of galaxies $n_g$ and the median redshift $z_m$. A wide survey spends the exposure time increasing the area, keeping $n_g$ and $z_m$ constant, while a deep survey increases them. Therefore, for a given observation time, we should always prefer a wide survey rather than a deep survey. Heavens *et al.* (2006) optimised 3D weak lensing surveys, and found that for a fixed time, larger surveys provide better cosmological parameter constraints, in particular for dark energy, once a median redshift of $z \approx 1$ is reached. The right panel of Fig. 5.1 shows that the dark energy is best constrained as soon as the median redshift $z_m \approx 1$. They showed that it gives comparable, but a little better, constraints on dark energy than weak lensing tomography. Taylor *et al.* (2007) have adapted and optimised the method around galaxy clusters to constrain dark energy and its evolution in large surveys. They showed that a large survey yields better constraints on dark energy than a targeted survey of 60 of the largest clusters.

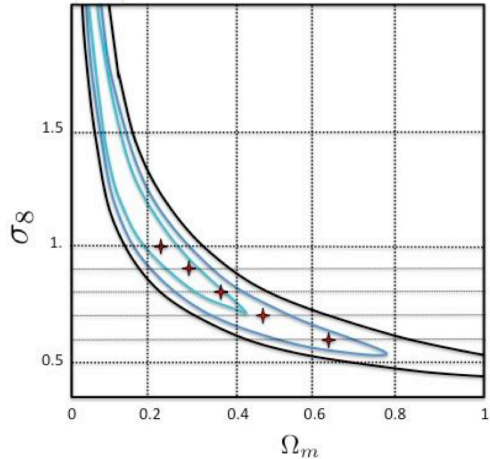

Figure 5.2: $\Omega_m$-$\sigma_8$ degeneracy left over by the weak lensing power spectrum (from Pires *et al.* 2009). The pluses represent the models that they considered (see main text).

To optimise dark energy measurement, we will survey the entire extragalactic sky (20,000 deg$^2$) at an intermediate median redshift ($z_m \approx 1$), providing us with $n_g \approx 40$ galaxies per square arcminute.

## 5.3  Optimising constraints on cosmological parameters

Besides optimising the survey configuration, one must also optimise the methods with which cosmological information is extracted. For instance, one must investigate how the well-known degeneracies left over by the statistics introduced in the previous section, as shown by Fig. 5.2 for the matter density $\Omega_m$ and matter power spectrum normalisation $\sigma_8$, can be broken. This can be done with observables sensitive to non-Gaussianity, such as third-order statistics and galaxy cluster counts.

Third-order statistics are the lowest order statistics sensitive to non-Gaussianity. Hence, they can be combined with the aforementioned 2-points statistics to break degeneracies left over by the latter. For instance, the weak lensing bispectrum is the Fourier transform of the convergence 3-point correlation function. In the flat-sky approximation, it is defined as

$$B_{\kappa,(ijk)}(\mathbf{l_1}, \mathbf{l_2}, \mathbf{l_3}) = \int_0^{\chi_h} d\chi g_i(\chi) g_j(\chi) g_k(\chi) \chi^{-4} B_\delta(\mathbf{k_1}, \mathbf{k_2}, \mathbf{k_3}; \chi) \qquad (5.1)$$

where $g_i(\chi)$ is the weak lensing efficiency function, $i$, $j$ and $k$ denote redshift bins, and $B_\delta(\mathbf{k_1}, \mathbf{k_2}, \mathbf{k_3}; \chi)$ is the matter bispectrum, which can be estimated e.g. with the halo model or related to the matter power spectrum with scaling relations from the hyper-extended theory perturbation (e.g. Scoccimarro & Couchman 2001). We have developed and compared independent codes to assess the constraining power of the weak lensing bispectrum, with Fisher matrices analyses. Joachimi *et al.* (2009) have derived the bispectrum covariance in the flat-sky approximation. We found that the bispectrum gives constraints similar to those provided by the power spectrum.

Table 5.1: Discrimination efficiencies (in percent) achieved on wavelet-filtered convergence maps with peak counts (from Pires *et al.* 2009).

|         | Model 1 | Model 2 | Model 3 | Model 4 | Model 5 |
|---------|---------|---------|---------|---------|---------|
| Model 1 | x       | 86      | 100     | 100     | 100     |
| Model 2 | 87      | x       | 94      | 100     | 100     |
| Model 3 | 100     | 92      | x       | 94      | 100     |
| Model 4 | 100     | 100     | 93      | x       | 99      |
| Model 5 | 100     | 100     | 100     | 100     | x       |

Clusters of galaxies are the most massive structures evolved from the non-linear gravitational evolution of the density field. Assessing their abundance is thus a direct probe of non-Gaussianity, and has been shown to be strongly dependent on cosmology (e.g. Oukbir & Blanchard 1992; Wang & Steinhardt 1998; Horellou & Berge 2005). Galaxy clusters can be selected in different ways. For instance, Euclid will allow us to detect them thanks to their optical signal (e.g. Adami *et al.* 2009) and thanks to their weak lensing signal (e.g. Marian & Bernstein 2006; Bergé *et al.* 2008). In the following, we show how weak-lensing selected cluster can break the degeneracies between cosmological parameters left with the power spectrum.

Pires *et al.* (2009) and Bergé *et al.* (2009) have shown how weak lensing alone can measure dark energy at the percent level by combining non-Gaussian measures with the power spectrum. Using numerical simulations, Pires *et al.* (2009) have compared how higher order weak lensing statistics break the degeneracy between $\Omega_m$ and $\sigma_8$. They showed that while third-order statistics, such as the skewness, are efficient at this task, counting galaxy clusters provides the best way to discriminate between several models. They considered five different cosmological models lying along the $\Omega_m$-$\sigma_8$ degeneracy, shown by pluses in Fig. 5.2. Table 5.1 presents the discrimination efficiency of peak counts on a wavelet-filtered weak lensing convergence map (where peaks correspond to galaxy clusters).

Bergé *et al.* (2009) used Fisher matrices to generalise Pires *et al.* (2009)'s results. They showed that, when counting galaxy clusters as signal-to-noise peaks, non-Gaussianities are captured as well as when including the redshift and/or mass information in clusters counts. This makes it an easy measurement, that does not require any additional information nor nuisance parameters to allow for the lack of knowledge in how the observable one uses to estimate a cluster's mass is related to its real mass (e.g. Lima & Hu 2004). Furthermore, they showed that the weak lensing bispectrum is as efficient as cluster counts to capture non-Gaussianity. Figure 5.3 shows the 1-$\sigma$ constraints that will be obtained with Euclid on the varying dark energy equation of state parameters $w_0$ and $w_a$, when marginalising over the other six parameters. The blue ellipse shows the forecasted constraints provided by the power spectrum. Non-Gaussianities are captured when combining it with the bispectrum (red ellipse) and with weak-lensing-selected galaxy clusters (green ellipse), hence tightening the constraints.

Besides these weak-lensing-only constraints, one can combine weak lensing measurements with other probes, such as the Cosmic Microwave Background (CMB) currently surveyed by Planck, Baryon Acoustic Oscillations (BAO), or type Ia supernovae (SNIa). In particular, the shear-ratio test provides constraints that are orthogonal to those from CMB : the shear-ratio test constrains $w_0$, and its combination with the CMB constrains $w_a$. Heavens *et al.* (2006) and

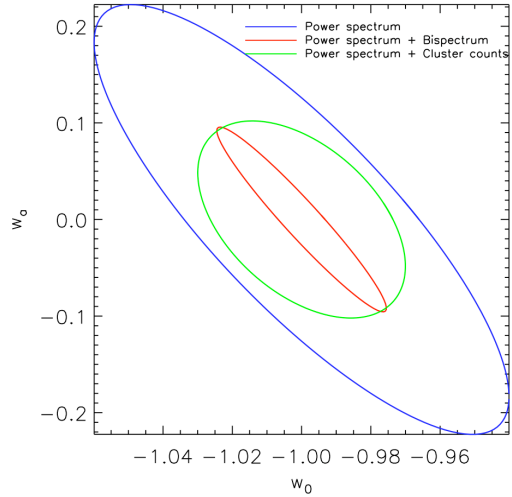

Figure 5.3: Forecasted marginalised 1-$\sigma$ constraints on the varying dark energy equation of state provided by Euclid. The blue ellipse shows the constraints from the weak lensing power spectrum. The red and green ellipses show the constraints when the weak lensing bispectrum and weak-lensing-selected clusters counts are combined with the power spectrum, respectively (from Bergé *et al.* 2009).

Taylor *et al.* (2007) have shown how 3D weak lensing and the shear-ratio geometric test constrain dark energy when combined with observations of the CMB, of BAO, and of SNIa. They found that those combinations will yield percent level constraints on the evolving dark energy.

### 5.3.1 Towards a more precise description of dark energy

It is usual to the time evolution of the dark energy equation of state by the first order expansion $w(a) = w_0 + (1 - a)w_a$. However, higher order expansions, using different data sets, will better represent the reality. Kitching & Amara (2009) have investigated how we must use Fisher matrices when dealing with such high order expansions. They showed that while finding the eigenvalues of $w(a)$ from Fisher matrix depends on the basis functions choice and on the order of the expansion, Euclid will allow us to put sub-percent errors on the best constrained eigenvalues. They also recommend that physical motivated functional forms should be used as basis sets.

## 5.4 Conclusion

To optimally measure dark energy using cosmic shear, the Euclid survey has been optimised to be as wide as possible : we will survey the entire extragalactic sky (20,000 deg$^2$), with a median redshift $z_m \approx 1$. The cosmological information will be optimally extracted by first measuring the weak lensing power spectrum, then combining it with probes of non-Gaussianities. For instance, counting clusters of galaxies, defined as signal-to-noise ratio peaks, will optimally break degeneracies. Combining the power spectrum with the bispectrum will then give similar constraints, as well as a check on the constraints obtained when combining the power spectrum with clusters counts. Combining weak lensing measurements, such as 3D weak lensing or the shear-ratio geometric test, with other cosmological probes, like the CMB or BAO, will provide additional constraints. With these available combinations, Euclid will provide percent level accuracy on dark energy.

# Modified Gravity with Lensing

*Authors: David Bacon (ICG Portsmouth), Adam Amara (ETH Zurich), Filipe Abdalla (UCL), Luca Amendola (Rome), Emma Beynon (ICG Portsmouth), Sarah Bridle (UCL), Alan Heavens (Edinburgh), Tom Kitching (Edinburgh), Kazuya Koyama (ICG Portsmouth), Martin Kunz (Geneva), Shaun Thomas (UCL), Jochen Weller (U. Munich/MPE) + EIC Weak Lensing Working Group.*

**Abstract**

Here we describe the work that has been carried out by EIC Weak Lensing group members to characterise modified gravity theories, particularly with a view to predicting the capabilities of weak gravitational lensing with Euclid to detect and measure deviations from Einstein's theory.

## 6.1   Introduction

As has been discussed in Chapter 3, it is possible that the indications of dark energy which we have observed are actually the signature of something even more radical: the departure of the law of gravity from Einstein's description. In this view, General Relativity is seen as a limiting case of a more general theory of gravity. Note that we still want GR as a limit at small scales, due to the very stringent tests which GR has passed in our Solar System.

In this chapter we describe recent work which members of our group have carried out, seeking to understand the extent to which these theories can be excluded or distinguished with Euclid lensing analyses.

It should be noted that various group members (Amara, Amendola, Heavens, Kitching, and Thomas) have compared their cosmology codes for calculating weak lensing predictions for modified gravity, particularly for the $\gamma$ growth parameter. Careful notice has been taken of the varying assumptions between models, summarised at the end of this chapter in Table 6.1.

## 6.2   Distinguishing between Modified Gravity and Dark   Energy

A key concern is how one can distinguish between modified gravity models with a particular set of parameters, and dark energy models with a different set of parameters - and indeed a different number of parameters. Members of the weak lensing group have addressed this in Heavens *et al.* (2007).

This paper demonstrates how the Bayesian evidence can be used to distinguish between models, using the Fisher matrices for the parameters of interest. In particular, the authors calculate the ratio of evidences $B$ for a 3D weak lensing analysis of the full Euclid survey, for a dark energy model with varying equation of state, and modified gravity with additionally varying growth parameter $\gamma$; the survey parameters are chosen as in the Yellow Book. The work focusses on models such as DGP, for which the relation between the lensing potential and density fluctuations is the same (i.e. $k^2(\phi + \psi)$ is the same).

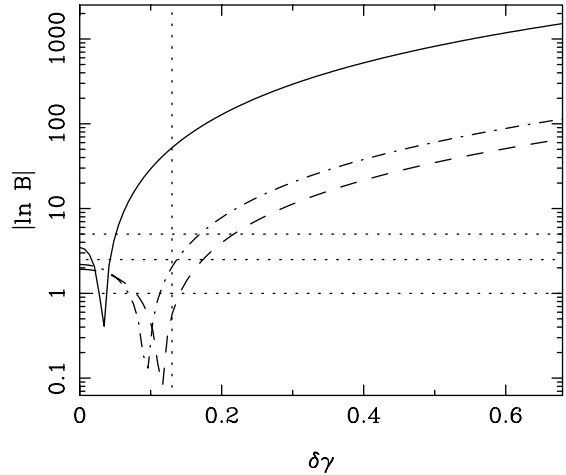

Figure 6.1: The value of $|\ln B|$ for Euclid (solid), Pan-STARRS (dot-dashed) and DES (dashed), in combination with CMB constraints from Planck, as a function of the difference in the growth rate between the modified gravity model and GR. The dotted vertical line shows the offset of the growth factor for the DGP model. The horizontal lines mark the boundaries between 'inconclusive', 'significant', 'strong', and 'decisive' in Jeffrey's (1961) terminology.

The results are shown in Figure 6.1. We see that Euclid can decisively distinguish between e.g. DGP and dark energy using the methods of this paper. Moreover, we will be able to distinguish any departure from GR which has a difference in $\gamma$ greater than $\simeq 0.03$.

As has been described in Theory, it is possible to construct rather peculiar dark energy models which could mimic the modified gravity models considered. So when we refer to the discriminatory power of Euclid for modified gravities such as those above, it should be understood that we are distinguishing between these and standard dark energy quintessence models. However, the Bayesian approach suggested here will disfavour artificial dark energy models which require more parameters to fit the data.

## 6.3 Constraining Modified Gravity with Weak Lensing

### 6.3.1 Parameter constraints

Going beyond the above discriminatory test, we are interested in the degree to which we can constrain modified gravity parameters using Euclid lensing. This is an active topic of research, especially for predictions in the linear regime (e.g. Knox *et al.* 2006; Yamamoto *et al.* 2007; Afshordi *et al.* 2008; Schmidt 2008; Song & Dore 2008; Tsujikawa & Tatekawa 2008; Zhao *et al.* 2009a,b); a key study has been carried out by members of our group in Thomas *et al.* (2009).

In this paper, the authors initially consider a set of models which they call 'mDGP' with $\Lambda$CDM and DGP as extremes, using a Friedmann equation with new parameter $\alpha$:

$$H^2 - \frac{H^\alpha}{r_c^{2-\alpha}} = \frac{8\pi G\rho}{3} \tag{6.1}$$

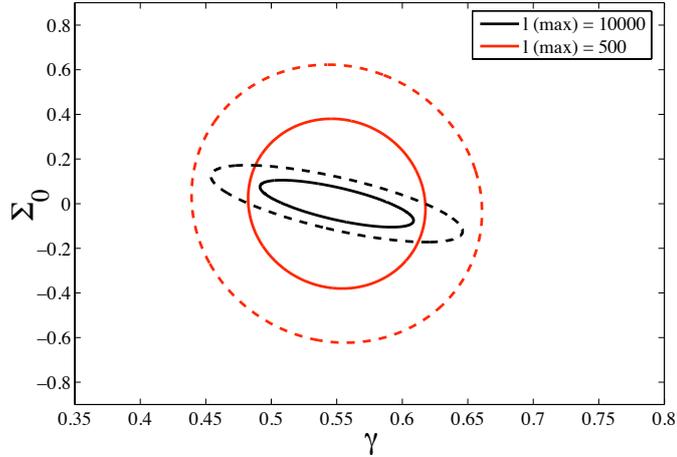

Figure 6.2: Marginalised $\gamma - \Sigma_0$ forecast for weak lensing only analysis with Euclid. Black contours correspond to $l_{max} = 10000$, demonstrating an error of $0.069(1\sigma)$ on $\Sigma_0$, whereas the red contours correspond to $l_{max} = 500$ giving an error of 0.25. In both cases, the inner and outer contours are $1\sigma$ and $2\sigma$ respectively. General Relativity resides at $[0.55, 0]$, while DGP resides at $[0.68, 0]$.

Using the CFHTLS lensing dataset, together with 2dF and SDSS data for BAO and SNLS data for supernova information, they show that DGP ($\alpha = 1$) can already be excluded, while $\gamma$ is still unconstrained.

On the other hand, by constructing Fisher matrix forecasts for Euclid, with survey parameters as given in the Yellow Book, they find that $\gamma$ will be accurately measured, together with the $\Sigma_0$ parameter; the resulting constraints from Euclid weak lensing alone are shown in Figure 6.2.

We see that the constraints obtained are dependent upon the maximum wavenumber $l_{\max}$ probed by the survey; $l_{\max} = 500$ corresponds to looking at the linear regime of the power spectrum, while $l_{\max} = 10000$ corresponds to including non-linear power. The latter requires the use of the Smith *et al.* (2003) fitting function, and results for this should be considered as indicative only (see below for further progress on this issue). We see that $\gamma$ is not found to be very sensitive to $l_{\max}$, while $\Sigma_0$ is measured much more accurately in the non-linear regime. In both cases, DGP is strongly excluded if the true model is $\Lambda$CDM.

### 6.3.2 Evolution of parameters

Members of our team considered an expansion of this parameterisation in Amendola *et al.* (2008). After considering predictions for Euclid accuracy of measurement for $\gamma$ and $\Sigma = 1 + a\Sigma_0$ in a similar fashion to above, the evolution of $\Sigma(z_i)$ was investigated by dividing the Euclid weak lensing survey into three redshift bins with equal numbers of galaxies in each bin. $\Sigma(z_i)$ was then approximated by $\Sigma(z_i) = \Sigma_i$, a piece-wise constant in the three bins. $\Sigma_1$ is set to 1, as there is a degeneracy between the overall amplitude of $\Sigma$ and the amplitude of matter fluctuations. The results are shown in Figure 6.3.

We see that the $\Sigma$ evolution is well characterised by Euclid lensing; a deviation from unit $\Sigma$ (i.e. General Relativity) of 3% can be detected in our second redshift bin, and a deviation of 10% is still detected in our furthest redshift bin.

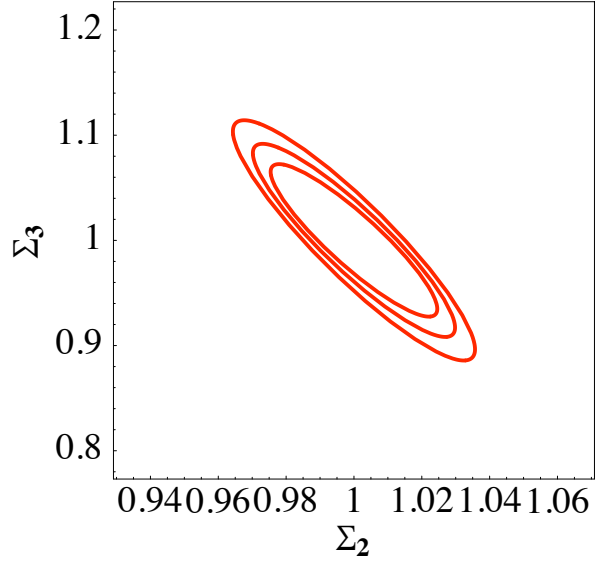

Figure 6.3: Forecast for Euclid constraints for $\Sigma$ evolution, with $z_{\mathrm{mean}} = 0.9$ and number density of galaxies 35 per square arcmin (outer contour), 50 per square arcmin (middle contour), 75 per square arcmin (inner contour).

### 6.3.3 Optimising the Survey for Modified Gravity

We have also examined whether the optimization of the survey for Dark Energy constraints also holds for modified gravity studies. Initial indications are positive; using the Beynon *et al.* (2009) models described below, the fiducial Euclid survey with $n_{\mathrm{gal}} = 35$ per square arcmin, $A$=20000 sq deg, $z_m = 0.9$ provides constraints on the growth factor $\gamma = 0.55 \pm 0.12(1\sigma)$ for a $\Lambda$CDM model. On the other hand, an equal-time deep survey with $n_{\mathrm{gal}} = 130$ per square arcmin, $A$=1000 sq deg, $z_m = 1.1$, we obtain $\gamma = 0.55 \pm 0.34$. These results are using the linear regime only, marginalising over $\Omega_m$ and $\sigma_8$. We see that the wide survey strategy is much preferred for obtaining constraints on $\gamma$.

## 6.4 Using the Non-Linear Power

A limitation in using weak lensing for modified gravity studies has been that, while much of the lensing signal on small scales arises from the non-linear matter power spectrum, we have not had good predictions for modified gravity in this regime. This is a crucial area where much theoretical work is being carried out, looking at how to probe non-linear power perturbatively (e.g. Koyama *et al.* 2009), via N-body simulations (Oyaizu *et al.* 2008; Schmidt 2009), halo statistics (Schmidt *et al.* 2008) and fitting functions (Hu & Sawicki 2007). A way of making progress by combining several of these insights has been described by members of our group in Beynon *et al.* (2009).

This paper has used the fact that the power spectrum must asymptote towards General Relativity predictions on small scales, for the particular expansion history in question. We can therefore approximate the power spectrum using an interpolation suggested by Hu & Sawicki

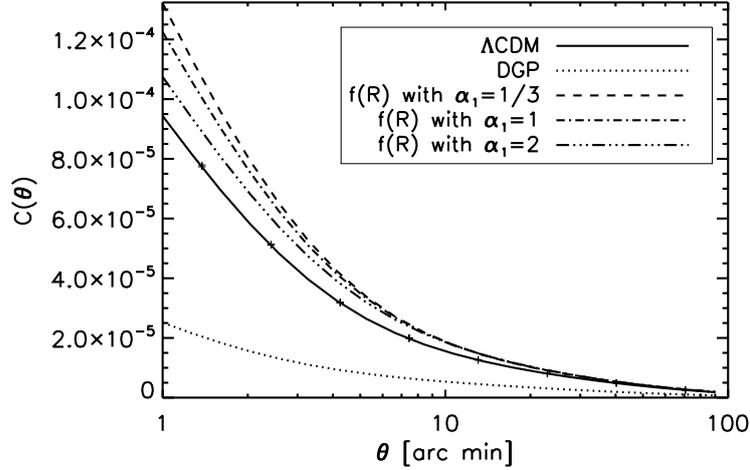

Figure 6.4: Correlation function predicted for $\Lambda$CDM, DGP and $f(R)$ with error estimates for Euclid. Models are for the central cosmological parameter values fitting WMAP+BAO+SNe, using a $\Lambda$CDM background (for $\Lambda$CDM and $f(R)$) and a DGP background (for DGP).

(2007),

$$P(k,z) = \frac{P_{non-GR}(k,z) + c_{nl}(z)\Sigma^2(k,z)P_{GR}(k,z)}{1 + c_{nl}(z)\Sigma^2(k,z)} \tag{6.2}$$

with

$$\Sigma^2(k,z) = \left(\frac{k^3}{2\pi^2}P_{lin}(k,z)\right)^{\alpha_1}, \quad c_{nl}(z) = A(1+z)^{\alpha_2} \tag{6.3}$$

i.e. with three new parameters, $A$, $\alpha_1$ and $\alpha_2$. By fitting these parameters to N-body simulations of DGP and $f(R)$ models (Oyaizu *et al.* 2008; Schmidt 2009), this study obtains weak lensing predictions for these models down to small scales, and shows that the discriminatory power of Euclid lensing is (as might be expected) stronger than in the linear regime alone. This paper restricts itself to the $\Sigma_0 = 0$ case, and measures the covariance on non-linear scales from the Horizon simulations of Teyssier *et al.* (2009).

An example of the shear correlation functions that are expected for Euclid following this method are shown in Figure 6.4; note the extremely small errors expected on this quantity for our survey. This translates into constraints on modified gravity parameters exemplified by Figure 6.5; here $A$, $\alpha_1$ and $\alpha_2$ have not been fit but instead are constrained by the lensing data itself. We see that, even with this larger range of parameters to fit, Euclid provides a measurement of the growth factor $\gamma$ to within 10%, and also allows some constraint on the new $\alpha_1$ parameter, probing the physics of nonlinear collapse in the modified gravity model.

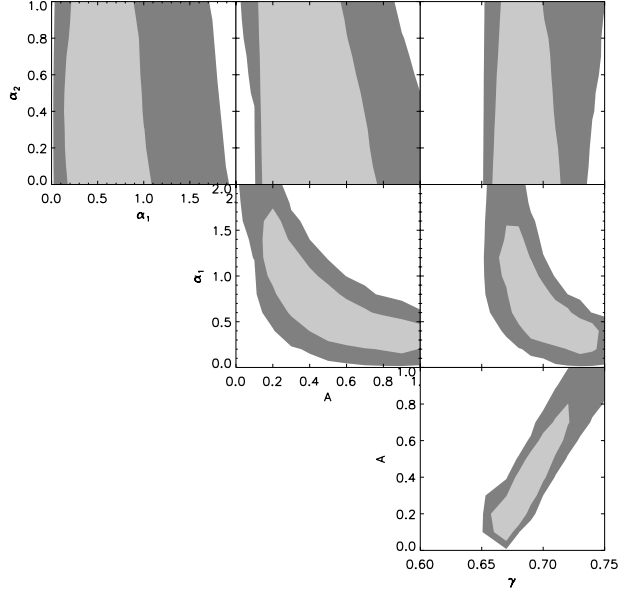

Figure 6.5: Constraints on $\gamma$, $\alpha_1$, $\alpha_2$ and $A$ from Euclid, using a DGP fiducial model and 0.4 redshift bins between 0.3 and 1.5 for the central cosmological parameter values fitting WMAP+BAO+SNe.

|  | Includes Planck? | Marginalised over # parameters | Assumes flatness? | Nonlinear approach |
|---|---|---|---|---|
| Heavens et al. | Yes | 10 | No | Smith et al. |
| Thomas et al. | No | 7 | No | Smith et al. |
| Amendola et al. | No | 8 | No | Smith et al. |
| Beynon et al. | No | 2 | Yes | Smith et al., Hu & Sawicki |

Table 6.1: Summary of assumptions in our models and analyses.

# Illuminating Dark Matter with Weak Lensing from Euclid

*Authors: T. D. Kitching (University of Edinburgh), A. Amara (ETH, Zurich), D. J. Bacon (University of Portsmouth), S. L. Bridle (University of London), K. Markovic (University of Munich), R. Massey (University of Edinburgh), J. Read (University of Zurich) + EIC Weak Lensing Working Group.*

**Abstract**

In the EIC we have shown that Euclid's low level of systematics and high precision statistics, available only from a space based experiment, can be used to improve our knowledge of all aspects of dark matter. Here we focus on the work of this group in during the Euclid Phase A activities. The power spectrum of fluctuations in the dark matter distribution, measured through the weak lensing power spectrum, can be used to constrain both the growth of large scale structure and the geometry of Universe. The distribution and halo shape of dark matter from galaxy to cluster scales will be mapped using shear and flexion statistics. The particle properties and nature of dark matter will be constrained using cosmic shear observations. Euclid will be a unique experiment for the investigation dark matter, and complementary to advanced particle physics and direct detection experiments. The large scale distribution, self-interaction cross-section and confirmation of any dark matter candidate as being the actual dark matter can only be achieved using high precision cosmological observations.

## 7.1 Introduction

The matter content of the Universe is dominantly in the form of a cold non-baryonic component that interacts with ordinary baryonic matter only via the gravitational force on cosmic scales. This astonishing realisation has been proved by a number of independent and disparate experiments. The first evidence for some "missing" (non luminous) mass came from observations of the Coma cluster (Zwicky 1933, 1937) and Virgo cluster (Smith 1936) where the velocity dispersion of the galaxies could not be accounted for by the mass implied by the luminous matter. On galactic scales a similar line of evidence was discovered when stars in Andromeda (Babcock 1939; Rubin & Ford 1970; Roberts & Whitehurst 1975), NGC3115 (Oort 1940) and other spiral galaxies (Rubin *et al.* 1985) were found to be rotating about the galactic enters at velocities that could not be explained by the luminous matter alone.

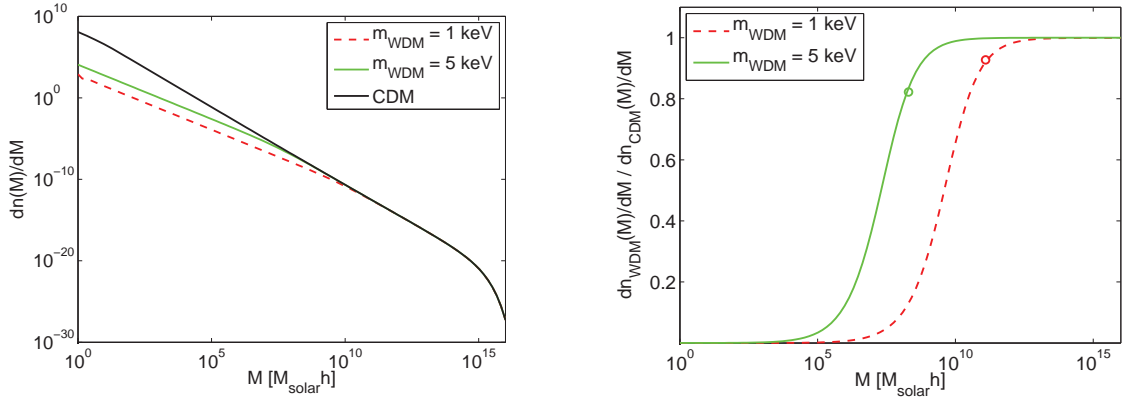

Figure 7.1: Left: the halo mass functions for Warm (1KeV or 5KeV) and cold dark matter particles. Right: the ratio of the mass function in a universe containing warm dark matter particles with masses of 1 keV (dashed line) and 5 keV (solid line) to the mass function in a universe with standard cold dark matter. The circles mark the halo mass that corresponds to the scales suppressed by free-streaming of these two types of warm dark matter particles. particles. From Markovic *et al.* (2009).

The modern ΛCDM model of cosmology also requires a large reservoir of (cold) dark matter to account for the observed energy density of the Universe, and to explain structure formation. Slowing the expansion after the Big Bang required much more mass than the baryons could provide (Coc *et al.* 2004). However, the dark matter must have stopped interacting with photons very early to preserve the uniformity of the primordial Cosmic Microwave Background radiation, in which fluctuations reach only one part in $10^{-5}$ (Peebles 1982). Indeed, after decoupling from other particles, dark matter could collapse under its own gravity before ordinary matter and these dark matter potential wells laid the initial seeds about which structure formation grew (Davis *et al.* 1985; Efstathiou *et al.* 1985, 1990; Springel *et al.* 2006). The latest measurements of the Cosmic Microwave Background and Large-Scale Structure (Dunkley *et al.* 2009; Lesgourgues *et al.* 2007) indicate that the Universe contains approximately one hydrogen atom per cubic metre, but five times that in the form of dark matter.

Most significantly gravitational lensing (see Massey *et al.* 2009, for a recent review by EIC members) probes the total matter content along the line of sight so in our Universe, dominated by dark matter, gravitational lensing is effectively a direct probe of the dark matter distribution. On cluster scales the "bullet cluster" (Clowe *et al.* 2007) represents the first empirical observation of dark matter and has been used to constrain the dark matter cross-section from lensing measurements. We review other dark matter constraints from lensing measurements in the following sections.

Despite the overwhelming abundance of dark matter we do not know what it is, how much there is over all scales or how it is distributed in detail. Euclid's exquisite imaging surveys will help to illuminate the nature of this unknown but dominant component of our Universe.

## 7.2 Particle Properties

One explanation for dark matter is that there is some as yet undetected non-baryonic sub-atomic particle beyond our standard model of particle physics that, once created in the early Universe exists and pervades the Universe until the present time. This Weakly Interacting Massive Particle (WIMP) picture has emerged as being preferable over an alternative baryonic explanation, that undetected MAssive Compact Halo Objects (MACHOs) surround galaxies and clusters, due to substantial evidence from gravitational lensing microlensing events. The MACHO survey (Popowski *et al.* 2003; Alcock *et al.* 2000), OGLE (Wyrzykowski *et al.* 2009; Udalski 2009) and POINT-AGAPE (Calchi Novati *et al.* 2005) all rule out MACHOs as being the dominant dark matter component, in fact after collectively monitoring may tens of millions of stars for over 10 years only a handful of microlensing events have been observed - whilst many hundreds were expected. There are two extensions to the standard model of particle physics both of which provide natural dark matter candidates. The axion is a Nambu-Goldstone boson produced by the addition of an extra symmetry (the U(1) Peccei-Quinn symmetry) that has been proposed to solve the CP problem. These particles may exist in the range of a few eV to a a few GeV. The supersymmetric (SUSY) extension to the standard model implies that any fermion/boson superpartners to standard boson/fermion particles must obey a condition called R-parity. Lepton and baryon number are not conserved in general SUSY models but these conservation rules are very well constrained by observations so R parity was introduced to suppress any coupling between SUSY particles and any interaction that may violate lepton or baryon number (e.g. a SUSY particle decaying into photons). This implies that the lightest SUSY particle cannot decay even if it is heavier that most (or all) standard model particles. Lightest SUSY particle include the neutralino (which is in fact an admixture of many SUSY particles), the axino and the gravitino. These particle are expected to have masses anywhere between a few eV-GeV (for the neutralino) and a few eV-TeV (for the gravitino).

Candidates for particle nature of dark matter may appear in future particle physics experiments, such as the LHC, or be detected in direct detection experiments. However to prove that these candidates are indeed the dark matter that constitutes the majority of mass in the Universe we will have to use cosmological observations.

Euclid will be extremely sensitive to any dark matter particle in the range of a few eV to a few hundred eV. A direct effect of dark matter is to change the relative abundance of sub-structure in galaxy clusters. For dark matter with a particle mass of 1 keV the majority of substructure below a mass of $10^{11}$ $M_{\odot}$ will be diffused at 5 keV; this limit is reduced to $10^9$ $M_{\odot}$. Using a weak lensing flexion variance (Bacon *et al.* 2009) such a signature could be distinguished from LCDM at a 10-$\sigma$ confidence.

There exists a discrepancy between observations of structure on sub-galactic scales and high resolution N-body simulations assuming the $\Lambda$CDM cosmological scenario. In particular, the observed numbers of dwarf galaxies seem to differ from the theoretical prediction by up to an order of magnitude. Observations can potentially be reconciled with theory by replacing the standard cold dark matter with particles that are mildly relativistic (or 'warm') at their decoupling. Such particles suppress the formation of structure by dampening the density fluctuations on scales below their free-streaming length. A fortunate consequence of such a scenario is that measuring the precise characteristics of the matter distribution on sub-galactic scales can provide constraints on the properties of the dark matter particle. Weak gravitational lensing traces the distribution of matter in the universe and so it can be used to calculate the power spectrum of matter structure. With the cosmic shear power spectrum from an upcoming survey like Euclid one can calculate the matter power spectrum on scales small enough to be of importance in warm dark matter calculations. Markovic *et al.* (2009) calculate a prediction for the upper limit detectable with such cosmic shear data providing a lower limit on the warm dark matter particle mass that would be allowed if the Euclid survey measured an apparently $\Lambda$CDM shear power spectrum. In this case, the least energetic, but still detectable neutrino-like, early decoupled dark matter particle would have a mass of 2keV. Figure 7.2 shows how the halo mass distribution changes with respect to the mass of the warm dark matter particle. These are Sheth-Tormen mass functions in which the WDM effect was simulated by suppressing the linear power spectrum with the warm dark matter transfer function from Viel *et al.* (2005).

Dark matter is most likely to contain multiple particle components, one of these components which is already known to exist will be massive neutrinos. Euclid will be extremely sensitive to massive neutrino properties through the effect that such particles have on structure formation. Massive neutrino tend to damp small scale structure by providing an effective pressure at small scales, this effect will be detectable by measuring the matter power spectrum through cosmic shear. Kitching *et al.* (2008) have shown that Euclid will be able to constrain the sum of neutrino masses and number to 3% and 10% respectively in combination with Planck. De Bernardis *et al.* (2009) have shown that Euclid will be able to distinguish the mass of individual neutrino species to percent accuracy and will be able to decisively (Jeffrey's scale of Bayesian evidence) distinguish between the neutrino hierarchies.

## 7.3 The Large-Scale Distribution of Dark Matter

The large-scale distribution of mass can be described statistically in terms of its power spectrum $P(k)$, which is shown in Figure 7.2. This is the Fourier transform of the correlation functions and measures the amount of clumping on different physical scales. The power spectrum is commonly decomposed into several components: the primordial power spectrum, the transfer function that maps the primordial fluctuation onto the dark matter power spectrum, and the growth factor that evolves the power spectrum as a function of time (or redshift)

$$P(k) \propto k^{n(k)} T^2(k) \text{Growth}^2, \tag{7.1}$$

where the primordial power spectrum $n(k) = n + \alpha \frac{d\ln n}{d\ln k}$ can be written as a power law parameterised by a slope $n$ and some running $\alpha$. A slope of unity is the Harrison-Zeldovitch approximation. If the density field is Gaussian, the same information can also be expressed as the mass function $N(M, z)$, i.e. the density of haloes of a given mass as a function of that mass and cosmological redshift. There has been a substantial amount of work on analysing the properties of the dark matter power spectrum, including its growth over time from a power spectrum of primordial density fluctuations. In order to compare theoretical models to data, the semi-analytic approach of Eisenstein & Hu (1998) is used, predominantly for its simple fitting functions to the linear power spectrum, which can be extended (at 5-10% accuracy) into the mildy nonlinear high-$k$ regime (Peacock & Dodds 1996; Smith *et al.* 2003).

The matter power spectrum thus encodes a wealth of information on the statistical distribution and growth of dark matter on all scales. The cosmic shear power spectrum combines measurements of the geometry of the Universe through the lensing effect with measurements of the dark matter power spectrum directly. For example the observed strength of the cosmic shear signal as a function of distance and angle can be written like

$$C_\ell^{\kappa\kappa} = \frac{9}{4} \left(\frac{H_0}{c}\right)^4 \Omega_m^2 \int_0^{r_H} dr P_\delta \left(\frac{\ell}{r}, r\right) \left[\frac{\overline{W}(r)}{a(r)}\right]^2 \tag{7.2}$$

where $P_\delta$ is the dark matter power spectrum. This has the advantage over using the galaxy distribution to infer the dark matter power spectrum since the direct measure does not include any assumptions about the way that galaxies trace the underlying dark matter structure. The EIC has shown (see Chapter 4) that using the cosmic shear power spectrum the clumpiness of dark matter matter over a wide range of scales can be used to constrain cosmological parameters to unprecedented accuracy. The growth of structure contains information on how the dark matter structure grows over cosmic time as well as the dark energy component or signatures of modified gravity (see Chapter 6), weak lensing observations been used to detect the growth of these perturbations (e.g. Massey *et al.* 2007; Bacon *et al.* 2005).

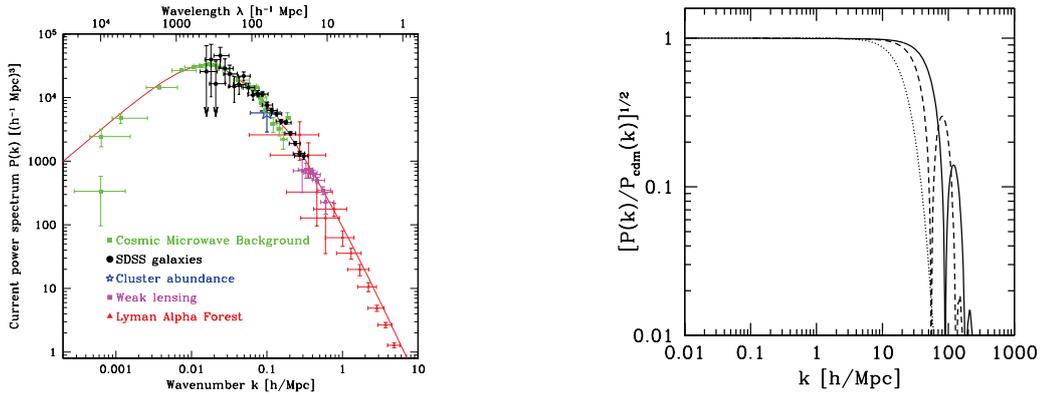

Figure 7.2: The statistical distribution of dark matter in Fourier space. *Left:* The mass power spectrum, showing the clumping of dark matter as a function of large scales (low $k$) to small scales (high $k$) (Tegmark *et al.* 2004), including some very early weak lensing constraints (Hoekstra *et al.* 2002). Current constraints are tighter and extend over a wider range of scales in both directions. *Right:* The effect on the power spectrum of 1 MeV mass WIMP dark matter that remains coupled to other particles in the early universe (Hooper *et al.* 2007); the solid (dashed) line assumes a 10 keV (1 keV) kinetic decoupling temperature. The dotted line illustrates the similar small-scale damping effect of a component of warm dark matter.

Using gravitational lensing Euclid will map the dark matter distribution. Pires *et al.* (2009) have reviewed the state-of-the-art in dark matter mass map reconstruction within the context of Euclid and discussed how to analyse this wealth of information to maximise the cosmological information that can be constructed.

The dark matter distribution can also be mapped in three dimensions using weak lensing techniques coupled with photometric redshifts. Simon *et al.* (2009) has recently shown how the dark matter density map in three dimensions can be reconstructed, an extension of the work of Taylor *et al.* (2004) that showed how the gravitational potential of the dark matter field can be mapped in three dimensions. Such techniques will allow Euclid to map the three dimensional dark matter distribution over the entire extra-galactic sky.

The large number density of high resolution galaxy images from Euclid will also allow non-linear features in the dark matter cosmic web to be observed such as filamentary structures that should connect dark matter haloes on large scales. Mead *et al.* (2009) have developed a number of weak lensing techniques that could be used to detect filaments in individual clusters or by stacking clusters.

## 7.4 Dark Matter Sub-Structure

The relative and absolute abundance of dark matter sub-structure as well as the distribution of dark matter sub-structure contains information on the nature of dark matter. For example if the temperature of dark matter was high then the effective pressure created on small scales will act to damp the growth of small sub haloes. In conjunction with large N-body simulations, that can be used to predict the abundance of small dark matter haloes for a given dark matter candidate, lensing measurements are already testing the CDM paradigm.

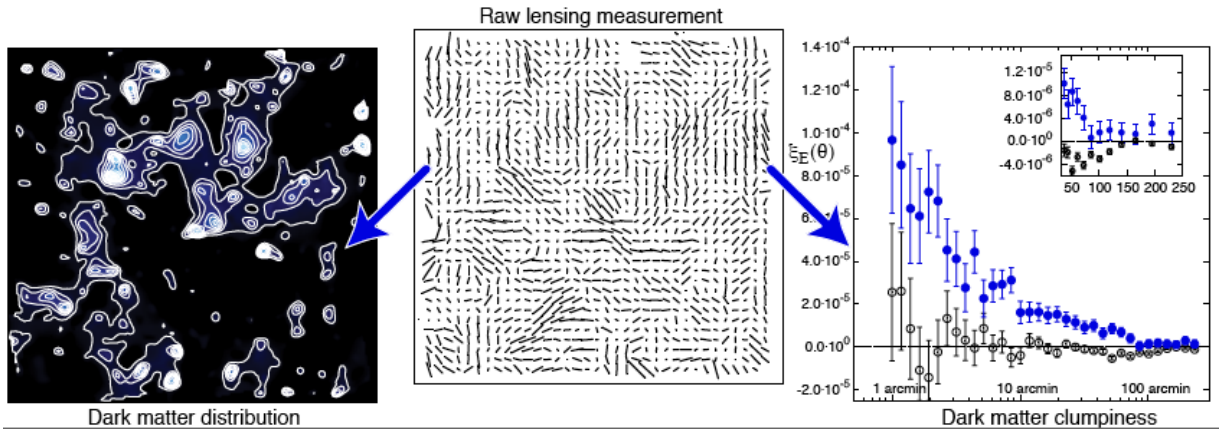

Figure 7.3: Plot adapted from Massey *et al.* (2007) and Fu *et al.* (2008). This central panel shows the raw ellipticity measurement from the weakly lensing galaxies, the length of the stick os proportional to the total ellipticity. Right : this measurement can be used to create dark matter maps where the colour represents the dark matter density. Left : this measurement can be used to construct a correlation function (or an equivalent power spectrum) from which global dark matter properties can be inferred.

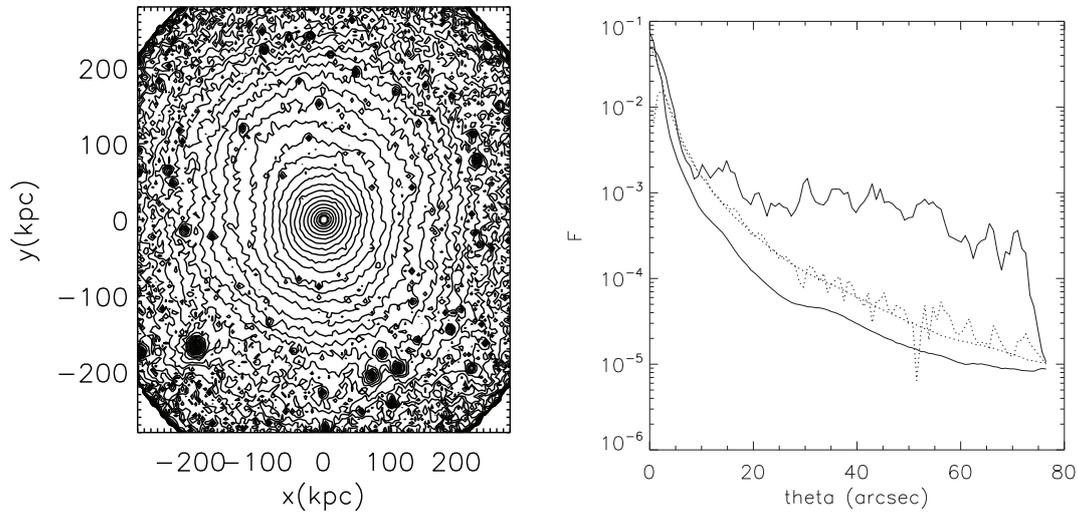

Figure 7.4: Images from Bacon *et al.* (2009). Left : a simulated cluster, contours show lines of constant density, this exhibits a large amount of substructure on all scales and at all radii. Right : Dashed are the mean flexion signal for simulations that include sub-structure (upper line) and for those that do not (lower line), solid are the flexion variance signal for simulations with (upper line) and without (lower line) sub-structure. Using flexion variance sub-structure is detected at all scales.

We here briefly review weak lensing flexion signal, where the galaxy distortion is measured at second order (a small 'arcness' in the image), which is particularly sensitive to small scale mass distributions including dark matter sub-structure. To measure this effect, even smaller in signal-to-noise that cosmic shear, a stable and well characterised PSF is crucial, hence Euclid is the ideal experiment to take advantage of this next-level of weak lensing information.

Gravitational lensing maps a coordinates on the sky $\theta'_i$ onto observed coordinates $\theta_i$ for $i \in \{1, 2\}$ via

$$\theta'_i = A_{ij}\theta_j + \frac{1}{2}D_{ijk}\theta_j\theta_k + \dots, \tag{7.3}$$

where the usual first-order matrix

$$A_{ij}(\theta) \equiv \frac{\partial\theta'_i}{\partial\theta_j} = (\delta_{ij} - \partial_i\partial_j\psi(\theta)), \tag{7.4}$$

$$\mathbf{A} = \begin{pmatrix} 1 - \kappa - \gamma_1 & -\gamma_2 \\ -\gamma_2 & 1 - \kappa + \gamma_1 \end{pmatrix}$$

encodes shear and magnification, but the next terms in a Taylor series $D_{ijk} = \partial_k A_{ij}$ include flexion.

This extension of the lensing formalism is most efficiently explored by first defining a complex gradient operator $\partial = \partial_1 + i\partial_2$. The lensing deflection angle is obtained by applying this operator to the lensing scalar potential $\psi$

$$\alpha = \alpha_1 + i\alpha_2 = \partial\psi. \tag{7.5}$$

The first order lensing observables are combinations of applying the operator twice

$$\kappa = \frac{1}{2}\partial^*\alpha = \frac{1}{2}\partial^*\partial\psi \tag{7.6}$$

$$\gamma = \gamma_1 + i\gamma_2 = \frac{1}{2}\partial\partial\psi, \tag{7.7}$$

where the gradient and its complex conjugate gradient $\partial^*$ commute. As shown in figure 7.5, these respectively produce spin-zero (magnification) and spin-two (shear) distortions to an image. Taking the third derivative of the lensing potential we find the unique combinations

$$\mathcal{F} = \frac{1}{2}\partial\partial^*\partial\psi = \partial\kappa = \partial^*\gamma,$$

$$\mathcal{G} = \frac{1}{2}\partial\partial\partial\psi = \partial\gamma. \tag{7.8}$$

These are known as the first and second flexions. They introduce spin-one and spin-three perturbations to an initially round image. These can be measured in a similar way to weak lensing shear methods, and yield constraints on the local derivative of the projected mass distribution.

The use of flexion in the effort to reconstruct sub-structure was first suggested by Massey *et al.* (2007). More recently Bacon *et al.* (2009) have developed the analysis method by investigating the variance of the flexion signal around clusters. The key realisation in this work is that for a smooth matter profile the flexion signal increases towards areas of high density but the variance of the signal in azimuthal bins is small (for a smooth profile the flexion amplitude is the same in all directions). If sub-structure is present then the azimuthal symmetry is broken and the variance of the flexion signal increases. Using this technique Bacon *et al.* (2009) have shown that dark matter haloes as small as $10^9 \, M_\odot$ could be distinguished and that the signature of dark matter particles with masses of up to 5 keV could be detected.

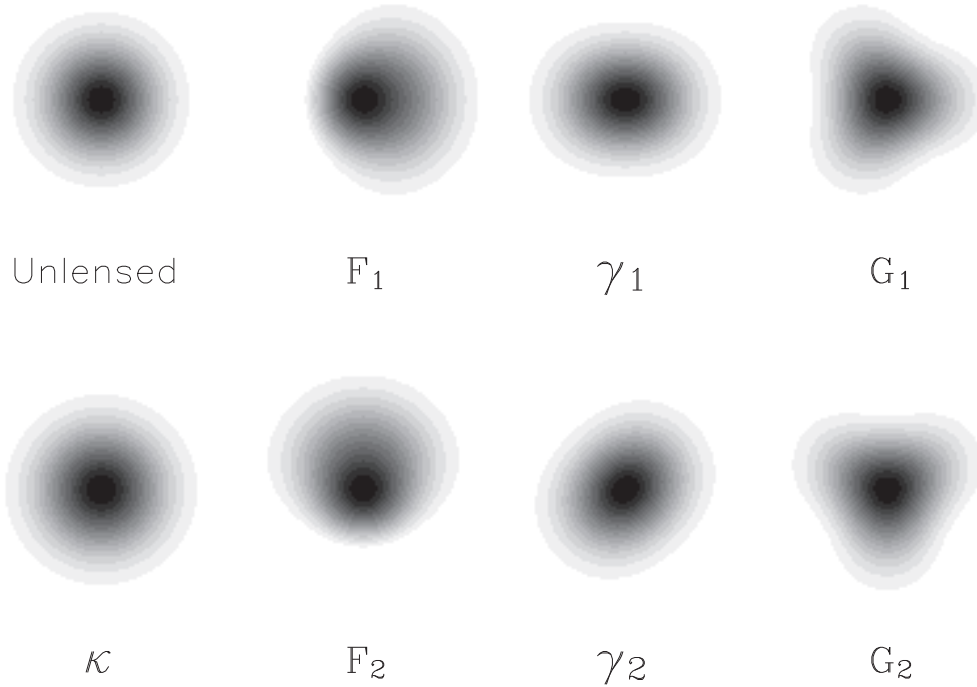

Figure 7.5: Weak lensing distortions with increasing spin values. Here an unlensed Gaussian galaxy with radius 1 arcsec has been distorted with 10% convergence/shear, and 0.28 arcsec$^{-1}$ flexion. Convergence is a spin-0 quantity; first flexion is spin-1; shear is spin-2; and second flexion is spin-3. From Bacon et al. (2006).

In addition to the flexion signal there are also extra higher order moments of weak lensing that can be used as a systematic check. These are known as 'twist' and 'turn' and are described in Bacon & Schäfer (2009).

## 7.5 Dark Matter Halo Properties

In the standard cosmological model, dark matter haloes are expected to be significantly non-spherical. Measurements of weak gravitational lensing in the Sloan Digital Sky Survey confirm that the axis ratio of haloes around isolated galaxy clusters (projected onto the 2D plane of the sky) is $0.48^{+0.14}_{-0.19}$ (Evans & Bridle 2008). This rules out sphericity at 99.6% confidence and is consistent with the ellipticity of the cluster galaxy distribution. However, the dark matter haloes around individual galaxies are slightly rounder than the light profiles. By creating a 1D analogue of some computationally intensive 2D integrals Hawken & Bridle (2009) have shown that constraints on the ellipticity of dark matter haloes should be two orders of magnitude tighter from gravitational flexion than those from shear.

Schneider & Bridle (2009) have extended the standard dark matter halo occupation to include the effects that these haloes have on the alignment of galaxies. This new model is based on the placement of galaxies into dark matter halos. The central galaxy ellipticity follows the large scale potential and, in the simplest case, the satellite galaxies point at the halo center. The two-halo term is then dominated by the linear alignment model and the one-halo term provides a motivated extension of intrinsic alignment models to small scales. Such effects will need to be understood to characterise the intrinsic alignment systematic in weak lensing as well as being a direct probe of the local dark matter environment.

Accurate measurement of the cluster mass function is a crucial element in efforts to constrain dark matter structure formation models. Euclid will observe hundreds of thousands of clusters and groups from $10^8$–$10^{15}$ $M_\odot$. By measuring the averaged mass profiles of these clusters over a wide range of masses and redshifts, the characteristic 'clumpiness' and 'cuspiness' of dark matter structures can be tightly constrained as a function of redshift and mass. Constraining the evolution of cluster mass profiles in this way places limits on the overall dark matter content of the Universe as well as the temperature and mass properties of the dark matter particles. Corless & King (2009) have shown that the effect of ignoring potential tri-axiality of dark matter haloes when stacking the lensing signals of many clusters and groups, binned by mass-correlated observables such as richness and luminosity, bias 3D virial mass estimates by a few percent. This bias affects not only direct lensing constraints on the mass function but can also affect the scatter and normalisation of the mass-observable relations derived from lensing that are so crucial to constraining the cluster mass function with large samples.

## 7.6 Summary

Euclid is poised to revolutionise the study of the dominant dark matter component of our Universe. Euclid will map the universal distribution of dark matter on large scales over the entire extra-galactic sky and on cluster and galaxy scales. Euclid will provide the data with which galaxy clusters and galactic haloes can be used as laboratories for the study of dark matter. Using shear and flexion statistics dark matter haloes will be detected from $10^8$–$10^{15}$ $M_\odot$. The constraints on the shapes of dark matter haloes including the cuspiness and tri-axiality will be improved by two orders of magnitude over current results. The nature of dark matter will be tested, using cosmic shear and flexion Euclid will constrain any $eV$–$KeV$ dark matter models, and in particular constrain the neutrino mass and hierarchy to percent accuracy.

# Point-Spread Function and Instrument Calibration

*Authors: Stéphane Paulin-Henriksson (CEA - Paris), Adam Amara (ETH - Zurich), Sarah Bridle (UCL - London), Eduardo Cypriano (UCL - London), Alexandre Réfrégier (CEA - Paris), Jason Rhodes (JPL - Los Angeles), Michael Seiffert (JPL - Los Angeles), Mario Schweitzer (MPI - Munich), Lisa Voigt (UCL - London) + EIC Weak Lensing Working Group.*

### Abstract

The Point-Spread-Function (PSF) and instrumental calibrations have been identified as critical issues to achieve the required accuracy in shape measurements, characterised by the level-1-requirement $\sigma_{\mathrm{sys}}^2 < 10^{-7}$. A simple analytical model allows us to estimate that one needs typically 50 stars to calibrate the PSF. After presenting this, we discuss three effects that could degrade the accuracy of shape measurements, but are not taken into account in the analytical model: the pixellation, whose impact is correlated with the PSF shape; the colour dependence of the PSF, and the Intra-Pixel Sensitivity Variations (IPSV). The latter effect is also discussed in the framework of infra-red detectors. First, to investigate the impact of pixellation, we have implemented a PSF assessment pipeline and derived preliminary level-2 requirements on the PSF shape. Second, we find the PSF colour dependence is compatible with the required accuracy. Third, we investigate the impact of IPSV in a general case for shape measurements in the visible and photometry in the near infra-red.

## 8.1 Introduction

Since the Point-Spread-Function (PSF) of an instrument varies on all scales, measuring the shape of a galaxy requires the PSF being calibrated using the stars that surround the lensed galaxy. Any error on the estimated PSF propagates in a systematic error on the measure of the galaxy shape. The required accuracy of the PSF calibration is fixed in the level-1-Top-Level-Science-Requirements (EIC.LEV1.4), in terms of an upper limit of $10^{-7}$ on the variance of the systematic effects $\sigma_{\mathrm{sys}}^2$ when estimating the galaxy shear. $\sigma_{\mathrm{sys}}^2$ is defined in Amara & Réfrégier (2008).

In Sect. 8.2, we estimate the number of stars that must be considered during the PSF-calibration process, to achieve this accuracy, in the simple framework of an analytical propagation of uncertainties, for a given wavelength, and for infinitely small pixels. We then review 3 effects that could significantly degrade this accuracy, and study them in details through simulations.

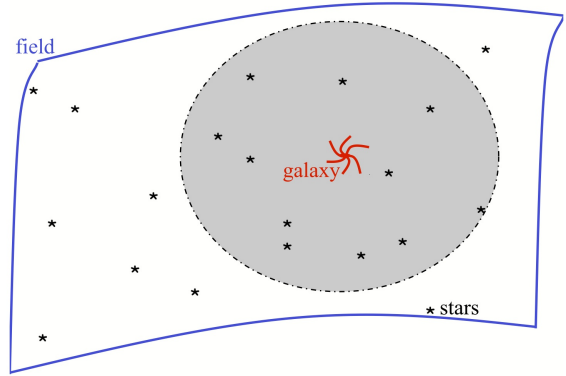

Figure 8.1: Illustration of the required number of stars $N_*$ to calibrate the PSF: when measuring the shape of a galaxy (indicated by the red spiral), the PSF needs to be calibrated with at least $N_*$ nearby stars (black asterisks) contained in the shaded region, to achieve the requirement $\sigma_{\mathrm{sys}}^2 < 10^{-7}$. In this example $N_* = 11$. Therefore, on scales smaller than the shaded region, there is not enough information coming from the stars to measure the PSF variations with the required accuracy.

Each of these 3 effects have a dedicated section in the following, that present the current status and point out that there is on going work to further develop our requirements:

- **The pixellation and PSF shape**. For shape measurements, pixellation induces biases and larger statistical errors than in the analytical case. The impact of pixellation is highly correlated with the PSF shape. This is discussed in Sect. 8.3.

- **The wavelength dependence of the PSF**: In general the PSF depends on the Spectral Energy Distribution (SED) of the object, and therefore will not be the same for stars and galaxies. Thus, calibrating the PSF on stars to deconvolve galaxies, is a biased approach. This point is discussed in Sect. 8.4.

- **Intra-Pixel Sensitivity Variations (IPSV):** The IPSV is discussed in the framework of the photometric observations performed in the Near Infra-Red (NIR) bands. This is presented in Sect. 8.5.

This chapter does not discuss of the finite accuracy of the analysing, which can be considered as another effect that may degrade the accuracy of measures. Indeed, as discussed in Paulin-Henriksson *et al.* (2009), this is a software issue that concerns the data analysis, while this chapter is dedicated to instrumental effects.

## 8.2 The number of stars required to calibrate the PSF

Each star provides an image of the PSF that is pixellated and noisy, which means that to reach a given accuracy in the knowledge of the PSF, a number of stars is required. This is illustrated in Fig. 8.1. But increasing the number of stars does not necessarily imply that the accuracy of the PSF calibration will be better. There is a trade-off between the number of stars, their density, the stability of the PSF, and the accuracy of the interpolation scheme (i.e. the way the PSF is interpolated between stars).

The required number of stars to calibrate the PSF at a given accuracy was first issued in Paulin-Henriksson *et al.* (2008), in the simple case of an analytical propagation of uncertainties, for infinitely small pixels and assuming the shear estimator is function of the second order moment of the flux. In this paper we find the scaling relation between the number of stars $N_*$ involved in the PSF calibration, the value of $\sigma_{\mathrm{sys}}^2$, the complexity $\Psi$ of the PSF model (characterising the number of degrees of freedom in the model), the Signal-to-Noise Ratio (SNR) of stars, and the dilution factor $D = \frac{R_{\mathrm{galaxy}}}{R_{\mathrm{PSF}}}$, which is the ratio between the radius of the smallest galaxy and the radius of the PSF:

$$N_* \approx 50 \times \left(\frac{\mathrm{SNR}}{500}\right)^{-2} \times \left(\frac{D}{1.5}\right)^{-4} \times \left(\frac{\Psi}{3}\right)^{2} \times \left(\frac{\sigma_{\mathrm{sys}}^2}{10^{-7}}\right)^{-1} \tag{8.1}$$

For the simplistic case of a nearly circular Gaussian-like PSF, we have $\Psi \approx \sqrt{2}$. In a realistic case, $\Psi$ is at least of few. The accuracy of this scaling relation is limited by two main approximations:

- first, it is pessimistic in the sense that it assumes the shear estimate is a function of unweighted $2^{\mathrm{nd}}$ order moments of the flux. This assumption was made to allow analytical calculations but in practice we expect the required number of stars to be lower;

- second, it is optimistic in the sense that it does not take into account a number of significant factors, such as the pixellation. In other words, the scaling relation 8.1 holds for infinitely small pixels. In practice, for big pixels or wide observation bands, the required number of stars may be larger. Moreover, radiation damages during the mission on the charge transfer efficiency may require this number of stars to increase during the mission time.

To investigate the required number of stars in practice, we performed a number of simulations summarised in the following.

## 8.3 Impact of pixellation and PSF shape

The impact of pixellation is highly correlated with the shape of the PSF. That is why it is necessary to investigate together pixellation and PSF shape. For this, we have implemented a PSF shape assessment pipeline, that is described below (Sect. 8.3.1), and that is used to perform large simulation sets. In particular, we focused our attention on a realistic system PSF, noted nominal' PSF because it is derived from the optical design proposed by industry. The nominal PSF is described in Sect. 8.3.2, and our current results on the impact of pixellation and PSF shape are summarised in Sect. 8.3.3.

### 8.3.1 PSF shape assessment simulation

An overview of the PSF shape assessment simulations is shown in Fig. 8.2. The main steps are the following:

- Simulation of star fields and galaxies: the system PSF is used to simulate N star fields, and M 2D-exponential profile galaxies. We currently take into account the dithering strategy and the Intra-Pixel Sensitivity Variations.

- Calibration of the PSF on a star field: we consider the system PSF with few arbitrary chosen degrees of freedom (e.g. a rotation, a dilation/compression and a distortion) and fit these degrees of freedom on one star field, with a $\chi^2$ minimisation.

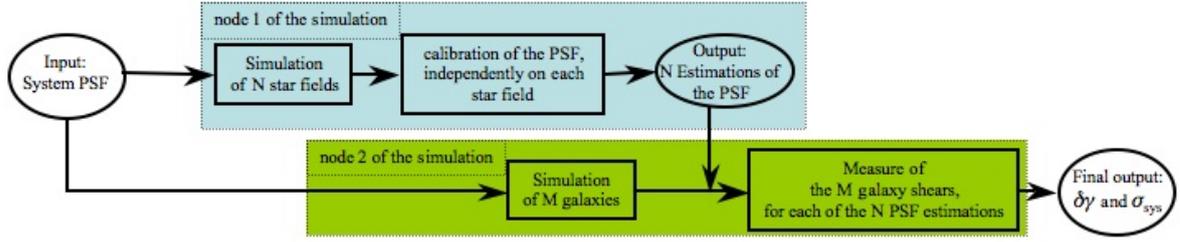

Figure 8.2: Summary of the PSF assessment simulations that estimates the bias on the shear measurements, for a given system PSF and galaxy population. The system PSF, in input, is used to simulate $N$ star fields and $M$ galaxies. Then the simulations can be sorted into 2 nodes. First, we simulate $N$ times the PSF calibration on the star fields. In output, each star field leads to an estimation of the PSF. Second, we estimate the galaxy shears, considering the PSF estimations instead of the true PSF. Each calibration of the PSF leads to a bias on the galaxy shear coming from the fact that the estimated PSF is not exactly equal to the true one. From the N biases, we derive $\sigma_{\text{sys}}$.

- Measure of the M galaxy shears, for a given estimated PSF: For a given estimation of the PSF, we fit a 2D-exponential profile on each galaxy, with a chi2-minimisation, and derive the corresponding fitted ellipticity. There is a bias in the fitted ellipticity, due to the fact that the estimated PSF is not infinitely accurate. Each of the N estimations of the PSF leads to a different bias.

- From the N biases, we compute the corresponding $\delta\gamma$ and $\sigma_{\text{sys}}$.

These simulations allow us to investigate the impact of number of effects, such as: the PSF and the galaxy shapes, the pixel size, the SNR of stars and galaxies, the number of stars involved in the PSF calibration, the interpolation scheme between stars, the dithering strategy, and some detector effects. We are currently including a detailed simulation of Radiation Damages on the Charge Transfer Efficiency.

### 8.3.2 The nominal PSF

The 'nominal' PSF is derived from the optical design proposed by industry. We calculated it by the convolution of 3 components of the PSF:

- **The optical PSF:** This is the component that comes from the optical configuration. For instance this would include diffraction effects. We have simulated the optical PSF (illustrated in Fig 8.3) that corresponds to the optical design proposed by industry with a resolution of $5 \times 10^{-3}$ arcsec per step, over a postage stamp of $5.12 \times 5.12$ arcsec$^2$.

- **The AOCS PSF:** This component comes from the movement of the instrument during an exposure. We adopted a realisation of the AOCS pattern.

- **The detector PSF:** This component comes from the detectors effects such as diffusion (but this does not include pixellation, which can not be modelled as a convolution). We adopted a simple Gaussian description of the detector PSF.

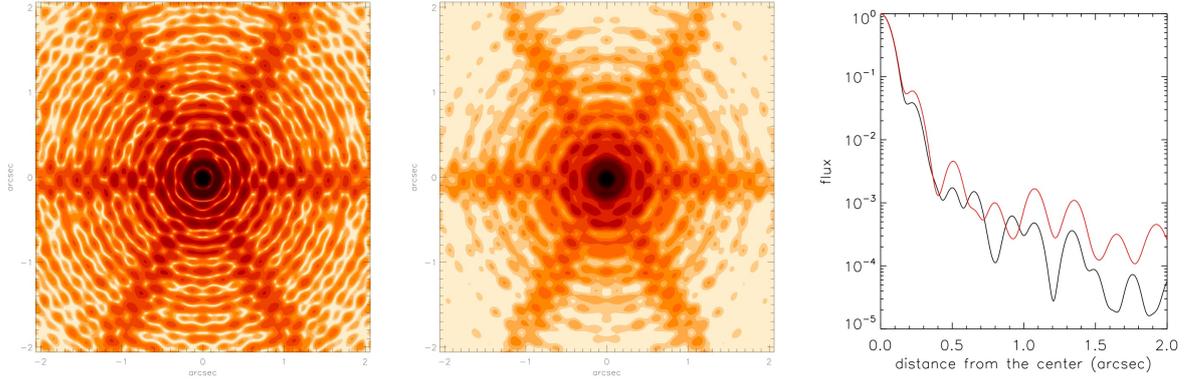

Figure 8.3: PSF at 800nm generated by the nominal optical design. The left-hand panel is the optical PSF (with a 3-arm spider of the structure maintaining the central obstruction) in logarithmic scale. The middle shows the system PSF (ie. the convolution of the optical PSF with the AOCS and the detector patterns). The right-hand panel shows the profile of the system PSF. Black line: from the centre toward the top (ie. far from any diffraction spike). Red line: from the centre toward the right-hand side (ie. inside a diffraction spike).

It should be noted that the system PSF under discussion here is the *un-pixellated* realisation of the PSF. The final image that a point source would form is a pixellated version of the system PSF, that depends on the position of the point source inside a pixel. The requirements and verifications are performed on the un-pixellated system PSF.

The properties of the nominal PSF are listed in Tab. 8.1. The level-2-requirements are derived from the level-1-requirement $\sigma_{\mathrm{sys}}^2 < 10^{-7}$, using the PSF shape assessment simulations presented in Sect. 8.3.1. As explained in Sect. 8.3.3, these level-2-requirements are conservative: it is possible to achieve the level-1-requirement $\sigma_{\mathrm{sys}}^2 < 10^{-7}$ even if the level-2-requirements on the PSF shape are marginally achieved. This is the case for the nominal PSF, that is acceptable although the FWHM is too small compared to the corresponding level-2-requirement.

### 8.3.3 Current results about the impact of pixellation and PSF shape

Fig. 8.4 shows $\sigma_{\mathrm{sys}}^2$ as a function of the PSF FWHM, for pixels of $0.1 \times 0.1 \, \mathrm{arcsec}^2$, in 3 cases:

1. **Analytical pixellation-free model for a Gaussian-like PSF (dotted line):** This is by far the most simple case, that requires no simulation. When applying the analytical calculations of Paulin-Henriksson et al. 2008, already mentioned above, in the case of a nearly circular Gaussian PSF, one obtains:

$$\sigma_{\mathrm{sys}}^2 \approx \frac{1.33}{D^4 \, \mathrm{SNR}^2 \, N_*} \qquad (8.2)$$

We can see that, for a given star population, $\sigma_{\mathrm{sys}}^2$ depends only on the dilution factor $D$ (the ratio between the radius of the galaxy and the radius of the PSF): ie. larger the galaxy, smaller $\sigma_{\mathrm{sys}}^2$. In this example, the dotted line corresponds to the case $R_{\mathrm{quadrupoles}} = 2 \, \mathrm{pixels}$.

2. **Semi-Analytical model for a Gaussian-like PSF, taking into account the pixellation (solid line):** We multiply the previous analytical prediction of $\sigma_{\mathrm{sys}}^2$ (ie. the relation 8.2 for $R_{\mathrm{quadrupoles}} = 2 \, \mathrm{pixels}$), with a corrective factor that corresponds to the impact of pixellation, that was estimated with the node 1 of the PSF shape assessment pipeline (see Fig. 8.2). One can see 2 regimes: for a large PSF (right hand side of the plot, that corresponds to small pixels compared to the PSF size), the solid line is close to the dashed one; one the other hand, when the PSF size decreases (and therefore the pixel size increases compared to the PSF size), the impact of pixellation increases and the solid line moves away from the dashed line, the optimal PSF size being around 0.18 arcsec, with a nearly flat plateau between 0.15 arcsec and 0.18 arcsec. Note that this plateau holds for the assumption of a Gaussian PSF. We are not sure of this plateau for a realistic PSF. That is why level-2-requirements recommend the FWHM to be between 0.18 arcsec and 0.23 arcsec (see Tab. 8.1).

| | FWHM (arcsec) | $\epsilon_{PSF}$ | EE50 (arcsec) | $\frac{EE90}{EE50}$ | averaged shear error $|<\delta\gamma>|$ | $\sigma^2_{sys}$ |
|---|---|---|---|---|---|---|
| level 1 requirement | | | | | | $< 10^{-7}$ |
| level 2 requirements (goal) | $0.18 - 0.23$ | $< 0.1$ $(< 0.05)$ | none | $< 5$ $(< 4.5)$ | none | |
| nominal PSF | 0.16 | 0.034 | 0.22 | 5.0 | $2.7 \times 10^{-4}$ | $7.3 \times 10^{-8}$ |

Table 8.1: Properties of the nominal PSF properties, at 800nm, at the centre of the field. The nominal PSF is the system PSF derived from the current optical design (illustrated above), after convolution with the AOCS and detector patterns. FWHM: Full Width at Half Maximum. $\epsilon_{PSF}$: ellipticity of the PSF. EE50 and EE90: diameters containing 50 and 90 percents of the flux, respectively. The level-1-requirement $\sigma_{sys} < 10^{-7}$ is the criteria that drives the level-2-requirements. We can see that it is achieved, implying that the nominal PSF is acceptable, even if the conservative level-2-requirement on the FWHM is not achieved. This is discussed in the text.

3. **Complete simulations with a realistic PSF (colour-shaded region and star):** We use nodes 1 and 2 of the PSF assessment pipeline to fully simulate $\sigma^2_{sys}$ for a given PSF, and characterise the PSF shape with 4 parameters: its FWHM, the diameters EE50 and EE90 containing 50 and 90 percents of the flux, respectively, and its ellipticity $\epsilon_{PSF}$. To investigate the impacts of these 4 shape parameters, we perform thousand of times the whole simulation set, each time with a different system PSF. For the moment, to focus on realistic PSFs, we restrict our study to small variations around the nominal PSF (presented in Sect. 8.3.2). And we adopt a PSF model with 5 degrees of freedom: a distortion; a rotation; a dilatation; 1 parameter ruling the wings, correlated with the centre; and a $2^{nd}$ wing parameter uncorrelated with the centre. For small galaxies of size $R_{quadrupoles} = 1$ pixel, the realistic PSFs we have investigated so far, lie in the colour-shaded area of the Fig. 8.4. This area is divided in 3 regions: the red one that is not acceptable, because the $\sigma^2_{sys} < 10^{-7}$ requirement is not achieved; the green one that is acceptable, and an orange one at the interface. The star shows the position of the nominal PSF, that is acceptable. We find that the following preliminary requirements can be adopted to achieve $\sigma^2_{sys} \lesssim 10^{-7}$: $0.18 \lesssim$ FWHM $\lesssim 0.23$ arcsec, $\epsilon_{PSF} < 0.1$, and EE90 / EE50 $< 5$. We emphasise that these preliminary requirements were derived for some small variations around the nominal PSF. On one hand, they are conservative: it possible to achieve the level-1-requirement $\sigma^2_{sys} < 10^{-7}$, even if these level-2-requirement are marginally achieved.

This is the case, for instance, of the nominal PSF (see Tab. 8.1). On the other hand, these requirements are probably not sufficient to ensure $\sigma_{\rm sys}^2 < 10^{-7}$ in a less specific case.

## 8.4 Impact of the Point-Spread-Function Colour Dependence

In general, the PSF depends on the spectral energy distribution of the object, and therefore will not be the same for stars and galaxies. Cosmic shear analyses of Hubble Space Telescope data do often use a PSF found from the telescope model due to the stability of the instrument. However even in that case the spectral energy distribution of the galaxy and the wavelength dependence of the PSF must be taken into account. The simulations and results are shown in more detail in Cypriano et al. (2009).

We use realistic galaxy and stellar populations to quantify the amount by which the PSF FWHM is likely to be misestimated. Galaxy SEDs are generated using mock catalogues designed to simulate the distribution of redshifts, colours and magnitudes of galaxies in GOODS-N (Kinney *et al.* 1996; Cowie *et al.* 2004). Template spectra are taken from Kinney *et al.* (1996) and intermediate types obtained by linear interpolation of these templates (see Abdalla *et al.* (2008) for further details). The PSF sizes for stars are estimated using stellar SEDs from the Bruzual-Persson-Gunn-Stryker Spectro-photometric Atlas. The catalogue contains 175 different SEDs covering a broad range of spectral types.

Fluxes are obtained for each object in the mock catalogues for the Euclid on-board filters F1 and Y up to the limiting magnitudes 24.5 and 22.80 respectively (AB magnitudes, $10\sigma$ detections). In addition, fluxes are measured in the g, r, i, z and y filters up to the magnitudes 24.45, 23.85, 23.05, 22.45 and 20.95. These configurations are chosen to simulate data from future ground based cosmology surveys such as DES and Pan-STARRS. Such observations could be exploited to optimally correct the wavelength-dependence of the PSF.

We used a template fitting method to predict the PSF FWHM of a galaxy by using all the colours available. Using ANNz (Collister & Lahav 2004) we trained a neural network using approximately one fourth of the simulated galaxies available, to predict the redshifts, spectral types and reddening of each galaxy, given the fiducial multi-colour information. With this information we compute the SED of each object and use a model of the wavelength dependence to predict the PSF FWHM for this galaxy. A telescope model and stars will be used to build this model for the wavelength dependence, and the accuracy of the model will depend on the stability of the wavelength dependence with telescope properties, and the number of stars available to calibrate which model to use. In the case of the Hubble Space Telescope, a PSF model taken from the telescope design is routinely used in conjunction with calibration from any stars in the field to assess the telescope configuration in a given observation. Therefore in this paper we use the exact model as given in its Eq. 9 and propagate the noisy and potentially biased galaxy SED estimates through to PSF biases and cosmological parameter biases. The comparison between this predicted PSF FWHM and the truth for the exact galaxy SED and redshift can be seen in the right hand panel of Fig. 8.5.

We propagate the biases as a function of redshift through to biases on cosmological parameters assuming a tomographic cosmic shear survey with the usual seven free parameters and find that the residual biases meet our requirements.

## 8.5 Intra-Pixel Sensibility Variations

Intra-Pixel Sensibility Variations (IPSV) are variations of the detection efficiency within pixels. These small variations are the result of wavelength dependent absorption depth of CCDs. As such, they are amenable to the same techniques described to deal with the wavelength dependent PSF (see Sect. 8.4). The size of the effect depends on both the resistivity and thickness of the CCD (with thickness and resistivity being mitigating factors). Future work will include a study of exactly how large this effect is for the chosen Euclid configuration and the development of mitigating techniques. Below, we describe a preliminary estimate of the impact of IPSV for the infra red photometry, and shape measurements in the visible channel.

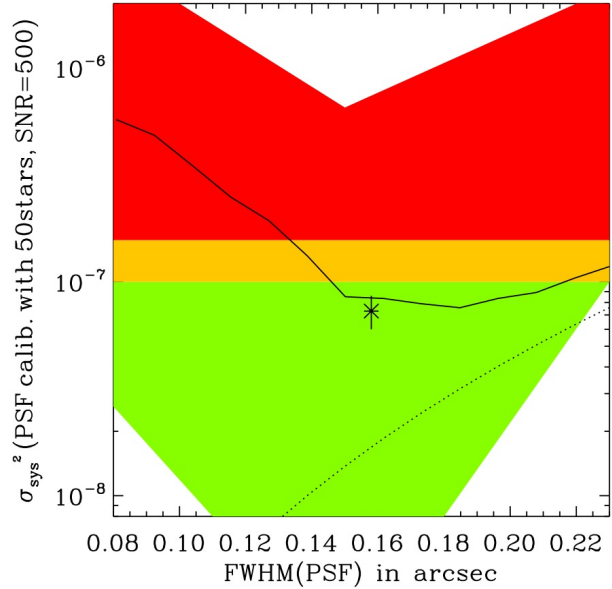

Figure 8.4: $\sigma^2_{sys}$ according to the PSF size, for a fixed pixel size of 0.1 arcsec. Dotted line: analytical pixellation-free model for a Gaussian-like PSF, for galaxies of size $R_{quadrupoles} = 2$ pixels. Solid line (semi-Analytical model for a Gaussian-like PSF, taking into account the pixellation): the previous analytical pixellation-free model is corrected for the impact of pixellation, estimated with the node 1 of the simulations (see Fig 8.2). Note that the nearly flat plateau, for 0.15 <FWHM< 0.18 arcsec, holds for the simple assumption of a Gaussian-like PSF. This plateau depends on the PSF shape. In particular, for a realistic PSF this region 0.15 <FWHM< 0.18 arcsec is unstable: there can be a large increase of $\sigma^2_{sys}$ when decreasing the PSF size below 0.18 arcsec. That is why level-2-requirements, that are conservative, recommend the FWHM being larger than 0.18 arcsec. Colour-shaded area: region where the systematics lay, according to the current results of the PSF assessment simulations (see Fig. 8.2), for a realistic PSF close to the nominal one, a PSF model with 5 degrees of freedom described in the text, and for a galaxy size of $R_{quadrupoles} = 1$ pixel. Red corresponds to $\sigma^2_{sys} > 10^{-7}$ and is not acceptable, green corresponds to $\sigma^2_{sys} < 10^{-7}$, and orange corresponds to the interface. star: position of the nominal PSF.

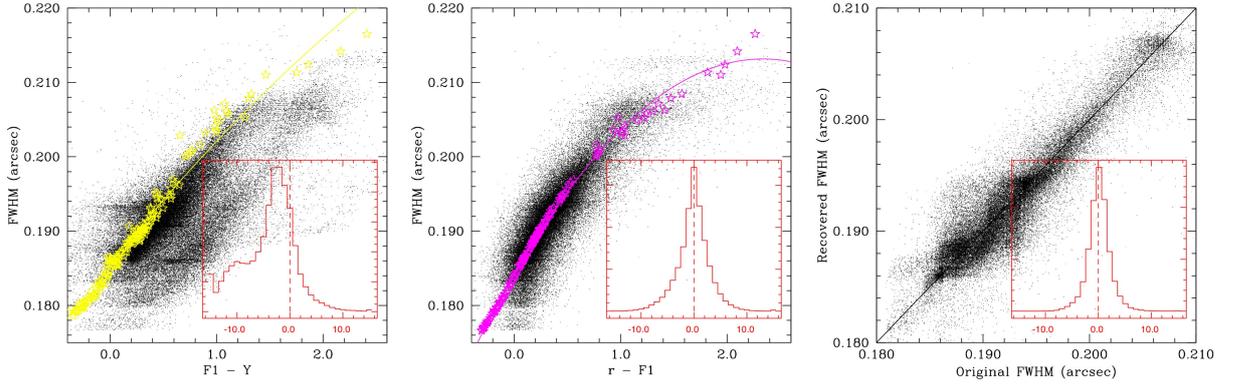

Figure 8.5: PSF FWHM for galaxies (black dots) and stars (yellow stars) as a function of object colour. The left hand panel shows calibration using space-data alone and the middle panel shows the result of using a colour composed of the wide Euclid optical filter and a ground based r band filter such as that used for photometric redshift estimates. The insets show the dispersion about the best fit first or second order polynomial to the stars (yellow line). We see the results using r-F1 are significantly better than using F1-Y. The right hand panel shows the results of our template fitting method.

### 8.5.1 Impact of Intra-Pixel Sensibility Variations in the near infrared

#### 8.5.1.1 Discussion

The required relative photometric accuracy of NIP for photo-z (as broken down from science requirements) has been defined as follows: Relative Photometric Accuracy for photo-z: the final relative photometric accuracy (i.e. post-calibration) $\delta F < 0.5$. This number applies to the uniformity within a field and among areas on the sky. The final calibration includes the off-line data processing of the fields, which assumes additional ground based information (secondary standard stars) and the possibility of self calibration based on consistencies in the (redundant) Euclid data. As long as the relative accuracy is met, the absolute photometric accuracy can be constructed from the data afterwards. Enough statistics will be available from standard stars to obtain a consistent absolute photometric calibration of TBD final accuracy.

The size of the PSF FWHM relative to the detector pixel size plays an important role w.r.t. to the impact that the detector Pixel Response Function (PRF) will have on the final photometric accuracy. In case the PSF FWHM becomes significantly smaller than the detector pixel size the PRF may lead to a degradation of the photometric accuracy. To meet the aforementioned relative photometric accuracy the following requirement on the PSF sampling has been defined on the Near Infra-Red (NIR) spatial resolution: the number of pixels per system PSF FWHM in the J band must be greater than 1 (TBC). The PSF FWHM of the current (EIC) optical design is less ($< 13 \, \mu m$) than the NIR detector pixel size of $18 \, \mu m$. In this case the PRF may lead to a degradation of the photometric performance.

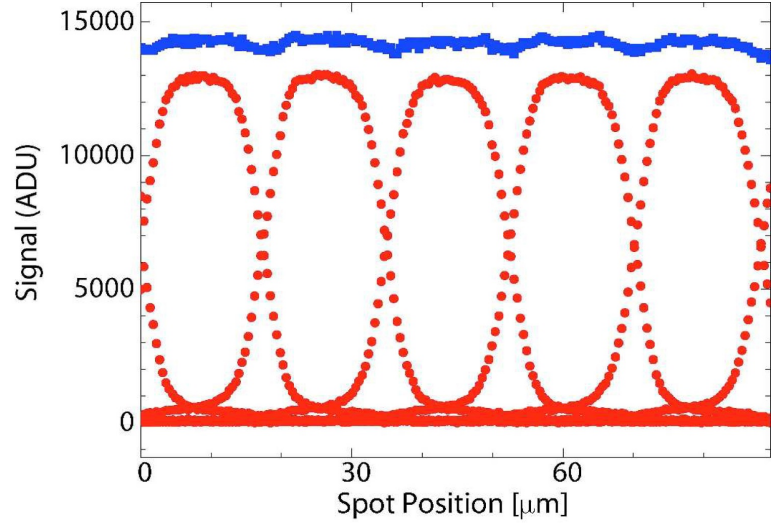

Figure 8.6: PRF from Barron *et al.* (2007).

Inter Pixel Capacity (IPC), lateral charge diffusion, optical cross talk and intra pixel variations are shaping the PRF. The PRF has been measured for the H2RG 1.7 $\mu$m cut-off detector by Barron *et al.* (2007), see Fig. 8.6. Based on this data the impact of the PSF size on the photometric accuracy has been investigated. Instead of using the PSF profile from the ZEMAX optical simulation, a Bessel function J1 of first order and of first kind has been used to simulate the PSF (IJ_actual) in J band :

$$\text{IJ}_{\text{actual}}(x, \delta) := \text{if} \left[ |x - \delta| < 10^{-48} \, , \, 1 \, , \, \left[ \frac{2\text{J1}\left[ \pi \frac{(x-\delta) \times 10^{-6}}{\lambda 2 \times N} \right]}{\pi \frac{(x-\delta) \times 10^{-6}}{\lambda 2 \times N}} \right]^2 \right] \, , \tag{8.3}$$

with f-Number N=10. The wavelength lambda2 has been chosen such that the PSF FWHM equals 12.6 $\mu$m (as determined by the optical simulation). The resulting flux (FJ_actual(delta)) has been calculated by integrating the product of the PSF (IJ_actual) and the normalised PRF (PRFtotal6) from Fig. 8.6 (blue curve) over a photometric aperture of 18 $\mu$m:

$$\text{FJ}_{\text{actual}}(\delta) := \int_{-9+\delta}^{9+\delta} dx \, \text{PRFtotal6}(x) \, \text{IJ}_{\text{actual}}(x, \delta) \tag{8.4}$$

The resulting flux FJ_actual($\delta$) is a function of $\delta$, which determines the relative shift between the centre of the PSF and the PRF. Further, it was assumed that the Euclid observation strategy will involve four dither positions so that the total final flux (Ftotal) will be the sum of four individual flux measurements resulting from four randomly shifted PSFs w.r.t. to the PRF:

$$\text{Ftotal}(d, \text{in}) := \frac{(\text{FJ}_{\text{actual}}(\text{r}_{\text{in}} + d) + \text{FJ}_{\text{actual}}(\text{rb}_{\text{in}} + d) + \text{FJ}_{\text{actual}}(\text{rc}_{\text{in}} + d) + \text{FJ}_{\text{actual}}(\text{rd}_{\text{in}} + d))}{4} \, . \tag{8.5}$$

$d$ is the common shift of the individual PSF's w.r.t. the PRF; r,rb,rc,rd are random shifts (in a range of $\pm 10 \, \mu$m) from a set of random numbers with index 'in'. It was found (see Fig. 8.7) that the resulting flux variations are significantly reduced (averaged out) when the actual observing scheme is taken into account.

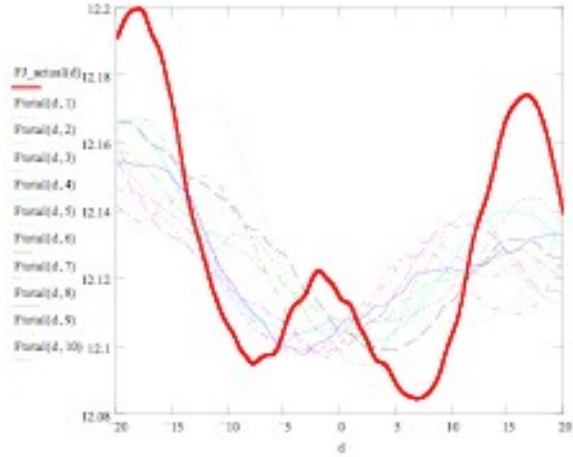

Figure 8.7: Comparison of flux variations between FJ_actual and Ftotal (for 10 different sets of random shifts).

Using 10 different sets of random shifts for Ftotal on average it was found:

$$\frac{\text{stddev}(\text{Ftotal}(d))}{\text{mean}(\text{Ftotal}(d)))} = 0.0014 \tag{8.6}$$

#### 8.5.1.2 Conclusion

It can be concluded that the actual size of the PSF FWHM is not expected to be critical w.r.t. the photometric calibration requirements. Dithering allows to reduce (averaging out) the photometric error caused by the detector effects to a level which is well acceptable and in line with the photometric requirements. We caution that to our knowledge no PRF has been measured for the 2.5 $\mu$m detector. In addition it is known that single detector devices can show untypical PRF profiles that may seriously degrade the photometric accuracy. For this reason it is assumed that each single detector has to be characterised w.r.t. to its PRF performance. In addition it may be uncertain to which level the PRF measurements of the intrinsic detector PRF effects may be biased by the precision of the measurement itself. If the measurement errors are significant there is further room for improvement. If the above mentioned result holds true also for the 2.5 $\mu$m cut-off detector at operational conditions it is expected that no PSF enlargement for the NIP is required.

### 8.5.2 Impact of Intra-Pixel Sensibility Variations for shape measurements

We implemented the IPSV in the simulations presented in Sect. 8.3.1, and performed an initial estimation of their impact on $\sigma_{\text{sys}}$ in case of a miscalibration of the IPSV pattern. As we do not know yet the IPSV pattern of pixels in Euclid CCDs, we have restricted our study to a 2D elliptical Gaussian pattern and we have investigated the bias on ellipticity measurements, implied by such a miscalibration, with respect to the pixel size, with the assumption that all the pixels have the same IPSV pattern. To do that, we simulate galaxies with a given IPSV pattern of ellipticity $\epsilon_p$, but we analyse them assuming perfect top-hat pixels. $\epsilon_p$ is interpreted as the miscalibration of the IPSV pattern (noted 'accuracy of the pixel ellipticity calibration' on the plot). The result is shown in Fig. 8.8. We find that, as expected, the systematic effects induced by IPSV for shape measurements, increase rapidly with the pixel size: For 0.1 arcsec pixels, the accuracy of the calibration must be around 0.01, while it must be lower than $5 \times 10^{-3}$ for 0.2 arcsec pixels. The study of correlations between IPSV and wavelength dependence of the PSF will be subject of further work.

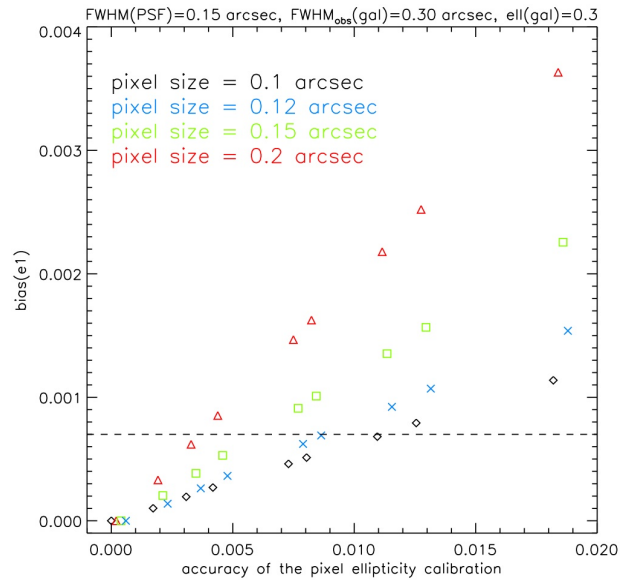

Figure 8.8: Impacts of Intra Pixel Sensitivity Variations (IPSV) for shape measurements. The y-axis shows the bias on one component of the ellipticity. The horizontal dashed lines show the acceptable upper limit (corresponding to $\sigma_{sys}^2 = 10^{-7}$). The x-axis shows the ellipticity $\epsilon_p$ of the 2D-Gaussian IPSV pattern used to simulate galaxies. These galaxies are analysed assuming perfect top-hat pixels, and $\epsilon_p$ is interpreted as a miscalibration of the IPSV pattern, noted 'accuracy of the pixel ellipticity calibration' in this example.

# Mitigating Charge Transfer Inefficiency

*Authors: Richard Massey (University of Edinburgh), Olivier Boulade (CEA, Saclay), Mark Cropper (Mullard Space Science Laboratory, University College London), Jason Rhodes (NASA Jet Propulsion Laboratory).*

## Abstract

Charge Transfer Inefficiency (CTI) due to radiation damage above the Earth's atmosphere creates spurious trailing in imaging with any Charge Coupled Device (CCD) detectors. We have examined the specific effects of CTI on the demanding galaxy shape measurements required by studies of weak gravitational lensing, and constraints on cosmology. We have developed an end-to-end method of simulating these effects for any specific CCD model and mission design (years at L2, operating temp, shielding, etc), plus a software post-processing method to correct images pixel by pixel. These are the tools that will be required during the development phase to perform a thorough analysis of the CTI effects on *Euclid* images to develop a mission design that meets requirements. A first pass using these tools has demonstrated that existing technology can be configured to implement a five-year *Euclid* mission throughout which the requirements on galaxy shape measurement are always met.

## 9.1 Origin of CTI

*Euclid* will image the sky using Charge-Coupled Device (CCD) detectors. During an exposure, these convert incident photons into electrons, which are stored within a silicon substrate, in a pixellated grid of electrostatic potential wells that gradually fill up. At the end of the exposure, the electrons are shuffled, row by row, to a readout register at the edge of the device as illustrated in figure 9.1. The electrons that emerge are then counted and converted into a digital signal.

Above the Earth's atmosphere, a continuous bombardment of high energy particles makes a harsh environment for sensitive electronic equipment. *Euclid* will operate at the Earth-Moon Lagrange point L2, outside the Earth's protective Van Allen belts, where the dominant concern is solar radiation (Barth *et al.* 2000). While high-energy, ionising particles generally create spurious streaks within a detector that impact individual exposures, low energy protons and heavy ions create longer-lasting bulk damage by colliding with and displacing atoms from the silicon lattice. The dislodged atoms can come to rest in the interstitial space, and the vacancies left behind move about the lattice until they combine with interstitial impurities, such as Phosphorus, Oxygen or another vacancy (Janesick 2001). Such defects degrade a CCD's ability to shuffle electrons, known as its Charge Transfer Efficiency (CTE; Charge Transfer *Ine*fficiency CTI=1-CTE). Electrons can become temporarily trapped in the local potential, then released after a delay $\tau$ that depends upon the properties of the lattice and impurities, and the operating temperature of the device (Shockley & Read 1952; Hall 1952). Several different species of charge trap may be present in any given device, with different characteristic release times.

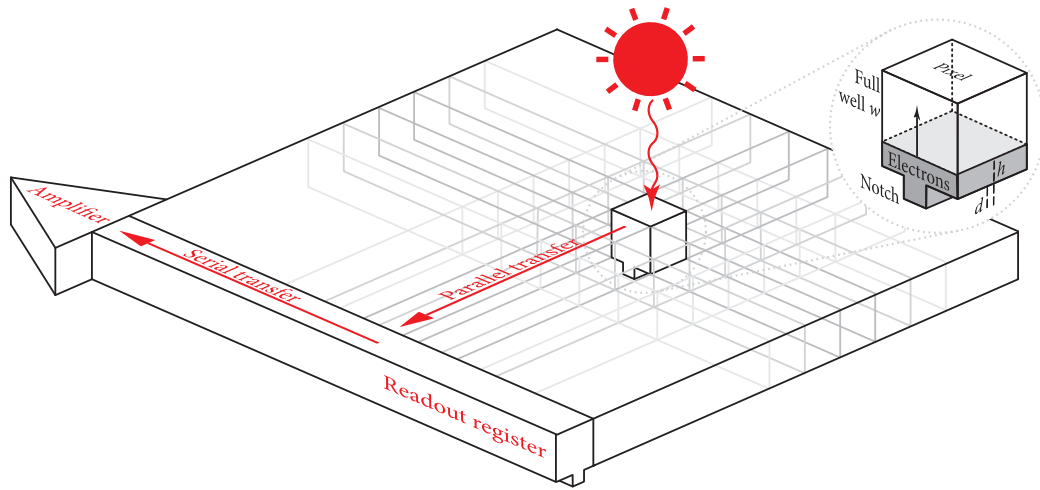

Figure 9.1: Cartoon illustrating imaging with a CCD device, as will be used for *Euclid*. Incident photons generate photoelectrons that are stored in and gradually fill up electrostatic potential wells in a grid of pixels. At the end of the exposure, they are shuffled through the device in the parallel and serial directions, to a single amplifier and analogue-to-digital converter. Reproduced from Rhodes & et al. (2009).

The trapping and delayed release of charge mainly causes problems during CCD readout. As demonstrated in figure 9.2, if a few electrons are trapped and held while others are moved along, the trapped electrons are released as a spurious trail. Regions of the image furthest from the readout register are worst affected, because electrons starting there encounter the most charge traps during their journey across the device, and the effect gets worse over time, as cumulative radiation damage creates more charge traps.

## 9.2 Observational consequences of CTI

The trailed electrons affect weak lensing science by altering the photometry and astrometry of faint galaxies, and by adding a spurious, coherent ellipticity in the readout directions. It is important to note that the effect is non-linear: faint galaxies are most affected, as a higher fraction of their electrons are trailed by a fixed number of charge traps, compared to bright stars. It thus closely mimics the effect of cosmic shear, in which distant galaxies are coherently elongated and, since the effect is not a convolution, it cannot be dealt with by traditional weak lensing measurement methods. **CTI was the dominant systematic in the weak lensing analysis of the largest survey by the *Hubble Space Telescope*** (Massey & et al. 2007), and is one of the main concerns in the ESA *Gaia* mission.

To investigate the level of CTI damage in *Euclid*, and its effect on galaxy shape measurement, we have simulated the process of image collection and readout in a software model (Massey & et al. 2009). The model is built around physical parameters, and has been verified and calibrated using real data from both in-flight *Hubble Space Telescope* observations and p-channel CCDs irradiated in a laboratory. By running this software readout model on the simulated *Euclid* images described in part V, we have demonstrated that **not all CTI is equally bad for the purposes of weak lensing measurement**. As shown in figure 9.3, only charge traps with characteristic release times of the same order as the clock cycle during readout affect galaxy shape measurements. Charge traps with much longer release times steal flux from sources and primarily affect photometric redshift estimation.

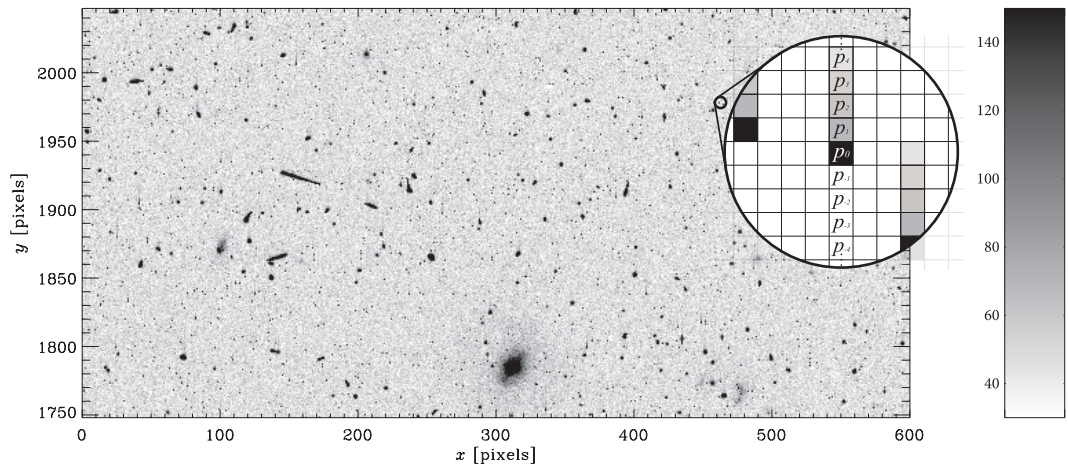

Figure 9.2: A typical, raw *Hubble Space Telescope/Advanced Camera for Surveys* image, in units of electrons. This $30''\times15''$ ($600\times300$ pixels) region is at the far side of the CCD to the readout register, which lies towards the bottom of the page. It was obtained in May 2005, 1171 days after the launch of ACS. Upon close inspection, as illustrated in the zoomed inset, spurious trailing of charge is manifest behind (above) every object. Reproduced from Massey & et al. (2009).

Some CTI can be traded off again other. An example of this is presented in figure 9.4, in which the operating temperature (which governs the traps' characteristic release times) and CCD clock speed can be adjusted to optimise the operating regime of *Euclid* to avoid particularly problematic trap species. The EIC has now established a comprehensive analysis programme tailored to weak lensing measurement from hardware development to shape measurement. We have developed the necessary tools and are now working closely with manufacturers e2v to enforce specifications beyond the usual single number for CTI.

## 9.3 CTI mitigation by hardware design

Various schemes can be used to mitigate the damaging effect of the radiation flux on the CCD hardware:

- optimising CCD operating temperature and clock speed to avoid "resonance" with the trapping speeds of common charge trap species (see figure 9.4);

- building in radiation shielding;

- using radiation-tolerant p-channel CCDs;

- adding more readout registers or altering the CCD's aspect ratio to decrease the distance travelled by electrons within a CCD;

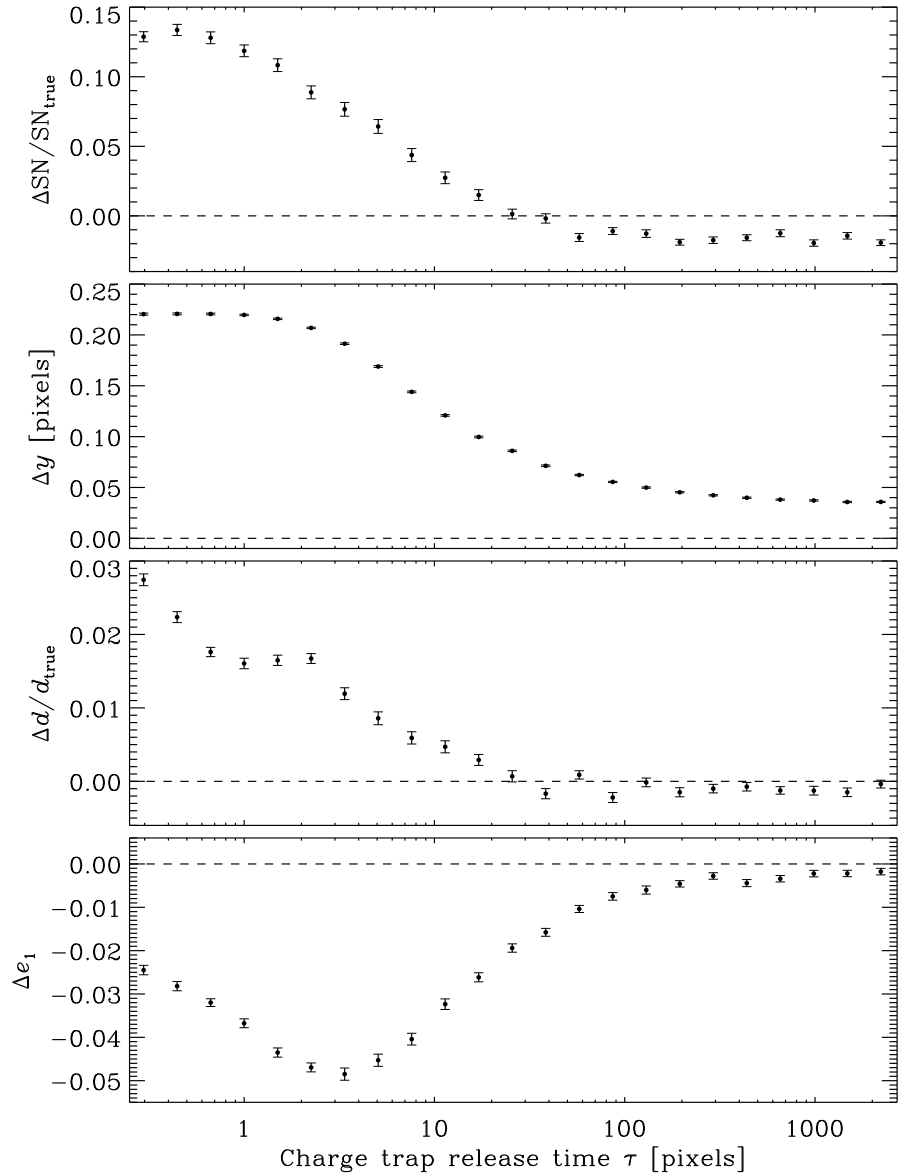

Figure 9.3: The effect of charge trap characteristic release time $\tau$ on a measurement of photometry, astrometry, size and ellipticity of a typically small, faint galaxy degraded by CTI in *Euclid*. Each $y$ axis represents the fractional change in that quantity. The absolute values of the $y$ axes are largely irrelevant, depending upon the assumed density of charge traps, CCD well filling model, galaxy SN (13), size (FWHM=3.8 pixels) and morphology (circularly symmetric De Vaucouleurs profile). However, the trends reveal several lessons for future hardware: notice particularly the local maximum in $\Delta e_1$, which implies a worst case clocking time, or readout cadence, for CCDs. Reproduced from Rhodes & et al. (2009).

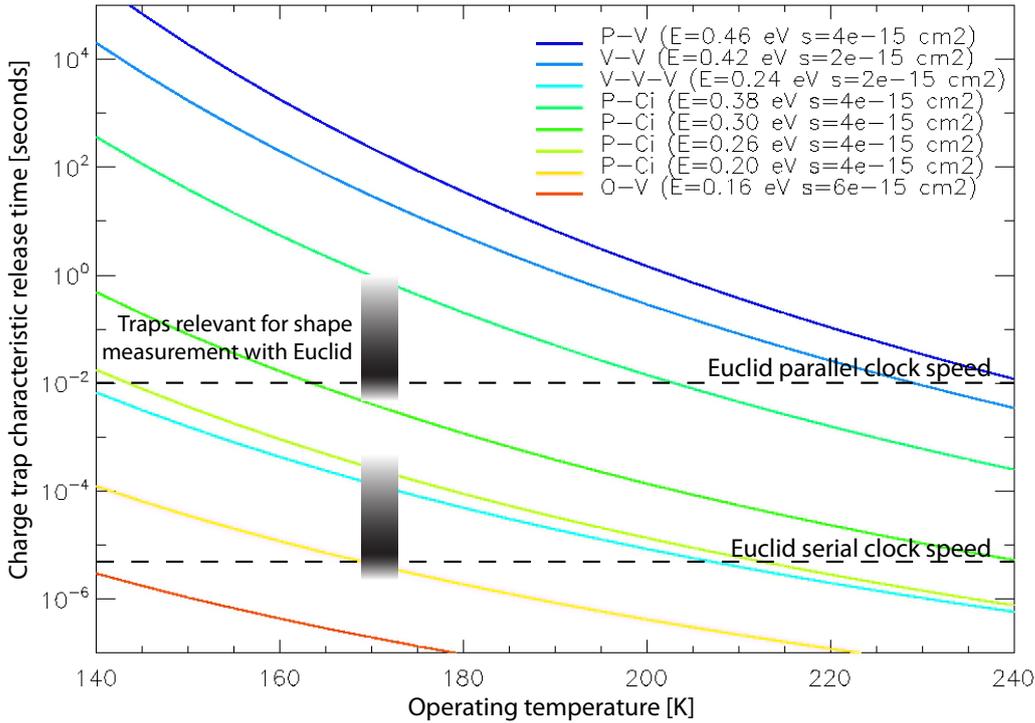

Figure 9.4: The characteristic release times of charge trap species in n-channel CCDs, which depend upon operating temperature. By adjusting the temperature and clock speed, it is possible to avoid the effects of several charge trap species, as shown by the target greyscale boxes (whose shading reflects the bottom panel of figure 9.3).

- implementing additional calibration diagnostics, such as pocket pumping, which allow the locations and properties of existing traps to be accurately measured on-orbit.

The use of additional shielding is limited in the *Euclid* case because of the size of the focal plane, and the penetrating nature of the damaging radiation. Nevertheless, shielding is naturally available from the surrounding equipment, and some progress may be possible from the payload layout.

There is an ongoing programme of discussions and workshops within the *Euclid* visible imager team in coordination with leading CCD manufacturers e2v and with ESA. An important consideration is to maintain a high level of technology readiness level, and the baselined e2v CCD203/204 family provides essentially all of the required characteristics. The CCD203/4 optimisation programme includes the structural options noted above (such as different aspect ratios), and in addition considers the reduction of the size of the readout register to improve CTI in the serial direction: larger readout registers are used in case CCD rows are summed before readout, which will not be the *Euclid* case. Additional work centres on the clock timing, shaping and sequencing, how many phases are used for the integration, voltage bias levels and using different readout speeds. This work is being done with flight representative readout electronics, which have already been developed in the Phase A, since the *Gaia* experience has shown how critical it is to treat the detector and electronics as a coupled system. The CCD204 includes a charge injection structure, which can be used to fill traps on the CCD readout, but with the drawback that additional lines of charge are imprinted on the image. The usefulness of such a structure in the *Euclid* case is still under investigation: in ESA's *Gaia* mission charge injection is baselined, but there the operation is in drift scan (TDI) mode.

In addition to these initiatives, there is an ESA-funded test programme on CCD204s irradiated with the worst case *Euclid* mission dose under way at SSTL-Astrium and institutes in the UK (Open University, MSSL-UCL) and France (CEA Saclay). The test plan (0128465_SSTL) has been constructed using the experience gained from the extensive *Gaia* campaign. Earlier tests have also been conducted on e2v-provided devices at Open University, and are the subject of several reports (Open-Euclid-TN-004-02; Open-Euclid-TN-003-01; Minutes of Meeting for Euclid CCDs 23 July 2009).

Alternative p-Channel CCDs are being actively considered for the *Euclid* programme. These devices collect holes rather than electrons and are not affected by the dominant electron trapping caused by Phosphorus-vacancy displacement damage. They have been shown to have charge transfer efficiency up to an order of magnitude better than several models of n-channel CCDs designed for space applications (Marshall *et al.* 2004; Lumb 2009; Gow *et al.* 2009). Such devices have been developed at LBNL in the USA (reported more fully in the next section) and by e2v, but in neither case is their TRL as high as for well-proven n-channel CCDs. A UK-funded test programme on irradiated e2v p-channel devices is under way at Open University (Open-P-Chan-TP-002.01; Minutes of Meeting for Euclid CCDs 23 July 2009), and ESA have funded a new fabrication run of e2v CCD204 p-channel CCDs, which should become available for investigation and characterisation in early 2010.

Survey strategy can also be used to partially mitigate radiation damage. Since the effects of CTI depend on the position of a source on the CCD relative to the readout register, sequences of dithered images can be used in principle to decouple its intrinsic shape from spurious CTI degradation. Four-step dithering is now established in the *Euclid* mission baseline.

## 9.4 CTI mitigation by software post-processing

The spurious ellipticity induced in faint galaxies can be measured in large weak lensing surveys as a shear signal that depends on chip position, after many pointings are stacked, to average away any cosmological signal. Parametric schemes have been developed for several instruments to correct this statistically at the catalogue level (Rhodes & et al. 2007). However, they fail to account for the "shielding" of one galaxy by another one nearby, which prefills charge traps during readout, and may not account for variation of the spurious ellipticity as a function of galaxy size, concentration and morphology.

Fortunately, the same software algorithm that we developed to mimic CCD readout and *add* charge trailing can also be used to *remove* it, via an iterative procedure illustrated in table 9.1. This attempts to find to find an image that, when read out through software, results in the image obtained from the telescope. This is the unblemished image, corrected to appear as it would have without CTI. Since charge transfer is one of the last process to occur during data acquisition, correction for CTI should be one of the first operations during data analysis.

To demonstrate the technique, we have removed charge transfer trailing from real imaging with the *Hubble Space Telescope/Advanced Camera for Surveys*. We used warm pixels in in-flight science exposures to measure the eight model parameters that encode the densities and release times of two charge trap species, and the rate at which potential wells in pixels are filled by electrons. Warm pixels also happen to be created by radiation damage. They are short circuits in the detector that create perfect delta-functions in images, and deviations from single-pixel spikes clearly show trails from charge traps with release times similar to the clock speed. The trap density was found to be about the same as that expected in a modern p-channel CCD after approximately 30 years at L2. The correction was then applied to the HST COSMOS survey.

| True image | $I$ | Not available |
|---|---|---|
| Downloaded from HST | $I + \delta$ | (A) |
| After one extra readout | $I + 2\delta + \delta^2$ | (B) |
| (A)+(A)−(B) | $I - \delta^2$ | (C) |
| After another readout | $I + \delta - \delta^2 - \delta^3$ | (D) |
| (A)+(C)−(D) | $I + \delta^3$ | (E) |
| After another readout | $I + \delta + \delta^3 + \delta^4$ | (F) |
| (A)+(E)−(F) | $I - \delta^4$ | etc. |

Table 9.1: Iterative method to remove CTI trailing, using only a forward algorithm that *adds* trailing. The true image $I$ is desired but not available. Only a version with (a small amount of) trailing, $I + \delta$, can be obtained from the telescope. However, by running that image through a software version of the readout process, and subtracting the difference, deviations from the true image can be reduced to $\mathcal{O}(\delta^2)$. Successive iterations further reduce the trails until an image is produced that, when "read out", reproduces data arbitrarily close to those obtained from the telescope. This is the corrected image.

As shown in figure 9.5, the spurious shear in the faintest galaxies was reduced by an order of magnitude. The technique is currently limited by the accuracy of the charge integration and readout model. Even better model calibration (and hence CTI correction) should be possible with more data: especially if calibration diagnostics like pocket pumping can be enabled by the *Euclid* readout electronics, to bring tailored CTI measurements in flight.

Other more dedicated software mitigation and calibration schemes are also being considered for the *Euclid* survey. These may use developments of the CDM02 model (GAIA-CH-TN-ESA-AS-015-1) baselined in the *Gaia* programme and modified for the standard exposure mode of operation. CDM02 is an analytical charge distortion model based on physical processes within the pixel, which predicts the effect on the detected charge distribution as the CCD is read out. Operating on an aggregated readout rather than a pixel-pixel transfer level, the model is fast enough to be applied to all *Euclid* visible imager data in real time. The parameters within the model are determined from calibrations (stellar sources) and the effect of prior sources read out ahead of the profile of interest is implicit. The investigation is ongoing.

## 9.5 A feasible Euclid mission design unimpeded by CTI

Our CCD readout software and end-to-end characterization methods are general; they can be applied to any CCDs, once the relevant density and characteristic release times of charge trap species are known. Our comprehensive software suite then allows us to couch our results in terms of the spurious galaxy ellipticities $\Delta e$ added to the faintest *Euclid* galaxies (which dominate the weak lensing signal), and hence in terms of the resultant errors on dark energy parameters. To benchmark one possible configuration for *Euclid*, we have modelled the properties of fully-depleted LBNL p-channel CCDs. Dawson & et al. (2008) irradiated several of these CCDs at the LBNL 88-Inch Cyclotron with 12.5 MeV protons to characterise their CTI properties. Although a variety of irradiation levels were tested, we have confirmed (Rhodes & et al. 2009) that the spurious ellipticity induced in faint galaxies depends linearly upon the density of relevant trap species, so we only considered data with the maximum irradiation levels of $2 \times 10^{10}$ protons/cm$^2$.

For a baseline *Euclid* mission using p-channel CCDs, and averaging over the solar cycle, we calculate that bias in the measurement of galaxy shapes furthest from the readout registers will increase at a rate of $d\Delta e/dt = (2.65 \pm 0.02) \times 10^{-4}$ per year at L2. If uncorrected, this would consume the entire shape measurement error budget $\sigma_{sys}^2 \sim 10^{-7}$ of a dark energy mission surveying the entire extra-galactic sky within about 4 years of accumulated radiation damage (Amara & Réfrégier 2008). However, software mitigation techniques at an accuracy already demonstrated on *Hubble Space Telescope* data in §9.4 can reduce this by a factor of $\sim 10$, bringing the effect well below mission requirements.

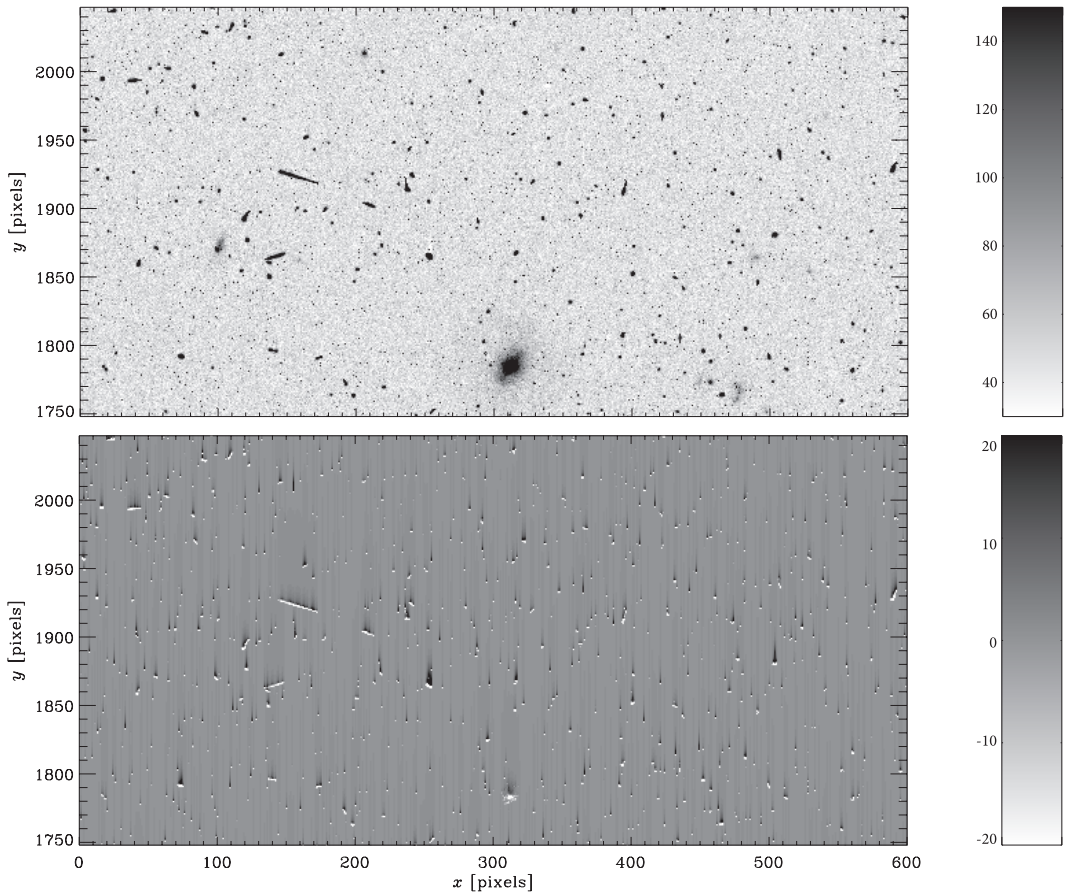

Figure 9.5: *Top*: The *Hubble Space Telescope* image from figure 9.2, after CTI correction, in units of electrons. *Bottom*: Difference image. Reproduced from Massey & et al. (2009).

We have thus demonstrated that an implementation of the *Euclid* mission using existing technology and proven software post-processing techniques can achieve galaxy shape measurement accuracy that is unimpeded by CTI. This conclusion is currently valid only for the radiation-tolerant p-channel CCDs we have modelled; CCDs with worse CTI will fare worse and may not meet the mission requirements. The EIC is therefore continuing to develop mitigation strategies and extend them to detector technologies with higher TRL. All the tools are in place to quantitatively test alternative mission configurations during the *Euclid* development phase.

# Intrinsic Alignments of Galaxies

*Authors: Michael Schneider (ICC Durham), Filipe Abdalla (UCL), Adam Amara (ETH Zurich), Sarah Bridle (UCL), Julien Carron (ETH Zurich), Benjamin Joachimi (U. Bonn / UCL), Tom Kitching (U. Edinburgh), Alexandre Réfrégier (CEA Saclay), + EIC Weak Lensing Working Group.*

## Abstract

Intrinsic alignments (IAs) of galaxy shapes are likely to be a significant source of systematic error in cosmic shear measurements. EIC members have pursued two approaches to mitigate this systematic error: "nulling" methods that subtract certain combinations of tomographic redshift bins that are systematics dominated, and "modelling" methods that fit parameterised models for IAs that are treated as nuisance parameters when constraining cosmology. In this section we describe the relative merits and drawbacks of each method for removing IA systematics from cosmic shear measurements. Simulations done by EIC members of the intrinsic alignments of galaxy discs and dark matter halos with the large-scale matter distribution are described in the section Numerical Simulations.

## 10.1 Introduction

Based on recent low-redshift measurements Brown *et al.* (2002); Mandelbaum *et al.* (2006); Hirata *et al.* (2007), IAs are expected to be a non-negligible systematic error for cosmic shear (and potentially dominant for constraining the dark energy equation of state).

Two types of intrinsic alignments can contaminate the shear two-point functions. So-called intrinsic-intrinsic (II) alignment occurs when two galaxies *at the same redshift* are aligned with each other because of correlations in the gravitational potential in which they formed and reside. On the other hand, shear-intrinsic (GI) alignments occur between galaxies *at widely separated redshifts*. This is caused when one galaxy resides in the lensing potential at intermediate redshift that is lensing a background galaxy at even larger redshift. This causes an anti-correlation of galaxy shapes because the galaxy in the lensing potential is preferentially aligned parallel to the gradient of the potential while the sheared galaxy is flattened tangentially to the gradient Hirata & Seljak (2004).

Because the lensing amplitude increases with increasing distance between the lens and the source, II alignments dominate shear measurements at low redshift. Based on both theory Crittenden *et al.* (2002); Jing & Suto (2002) and observations Brown *et al.* (2002); Mandelbaum *et al.* (2006) II alignments are expected to contribute 10% of the observed shear correlations for surveys with median redshift $\sim 1$. The GI alignments, which include lensing effects, become more dominant at higher redshifts. Heymans *et al.* (2004) modelled the GI alignments for central galaxies in low-mass halos using N-body simulations and predicted the GI signal to also decrease the shear power spectrum amplitude by $\sim 10\%$. The analytical models of Hirata & Seljak (2004); Schneider & Bridle (2009) are consistent with this level of contamination.

## 10.2 Nulling

Both the II and GI alignments have particular geometric properties that can be exploited to remove the IA contamination from cosmic shear observations. The II alignments occur between galaxies that are located in the same gravitational potential, which means the galaxies must have the same redshift. By downweighting galaxy pairs that are close in redshift and in angular separation on the sky, the II contamination can be largely removed (depending on the accuracy of the galaxy redshifts) King & Schneider (2002); Heymans & Heavens (2003); King & Schneider (2003).

Joachimi & Schneider (2008) introduced a nulling technique to remove the GI signal from the cosmic shear signal. This method is purely geometric in the sense that it only takes the characteristic and well-known dependence of the intrinsic alignment signal into account and does not rely at all on the existing models of the underlying correlations between matter and the intrinsic orientation of galaxies. In the limit of perfect knowledge of galaxy redshifts, nulling is capable of eliminating the GI effect completely. Joachimi & Schneider (2009) have demonstrated that this technique reduces the GI term to about 10% in presence of photometric redshifts whose quality corresponds to the requirements of Euclid. However, the price to pay for the complete model independence of this approach is a reduction in the dark energy Figure of Merit of roughly an order of magnitude. The degradation of the constraints on a 6-parameter cosmological model using this technique are shown in fig. 10.1. The 1-$\sigma$ errors for the example data set used in fig. 10.1 are increased by factors of 2-3 after nulling. The biases in the mean values of the parameters are simultaneously reduced by factors of 2-10 so that in all cases the biases are much less than the size of the 1-$\sigma$ errors after nulling.

## 10.3 Modelling

A second approach for removing the IAs as a systematic error is to construct a parameterised model for the IA correlations and marginalise over the model parameters when constraining cosmology. An advantage of this approach is that, with an accurate model, all the information in the data is retained. However, without an accurate model for IAs two issues must be considered. An insufficiently flexible IA model will leave a systematic bias in the cosmic shear data while a model with too many parameters will degrade the cosmological parameter constraints to a sub-optimal level.

Building on the model of Hirata & Seljak (2004), Bridle & King (2007) introduced a tunable set of parameters for describing the power spectrum of IAs. By marginalising over the amplitude of the IA power spectra binned in wavenumber and in redshift (with varying bin sizes) Bridle & King (2007) explored the degradation in cosmological parameter constraints from cosmic shear as the number of IA parameters is increased. Figure 10.2 shows the change in the dark energy Figure of Merit as a function of the number of wavenumber and redshift bins. For such flexible IA models, the Figure of Merit can be degraded by up to a factor of 4 compared to the prediction assuming no IA contamination. The Figure of Merit for the maximally flexible IA model is a factor of 3 times smaller than the minimally flexible model.

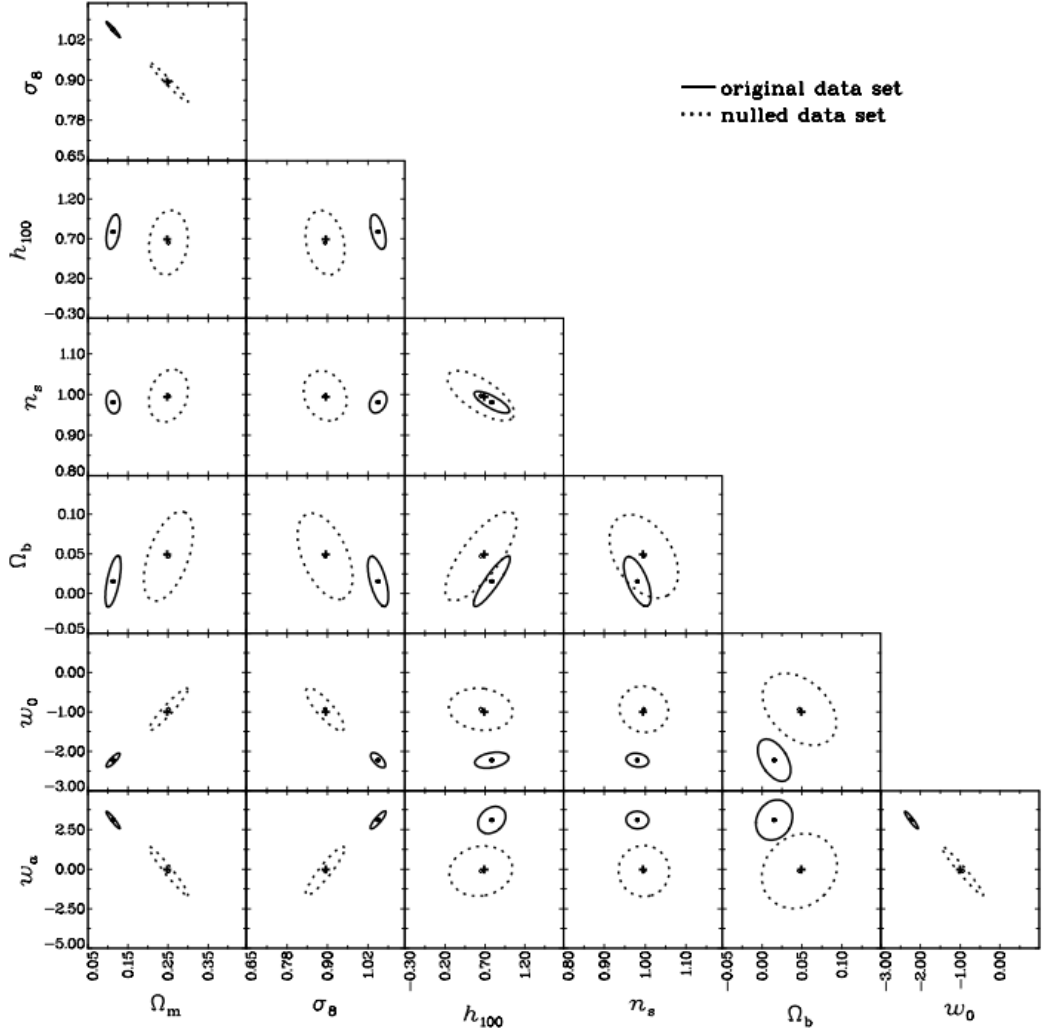

Figure 10.1: Parameter constraints before and after nulling the GI contamination from cosmic shear tomographics power spectra. Shown are the two-dimensional marginalised $2\sigma$-errors for the original data set (with IA contamination) as solid curves and for the nulled data set as dotted curves. The fiducial parameter values are marked by the crosses. The survey has been divided into 10 photometric redshift bins. Photometric redshift errors are characterised by a dispersion in $z$ of 0.05, catastrophic error fraction 0.05, and bias in the mean of the catastrophic error population of $\Delta z = \pm 1.0$. The linear alignment model, downscaled by a factor of five, was used to model the IA contribution. (Taken from Joachimi & Schneider (2009).)

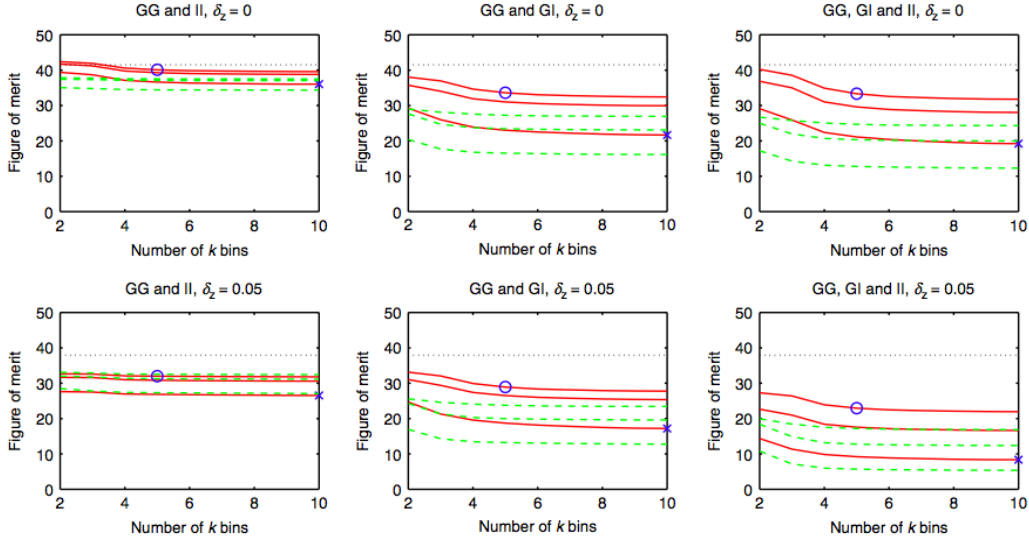

Figure 10.2: Degradation in the dark energy Figure of Merit as the number of IA model parameters is increased. Top row: assuming perfect photometric redshifts. Bottom row: more realistic photometric redshifts (with dispersion in $z$ of 0.05). Left column: including just the II terms. Middle column: including just the GI terms. Right column: including both II and GI terms. Solid lines: Figure of Merit as a function of number of wavenumber bins in the IA power spectra. Lines from top to bottom within one panel: the number of bins in the redshift direction increases 1, 2, 5. Dotted line: Figure of Merit for GG alone. Dashed line: using the linear alignment model with the linear theory matter power spectrum, instead of the nonlinear theory matter power spectrum. In every case a free amplitude parameter and unknown power law evolution was marginalised over (allowed to be different for each of GI and II). The default intrinsic alignment model is shown by a circle and the maximally flexible model is marked by a cross. (Taken from Bridle & King (2007).)

In an effort to build a simultaneously more descriptive and minimally flexible IA model, Schneider & Bridle (2009) constructed a model for IAs on small scales by considering the alignments of galaxies within individual dark matter halos. This modelling showed that a range of assumptions about the small-scale alignments of galaxies have a limited range of effects on the IA power spectra. Within the framework considered in Schneider & Bridle (2009) the IA spectra can be described by just 2 effective amplitude parameters: one for the large-scale alignments and one for the small-scale alignments within halos. This result illustrates the potential for future physical modelling of IAs to yield constrained parameterisations of the systematic contamination to cosmic shear.

Using alternate cosmic shear statistics Kitching *et al.* (2008) showed that IA, photometric redshift and shape measurement systematics models can be simultaneously constrained. They found that the combination of 3-D cosmic shear and the shear ratio test are very complimentary and that with only modest priors on the systematics parameters (10 to a few percent) the degradation of the dark energy Figure of Merit is at most a factor of 2.

Further progress has been made by considering the additional information available in the Euclid imaging survey in the form of the galaxy angular positions. In addition to shear correlations, one can consider galaxy number density correlations and number density-shear cross correlations. The number density of galaxies is determined by the intrinsic galaxy clustering and the changes due to lensing magnification effects. However, uncertainties in the clustering bias of galaxies with respect to the dark matter add additional astrophysical systematics beyond the IAs in the cosmic shear signal.

Joachimi & Bridle (in prep.) have undertaken a joint analysis of the complete set of galaxy shape and position correlations, choosing a general parameterisation for both intrinsic alignment and galaxy bias with very low model dependence. In their most flexible model over 200 nuisance parameters have been marginalised over accounting for uncertainties in galaxy bias, intrinsic alignments, galaxy luminosity function, and photometric redshift information. Making the further conservative assumption that even this large degree of freedom is not sufficient to describe galaxy bias on small non-linear scales, all clustering signals in the substantially non-linear regime were discarded.

With this setup Joachimi & Bridle showed that the total information content (expressed in terms of the determinant of the Fisher matrix) of the pure lensing signal and the combined set of shape and position signals after marginalisation is the same to good accuracy. Dark energy parameters are slightly more affected by the marginalisation process, so that the Figure of Merit is reduced by about 40% with respect to the pure lensing signal. However, to obtain these results, all priors on the nuisance parameters were increased to non-informative values. Any substantial prior information, in particular on the galaxy bias and intrinsic alignments, readily brings the Figure of Merit back to pure lensing values and possibly beyond.

# Part IV

# ADDITIONAL COSMOLOGICAL PROBES & LEGACY SCIENCE

# Cosmology with Galaxy Clusters Counts

*Authors: Jochen Weller (Ludwig-Maximilians-Universität München and Excellence Cluster 'Origin and Structure of the Universe'), Filipe Abdalla (University College London), Joel Bergé (Jet Propulsion Laboratory, California Institute of Technology), Lauro Moscardini (Bologna University), Alexandre Réfrégier (CEA Saclay), Stella Seitz (Ludwig-Maximilians-Universität München), Adam Amara (ETH Zürich), Nabila Aghanim (IAS, Orsay), Thomas Kitching (ROE, Edinburgh), Marian Doupsis (IAS, Orsay).*

### Abstract

We discuss how the imaging survey of Euclid will be able to identify galaxy clusters and how they can be exploited to constrain cosmological parameters. We show the extreme sensitivity of galaxy cluster counts on the growth of structures and hence their ability to constrain dark energy scenarios and modified gravity models. We discuss how uncertainties in the selection function degrade this ability. These uncertainties come in two ways: First the scaling of the mass-observable relation and second the scatter in this relation. We show how a self-calibration approach can address the first problem and discuss how much prior information is required on the scatter in order to obtain competitive constraints from galaxy cluster counts.

## 11.1 Introduction

Clusters of galaxies are the largest over-dense objects in the Universe and bear imprints of the history of their formation because they trace: a) the spectrum of initial fluctuations; b) the growth of these structures over time and c) the dynamics of the collapse of halos. This threefold dependency makes clusters an excellent probe of the growth of structure in the Universe. Clusters of galaxies have long been suggested as cosmological probes. They are sensitive to the cosmological parameters in three ways: Firstly their distribution in redshift is sensitive to the above mentioned processes and hence cosmology, secondly their spatial distribution is a biased tracer of the underlying dark matter power spectrum (see Evrard 1989; Bahcall & Bode 2003) and thirdly individual clusters can be viewed, under certain conditions, as fair samples of the matter content in the Universe as pointed out by White *et al.* (1993). We are not pursuing the last approach, since this requires a resolved observation of a cluster in order to only select relaxed galaxy clusters. However, the distribution of galaxy clusters in redshift and space will be accessible to Euclid. Currently Vikhlinin *et al.* (2009) have performed the most detailed study of discovered galaxy clusters with x-ray observations. These will be complemented in the near future by observations of galaxy clusters via the Sunyaev-Zel'dovich decrement in the cosmic microwave background, as first shown by Sunyaev & Zeldovich (1972), for example with the South-pole Telescope (see Ruhl *et al.* 2004) and the Planck satellite (see Planck Collaboration 2006). However, there have been already promising results for optically selected clusters as well. For example, the Sloan Digital Sky Survey (SDSS) cluster sample of about 130,000 objects has been exploited (see Bahcall & Bode 2003; Koester *et al.* 2007; Rozo *et al.* 2007b; Rozo *et al.* 2009a) and provided interesting constraints. Further the Red-Sequence Cluster Survey (RCS) has also provided cosmological information, mainly on the amplitude of density fluctuations $\sigma_8$ as presented by Gladders *et al.* (2007). While these constraints are mainly exploiting a relation between the mass of a galaxy cluster and the number of member galaxies, a good way to calibrate the mass-observable relation is by facilitating the weak lensing properties of clusters as shown in a recent series of papers by Sheldon *et al.* (2009b,a). Dahle (2006) has shown the first promising indications that the distribution of clusters as measured from weak lensing alone yield meaningful cosmological constraints.

The rest of the chapter is organised as follows: We first discuss how the redshift distribution of galaxy clusters is sensitive to cosmological models, then we discuss the different selection methods, which are relevant for the Euclid imaging survey and self- and cross-calibration techniques. We then discuss the prospects for Euclid imaging to constrain cosmological parameters with galaxy clusters.

## 11.2 Cosmology with the Redshift Distribution of Galaxy Clusters

Recent years have seen a plethora of papers forecasting, mainly x-ray and Sunyaev-Zel'dovich, galaxy cluster counts and their ability to constrain cosmology (see Holder *et al.* 2001; Haiman *et al.* 2001; Majumdar & Mohr 2003; Battye & Weller 2003; Hu 2003; Majumdar & Mohr 2004; Lima & Hu 2004, 2005). In order to forecast the number of galaxy clusters in a particular redshift bin we have to calculate

$$\Delta N(z) = \int\limits_{z-\Delta z/2}^{z+\Delta z/2} \Delta\Omega \frac{d^2 V}{dz d\Omega} \int_{M_{\lim(z)}}^{\infty} \frac{dn}{dM} dM \ , \tag{11.1}$$

with $\Delta z$ the width of the redshift bin, $\Delta\Omega$ the sky coverage, $d^2 V/dz/d\Omega$ the volume element and $dn/dM$ the mass function of dark matter halos. $M_{\lim}(z)$ encodes the selection function of the particular survey. Analytically derived mass functions, like Press & Schechter (1974) or its extension by Sheth & Tormen (2002), have the advantage that they capture the entire cosmology dependence of the mass function. However, fits to numerical simulations, are generally more accurate. Currently the most widely used mass function is by Jenkins *et al.* (2001), although this has been surpassed by more precise versions by Warren *et al.* (2006) and most recently by Tinker *et al.* (2008). In the analysis presented here we use the fit by Jenkins *et al.* (2001), which is given by

$$\frac{dn}{dM}(M, z) = -0.316 \frac{\rho_{m,0}}{M} \frac{d\sigma_M}{dM} \frac{1}{\sigma_M} \exp\left\{-|0.67 - \log\left[D(z)\sigma_M\right]|^{3.82}\right\} \ , \tag{11.2}$$

where $\sigma_M$ is the variance of a fluctuation of mass $M$, $\rho_{m,0}$ is the density in mass today and $D(z)$ is the growth of linear matter fluctuations with redshift. From equations (11.1) and (11.2) we see that the strongest cosmology dependence of the number counts actually comes from the volume element and the growth of structures. There is an almost exponential dependence on the growth of structures and this becomes an important feature to investigate modified gravity scenarios as described in Chapter 6. In order to get an understanding of the cosmology dependence of the number counts we show in Fig. 11.1 the redshift distribution of clusters for three different cosmologies. The base line is a concordance $\Lambda$CDM model (solid line). The dotted line is for a model with an equation of state of $w = -0.9$ compared to the concordance model with $w = -1$. From the plot we see a different behaviour at low redshift ($z \leq 1$) compared to high redshift. This is because the difference from a $\Lambda$CDM model is driven at low redshifts by the different volume factor, while at high redshift from the difference in the growth of structures. Although, the difference between the two models seems miniscule, one has to keep in mind that the statistical error for number counts is driven by the Poisson noise. The overall number of clusters for this setup is over half a million, so the Poisson noise is tiny, and of the order of a few hundred per bin, compared to the overall number of over 10,000 per bin. Hence in order to study the statistical significance of the difference we need to understand the systematics, and we will come to this later. The dashed line corresponds to a modified gravity model, which we parameterised with $\gamma = 0.68$ as described in Chapter 6. This is a huge difference, which demonstrates the power of galaxy cluster counts to constrain the growth of structures.

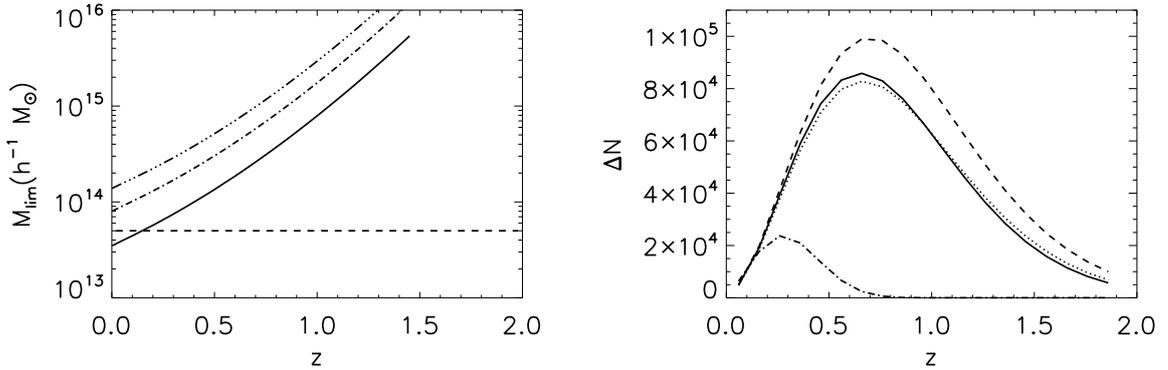

Figure 11.1: Left: Mass limits expected from Euclid optical cluster selection (dashed) and weak lensing selection thresholds for a 3-$\sigma$ detection (solid), 5-$\sigma$ (dash-dotted) and 7-$\sigma$ (dash-triple dotted). Right: Distribution of galaxy clusters in redshift bins of width $\Delta z = 0.1$ for different cosmologies observed on 20,000 $deg^2$. All clusters above a mass limit of $5 \times 10^{13} h^{-1} M_\odot$ are selected. Lines show a $\Lambda$CDM model (solid), a $w = -0.9$ model (dotted) and a modified gravity model ($\gamma = 0.68$; dashed). The dot-dashed line is for a $\Lambda$CDM model with the mass limit of the weak lensing 3-$\sigma$ detection limit. Note that the Poisson errors $\sim \sqrt{N}$ are of the order of a few hundred in most bins for Euclid and hence negligible on this plot.

## 11.3 Selection of Galaxy Clusters with the Euclid Imaging Survey

The Euclid imaging survey can target clusters with three methods. The first is to count all the member galaxies of the over-dense structures, the second is to look at the weak lensing signal imposed by the massive galaxy clusters on the background galaxies, and the third is by the strong lensing signal. The last is discussed in detail in Chapter 15. Here we concentrate on the first two possibilities. In recent years the maxBCG method by Koester *et al.* (2007) has been put forward. maxBCG assumes that the Brightest Cluster Galaxy (BCG) sits in the centre of every cluster. The tight colour-magnitude relation of these objects is used to select candidate BCGs. In addition to identifying the BCG, so called ridge-line galaxies are identified with a model, by finding them with their radial and color distribution. The two models are *maximised* as a function of redshift and hence provide an estimate of the photometric redshift of the cluster. In an iterative scheme the most likely clusters and their satellites are removed from the sample. A chain of probabilities, which have been calibrated with mock catalogues, is then applied to infer the mass of the cluster. The free parameters in the models, besides the cosmological parameters, are the halo model parameters. In terms of systematics or noise the biggest problems are the completeness and purity of the sample. For example, projection effects along the line of sight can lead to a misestimate of cluster members for galaxies along this line. Euclid's good photometry can in part avoid this problem. For more details see the paper by Rozo *et al.* (2007a). This method has been successfully applied to the SDSS cluster sample by Rozo *et al.* (2007b); Rozo *et al.* (2009b), where they analysed the likelihoods of over 10,000 galaxy clusters and could place tight constraints on $\sigma_8$ and $\Omega_m$.

The second possibility to select clusters with the Euclid imaging survey is weak lensing. For example, Dahle (2006) describes how the mass function can be estimated entirely from lensing masses, albeit for an x-ray selected sample. A generalisation of this for a blind survey was presented by Bergé *et al.* (2008) for the XMM and CFHTLS survey in combination. Also the cross-correlation of the SDSS photometric cluster sample with weak lensing, for 130,000 objects, allowed a calibration of the signal to the $\sim 10\%$ level (see Sheldon *et al.* 2009b; Johnston *et al.* 2007; Sheldon *et al.* 2009a).

A large problem for estimating cosmological parameters with galaxy cluster counts is not just the statistical uncertainty in the *mean* mass-observable relation, but even more so the scatter in this relation. This in general leads to a shift to larger cluster numbers in the sample, due to the shape of the mass-function. In order to analyse galaxy cluster counts in a general set-up we adopt the approach proposed by Hu (2003) and Lima & Hu (2004, 2005). They suggest to study the galaxy cluster counts in cells of redshift *and* observable bins, i.e.:

$$\bar{n}_i = \int_{\theta_i^{\mathrm{obs}}}^{\theta_{i+1}^{\mathrm{obs}}} \frac{d\theta^{\mathrm{obs}}}{\theta^{\mathrm{obs}}} \int \frac{dM}{M} \frac{d\bar{n}}{d\ln M} p(\theta^{\mathrm{obs}}|M) \,, \tag{11.3}$$

where the index 'i' refers to bins in redshift and the observable $\theta^{\mathrm{obs}}$ should be viewed as a generic observable. This could be x-ray flux, SZ flux, number of member galaxies (richness), etc. The selection function is encoded in the probability of a cluster with true mass $M$ having the observed "mass" $\theta^{\mathrm{obs}}$ and we choose following Lima & Hu (2005):

$$p(\theta^{\mathrm{obs}}|M) = \frac{1}{\sqrt{2\pi\sigma_{\ln M}^2}} \exp\left[-x^2(\theta^{\mathrm{obs}})\right] \,, \tag{11.4}$$

where

$$x(\theta^{\mathrm{obs}}) \equiv \frac{\ln\theta^{\mathrm{obs}} - \ln M - \ln M^{\mathrm{bias}}}{\sqrt{2\sigma_{\ln M}^2}} \,. \tag{11.5}$$

Note that $M^{\mathrm{bias}}$ encodes the mean scaling relation between the observable and the true underlying mass. We assume here Gaussian scatter with a variance of $\sigma_{\ln M}^2$. Note that this can be easily extended to include non-Gaussian likelihoods as discussed by Shaw *et al.* (2009). In this approach by Lima & Hu (2005), the scatter and bias can be allowed to take any form, even changing completely freely from bin to bin.

Rather than knowing the exact scaling relation and scatter a priori, the survey can self-calibrate for the unknowns in the mass-observable relation (see Majumdar & Mohr 2003; Lima & Hu 2005). However if there is too much freedom there are of course no meaningful cosmological constraints. For the analysis presented here we parameterise the bias with $\ln M^{\text{bias}} = A_b + n_b(1 + z_i)$ and the scatter quadratic in redshift, meaning an additional three parameters. All these parameters are fitted for by the survey, in addition to the cosmological parameters. In order to judge if this is a meaningful parameterisation, we would require mock catalogs and analyse them with an analysis pipeline both adapted and tailored for Euclid's case. However a quadratic evolution in redshift of the scatter is already quite general. As Lima & Hu (2005) have shown it is actually not the scatter itself, it is the prior uncertainty on this quantity, which drives cosmological uncertainties. This uncertainty can be calibrated by observing clusters with different methods. For example Rozo *et al.* (2009) have constrained the scatter in the mass-richness relation of the SDSS maxBCG cluster with complementary weak lensing and x-ray observations and find $\sigma_{\ln M | N_{200}} \approx 0.45 \pm 0.20$. It is important to note that the power spectrum of clusters in addition to the number counts is also sensitive to these uncertainties and plays an important role for the self-calibration of the galaxy cluster observation. We have not yet implemented this in our forecast pipeline, which would further tighten our error forecasts. So from this point of view our forecasts should be considered conservative.

## 11.4 Forecasting Cosmological Parameters with Clusters

In this section we concentrate on cosmological constraints from Euclid imaging survey clusters alone. We want to emphasise that, in particular the weak lensing information on clusters from Euclid imaging can play a pivotal role in calibrating scaling relations of galaxy clusters in surveys such as Planck or eRosita. This potential has been already indicated by Dahle (2006) and Bergé *et al.* (2008). We also want to stress that Euclid is ideally complementary for the aforementioned surveys in a sense that it provides a full sky coverage like these surveys. We have performed a preliminary forecasting analysis for calibrating SZ selected clusters of the Planck survey, by stacking these clusters and constraining the mass - SZ flux relation with Euclid weak lensing. This improves the uncertainty in the scaling relation typically by 50% and in turn improves the cosmological constraints, compared to a Planck alone analysis of SZ selected clusters (see also Cunha 2009). A particular problem, which arises for this type of calibration, is that one has to be careful how the different probes are correlated in a joint analysis as shown by Takada & Bridle (2007).

Here we concentrate on a maxBCG like analysis. However we want to stress that a range of cluster finding algorithms could be used in the Euclid data besides the more popular maxBCG method used extensively in the SDSS data. These would include Voronoi tesselation, matched filter techniques, hybrid filter techniques, counts in cells, percolation algorithms such as friends of friends in photo-z space, smoothing kernels, adaptive kernels, surface brightness enhancements, cut and enhance techniques, just to name a few. Each technique is sensitive to different types of structures and there is no consensus within the community which is the best technique for dark energy studies or for general cluster studies. Some techniques rely of the red properties of galaxy clusters where others could find bluer concentrations of objects or clusters which are not necessarily spherically symmetric.

In the SDSS photometric survey it has been shown that clusters with masses above $10^{13.5} h^{-1} M_\odot$ are detectable with high purity and completeness, with more than 10 bright red galaxies out to $z \sim 0.3$ (see Koester *et al.* 2007; Johnston *et al.* 2007). Note that we assume that this still holds out to a maximum redshift of $z_{\max} = 2$. There have been indications that clusters have a robust red sequence out to redshift $z = 1$, but this will eventually break down and detailed observations and simulations still have to show this beyond $z = 1$. However we should note that given the filter set and depth chosen for Euclid it will be possible to perform high redshift cluster selection with Euclid alone. Given the lack of optical filters the low redshift will not be optimally sampled by Euclid, but given the depth of the Y, J and H filters, Euclid will be able to probe the tail of the cluster redshift distribution accurately since it will be deep enough to find ridgeline galaxy photo-z at high redshift $1.3 < z < 2.0$ (cf. Chapter 20) and will also cover the entire sky, hence have as little cosmic variance as possible.

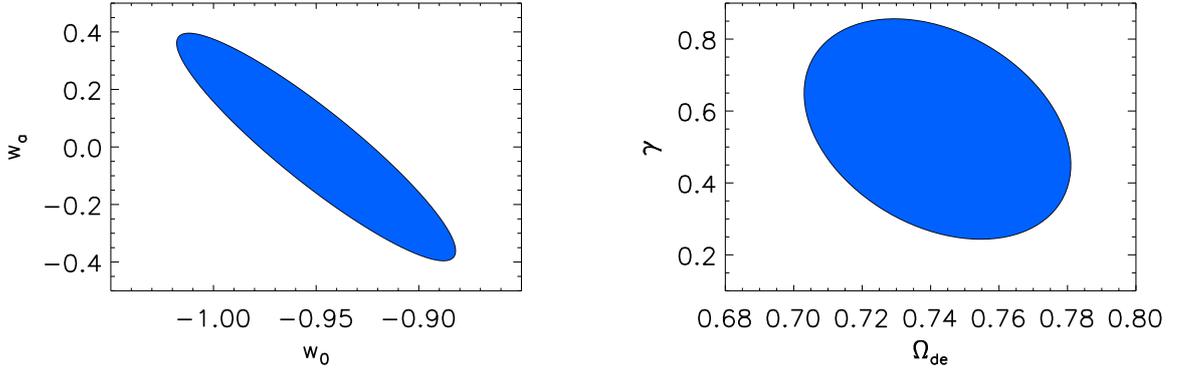

Figure 11.2: On the left the $1-\sigma$ joint likelihood contour in the $w_0 - w_a$ plane after marginalisation over all the cosmological and nuisance parameters with the priors mentioned in the text. On the right we show the joint likelihoods for constraining modified gravity vs. the 'dark energy' contribution.

As mentioned above we assume two parameters on the scaling relation we fit for and 3 for the scatter. In addition we vary the matter contents $\Omega_m$, the dark energy contents $\Omega_{de}$, the two dark energy parameters $w_0$ and $w_a$, the tilt of the primordial power spectrum $n$ and the amplitude of the fluctuations $\sigma_8$. Where indicated we also vary the modified gravity parameter $\gamma$. Note that in the way we set up the survey there is *no* dependence on the Hubble constant $h$ and only very weak dependence on the baryon contents $\Omega_b$. This is because the selection function is fixed and the cluster mass function depends only on the dark matter power spectrum. For the fiducial model we assume $\Omega_m = 0.25$, $\Omega_{de} = 0.75$, $w_0 = -0.95$, $w_a = 0$, $n = 1.0$, $\sigma_8 = 0.8$. The scaling bias is the mass-observable relation is assumed to be 10% constant, the intrinsic scatter in the logarithm of the mass 25%, we use four equally spaced bins in logarithmic mass for the self calibration between $10^{13.5} h^{-1} M_\odot$ and $2 \times 10^{15} h^{-1} M_\odot$, and redshift bins of width $\Delta z = 0.1$ between $z = 0 - 2$. For the sky coverage we assume half the sky with $\Delta \Omega = 20,000 \deg^2$. Finally we assume a prior uncertainty on the scatter and bias parameters of 25%, in line with observations and simulations. For the presentation of our results we add a prior on the cosmological parameters as expected from the Planck mission and described in Rassat *et al.* (2008).

In Fig. 11.2 we show the forecasted constraints for Euclid imaging survey clusters, on the left for a standard dark energy scenario and on the right for a modified gravity parameterisation. Note that for modified gravity we still use a Planck prior *without* the inclusion of the $\gamma$ parameter. A proper treatment is likely to tighten the errors even further. In addition the size of the errors is entirely driven by the large uncertainties in the scatter and bias in the mass-observable scaling relation. This relation is likely to be much tighter than assumed here, in particular if we calibrate with Euclid weak lensing clusters or x-ray observations from surveys such as eRosita. For example, Wu *et al.* (2009) show that an x-ray follow up for only 200 clusters can improve the figure of merit by 50%. Clearly eRosita will provide many more clusters, so a cross-calibration of Euclid with eRosita and vice versa will benefit both. In addition the spectroscopic part of Euclid bears potential to calibrate some of the clusters masses via velocity dispersion measurements.

# The Integrated Sachs Wolfe Effect

*Authors: Nabila Aghanim (IAS, Orsay, France), Jochen Weller (Ludwig-Maximilians-Universität München and Excellence Cluster 'Origin and Structure of the Universe), Marian Douspis (IAS, Orsay, France), Anaïs Rassat (CEA, Saclay, France), Alexandre Réfrégier (CEA, Saclay, France).*

## Abstract

The Integrated Sachs-Wolfe (ISW) effect is an independent probe of dark energy, curvature and deviations from General Relativity on large scales. Though difficult to detect directly in the cosmic microwave background (CMB), it can be measured using correlations of the CMB with tracers of large scale structure (LSS) and has recently been detected using several LSS surveys. The Euclid wide survey provides a galaxy survey which is naturally optimised to lead to a maximum detection of the ISW effect when correlated with future Planck data. As Euclid covers a large redshift range, it will also be possible to perform a tomographic analysis, which will tighten constraints on cosmological parameters. As it probes both geometry and growth of structure, it is complementary to the primary probes of Euclid. We find that in a flat Universe, we can expect to constrain the cosmological constant $w_0$ at the 8% level with the ISW and Planck, and that with strong priors on other cosmological parameters, we can expect to discriminate modified gravity theories by constraining the growth parameter $\gamma$. We summarise in this Chapter the work done by the Euclid Imaging Consortium ISW working group.

## 12.1  Introduction

To better constrain and understand the present acceleration of the expansion there is a crucial need for multiple and complementary observational probes. The Integrated Sachs-Wolfe (ISW) effect (Sachs & Wolfe 1967) imprinted in the CMB and its correlation with the distribution of matter at lower redshifts (through the galaxy surveys) is one of them. The ISW effect arises from the time-variation of the scalar metric perturbations and offers a promising new way of inferring cosmological constraints (for e.g. Corasaniti *et al.* 2005; Pogosian 2006). There exists both an early- and a late-time ISW effect. The early effect is only important around recombination when anisotropies can start growing and the radiation energy density is still dynamically important. The late ISW effect, on the other hand, originates much later after the onset of matter domination from the time derivative of the gravitational potential (Kofman & Starobinskii 1985; Mukhanov *et al.* 1992; Kamionkowski & Spergel 1994). It is to this latter effect that we refer to as being the ISW effect. The late-time ISW effect can be due to dark energy domination at low redshift, curvature, or modifications to the growth of structure on large scales. In flat universe with no modifications to gravity, detection of the ISW is a direct signature of the presence of dark energy.

The ISW effect dominates the largest scales of the CMB temperature power spectrum ($l < 30$). Its importance comes from the fact that it is sensitive to the amount, equation of state and clustering properties of the dark energy. Detection of the weak signal in the CMB power spectrum is highly limited by cosmic variance. However, since the time evolution of the potential giving rise to the ISW effect is probed by observations of large scale structure (LSS), it is also possible to detect the ISW effect through the cross-correlation of the CMB with tracers of the LSS distribution.

This idea, first proposed by Crittenden & Turok (1996), has been widely discussed in the literature. The recent WMAP data provide high enough quality measurements at large scales to be used in combination with LSS tracers to assess the ISW detection (for e.g. Giannantonio *et al.* 2006; Rassat *et al.* 2007, and references therein). The ISW effect has been detected with 2-3.9$\sigma$ significance which increases to 4.5$\sigma$ when all current surveys are used in combination (Giannantonio *et al.* 2008).

## 12.2 Formalism

The ISW effect is a contribution to the CMB anisotropies that arises in the direction $\hat{\mathbf{n}}$ due to variations of the gravitational potential, $\Phi$, along the path of CMB photons from last scattering until now. It adds power to the CMB temperature anisotropies as:

$$\frac{\Delta T_{\text{ISW}}}{T}(\hat{\mathbf{n}}) = -2 \int_{r_{LS}}^{r_0} dr \, \dot{\Phi}(r, \hat{\mathbf{n}}r) \tag{12.1}$$

where $\dot{\Phi} \equiv \partial \Phi / \partial r$ can be related to the matter density field $\delta$ through the Poisson equation. The variable $r$ is the conformal distance, defined today as $r_0$ and at the surface of last scattering as $r_{LS}$. In a flat universe, within the linear regime, the gravitational potential does not change with time if the expansion of the universe is dominated by matter. Therefore, for most of the time since last scattering, matter domination ensured a vanishing ISW contribution. Conversely, a detection of an ISW effect would indicate that the effective potential of the universe has changed. In fact, detection of an ISW effect can only be attributed to three causes: the presence of dark energy at late times, spatial curvature or a modification to General Relativity at large scales. The Euclid mission will probe LSS at redshifts $z = 0 - 2$, i.e. it probes the dark energy dominated era and is therefore ideal to detect changes in the potential due to dark energy. Euclid will also be able to detect signatures of spatial curvature and modifications of gravity on large scales (see Figure 12.1) by constraining the parameter $\gamma$ (defined in Chapter 3).

The 2-point angular cross-correlation of the ISW temperature anisotropies with the galaxy distribution field is defined as

$$C_l^{\text{ISW}-\text{G}} = \frac{2}{\pi} \int dk \, k^2 P_{\delta\delta}(k) I_l^{\text{ISW}}(k) I_l^{\text{G}}(k), \tag{12.2}$$

where $I_l^{\text{G}}(k) = \int_0^{r_0} dr \, W^{\text{G}}(k,r) j_l(kr)$, and the galaxy window function is given by $W^{\text{G}}(k,r) = b_{\text{G}}(k,r) n_{\text{G}}(r) G(r)$. The power spectrum of the over-density fluctuations is described by $P_{\delta\delta}(k)$. The spherical Bessel functions are denoted by $j_l$; $b_{\text{G}}(k,r)$ is the bias, and $n_{\text{G}}(z) = n_{\text{G}}(r)/H(z)$ is the normalised redshift distribution of the galaxies, for which we use the following parameterisation:

$$n_{\text{G}}(z) = n_{\text{G}}^0 \left(\frac{z}{z_{med}}\right)^\alpha \exp\left[-\left(\frac{z}{z_{med}}\right)^\beta\right]. \tag{12.3}$$

The variables $\alpha$ and $\beta$ provide a description of the galaxy distribution at low and at high redshifts respectively and $z_{med}$ is the median redshift. The variable $n_{\text{G}}^0$ is a normalising constant.

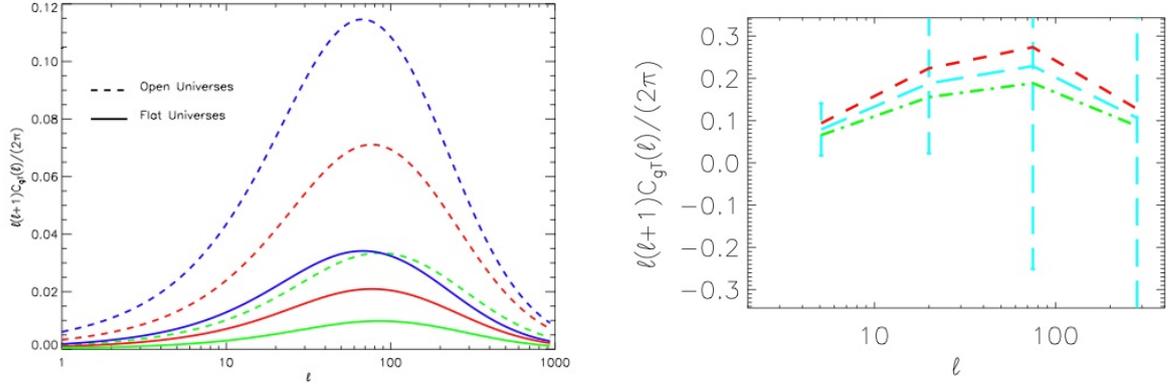

Figure 12.1: *Left Panel*: Prediction of the ISW cross-correlation signal for different values of the dark energy density ($\Omega_{DE} = 0.10$, green line; $\Omega_{DE} = 0.20$, red line; $\Omega_{DE} = 0.30$, blue line) for universes with flat geometry (solid lines) and universes with open geometry and no dark energy. The ISW signal for universes with the same matter density is larger in open universes than in flat universes. The signal is calculated for a Euclid-like photometric survey. *Right panel*: The ISW cross-correlation signal for different values of the growth parameter ($\gamma = 0.44$, green, dash-dotted line; $\gamma = 0.55$, blue dashed line; $\gamma = 0.68$, e.g. a DGP model, red short dashed). *Both figures are taken from Rassat (2007).*

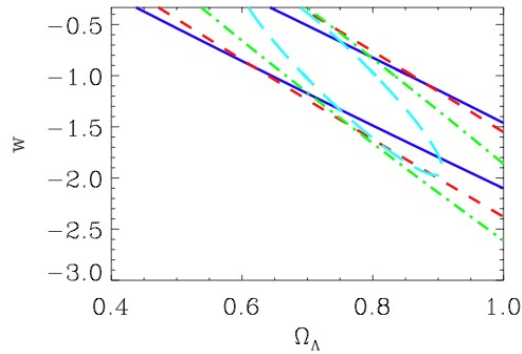

Figure 12.2: Contours for $w$ and $\Omega_{DE}$ from 4 redshift bins from the Euclid photometric survey with roughly equal number of galaxies per bin ($z = [0, 0.6], [0.7, 1.10], [1.1, 1.4], [1.5, 2.7]$). The direction of the degeneracy in the $w - \Omega_{DE}$ plane changes with the redshift of the galaxies considered. *Figure taken from Rassat (2007).*

As the Euclid survey will provide a galaxy survey over redshift $[0-2]$, it will be possible to measure the evolution of the ISW signal. This is of special importance to model the evolution of the equation of state $w(z)$. Degeneracies between different parameters also evolve with redshift (see Figure 12.2), and so a full tomographic analysis is expected to provide tighter constraints on cosmological parameters (see Rassat (2007) and Giannantonio *et al.* (2008)).

The uncertainty on the bias and most importantly on its redshift evolution might have a non negligible impact on the measure of the ISW. Consequently it will affect the cosmological parameter estimate, in particular the equation of state (Schaefer *et al.* 2009). Assuming a linear bias evolution, we find that bias evolution shifts the best fit position by an amount comparable to the statistical accuracy in the case of $\sigma_8$, $n_s$ and $w$. This comes from the fact that $\sigma_8$ affected due to the higher average bias at earlier times, $n_s$ due to the change in the shape of the cross-correlation spectrum and $w$ since it is not as well measured from primary CMB anisotropies. Ignoring bias evolution yields less negative values for $w$ introducing a bias towards dark energy models with respect to $\Lambda$CDM.

The galaxy distribution $n_G(z)$ will also be subject to redshift distortions (Kaiser 1987), which will affect both the projected galaxy field (Fisher *et al.* 1994; Padmanabhan *et al.* 2007) and the galaxy-temperature cross-correlation will be subject to redshift distortions (Rassat 2009). This will modify the ISW cross-correlation equation 12.2. To include redshift space distortions, the $I_\ell^G(k)$ term in equation 12.2 should be replace by the following term: $I_\ell^{G+z} = I_\ell^G(1 + \beta A_\ell) - \beta B_\ell I_{\ell-2}^G - \beta D_\ell I_{\ell-2}^G$, where $\beta = \frac{d \ln D}{d \ln a}$ is the growth factor and $A_\ell$, $B_\ell$ and $C_\ell$ all depend on the multipole $\ell$. The inclusion of redshift distortions (which will naturally be present in any redshift selected galaxy survey) adds a new dependence on the growth, which is useful for constraining both dark energy, curvature and modified gravity.

## 12.3 ISW requirements: Signal to Noise analysis

We explored what an ideal survey, optimised for detecting the ISW effect, would resemble, by considering the detection level of ISW measured through its signal-to-noise (SN) as in Douspis *et al.* (2008):

$$\left(\frac{S}{N}\right)^2 = f_{sky}^c \sum_{l=l_{min}}^{l_{max}} (2l+1) \times \frac{\left[C_l^{ISW-G}\right]^2}{\left[C_l^{ISW-G}\right]^2 + \left(C_l^{ISW} + N_l^{ISW}\right)\left(C_l^G + N_l^G\right)}, \qquad (12.4)$$

where $f_{sky}^c$ is the fraction of sky common to the CMB and the Euclid galaxy survey maps, and the cumulative SN is summed over multipoles between $l_{min} = 2$ and $l_{max} = 60$, which are the scales which contribute most to the ISW effect. The spectra $C_l^{ISW-G}$, $C_l^{ISW}$ and $C_l^G$ are the cross- and auto- correlation spectra of temperature (ISW) and the galaxy (G) fields. The noise contributions in the temperature signal and galaxy surveys are $N_l^{ISW}$ and $N_l^G$ with $N_l^{ISW} = C_l^{CMB} + N_l^{CMBexp}$ ($N_l^{CMBexp}$ is the CMB experimental noise for an experiment such as the Planck satellite). The galaxy survey noise is defined by the shot noise contribution: $N_l^G = \frac{1}{\overline{N}}$ where $\overline{N}$ is the surface density of sources per steradian used for the correlation with CMB. The noise part of the SN depends then, at first order, on the common sky fraction, on the surface density of sources, and on their median redshift through the amplitude of the ISW power spectrum $C_l^{ISW}$ and of the galaxy auto-correlation signal $C_l^G$.

We find that the SN plateaus once $\overline{N}$ reaches typically about 10 sources per arcmin$^2$. Above this limit for $\overline{N}$, the SN ratio is very sensitive to $f_{sky}^c$ and $z_{med}$ (see Figure 12.3). For a given $z_{med}$, the larger $f_{sky}^c$ the higher the detection level. Conversely, at a given $f_{sky}^c$, increasing $z_{med}$ significantly improves the ISW detection only up to $z_{med} \sim 1$. An optimal survey (with a detection level above 4 $\sigma$) should thus be designed so that it has a minimum number density of sources of around 10 galaxy per arcmin$^{-2}$, covers a minimum sky fraction of the order of 0.5 and is reasonably deep, with a minimum median redshift of about 0.8. Such a survey, optimised for the weak lensing science case, is being planned for the future and satisfies the ISW requirements. It is the wide survey of Euclid mission proposed to the ESA's Cosmic Vision.

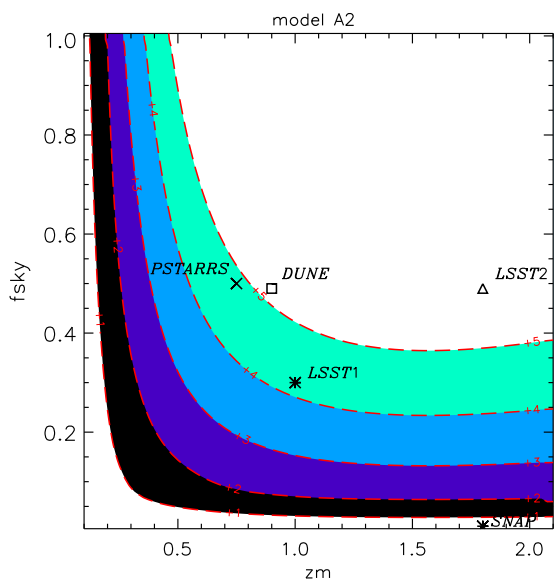

Figure 12.3: Total signal-to-noise for an ISW detection in the constant equation of state model with $w = -0.9$ as function of the galaxy survey parameters $f_{sky}^c$ and $z_{\mathrm{med}}$.

## 12.4 Fisher Matrix analysis

We calculate forecasts for the Euclid imaging survey, considering the same fiducial cosmology and survey details as in Chapter 13, and consider a tomographic calculation, which includes the effect

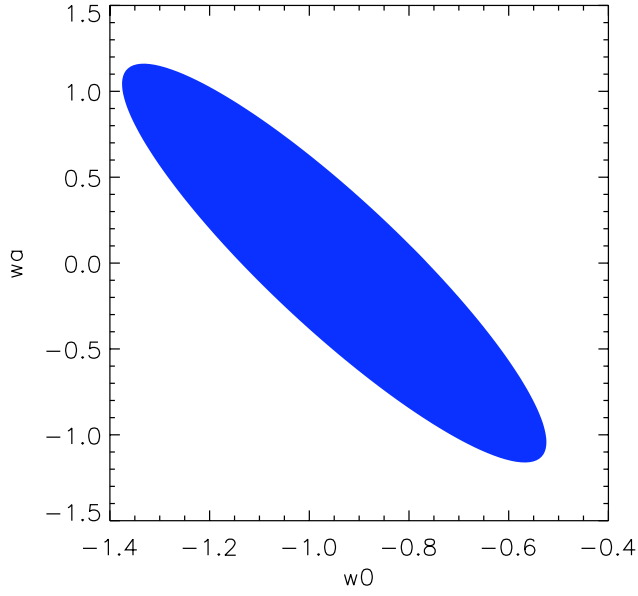

Figure 12.4: Expected constraints (marginalised over the other parameters shown in Table 12.1) from the ISW probe using the cross-correlation of the Euclid photometric galaxy survey with Planck data. Constraints include priors from Planck and allow for curvature.

of redshift distortions. The Fisher matrix can then be calculated using (see Pogosian 2006):

$$F_{\alpha,\beta}^{G-ISW} = f_{sky}^c \sum_{\ell} \sum_{i,j} \frac{\partial C_{\ell}^{Gi-ISW}}{\partial p_{\alpha}} COV^{-1} \left( C_{\ell}^{Gi-ISW} C_{\ell}^{Gj-ISW} \right) \frac{\partial C_{\ell}^{Gj-ISW}}{\partial p_{\beta}}, \qquad (12.5)$$

where $Gi$ and $Gj$ correspond to two separate galaxy redshift bins and $\alpha$ and $\beta$ to different cosmological parameters.

For the Fisher forecasts we consider two parameter sets, one which allows for curvature and a varying equation of state $w_0 = w_0(z)$:

$$\Theta_2 = (H_0, \Omega_b, \Omega_m, \sigma_8, n_s, w_0, w_a, \Omega_{DE}). \qquad (12.6)$$

and another which considers only a flat geometry and a constant equation of state $w_0$:

$$\Theta_1 = (H_0, \Omega_b, \Omega_m, \sigma_8, n_s, w_0), \qquad (12.7)$$

The cosmological constraints we obtain for the ISW signal are shown in Table 12.1 and the marginalised constraints in the $w_0 - w_a$ plane are shown in Figure 12.4.

Table 12.1: Cosmological constraints for Euclid's Imaging survey (i.e. photometric redshifts) using the Integrated Sachs Wolfe effect. The constraints are presented with Planck priors calculated as in Rassat *et al.* (2008). The marginalised constraints in the $w_0 - w_a$ plane are shown in Figure 12.4.

| Method | $w_0$ | $w_a$ | $\Omega_{DE}$ | $\Omega_m$ | $\Omega_b$ | $n_s$ | $h$ | $\sigma_8$ |
|--------|-------|-------|---------------|------------|------------|-------|-----|-----------|
| Full model | 0.59 | 1.60 | 0.067 | 0.065 | 0.010 | 0.0038 | 0.08 | 0.068 |
| Flat Universe & constant $w$ | 0.081 | - | - | 0.0072 | 0.00091 | 0.0038 | 0.0073 | 0.048 |

Despite the constraints from the cross-correlation itself are weak, they play a non negligible role in the combination with CMB. This is mainly due to the 6–dimensional shape of the likelihood and its degeneracies. The ISW effect does help (as compared to CMB alone) in breaking degeneracies among the parameters describing the dark energy model and the other cosmological parameters, primarily $\sigma_8$. The cross-correlation between Planck and Euclid allows to place a constraint of the order of 8% on $w$ for a model with a constant equation of state and reduces the errors on the estimation of the parameter $w_a$ in a linear model. However, it does not allow to distinguish among a constant and a dynamical equation of state for the dark energy.

## 12.5 Conclusions

The ISW effect offers a promising new way of inferring cosmological constraints (for e.g. Corasaniti *et al.* 2005; Pogosian 2006) and though it is weaker than other probes, it provide a complementary way to probe our cosmological model. In particular it can be used to probe the energy density, inhomogeneities and redshift evolution of dark energy, as well as curvature and modifications to the large scale growth of structure. By construction, the Euclid photometric survey is ideally optimised for an ISW survey, which requires maximum sky coverage, at least 10 galaxies per square arcmin, and a redshift range covering that of dark energy domination. The relative depth of the Euclid photometric survey will also permit a tomographic analysis, which will improve cosmological constraints. We find that with a Planck prior and the assumption of a flat Universe, an ISW survey from Euclid can expect to constraint the cosmological constant equation of state to 8%. In the case of dark energy and allowing for curvature, the ISW constraints from Euclid can serve as priors to weak lensing constraints.

# Galaxy Correlations with Euclid's Imaging Survey

*Authors: Francisco Castander (ICE, Barcelona), Anaïs Rassat (CEA, Saclay), Adam Amara (ETH, Zurich), Luca Amendola (INAF, Rome), Julien Carron (ETH, Zurich), Thomas Kitching (ROE, Edinburgh), Martin Kunz (University of Sussex, UK / University of Geneva, Switzerland), Alexandre Réfrégier (CEA, Saclay), Jochen Weller (University of Munich, Germany).*

## Abstract

In recent years, galaxy surveys have proved to be powerful probes with which to constrain cosmology and there exists various methods in the literature which use different features present in the galaxy power spectrum, including Baryon Acoustic Oscillations (BAOs). In this Chapter, we present the work of the Euclid Imaging Consortium's (EIC) working group on galaxy correlations, which has studied a spherical harmonic approach to the studying the galaxy power spectrum, as well as a 'wiggles only' method in Fourier space. We find the Euclid Imaging (photometric) Survey has the power to constrain different sectors of the cosmological model, which though weaker than weak lensing constraints are useful for cross-checking results.

## 13.1 Introduction

Galaxy surveys are often used to trace a feature in the galaxy power spectrum named Baryon Acoustic Oscillations (BAOs) - or baryon wiggles, which are considered a powerful standard ruler with which to constrain cosmology. However, BAOs are only one feature in a rich statistic - the entirety of the statistical distribution of galaxies encodes information on various characteristic scales about the underlying dark matter distribution and the nature of dark energy (e.g., Peebles 1980). As it also probes both the growth of structure as well as the expansion history of the Universe, the full statistical distribution of galaxies is a very powerful tool with which to probe the composition of the Universe, the nature of dark energy and the behaviour of gravity on large scales.

Large scale galaxy surveys have been used extensively to constrain our cosmological model both in the optical (for e.g., Geller *et al.* 1987; Tegmark *et al.* 2004; Peacock *et al.* 2001), as well as in the near-infrared (for e.g., Fisher *et al.* 1994; Heavens & Taylor 1995; Erdoğdu (a) *et al.* 2006), and provided early evidence for a cosmological constant (Efstathiou *et al.* 1991).

There exists several methods to measure the statistical distribution of galaxies. The first variable is the statistics used in the analysis: the measurement can be done in real space (Eisenstein *et al.* 2005, $\xi(r)$), configuration space (Loverde *et al.* 2008, $w(\theta)$), Fourier space (Seo & Eisenstein 2003; Amendola *et al.* 2005, P(k)) or in Spherical Harmonic space (Dolney *et al.* 2006, $C(\ell)$). For a given statistic, it is also possible to use different methods which use information from specific scales. In Fourier space, some measure the full power spectrum (Seo & Eisenstein 2003, i.e. including information from the BAOs), whereas others subtract the smooth part of the spectrum and focus only on the oscillations (Blake *et al.* 2007; Seo & Eisenstein 2007) - the latter uses less information but may be more robust with respect to systematic uncertainties.

Knowledge of the full three-dimensional galaxy distribution can yield very competitive constraints, even if only considering the baryon wiggles feature. However, partial radial information - e.g., with photometric redshifts - can still be used to compute a statistical measure of projected features in spherical harmonic space (Dolney *et al.* 2006; Rassat *et al.* 2008), and the radial degradation can also be included in Fourier space measurements (Seo & Eisenstein 2007). Furthermore, radial features such as redshift distortions will also have secondary effects that can still be measured in the projected distribution (Padmanabhan *et al.* 2006; Rassat *et al.* 2008).

In this Chapter we will present the different building blocks which carry information in the galaxy distribution and present different measurement methods in both Fourier and projected spherical harmonic space. We explore the cosmological constraints we can expect to achieve for the Euclid Imaging Survey (i.e., with photometric redshifts) for the three sectors of the cosmological model: dark matter, dark energy and initial conditions.

## 13.2 Building Blocks of the Galaxy Power Spectrum

The three-dimensional galaxy power spectrum is a rich statistic, where several features on different scales contain cosmological information specific to different sectors of the cosmological model (dark energy, dark matter and initial conditions). On linear scales we identify the following three main features in the galaxy power spectrum, which are:

- The broad-band power: Information is contained in the shape, normalisation and time evolution of the power spectrum.

- The Baryon Acoustic Oscillations (BAOs): Information is contained in the tangential and radial wavelengths, as well as the wiggles amplitude.

- The linear redshift space distortions: Even when considering the projected power spectrum, this radial information is partially present.

The Fourier space matter power spectrum $P(k)$ describes the fluctuations of the matter distribution and is defined by $\langle \delta(\mathbf{k})\delta^*(\mathbf{k}') \rangle = (2\pi)^3 \delta_D^3(\mathbf{k} - \mathbf{k}')P(k)$, where $\delta(\mathbf{k})$ represents the Fourier transform of the matter overdensities $\delta(\mathbf{r}) = \frac{\rho(\mathbf{r}) - (\bar{\rho})}{\bar{\rho}}$, and the mean density of the Universe is $\bar{\rho}$. The term $\delta_D^3(\mathbf{k} - \mathbf{k}')$ represents the Dirac delta function.

The *galaxy* power spectrum is related to the matter power spectrum by:

$$P^g(k) \propto (1 + \beta\mu^2)^2 b^2(k, z)P(k), \tag{13.1}$$

where the first term describes the effect of redshift space distortions, modulated by the distortion parameter $\beta = \frac{1}{b(z)}\frac{d\ln D(a)}{d\ln a}$ and $\mu$ (the cosine of the angle between the tangential and radial modes) and on linear scales a scale independent galaxy bias can be assumed, i.e. $b(k, z) = b(z)$.

The three building blocks of the observed galaxy power spectrum are plotted separately in Figure 13.1 (panels B, C and D), while the observed galaxy power spectrum is plotted in panel

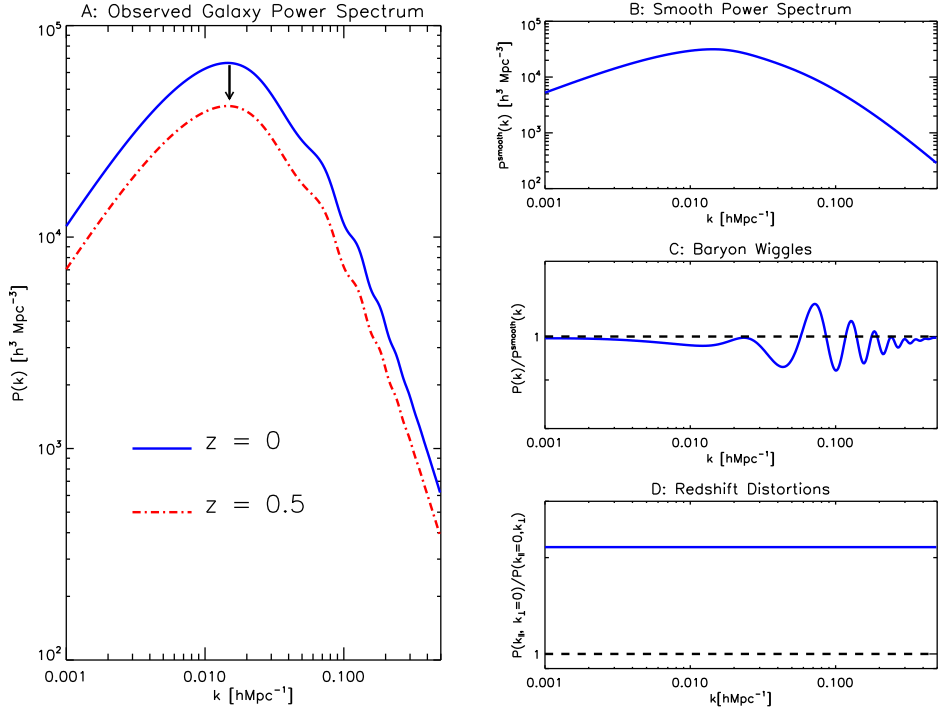

Figure 13.1: Illustration of the main building blocks of the linear Fourier space galaxy power spectrum. **Panel A:** The observed linear galaxy power spectrum. This includes the three building blocks of the matter power spectrum: the smooth power spectrum (broad band power), baryonic wiggles, linear redshift distortions as well as the linear galaxy bias $b(z)$. In linear theory the redshift evolution of the galaxy power spectrum depends solely on the linear growth factor ; we illustrate this by comparing the linear galaxy power spectrum at redshift $z = 0$ (solid blue line), and redshift $z = 0.5$ (dot-dashed red line). **Panel B, C and D** illustrate the different building blocks of the galaxy power spectrum. **Panel B:** The smooth part of the galaxy power spectrum contains information on the shape and normalisation of the power spectrum. **Panel C:** The ratio (blue solid line) of the full spectrum and the smooth part of the power spectrum reveals the residual baryonic wiggles. The dashed line corresponds to no baryonic wiggles. **Panel D:** The ratio (blue solid line) of the radial galaxy power spectrum to the tangential spectrum illustrates the scale-independent effect of linear redshift distortions.

A. Each feature depends on different cosmological parameters in complementary ways, and a detailed description of the cosmological dependence of each building block on the different sectors of the cosmological model is given in Rassat *et al.* (2008).

## 13.3 Galaxy correlations with photometric survey

It is possible to study the statistical distribution of galaxies using various basis sets. In this chapter we consider the projected spherical harmonic decomposition ($C(\ell)$) and a wiggles only method (BAO) in Fourier space.

### 13.3.1 Projected Spherical Harmonic Method [$C(\ell)$]

There are several advantages to using the projected spherical harmonic method. For data on a sphere, spherical harmonics provide a natural basis set in which transverse and radial modes are independent. In this basis, there exists an exact prescription for redshift distortions (Fisher *et al.* 1994; Heavens & Taylor 1995). This contrasts with the far field approximation necessary to describe redshift distortions in Fourier space (Kaiser 1987).

The bare observables provided by a galaxy survey ($\theta, \phi, z$) are also naturally translated into a single measurement $C(\ell)$, independently of cosmology, whereas to measure the Fourier space power spectrum, it is first necessary to relate these observables to the wavenumber $\mathbf{k}$ which requires an assumption of the cosmological model.

Finally, in the case of a photometric survey like the Euclid Imaging Survey, there is not as much information lost when using the projected galaxy information (as opposed to a three-dimensional decomposition), since the radial information is already weakened by redshift uncertainty. Furthermore, some radial information from linear redshift distortions can be measured in the projected spectrum.

In this section, we describe the tomographic method we use for the projected $C(\ell)$ method. This method draws on that described in Dolney *et al.* (2006), but includes the effect of redshift distortions as described in Padmanabhan *et al.* (2006) and Rassat *et al.* (2008). For full details of the tomographic method, the reader is referred to Rassat *et al.* (2008).

For a redshift selected galaxy survey, the spherical harmonic power spectrum can only be measured in redshift space; this can be theoretically modelled by adding a redshift distortion correction term to the real space window function:

$$
\begin{aligned}
C(\ell, z_i) &= \langle |a_{\ell m}|^2 \rangle \\
&= 4\pi b^2(z_i) \int dk \frac{\Delta^2(k)}{k} |W^r(k) + \beta W^z(k)|^2 \\
&= 4\pi b^2(z_i) \int dk \frac{\Delta^2(k)}{k} |W_\ell^{r+z}(k)|^2,
\end{aligned} \tag{13.2}
$$

where $a_{\ell m}$ represent the projected spherical harmonic coefficients and the galaxy bias is taken to be constant across the depth of each bin centred at redshift $z_i$. The power spectrum is expressed as: $\Delta^2(k) = \frac{4\pi}{(2\pi)^3} k^3 P(k)$ and is calculated using the publicly available code CAMB (Lewis *et al.* 2000).

Padmanabhan *et al.* (2006) showed that the final term in equation 13.2 could be re-written:

$$
\begin{aligned}
W_\ell^{r+z}(k) &= W^r(k) + \beta W^z(k) \,, \\
&= W_\ell^r(k) + \beta \left[ A_\ell W_\ell^r(k) + B_\ell W_{\ell-2}^r(k) + D_\ell W_{\ell+2}^r(k) \right] \,,
\end{aligned} \tag{13.3}
$$

where the terms $A_\ell$, $B_\ell$ and $D_\ell$ depend only on $\ell$. The real space window function $W_\ell^r(k)$ is given by $W_\ell^r(k) = \int dr \Theta(r) j_\ell(kr) D(r)$. The normalised galaxy distribution is denoted by $\Theta(r)$, the term $j_\ell(r)$ refers to the spherical Bessel function of order $\ell$ and $D(r)$ is the growth function.

For a galaxy survey with photometric redshifts, the different galaxy bins will necessarily overlap. The correlation functions between two bins $i, j$ can be taken into consideration for the Fisher matrix calculation by:

$$F_{\alpha,\beta}^{ij} = \sum_\ell \sum_{i,j} \frac{\partial C_\ell^{ij}}{\partial p_\alpha} \text{COV}^{-1} \left[ C_\ell^{ij} \right] \frac{\partial C_\ell^{ij}}{\partial p_\beta}, \tag{13.4}$$

where $C_\ell^{ij}$ is the correlation between two redshift bins $i$ and $j$ and the covariance matrix is given by: $\text{COV} \left[ C^{ij}(\ell) \right] = \sqrt{\frac{2}{(2\ell+1)f_{\text{sky}}}} \left[ C^{ij}(\ell) + \mathcal{N}_g \delta_{ij} \right]$, so that the shot noise $\mathcal{N}_g$ only intervenes in the diagonal elements of the covariance matrix.

### 13.3.2 Wiggles only BAO method

We also consider a 'wiggles only' method which uses a Fourier space decomposition. The main advantage of using Fourier space is that the measured power spectrum in is physical - not projected - coordinates. This means the scale $k$ of the wiggles is constant for every redshift bin considered, whereas in spherical harmonic the scale $\ell$ of the wiggles will shift as different redshift bins are considered. This method is also a three-dimensional method, where the radial degradation from photometric redshift is included.

In the case of the 'wiggles only' method, only the size of the radial and tangential BAO scale is used - all other information from the power spectrum $P(k)$ (see Figure 13.1) is discarded. This method omits all information from the shape and amplitude of the power spectrum, the redshift distortions and even the amplitude of the BAO wiggles.

We follow the method of Seo & Eisenstein (2007), which we briefly summarise here. The starting point is the expression to compute the Fisher matrix from the full Fourier space power spectrum $P(k)$. The non-linear power spectrum used is extracted from the linear power spectrum with a damping term representing the loss of information coming from the non-linear evolutionary behaviour. This damping term is approximated by an exponential computed using the Lagrangian displacement fields Eisenstein *et al.* (2006):

$$P_{\text{nl}}(k, \mu) = P_{\text{lin}}(k, \mu) \exp(-k^2 \Sigma_{\text{nl}}^2) \tag{13.5}$$

This non-linear damping can be decomposed into the radial and perpendicular directions by:

$$P_{\text{nl}}(k, \mu) = P_{\text{lin}}(k, \mu) \exp\left( -\frac{k_\perp^2 \Sigma_\perp^2}{2} - \frac{k_\parallel^2 \Sigma_\parallel^2}{2} \right), \tag{13.6}$$

where $\Sigma_{\text{nl}}^2$ varies with redshift, following $\Sigma_{\text{nl},\perp} = \Sigma_{\text{nl}} D(z)$ and $\Sigma_{\text{nl},\parallel} = \Sigma_{\text{nl}} D(z)(1 + f(z))$.

A delta function baryonic peak in the correlation function at the sound horizon scale, $s$, translates into a 'wiggles only' power spectrum of the form

$$P_b \propto \frac{\sin(ks)}{ks}. \tag{13.7}$$

This functional form is also obtained if the power spectrum is divided by a smooth version of it or if we only use the baryonic part of the power spectrum transfer function (e.g., Eisenstein &

Hu (1999)). However, the baryonic peak is widened due to Silk damping and non-linear effects. This Gaussian broadening translates into an exponential decaying factor in Fourier space:

$$P_b \quad \propto \quad \frac{\sin(ks)}{ks} \exp\left(\frac{-k^2 \Sigma^2}{2}\right), \tag{13.8}$$

$$= \quad \frac{\sin(ks)}{ks} \exp\left[-(k\Sigma_{\text{Silk}})^{1.4}\right] \exp\left(\frac{-k^2 \Sigma_{\text{nl}}^2}{2}\right). \tag{13.9}$$

We can further refine the calculation to 2 dimensions, separating the radial and tangential location of the baryon acoustic peak in the correlation function. Using the same notation as Blake *et al.* (2006) and Parkinson *et al.* (2007), the observables are $y(z) = r(z)/s$ and $y'(z) = r'(z)/s$ respectively, where $r(z)$ is the comoving distance to redshift z. Measuring the fractional errors on these two observables is equivalent to measuring the fractional errors on $H \cdot s$ and $D_A/s$.

We can now use the derivatives of the 'wiggles only' power spectrum by these quantities (see Seo & Eisenstein (2007) for all the details) to compute the Fisher matrix (equation 26 of Seo & Eisenstein (2007)). This equation includes the effect of redshift distortions on the baryon wiggles in the factor $R(\mu) = (1 + \beta\mu^2)^2$, which decreases the amount of shot noise.

The effect of photometric redshift errors can be included in this formalism as an additional exponential term in the redshift distortion factor, $R(\mu) = (1 + \beta\mu^2)^2 \exp(-k^2\mu^2\Sigma_z^2)$, where $\Sigma_z$ is the uncertainty in the determination of photometric redshifts. The effect of photometric redshift errors will increase the shot noise, leading to larger errors. We evaluate these integrals to obtain the Fisher matrix for the angular diameter distance, $D_A/s$, and the rate of expansion, $H \cdot s$.

To compute the errors on the parameters, we follow Parkinson *et al.* (2007), with some changes: Since we know the correlation between the errors on the vector $y_i = \{r(z_i)/s, r'(z_i)/s\}$, we form a small 2x2 covariance matrix for each redshift bin, which we invert to obtain the corresponding Fisher matrix, $F^{(i)}$. While Parkinson *et al.* (2007) compute the transformation to the cosmological parameters analytically, we perform it numerically.

The above is enough to provide constraints on cosmological parameters. We also use another equivalent implementation which adds another step. In this second implementation we form a Gaussian likelihood $L \propto \exp(-\chi^2/2)$ with

$$\chi^2 = \sum_i \sum_{a,b=1}^{2} \Delta_a F_{ab}^{(i)} \Delta_b \tag{13.10}$$

where $i$ runs over the redshift bins and $a, b$ over $r/s$ and $r'/s$, and $\Delta = X_{\text{data}}(z_i) - X_{\text{theory}}(z_i)$ is the difference between the data and the theory. We then compute numerically the matrix of second derivatives of $\chi^2$ at the peak of the likelihood, via finite differentiating in all parameters (where it is necessary to divide by 2 when using $\chi^2$ rather than $-\ln L$).

The accuracy to which one can constrain cosmological parameters from galaxy clustering depends on the accuracy one can measure the galaxy power spectrum. The error on the determination of the power spectrum comes from two factors: sampling variance and shot noise. The relative error can be written as (Feldman, Kaiser & Peacock 1984):

$$\frac{\Delta P}{P} \propto \frac{1}{\sqrt{V}}(1 + \frac{1}{nP}) \tag{13.11}$$

where the first term represents the sampling variance term and depends on the volume sampled. The second term is the shot noise term and depends on the effective number of galaxies used and the value of the power spectrum.

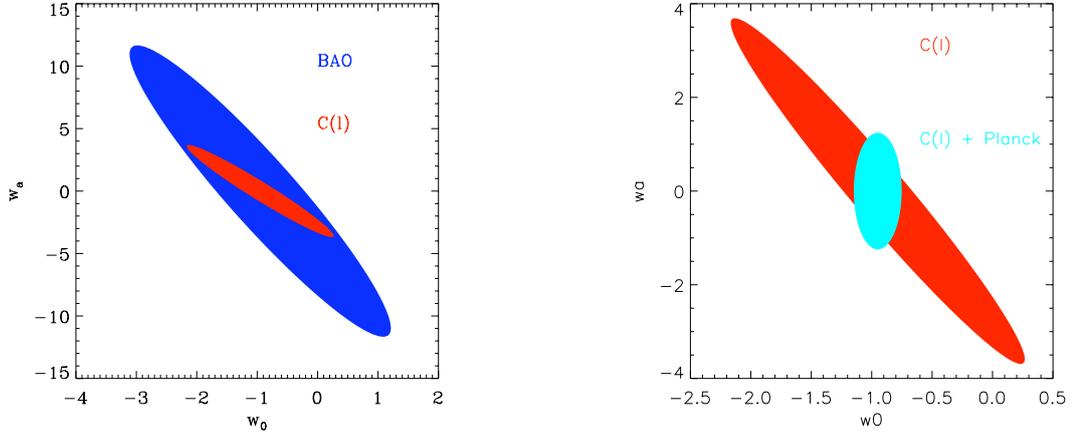

Figure 13.2: **Left Panel:** Constraints on $w_0$ and $w_a$ for the projected $C(\ell)$ and the 'wiggles only' BAO method using Euclid's Imaging (photometric) survey. **Right Panel:** Constraints on $w_0$ and $w_a$ for the projected $C(\ell)$ method using Euclid's Imaging (photometric) survey with and without Planck priors.

## 13.4  Constraints from the Euclid Imaging Survey

### 13.4.1  Implementation

In order to calculate forecasts for the Euclid Imaging Survey, we use a central fiducial model similar to WMAP5 results (Dunkley *et al.* 2009), i.e. flat universe (though we allow for curvature in the Fisher matrix calculations), with no running spectral index and no massive neutrinos. The central values are shown in Table 13.1, where we use the parameterisation of Chevallier & Polarski (2001) to quantify the redshift evolution of the equation of state $w(a)$:

$$w(a) = (1 - a)w_a = w_p + (a_p - a)w_a,\tag{13.12}$$

where $a = 1/(1 + z)$, and the subscript $p$ denotes the pivot point. We also consider the dark energy 'Figure of Merit', as defined by the DETF (Albrecht *et al.* 2006, Dark Energy Task Force):

$$\mathrm{FoM} = \frac{1}{\sigma(w_p) \times \sigma(w_a)}\tag{13.13}$$

For the 'wiggles only' method, we estimate the number of galaxies observed as a function of redshift integrating an evolving luminosity function down to Euclid's imaging survey magnitude limit. Together with the area, this also give us the sampled volume. We take the photo-z error from the simulations of Abdalla et al and Banerji et al. We take the bias as a combination of the VVDS determination and that obtained from numerical simulations.

Table 13.1: A flat universe (though we allow for curvature in the Fisher matrix calculations, with no running spectral index nor massive neutrinos.

| Parameter | Central fiducial Value |
|-----------|------------------------|
| $\Omega_m$ | 0.25 |
| $\Omega_{\rm DE}$ | 0.75 |
| $\Omega_b$ | 0.045 |
| $w_0$ | -0.95 |
| $w_a$ | 0 |
| $n_s$ | 1 |
| $h$ | 0.7 |
| $\sigma_8$ | 0.8 |

For the spherical harmonic method, we consider an imaging survey covering $20,000 \deg^2$, with median redshift $z = 1$, which we split into 10 tomographic bins with roughly the same number of galaxies in each bin (2 billion galaxies in total). We take the linear bias to be constant across each redshift bin and marginalise over these 10 extra parameters for the $C(\ell)$ method. As we assume linearity we do not consider $k$ modes above $k_{\max} = 0.25h\mathrm{Mpc}^{-1}$.

### 13.4.2   Combination with Planck priors

The constraints for both methods can be combined with constraints from other probes. Here we also consider priors we expect to achieve with the upcoming Planck experiment. We calculate Planck priors, using the Fisher matrix formalism and considering both the temperature and E-mode polarisation spectra (i.e., $TT$, $TE$ and $EE$ spectra; see Appendix B in Rassat *et al.* 2008, for details), and consider scales which are not correlated with the galaxy power spectrum (i.e., we do not use information from the Integrated Sachs Wolfe effect on large scales (see Chapter 12)). The Planck priors are included, using:

$$F_{\mathrm{Combined}} = F_{\mathrm{Planck}} + F_{\mathrm{X}}, \tag{13.14}$$

where $X$ is either $C(\ell)$ or BAO. Inclusion of Planck priors helps to break degeneracies between different cosmological parameters.

### 13.4.3   Constraints from the Euclid Imaging Survey

We consider cosmological forecasts for both the $C(\ell)$ and the wiggles only methods described earlier, with and without Planck priors. The constraints are summarised in Table 13.2.

As expected, the wiggles only method has less constraining power than the $C(\ell)$ method for all cosmological sectors. It cannot alone constrain parameters related to the primordial power spectrum ($n_s$). The parameters $\Omega_b$ and $h$ are also difficultly constrained though we find that using a different parameter set using $\Omega_b h^2$ allows this combination to be constrained - though weakly. The $C(\ell)$ method constrains dark energy parameters better than the wiggles only method, though still weaker than the weak lensing method for most parameters as expected (see Chapter 4), except $\Omega_b$ and $h$ which show similar constraints.. This is mainly due to the fact that the $C(\ell)$ ignores radial information, which is not present in this photometric survey.

Table 13.2: Cosmological constraints for Euclid's Imaging survey (i.e. photometric redshifts) using wiggles only and C(l) method. The constraints are presented with and without Planck priors.

| Method | $w_p$ | $w_0$ | $w_a$ | $\Omega_{DE}$ | $\Omega_m$ | $\Omega_b$ | $n_s$ | $h$ |
|---|---|---|---|---|---|---|---|---|
| No priors | | | | | | | | |
| BAO | 0.48 | 1.43 | 7.7 | 0.40 | 0.32 | - | - | - |
| $C(\ell)$ | 0.19 | 0.80 | 2.4 | 0.19 | 0.09 | 0.023 | 0.19 | 0.22 |
| with Planck priors | | | | | | | | |
| BAO | 0.063 | 0.55 | 1.65 | 0.047 | 0.046 | 0.0072 | 0.0038 | 0.055 |
| $C(\ell)$ | 0.13 | 0.36 | 0.82 | 0.039 | 0.030 | 0.0049 | 0.0037 | 0.038 |

By combining the results from the galaxy correlations with Planck priors, we can improve the cosmological constraints on all parameters. The right panel in Figure 13.2 shows the constraints on $w_0$ and $w_a$ with and without Planck priors, and we see the inclusion of priors greatly improves the constraints on both parameters. We also find that the difference between the wiggles only and the projected spherical harmonic methods is less important (see Table 13.4.2). A three-dimensional Fourier space analysis on spectroscopic data can provide competitive constraints, which will be possible using the Euclid Spectroscopic Survey (see the Euclid Yellow Book).

## 13.5 Conclusion

Over the past decades, galaxy surveys have proved to be powerful probes of our cosmological model. The Euclid Imaging survey naturally provides a photometric galaxy survey over the redshift ranges $z = 0 - 2$. Though lacking in radial precision, such as survey can be used to study the tangential modes of the statistical distribution of galaxy.

In this chapter we overviewed the work done by the Euclid Imaging Consortium's (EIC) galaxy correlations working group. We considered the different methods which exist in the literature regarding galaxy correlations, and decided to concentrate on a Fourier space wiggles only method (Seo & Eisenstein 2007) as well as a projected spherical harmonic method [$C(\ell)$] which is ideal to study tangential modes on a sphere.

We found constraints from the spherical harmonic method provided competitive constraints on all sectors of the cosmological method, though these constraints were weaker than those from weak lensing for this survey, as expected. Constraints from the wiggles only method - which only considers the modes of Baryon Acoustic Oscillations (BAOs), were much weaker. We also presented our constraints including Planck priors, for which many degeneracies are broken. The resulting constraints show that galaxy correlations are a useful and complementary probe for the Euclid Imaging Survey.

# Galaxy Evolution with Euclid's Imaging

*Authors: C. Marcella Carollo (Physics Department, ETH Zurich), R. Bender (Max Planck Munich), C. Lacey (Durham University), S. Lilly (Physics Department, ETH Zurich), J. Peacock (ROE, Edinburgh), H.-W. Rix (Max Planck Heidelberg), R. Saglia (Max Planck Munich).*

## Abstract

Euclid's Wide and Deep Imaging Surveys will achieve order of magnitude gains in the combined parameter space of area, depth, wavelength coverage and spatial resolution. The Wide Survey will span a transverse size of $\sim 1$ comoving Gpc at $z \sim 0.1$, up to more than 20 comoving Gpc at $z \sim 6$, with a total volume in excess of $\sim 1200$ Gpc$^3$ returning images for more than a billion galaxies. The Y, J and H wavelength coverage down to $AB \sim 24$ will probe the rest-frame optical light of high-z galaxies and enable photometric redshift measurements to an accuracy of $\Delta z/(1+z) \sim 0.05$, and reliable conversion of the observed galaxy properties to key physical parameters like stellar mass and star formation rate - out to epochs well beyond the peak of the integrated star formation rate in the universe (Lilly *et al.* 1995; Madau *et al.* 1996). The exquisite $< 0.2$ resolution of the $AB \sim 24.5$ optical images will provide detailed morphological and structural information e.g., bulges, disks, bars, spiral arms, lopsidedness out to $z \sim 2$, and fundamental structural information such as size, concentration, at any redshifts. At $AB \sim 26.5$-26, the Deep optical/NIR images will fill in the unexplored niche of area and depth between Euclid's Wide Imaging Survey and the extraordinarily deep but narrow-beam surveys of the James Webb Space Telescope (JWST), and will unveil galaxies and quasars (QSOs) at high redshifts that will remain undetected in ground-based surveys. The Euclid imaging surveys will therefore be a unique resource to study the co-evolution of galaxies and black holes from the reionisation era down to the present time.

## 14.1    Introduction

A crucial advantage of Euclid's Imaging Wide and Deep Surveys over similar past and planned surveys of galaxies will be the sheer number of galaxies ($> 10^9$) available with morphological information and accurate photometric redshifts. This massive database will be important for many areas of galaxy evolution studies, such as the discovery of:

- *The most luminous objects in the very early universe.* The first generations of galaxies were assembled around redshifts $z \sim 7 - 10$, in the heart of the reionization of the universe. Despite great efforts with space- and 8m-mirror ground-based telescopes, not many galaxies (or galaxy candidates) have been reliably detected so far at $z \sim 7$. Samples are too small to draw conclusions on the physical properties of galaxies near the end of the reionization epoch. Euclid's deep imaging survey will dramatically change this situation. Using the Lyman-dropout technique (Steidel 1990) in the near-IR, the survey will detect many thousands of the most luminous – and thus rarest but of key importance – objects in the early Universe at $z > 7$.

• *The largest scale signposts of structure formation at early epochs.* While at present only a few massive galaxy clusters at $z > 1$ are known, Euclid's near-infrared imaging will allow for red-sequence determinations, and thus the identification of hundreds of galaxy clusters at $z > 2$. The latter are the likely environments in which the peak of QSO activity at $z \sim 2$ takes place, and hold the empirical key to understanding the heyday of such activity, as well as its relationship with the formation and evolution of the massive galaxies that host the QSOs.

• *How the bias varies with galaxy type and cosmic epoch.* The bias function that links the galaxies to their host dark matter halos is a major unknown in galaxy evolution. Connecting Euclid's weak lensing maps for the total mass density with the stellar mass and luminosity of more than a billion galaxies out to $z \sim 3$ will solve this mystery.

• *Which physics shapes galaxies – and on what timescales?* Euclid's images will open up the study of the $z \sim 1$-2 universe, at the peak of cosmic activity, at a similar level of detail that the SDSS has achieved on the $z = 0$ universe.

## 14.2 Understanding reionization, and galaxy formation in the reionization era

Understanding galaxy formation at early times is a major challenge of today's astrophysical cosmology. The timescale for the formation of the first stars and first galaxies is unknown, with estimates ranging from $z \sim 30$ to $z \sim 15$ ($\sim 100 - 300$ Myr after recombination). Hints for the timescale for galaxy assembly are provided by e.g., WMAP5; these indicate that reionization extends from roughly $z \sim 11$ to $z \sim 6$ ($\sim 400 - 900$ Myr after recombination). This crucial period is virtually unexplored to date. The current state-of-the-art is set by the groundbreaking images delivered by the new Wide Field Camera-3, recently installed on the HST, which have provided of order 20 $z \sim 7$ galaxies, a handful of robust candidate $z \sim 8$ Ly-break galaxies, and possibly a few robust $z \sim 10$ candidates (Bouwens *et al.* 2009; Oesch *et al.* 2009; McLure *et al.* 2009; Bunker *et al.* 2009). Not much can be learned from such small numbers of galaxies at such fundamental epochs in our universe. Large samples are essential to understand the intrinsic physical properties of $z > 7$ galaxies – and the physics of reionization.

Studying early-epoch galaxies without star-formation, studying the highest redshifts ($z > 7$) and studying the faint and, at early epochs representative part of the luminosity distribution, is the realm of Euclid optical and NIR Deep Imaging Survey, whose four-colour imaging will provide galaxy populations statistics by itself. Using the Lyman-dropout technique in the near-IR, the survey will detect thousands of the most luminous objects in the early Universe at $z > 6$: for star formation rates of order and above 30 $M_\odot$/yr, about $10^4$ star-forming galaxies at $z \sim 8$ and up to $10^3$ at $z \sim 10$.

High redshift quasars (QSOs) may also play a crucial role in reionising the universe at epochs beyond $z \sim 7$. To date, no quasars have been found above $z = 6.4$, and presently planned ground-based programmes will struggle to push this beyond $z \sim 7$ because of the need for deep, wide area NIR imaging and spectra (see e.g., Venemans *et al.* 2007). Euclid's visible and NIR Wide Imaging Survey will identify likely QSOs by their colours out to redshifts $z \sim 12$ and above, if they exist. Many of these objects may be difficult to find by spectroscopic techniques, as their Ly$\alpha$ lines are often weak or even absent (e.g. Fan *et al.* 2006). The likely steep luminosity function of these objects implies that maximal imaging depth at $\sim 2\pi$ area is paramount: even a factor of two in flux limit improvement will provide a ten-fold increase in the expected number of objects. Such searches, and their follow-ups, may well hold the key on the rate of black hole growth at $z > 8$ and hence answer questions about the masses of likely seed-black holes, a fundamental open question in galaxy formation. Euclids deep imaging survey should furthermore add up to of order a thousand QSOs at $z > 7$, depending on the (unknown) details of the LF at such early epochs.

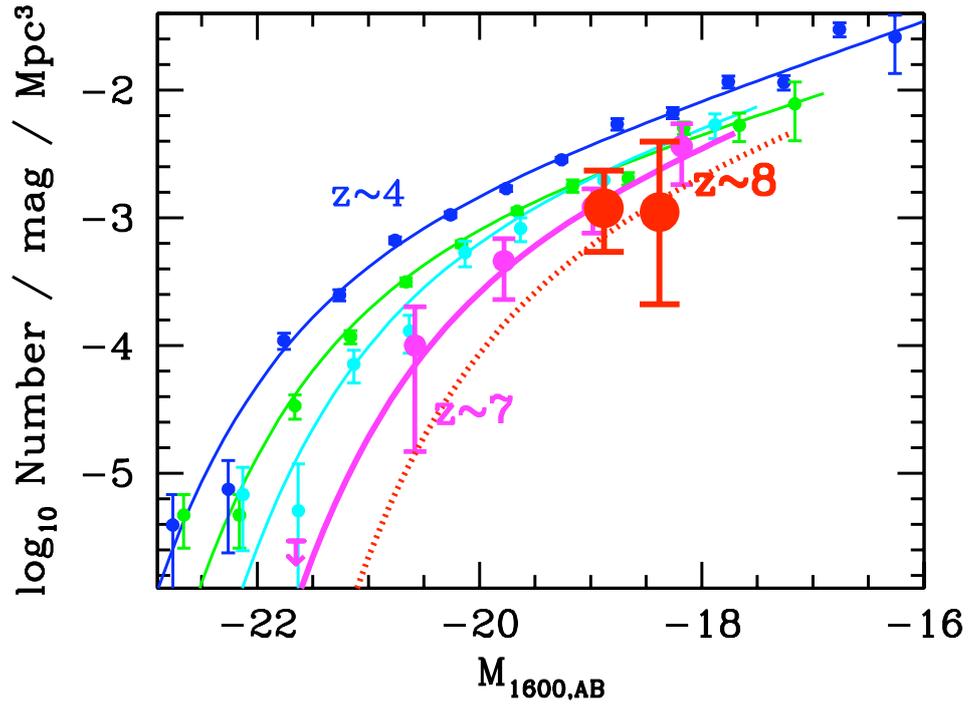

Figure 14.1: Figure reported from Bouwens *et al.* (2009), to show the state of the art knowledge on the *UV* Luminosity Function of galaxies at $z \sim 8$ from the present *Y*-dropout searches in the Hubble Ultra Deep Field WFC3/IR images (solid red circles and $1\sigma$ error bars). Also included are: (*i*) the *UV* LF determination at $z \sim 7$ from a *z*-dropout search over the same ultra-deep WFC3/IR field (magenta lines and points: Oesch *et al.* 2009); and (*ii*) the Bouwens *et al.* (2007) *UV* LF determinations at $z \sim 4$, 5 and 6 (shown in blue, green, and cyan lines and points). The dotted red line gives one possible Schechter-like LF that agrees with the observed data points at $z \sim 8$. Key is however to note that the $z \sim 8$ LF is totally unconstrained at the bright end. This will be determined with exquisite accuracy by Euclid's Deep Imaging Surveys - together with the physical properties of the $z \sim 8$ Lyman-break galaxy population.

Detection of these high redshift sources – Lyman-break galaxies and quasars – in Euclid's images will provide crucial targets for the European Extremely Large Telescope (E-ELT) and the JWST which will achieve greater depths but on much smaller areas, thus probing a fully complementary part of the luminosity functions.

## 14.3 Unveiling clusters and massive groups in the early universe

Euclid's near-infrared imaging will allow the identification, through their red sequence, of hundreds of galaxy clusters at $z \sim 2$-3 with $M > 10^{14} M_\odot$ i.e., comparable with the Virgo Cluster, and thousands of structures with $M > 1.0 - 3.0 \times 10^{13} M_\odot$, the mass scale of the massive groups. Current models of galaxy formation predict that many of the processes that lead to galaxy transformations should occur in such massive-group environment. First, massive groups should have an enhanced efficiency of tidal forces and dynamical friction. The in-spiral timescale from dynamical friction varies as $\sigma_{halo}^3/\rho_{halo}$ (with $\sigma_{halo}$ and $\rho_{halo}$ respectively the velocity dispersion and density of the dark matter halo), and it is thus minimised in relatively high densities but low velocity dispersion groups (Barnes 1990). Furthermore, state-of-the-art cosmological simulations indicate that ram pressure, which creates an induced pressure force proportional to $\propto \rho\sigma^2$, is already active at such intermediate mass scales (Rasmussen *et al.* 2008). Clear signatures of galaxy strangulation have also been observed at the intermediate mass scales of massive galaxy groups (Kawata & Mulchaey 2008). Finally, cold streams of 'unvirialised' gas persist in such potentials down to intermediate redshifts – and thus sustain enhanced star formation rates over substantially longer timescales than major mergers (Dekel *et al.* 2009).

Euclid's Wide Imaging Survey will thus dramatically advance our understanding of the evolution of galaxies in massive clusters and in the intermediate-mass potentials of massive groups. Euclid's optical morphologies at sub-kpc spatial resolution, and deep NIR galaxy images for robust stellar mass estimates, will be a fundamental ingredient in mapping the growth of stellar mass in the member galaxies, as a function of galactic structural properties and cluster-centric distance, over a wide span of galaxy- and cluster/group mass scales. The $> 1$ billion galaxies sample with morphological and spectrophotometric diagnostics, over the full range of galaxy properties, will allow to disentangle local effects of evolution from effects related to the global properties of the clusters, and to identify the physical processes responsible for differential galaxy evolution in high-, intermediate- and low-density environments.

The next generation of wide-area Sunyaev-Zeldovich (e.g., SPT, CCAT) and X-ray (e.g., eROSITA, WFXT) surveys will detect a huge number of distant clusters ($\sim 5000$ for eROSITA to $> 10^5$ for WFXT), but only above the $\sim 10^{14} M_\odot$ mass scale, and/or only up to $z \sim 2$. Capitalising on the large area covered down to $AB \sim 24$ in the NIR, the Euclid Wide Imaging Survey will be uniquely positioned to detect bound structures at higher redshifts, and down to one order of magnitude smaller scales.

## 14.4 Linking matter and light

Euclid's optical and NIR Imaging Surveys will provide rest-frame optical luminosities and stellar masses for the bulk of the galaxy population (to $\sim 3.5$ magnitudes below $M*$) out to $z \sim 3$, and the bulk of the unobscured star formation to $z \sim 6$ through rest-frame UV light. Comparison between the distribution of light and stellar mass – and weak lensing maps – will provide a direct mapping between total mass density and galaxies distribution and properties (stellar mass, luminosity, morphological type, etc), the *bias*, which is at the core of understanding galaxy formation. It will furthermore unveil or at least constrain the existence and abundance of 'naked' dark matter peaks.

The Euclid's images will also uncover a very large number of strong lensing systems: about $10^5$ galaxy-galaxy lenses, $10^3$ galaxy-quasar lenses and several thousand strong lensing arcs in clusters. Several tens of galaxy-galaxy lenses will be double Einstein rings (Gavazzi *et al.* 2008), which are exceptionally powerful probes of the cosmological model as they simultaneously probe several redshifts. Euclids large-area exploitation of Natures telescope, strong gravitational lensing, will assemble an unparalleled sample of intrinsically extremely faint, but highly magnified, objects at high redshifts. These are exciting targets for synergetic spectroscopic studies with the JWST of the seeds of the massive galaxies that are detected in the universe at later epochs.

Furthermore, the shapes of galaxies measured on Euclid's images, in addition to measuring the cosmic shear, will reveal possible alignments of galaxy ellipticities induced by tidal interactions. The presence of such intrinsic alignments is a source of contamination in Euclid weak lensing measurements, which must be carefully calibrated and subtracted from them (e.g., Hirata & Seljak 2004). Additional cross-correlation effects of the intrinsic ellipticities and the cosmic shear must also be calibrated and subtracted separately. Such alignments would however provide crucial information on the hierarchical assembly of structure in the universe. Recent measurements of intrinsic ellipticity correlations (Okumura *et al.* 2009; Brainerd *et al.* 2009) show the presence of a strong alignment that varies among different galaxy types. On large scales, the mean galaxy ellipticity alignment should behave linearly with the tidal field, and observations of the intrinsic alignments will provide a basic test for the assumed relationship between the observed galaxy density and the tidal field.

## 14.5 Identifying the physical drivers of galaxy evolution

The relative importance of internal vs. external drivers of galaxy evolution, and the relative timescales of star formation and mass assembly in galaxies, are still unknown. Galaxy properties at $z < 1$ are strongly dependent upon galaxy mass and current star formation rate. Massive galaxies (in dense environments) have old stars which passively evolve without further star formation. In contrast, less massive galaxies (in less dense environments) are still actively forming stars down to today. What physical mechanisms are responsible for quenching the star formation in the most massive galaxies (and thus driving to the Hubble sequence and the morphology-density relation)? Galaxy formation models can reproduce the 'red&dead' $z = 0$ massive galaxies by assuming that their host massive dark matter halos produce shock-heated, quasi-hydrostatic hot gas halos (e.g., Hopkins *et al.* 2008). Cold gas accretion from dense filaments in the low-mass halos prevents such form of 'heating' – and thus 'quenching' of star formation (Dekel & Birnboim 2006). Within this paradigm, the critical mass separating 'hot accretion' halos from 'cold accretion' halos should evolve with cosmic time (and was larger at earlier epochs). By providing detailed internal galactic structure (bars, spiral arms, lopsidedness, etc) for many millions of $z < 1$ galaxies as a function of cluster/group properties, and global structural, photometric and physical properties of central (and satellite) galaxies at earlier times, $z \sim 2$, when cosmic star formation rate and QSO activity peak their efficiencies, Euclid's images will directly test key predictions of galaxy formation models.

## 14.6 Conclusions

The last decade has seen unprecedented advances in our understanding of galaxy formation. Fundamental questions remain open: how is star formation in galaxies regulated, triggered, and quenched? How does the energy released by accreting black holes affect their host galaxies and their surroundings? How does the large scale environment affect galaxies? What is the relationship between dark matter halo mass and galaxy mass and properties? The next decade will be a critical period where these main open questions in astrophysical cosmology will be resolved. Euclid's Imaging Surveys will be instrumental in providing the answers by producing a massive legacy of deep images over at least half of the entire sky.

# Strong lensing with Euclid

*Authors: Matthias Bartelmann (Zentrum für Astronomie, Universität Heidelberg), Leonidas Moustakas (JPL, Pasadena), Massimo Meneghetti (Bologna Observatory), Frédéric Courbin (EPF Lausanne).*

## Abstract

Euclid will provide an excellent data set not only for weak, but also for strong lensing. The most important strongly lensing objects are individual galaxies, galaxies in groups, and galaxy clusters, whose strong lensing effects give rise to multiple or highly distorted images. Strong lensing sensitively probes the mass distribution in the innermost parts of these objects and thus allows to study mass profiles, concentrations, substructures, total masses and their relations to observable emission e.g. in the optical or X-ray regimes. Besides information on individual objects, strong lensing is cosmologically important because it constrains non-linear structure growth. Dedicated search algorithms exist which can scan huge amounts of data automatically and efficiently for the rare cases of strong lensing. Euclid is expected to find of order $10^5$ galaxies lensed by foreground galaxies, $\lesssim 10^3$ multiply-imaged quasars, and $\approx 5000$ clusters containing strongly distorted arcs.

## 15.1 Introduction

Strong and weak gravitational lensing are related but distinct phenomena. When the surface mass density in a "lens" is above a critical threshold, the appearance of an aligned background object can be both distorted as well multiply imaged into several images, arranged in a highly distinctive (and easy to identify) pattern. The specific appearance or arrangement of these multiple images is sensitive to the underlying cosmology through the relative distances, as well as to the distribution of the total lensing mass (meaning the combination of dark plus luminous matter). Thus, while weak lensing predominantly probes linear and weakly non-linear structures, strong lensing probes into the cores of non-linear structures and provides astrophysical information which cannot otherwise be obtained.

Strong lensing is a phenomenon that can occur across many scales of lensing mass. Individual galaxies, groups of galaxies, and clusters of galaxies, can all lens more-distant objects. The objects lensed are generally either distant quasars, or extended and faint background galaxies. In galaxy clusters, the predominant features are strongly distorted, arc-like images formed from background galaxies and multiple galaxy images. Multiple imaging of quasars by clusters does not play an important role because it is a very rare phenomenon.

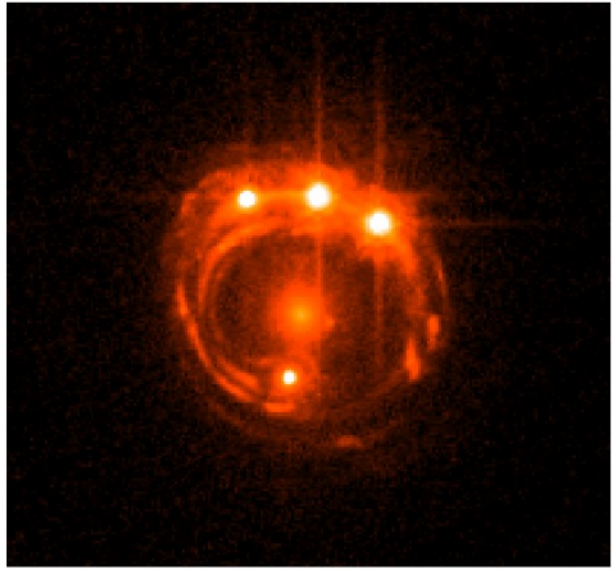

Figure 15.1: Example of a quasar, RX J1131−123, seen quadruply due to the strong lensing effect of a foreground galaxy. The host galaxy of the lensed quasar is seen as an almost complete Einstein ring that can be used to constrain the mass model of the lens, in the centre of the ring. The side length of this HST image from CASTLEs is about 10 ″.

Euclid will produce an ideal dataset for strong-lensing studies. Strong lensing is rare, thus surveying wide areas is essential, and many strongly lensed images are faint, thus depth is similarly crucial. Euclid combines both to the degree that will allow substantially increasing samples of strongly lensed systems. Equally importantly, these large samples will be homogeneously selected and thus statistically complete. The applications of these phenomena to astrophysics and fundamental physics are both diverse and powerful, and we explore some of the highlights here.

## 15.2 Objects of Interest

A broad variety of scientific topics can be addressed based on a large, homogeneously selected data base of strongly lensed objects. Searching for strong-lensing events in the Euclid imaging data targets three broad classes of lensing objects, in which strong lensing gives rise to a variety of different effects. We explore these in a general sense in this section, and concentrate on specific astrophysical applications in a subsequent section.

### 15.2.1 Individual, mostly massive galaxies

Individual lensing galaxies were the first in which gravitational lensing was detected, in the form of multiply imaged quasars (Walsh *et al.* 1979; Muñoz *et al.* 1998). Since the light-travel times between different images differ, time delays can be observed in systems whose source is variable.

For galactic lenses, the time delays are between a few and several hundred days. They have attracted considerable attention because they can be used to constrain the Hubble constant and thus the scale of the Universe without invoking a distance ladder (Refsdal 1964). Time delays are difficult to measure as they need frequent and long-term observations with high angular resolution. Nonetheless, they start to be measured with a few percent accuracy (e.g., Vuissoz *et al.* 2008; Oscoz *et al.* 2001). In combination with high-resolution imaging, as is currently done with the HST (or in the future with the JWST), time delays lead to measurements of the Hubble parameter that are fully competitive with, and complementary to, other cosmological probes (e.g., Suyu *et al.* 2009). Euclid will provide hundreds of lenses suitable for time delay measurements.

Background galaxies strongly lensed by foreground galaxies typically appear as highly curved arcs or partial rings (Hewitt *et al.* 1988; Warren *et al.* 1996; Impey *et al.* 1998). They constrain galaxy mass models substantially better than individual, point-like images because of their much larger number of image points. Some systems contain both point sources and arcs. For these, time delays can be measured thanks to the variability of the point source, and the lens model can be well constrained based on the arc feature (Fig. 15.1). Strong lensing occurs for sources located next to so-called caustic curves. Mathematically, these are catastrophes of the lens mapping. Peculiar types of image configurations can arise at catastrophes, more precisely metamorphoses, of critical curves (Blandford & Narayan 1986; Marshall *et al.* 2005). They are general in the sense that they do not depend on details of the lensing mass model and thus allow constraints independent of the precise mass distribution.

### 15.2.2  Galaxies in groups, whose environment is crucial for their strong lensing

These objects are interesting because their lensing effects are enhanced and can be largely modified by their environment. They can give rise to complicated configurations of multiple images (Keeton & Zabludoff 2004) or to arcs or (partial) Einstein rings around galaxy groups (Cabanac *et al.* 2007). Both types of image configuration are most useful for studying the mass distribution surround galaxies. Studying the effect of groups on the potential well of individual galaxies is also relevant for cosmology, as they modify time delays and thus the value inferred for the Hubble parameter (e.g., Momcheva *et al.* 2006). On even larger scales, it has been shown that large-scale structures influence the lensing effects of individual galaxies or by groups (Faure *et al.* 2009).

### 15.2.3  Galaxy clusters

Strong lensing by clusters typically causes spectacular effects. Most notable are the formation of large arc-like images (Fort & Mellier 1994; Fort *et al.* 1992), so-called (giant) arcs, which form either very close to the cluster cores and are then radially oriented, or they are located farther away and oriented tangentially to the cluster centre. Near higher-order catastrophes or metamorphoses in caustic curves, peculiar types of images can appear, such as straight arcs (Kassiola *et al.* 1992). Arc systems from sources at multiple redshifts (Kneib *et al.* 1993; Broadhurst *et al.* 2005) are very helpful in analysing the core mass distribution of galaxy clusters.

## 15.3   Detection Algorithms

Successful searches in the Euclid imaging data require automatic discovery techniques for these types of object. Various algorithms have been suggested and demonstrated to be useful:

### 15.3.1   Searching for clear identifying features in lenses

Although strong lens searches have seen enormous success by mining the spectroscopic portion of the Sloan Digital Sky Survey (SDSS; e.g., the Sloan Lens ACS Survey, SLACS, Bolton *et al.* 2006), searches with Euclid will be based on some combination of morphological and colour information from the comprehensive imaging it will acquire. There have been several pioneering explorations on the theory and application of designing such searches recently, in both ground- and space-based data, all of which will inform how the most effective search strategy should be built for Euclid. These include a) identifying multiple-image candidates based on large-separation objects in the SDSS (Oguri *et al.* 2006; Inada *et al.* 2003); b) searches for candidates based on a combination of multiple or highly distorted objects combined with colour information, as applied by the very successful CFHTLS-SL$^2$S survey (Cabanac *et al.* 2007); and c) a careful exploration of a probabilistic and automated search for lenses in space-quality data, which includes both detailed modelling of the candidates, but also inclusion of available colour information (Marshall *et al.* 2009).

### 15.3.2   Specialised algorithms

Such algorithms were recently proposed and tested to search for highly distorted images. The most relevant of these are the algorithm by Cabanac et al. and used within SL$^2$S (Cabanac *et al.* 2007) to detect arc-like images mainly around galaxies, the algorithm by Horesh et al. used for comparing real and simulated cluster lenses (Horesh *et al.* 2005), and the fast algorithm by Seidel et al. specifically designed for scanning huge data sets and minimising spurious detections (Seidel & Bartelmann 2007). These algorithms have their respective merits. The procedure by Cabanac et al. includes colour information into the image selection. Horesh et al. use the widely used image-analysis software packages IRAF and SExtractor and parameterise them such as to optimise the detection of faint, elongated images, and Seidel et al. devised a novel, highly efficient algorithm that avoids convolutions and is sufficiently fast for wide-area surveys.

## 15.4   Astrophysical Applications

Numerous astrophysical applications can be based on sufficiently large and homogeneously selected samples of strong lenses. We describe here a representative set.

### 15.4.1   Galaxy structure and its evolution over cosmic time

Modelling the image configuration, in some cases also their fluxes and their morphology, density profiles of the lensing galaxies and the haloes they are embedded in can be measured, whether the lens is an early-type galaxy or a spiral galaxy. Since gravitational lensing is sensitive to all forms of inhomogeneously distributed mass rather than the luminous matter only, strong lensing can thus constrain the balance between dark and luminous matter, when combined with galaxy scaling relations. Since lenses will be found across many redshifts, or epochs, this mapping will be done over some 10 billion years of cosmic time.

The general mathematical properties of imaging near caustics, their cusps and their higher-order metamorphoses, demand certain imaging properties that are independent of the detailed mass model. Perhaps the most prominent of those is a mathematical statement on the magnification ratios within image triplets, which is violated in almost all suitable observed cases. These so-called anomalous magnification ratios can help quantifying the amount of substructure in galaxies. Higher-order catastrophes and metamorphoses, such as umbilics, swallowtails, butterflies, in lensing caustics can be exploited in a similar manner, provided sufficiently many of such cases can be detected.

### 15.4.2 Cluster structure

Strong lensing by clusters, in particular the large arcs produced by it, allows inferences on the cluster density profiles, in particular by combining radial and tangential arcs with weak-lensing maps. Qualitatively, radial arcs constrain the local slope of the density profiles, while tangential arcs place limits on the mass enclosed by them. Density profiles of dark matter haloes are a firm prediction of the Cold Dark Matter paradigm, thus testing them observationally is cosmologically very important. While they are substantially modified on galactic scales by baryonic physics, they are expected to be conserved in clusters well into their cores. Core sizes, or equivalently concentration parameters, are determined in the CDM paradigm by the formation time of the most massive progenitor halo. They can effectively be constrained by strong lensing. Values derived so far indicate that strongly lensing clusters seem to be special because their concentrations are routinely measured to be much higher than expected theoretically. The morphology of observed large arcs helps quantifying the amount of cluster substructure. An obvious example is the "straight arc" in the cluster Abell 2390, which indicates the presence of a structure comparable in mass to the main cluster body, however much less luminous both in X-rays and in visible light. Automatic arc-detection algorithms are now efficient enough to be run over wide-area surveys to search for objects strongly lensed by underluminous or dark halos, allowing expectations on the mass-to-light ratio of galaxy clusters to be tested.

### 15.4.3 Cosmology

The possibility to test the CDM paradigm through substructure and halo density profiles was mentioned before but is of course cosmologically highly important. Similarly, reliable constraints on the total mass of dark-matter haloes cannot be obtained by other means than lensing without further simplifying and specialising assumptions. Halo masses are important to know on the galaxy scale in view of galaxy formation in a universe with low $\sigma_8$, and on the cluster scale because scaling relations between cluster temperature, luminosity and mass need to be externally calibrated. Mass determinations by methods other than gravitational lensing suffer from the necessity to adopt hydrostatic or virial equilibrium, which can hardly be independently tested.

Another important subject of cosmological relevance is the growth of non-linear structures throughout cosmic history. Since strong lensing relies on high surface mass densities, it constrains the growth of nonlinear structures through the occurrence of sufficiently compact and massive halos, which is particularly interesting at high redshift.

Table 15.1: Expected numbers of various strong-lensing events in the Euclid survey area

| | |
|---|---|
| Galaxies lensed by galaxies | $\approx 10^5$ |
| Quasars lensed by galaxies | $\lesssim 10^3$ |
| Clusters containing strongly lensed arcs (with a length-to-width ratio $\geq 10$) | $\approx 5000$ |

The geometry of the Universe is constrained by strong lensing through time delays between several of the images in multiply-imaged quasars. Given a model for the lensing mass distribution, such time delays measure the overall scale of the Universe through the inverse of the Hubble constant, as well as the overall expansion history of the universe through the angular diameter distances that are involved (e.g., Coe & Moustakas 2009).

Galaxy clusters containing multiple arc systems originating from sources at different redshifts also probe the cosmic geometry directly. Such sources experience lensing effects of different strength from the same mass distribution, where the difference depends solely on the cosmic geometry.

Statistically, the abundance of strong-lensing events place cosmological constraints through the comparison of survey results, in particular distributions of image separation, multiplicity and morphology, with expectations from simulations. Moreover, samples of strongly lensing galaxy clusters, selected by their lensing effects, provide cosmological information through their comparison to samples defined by the occurrence of a red cluster sequence, by their X-ray emission or their thermal Sunyaev-Zel'dovich effect.

## 15.5 Predictions

Euclid is expected to find thousands of interesting examples of strong gravitational lensing. The estimates given below are based on the following assumptions; The telescope will have a diameter of 1.5 metres; The exposure time is of order 1500 s; The FWHM of the PSF is $0.3''$; The pixel size is $0.1''$; and The approximate quantum efficiency $Q \approx 0.1$. This configuration is close to current Euclid baseline of 1.2m diameter, with an exposure time of 1800 seconds and a PSF FWHM of 0.18". The assumptions imply a 10-$\sigma$ point-source detection limit of $I_{AB} \approx 25$, and a 3-$\sigma$ detection limit for extended sources with $\approx 1''$ diameter of $I_{AB} \approx 25$.

Based on current optical-depth calculations for lensing of QSOs and galaxies and of galaxies by clusters, and on source counts from the HDFS, we estimate the numbers of lensing events on the expected 20,000 square degrees of the Euclid survey given in Tab. 15.1:

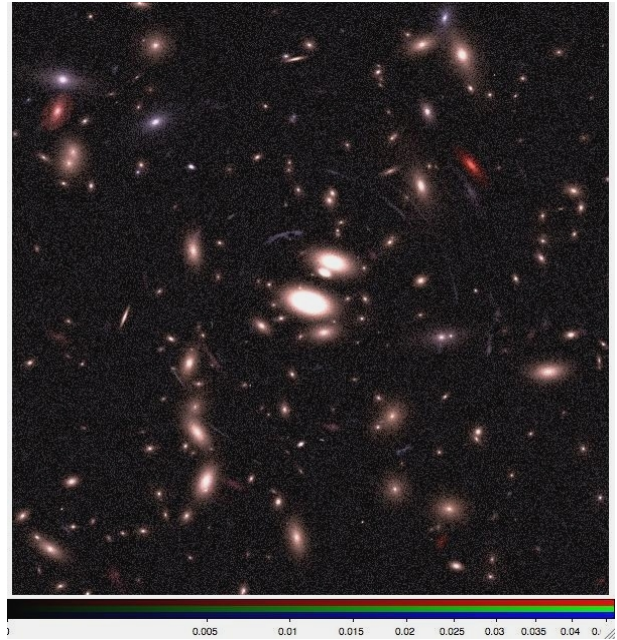

Figure 15.2: Simulated multicolour observation of a cluster core. The simulation assumes here to being able to observe the cluster in three bands (B,V,I), although Euclid will work with only one optical filter (a broad band RIZ filter, covering the wavelength range 550...920 nm)

## 15.6 Preparatory Work for Euclid

The SL working group has already developed tools which are being tested and particularised for a future usage with Euclid data. In particular, several studies have be undertaken, which make use of numerical simulations. In synergy with the Image Simulation WG, we have used state-of-the-art codes for image simulations for producing mock observations of galaxy cluster fields. The simulations take into account the optical design of Euclid, as well as its throughput and quantum efficiency of the camera. All the relevant noises have been included to make these simulations fully realistic. The galaxy morphologies are modelled using shapelet decompositions of real galaxies from the Hubble-Ultra-Deep-Field. Last but not least, the image simulations include lensing effects from galaxy cluster mass distributions, which are obtained from numerical and hydrodynamical cosmological simulations. An example of such images is shown in Fig. 15.2. These images are currently being used for the following applications:

- Train algorithms for arc detections. In particular, the code by Seidel & Bartelmann (2007) is being used and calibrated with such images. A comparison with other instruments is also done, highlighting the great capabilities of Euclid for SL studies (Seidel et al. in prep).

- In a paper by Meneghetti *et al.* (2008), we show that arcs arising from galaxies down to an apparent magnitude of 27 will be detectable with Euclid. We study how the arc properties change as a function of the morphology of the sources and we define some empirical methods to measure the length, the width and the curvature radius of each lensed image (see the left panel in Fig. 15.3).

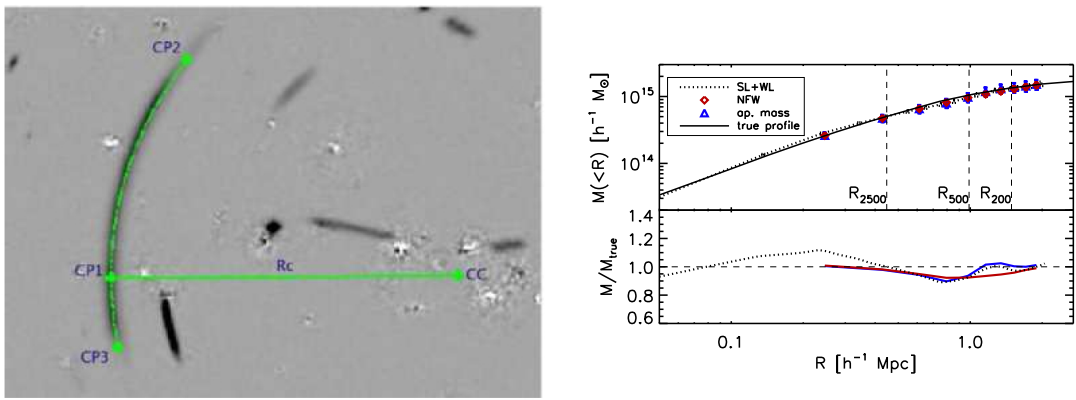

Figure 15.3: Right panel: detection and fitting of a gravitational arc in a simulated Euclid observations. The procedure involves fitting an arc through three characteristic points, measure the length and the curvature radius of the arc, and finally performing an azimuthal scansion of the arc to measure the width profile. Such properties can be used for arc statistics applications. Left panel: the measurement of a cluster mass profile. This is done analysing the images with several methods used for analysing real data. Among them, our method which combines strong and weak lensing constraints (dotted line). The solid line shows the true profile of the simulated cluster. The bottom panel shows the ratios between estimated and true masses.

- We have developed and tested with simulations some tools for reconstructing the mass of gravitational lenses using both parametric and non-parametric methods (Comerford *et al.* 2006; Cacciato *et al.* 2006; Merten *et al.* 2008). Using parametric methods we will be able to constrain the projected masses within the Einstein radii with an accuracy of order $\lesssim 5\%$. Our non-parametric code combines the strong lensing constraints in the cluster centres with the weak lensing signal in the external regions, and it allows to measure the mass profile from kpc to Mpc scales with an accuracy of order 10% (Meneghetti et al. in prep, see left panel of Fig. 15.3).

- We have developed an automatic image-deconvolution pipeline to unveil multiple point sources with separations as small as half the FWHM of the PSF. These algorithms have been tested on ground-based data such as the PQUEST survey and are currently used on SDSS to find small-separation lenses.

# Cosmology and Astrophysics with a Euclid Supernova Survey

*Authors: Isobel Hook (U. Oxford, UK and INAF Obs. Rome, Italy), Massimo Della Valle (INAF-Obs Capodimonte and International Centre for Relativistic Astrophysics, Pescara, Italy), Filippo Mannucci (INAF- Obs. Arcetri, Italy), see footnote.*[1]

## Abstract

During the repeat imaging of the deep survey fields, Euclid will find thousands of distant supernovae and other transients. As described below, supernovae have powerful uses as cosmological probes of dark energy (in a way that complements WL and BAO), as probes of the star formation history of the Universe and of stellar evolution.

## 16.1 Introduction

Supernovae are the violent explosions that occur at the end of the lives of some stars. The luminosity of these explosions can be comparable to that of an entire galaxy, allowing them to be observed at cosmological distances. In the sections below we describe how a sample of distant supernovae studied with Euclid could be used for cosmology and other applications in astrophysics.

## 16.2 Constraining Dark Energy with Type Ia Supernovae

Type Ia Supernovae (SNe Ia) provide accurate and well-understood cosmological distance indicators for cosmology, and were used to produce the first direct evidence for the accelerating expansion of the Universe (Riess *et al.* 1998; Perlmutter *et al.* 1999). Since then, several major surveys have amassed large samples of SNe Ia to further improve constraints on the properties of the dark energy driving that acceleration (e.g. see Figure 16.1). Recent results based on ∼250 distant SNe Ia from the Supernova Legacy Survey, in combination with WMAP-5 data and BAO data from SDSS and assuming a constant $w$ and a flat Universe, constrain the equation of state parameter $w$ to better than 5% statistical accuracy (Sullivan et al, in prep), and about 7% when systematic uncertainties are included. This is the tightest constraint on $w$ to date. Moreover, the cosmological constraints from SNe are complementary (i.e. in a different direction in the $w - w_a$ plane) to those from weak lensing and BAO.

---

[1]Much of this text was initially developed for the Euclid Yellow Book in collaboration with Anne Ealet, Ariel Goobar, Bruno Leibundgut, Bob Nichol and Pierre Astier, whose input is gratefully acknowledged. A summarised version appears in the Yellow Book.

The statistical uncertainty on $w$ from SNe Ia is now reduced to the level where systematic effects are comparable (e.g. see lower right panel of Figure 16.1, reproduced from Kowalski *et al.* 2008). To make major advances it will be necessary to control these systematic effects. The Euclid deep survey provides a unique combination of area, stability and depth, particularly in the infrared where the sky background is significantly lower than from the ground and where the effects of extinction by dust are reduced. This will allow us to create a larger and, more importantly, better controlled (in terms of systematic effects) SNe Ia sample for cosmology.

By the time Euclid enters operation, large, ground-based SNe surveys are expected to have collected of order 1000 well-measured SNe with redshifts up to $z \sim 1$. Perhaps a few hundred of those with redshift $z < 0.5 - 0.6$ will have near-IR measurements (for example from VISTA), which will provide control on systematics. In addition a smaller number (perhaps 50) at higher redshifts will have been studied from space using HST.

Euclid, by repeat imaging of 10-square degree patches within the deep survey, could detect and follow $1000 - 2000$ SNe Ia with $z < 0.7$ in the J band during the five-year mission lifetime (see Fig 16.2). This is the redshift range where the acceleration of the Universal expansion (and hence the direct effect of dark energy) is strongest. In addition, detections at peak magnitude will be available for a further $1000 - 2000$ SNe Ia up to $z \sim 1$.

The power of such a Euclid SN survey would be greatly enhanced if coordinated with a ground-based campaign obtaining optical light curves and spectroscopic redshifts and classification.

### 16.2.1 Using Euclid near-infrared photometry to improve the accuracy of cosmological measurements

The key aspect of any Euclid SN survey will be the availability of deep and temporally stable data (important when subtracting images) in the NIR. Such data are extremely difficult to obtain from the ground. There are two ways in which Euclid near-IR photometry would advance cosmological measurements from SNe Ia:

**(a) Better constraints on extinction by dust** Accurate cosmological measurements from SNe rely on correcting for extinction by dust, which is complicated by the fact that SNe Ia also show intrinsic colour variations that correlate with luminosity. In the near-IR not only are the effects of dust much reduced, but the combination with data at shorter wavelengths allows the two effects to be separated. Extinction measurements towards some well-measured SNe show that the dust extinction law (characterised by $R_V$) differs from that of the Milky Way (e.g. Krisciunas *et al.* 2004). Euclid near-IR photometry will allow us to calculate the extinction by dust along the line of sight towards each SN individually, thereby reducing the systematic error introduced by global assumptions about dust properties.

**(b) Lower intrinsic dispersion in the Hubble diagram** Traditional SN cosmology is based on the rest-frame B-band Hubble diagram. However, at longer rest-frame wavelengths, SNe Ia peak magnitudes are believed to have smaller intrinsic dispersion. The NIR imaging data will allow us to measure distances in the rest-frame I-band where the scatter is only 0.13 mag (Freedman *et al.* 2009).

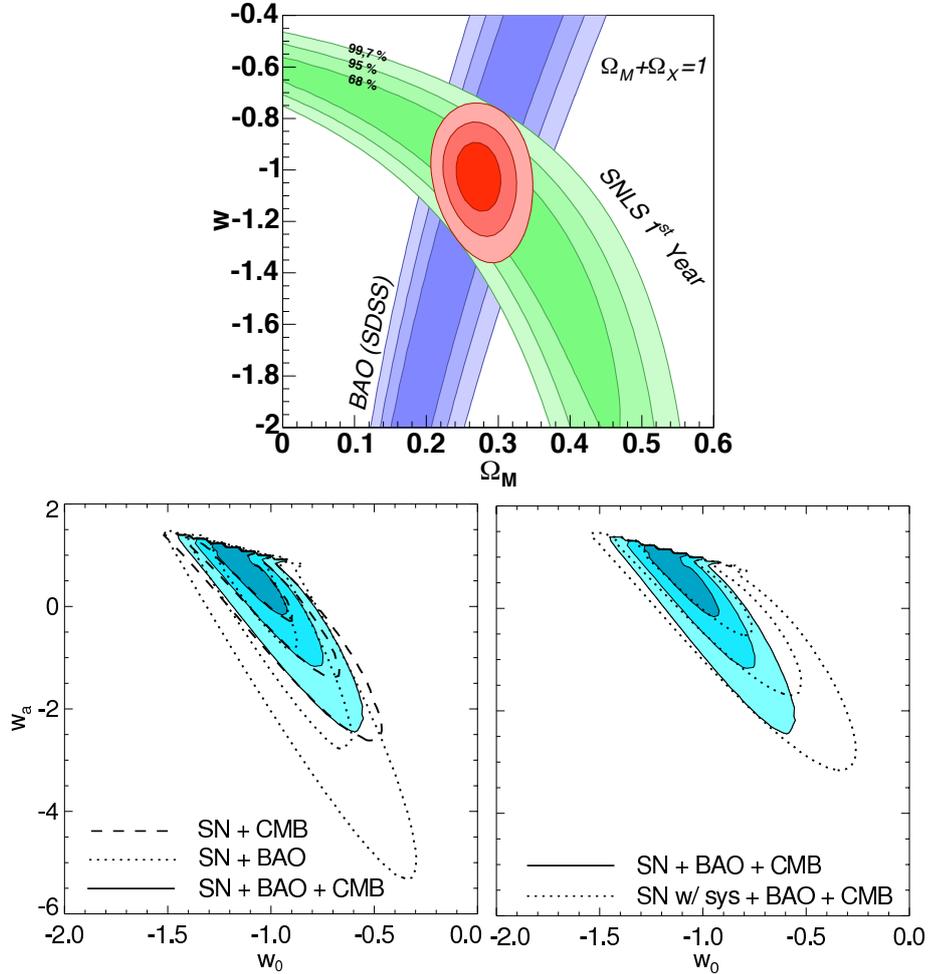

Figure 16.1: *Top*: (Reproduced from Astier *et al.* 2006) Constraints in the $\Omega_M - w$ plane from the first year Supernova Legacy Survey data, assuming a flat Universe and a constant $w$, are shown in green. The combined constraints with SDSS BAO results (Eisenstein 2005) are shown in red. The statistical error on $w$ of $\sim 9\%$ improves to better than 5% when the 3rd year SNLS data are included (Sullivan et al in prep). *Bottom left* (reproduced from Kowalski *et al.* 2008): Contours at 68.3%, 95.4% and 99.7% confidence level in the $w - w_a$ plane from the Union SNe Ia compilation (consisting of 307 SNe Ia compiled from multiple surveys) when combined with BAO and/or CMB constraints (from the 5-year WMPA data release, Dunkley *et al.* 2009) and assuming a flat Universe. *Bottom right*: as for previous plot but showing the effect of including systematic effects in the supernovae analysis.

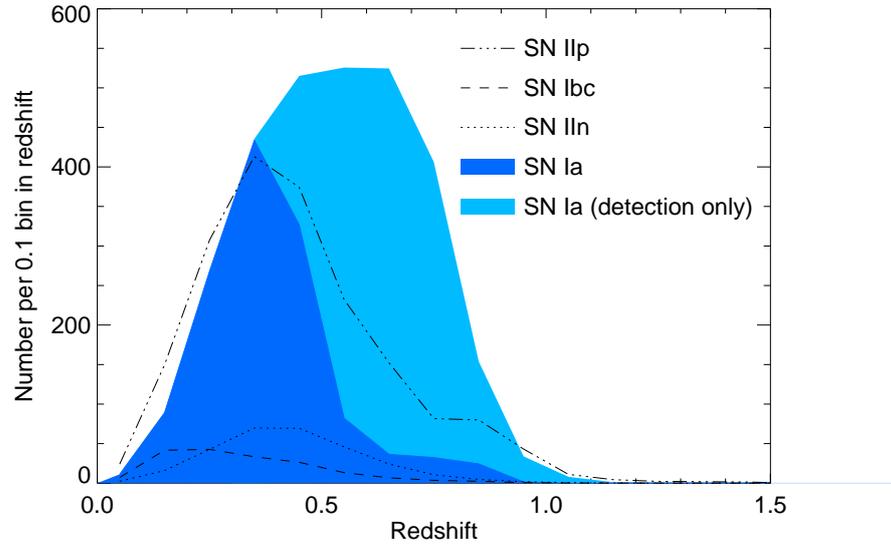

Figure 16.2: Number of SNe of various types that are expected to be detected by Euclid in the J band, as a function of redshift. Estimates for SNe of type Ia (dark blue shaded region), Ibc, IIn and IIp were provided by A. Goobar based on assumptions in Goobar *et al.* (2008), using SNe Ia rates from Dahlen *et al.* (2004) and assuming a 5 year survey that monitors a patch of 10sq deg at any time. These histograms represent the $N(z)$ for SNe with sufficient sampling to measure their lightcurve shapes (i.e. reaching 1 magnitude fainter than the peak brightness). The light-blue shaded region shows an independent estimate of the total number of SNe Ia detections including those only detected at peak luminosity, i.e. without full lightcurve measurements.

Overall we expect that the J-band photometry from the Euclid deep survey will be the most sensitive for supernovae, with the Y and H bands providing additional colour information. The optical component of the deep survey will also provide useful information (for example morphology of SN host galaxies and position of the SN within its host) but the single broad optical R+I+Z filter is difficult to calibrate for precision light-curve photometry, hence the benefit of a coordinated ground-based optical survey. When combined with ground-based data, 'standard' rest frame B-band distances and rest-frame I-band distances to the *same* supernovae could be compared. The large wavelength coverage can be used to study colour variations and in turn reduce the scatter in distance measurements. This would yield a high quality Hubble diagram in the rest-frame I-band with thousands of events to $z \sim 0.7$.

In summary, Euclid would provide much-improved extinction corrections up to $z \sim 0.8$ (plus additional objects with 5-sigma detections to $z \sim 1$), and a rest-frame I-band Hubble diagram with thousands of objects to $z \sim 0.7$.

### 16.2.2 Euclid spectroscopy of supernovae type Ia

Spectroscopy is needed in order to measure the redshift (usually from the host galaxy) and to determine the supernova type. Euclid itself, through the deep spectroscopic survey (depth TBD in the slitless case, or $H(AB) \sim 24$ in the DMD case), could provide spectra for the brightest

SNe as well as redshifts for most of the host galaxies below $z \sim 0.5$.

Repeated visits to a deep spectroscopic field (e.g. with a cadence of less than about 5 days), correlated with photometry, would provide time series of the brightest SN spectra "for free", which can provide typing information. The possibility of generating SN photometric measurements from spectroscopy in a slitless mode could also be investigated. This technique would require accurate positioning, dithering capability and high sensitivity and should be explored with simulations.

### 16.2.3 Cosmology from Core-collapse Supernovae

In addition to the type Ia SNe described above, the Euclid deep survey will discover several thousand (3000-6000) core-collapse SNe (i.e. SNe of types other than Ia) out to $z \sim 0.5$. For many years, various methods to derive the distance to type II SNe have been proposed based on the expanding photosphere effect which relates the SN luminosity to the geometry of the photosphere (e.g. Hamuy *et al.* 2001; Dessart *et al.* 2008). Now, tight Hubble diagrams are being constructed with the current scatter on the distance of $\sim 10\%$. This is somewhat larger than that can be currently derived with type Ia SNe, but it is considered to be subject to lower systematics because the luminosity is directly related to the geometry of the source. In addition metallicity is not expected to have a role as the hydrogen shell is optically thick. Type II SNe detected by Euclid will provide a complementary cosmological test compared to type Ia SNe and other cosmological probes.

## 16.3 Astrophysics from Supernovae of all Types

SNe have a central role in many other fields of astrophysics. Their study is important both to understand the exploding systems and to obtain a complete picture of galaxy formation and evolution, processes that are strongly affected by SN explosions. Euclid will have a central role in this study because, according to the current scanning strategy of the deep survey, it will discover a few thousands of well observed SNe in galaxies whose properties are well known. SN rates will be measured for both types of SNe, although the precision of these results will depend on details of the scanning strategy.

For such astrophysical use of the SNe, spectroscopic observations of the SNe themselves are not strictly required. We expect that a significant fraction of SNe will be discovered in galaxies with a spectroscopic redshift obtained by Euclid itself (depending on the final depth of the spectroscopic deep field, which remains TBD), while for most of the others a photometric redshift of the host will be available. In this situation, methods of SN classification based on repeated multi-wavelength photometric observations show promising results (Poznanski *et al.* 2007; Rodney & Tonry 2009), although simulations are required to demonstrate the application of this technique to the Euclid filter set. We note that for the purposes of measuring SN rates, precise optical photometry is not required.

### 16.3.1 Dust extinction and near-infrared searches for supernovae

Dust extinction is known to limit the number of SNe detected in the local Universe, especially in star-forming galaxies (Maiolino *et al.* 2002; Mannucci *et al.* 2003; Cresci *et al.* 2007; Mattila & Meikle 2001). This is even more important because the majority of the SNe exploding inside starburst regions are hidden (Lonsdale *et al.* 2003, 2006; Smith *et al.* 1998; Ulvestad 2009). The properties of the SNe exploding in starburst regions, which are directly related to dynamical feedback, remain unknown. At redshift larger than $\sim 1$, dust extinction becomes so important that up to half of the SNe could be hidden from optical searches (Mannucci *et al.* 2007). From the ground, near-IR searches are limited by sky background and detector sizes. Euclid will be the first large-scale near-IR search for SNe from space. At the long-wavelength end of Euclid's range (2 micron), the near-IR extinction magnitude $A_H$ is expected to be ~7 times smaller than in the optical, $A_V$. For example, trSNe hidden by 12 mag of optical extinction (a factor ~ 60000) and therefore out of reach of any optical SNe search, in the near-IR would lose less than 2 mags only (less than a factor of 6). As a consequence, much deeper parts of dusty regions can be investigated.

### 16.3.2 Supernova rate and galaxy evolution

Several effects link galaxy evolution with the number and the properties of the SNe they host. Both core-collapse and thermonuclear SNe are the main producers of heavy elements in the Universe. Chemical enrichment within a galaxy is determined by the SN rate as a function of galaxy age, while the cosmic SN rate as a function of redshift determines the chemical evolution of the Universe as a whole. Core-collapse SNe are believed to be the main producer of dust at high redshifts (Maiolino *et al.* 2004; Bianchi & Schneider 2007). Therefore understanding dust extinction at high redshifts requires a good knowledge of at least this type of SNe. Both types of SNe could contribute and even dominate the feedback processes in galaxy formation and could explain the ubiquitous presence of outflows in star forming galaxies. With the measurement of SN rates, Euclid will give an important contribution in all these areas.

### 16.3.3 Type Ia supernova progenitors and stellar evolution

SN rates are also an important tool to investigate the nature of the exploding systems. While the evolution of SN photometry and spectra and the stratification of the chemical elements can constrain the explosion mechanism, SN rates are crucial to constrain the progenitors. SNe Ia are believed to be due to the thermonuclear explosion of a C/O white dwarf (WD). Considerable uncertainties about the explosion model remain within this broad framework, such as the structure and the composition of the exploding WD (He, C/O, or O/Ne), its mass at explosion (at, below, or above the Chandrasekhar mass) and flame propagation (detonation, deflagration, or a combination of the two). Large uncertainties also remain as to the nature of the progenitor system. Usually, a binary system is considered, with the WD dwarf accreting mass either from a nondegenerate secondary star (single-degenerate model, SD) or from a secondary WD (double-degenerate model, DD). The evolution of the binary system through one or more common envelope phases, and its configuration at the moment of the explosion are not known (see Yungelson 2005 for a review). Single-star models are also possible (Tout 2005; Maoz 2008) and current observations are unable to solve the problem (Maoz & Mannucci 2008).

The key quantity to relate Type Ia SN rate to the parent stellar population is the delay time distribution (DTD), i.e. the distribution of the delay time between the formation of the progenitor system and its explosion as a SN. In general, an expected DTD can be derived from each progenitor model, and this can be compared with observations. Several different estimates of the DTD have been proposed, but severe limitations on the SN sample and on the knowledge of the parent galaxies do not allow these to be distinguished reliably. Euclid will allow us to measure the SN rate accurately. From this data DTD can be estimated in two ways, i.e. either by comparing the evolution of type Ia rate with with the cosmic star formation history (SFH), or by comparing the number of SNe in each galaxy with the SFH of that particular host, without the need of making averages among different galaxies.

### 16.3.4 Core Collapse supernova progenitors

Core Collapse (CC) SNe are considered to be due to the gravitational collapse of very massive stars (M>8 solar masses), although exactly how massive is still to be defined. This uncertainty in the mass range can be reduced by measuring accurate SN rates and the initial mass function (IMF) of the parent population. The high spatial resolution of Euclid in the optical will allow us to study the position of the exploding SN within its host galaxy. For example, it will be possible to study if and how much the different classes of CC SNe are associated with the spiral arms of the disk galaxies, or what is the radial distribution of the Ia SNe when compared with the host galaxies. This can be done up to a significant distance, although simulations are needed to quantify this. This method has proved to be able to put constraints on the mass of the progenitor stars. For example, it has been demonstrated that the progenitors of Ib and Ic SNe are more massive than for type II SNe (Hakobyan *et al.* 2009; Georgy *et al.* 2009; Raskin *et al.* 2009; Anderson & James 2009).

### 16.3.5 Core collapse supernova rates

The CC rates are an independent probe of the cosmic SFH (Botticella *et al.* 2008; Dahlen *et al.* 2008). Comparing this with the other probes (UV, H$\alpha$, far-IR) will help understand the relative biases and constrain the IMF. This is true, in particular, if dust effects can be controlled, as explained above. When using SNe, the SFH is measured by counting point-sources instead of measuring the luminosity of diffuse objects, and this have several advantages, as explained by (e.g. Botticella *et al.* 2008). Euclid is expected to detected a few thousand CC SNe in a well controlled sample of galaxies. These SNe will reach $z \sim 0.5$, i.e. will cover the about 1/3 of cosmic history. Therefore it will give a fundamental contribution in this field.

# Euclid Microlensing Planet Hunter

*Authors: Jean-Philippe Beaulieu (Institut d'Astrophysique de Paris, France), Jordi Miralda-Escudé (Institut de Cincies del Cosmos, Barcelona) and Joachim Wambsganss (Astronomisches Rechen-Institut (ARI), Heidelberg).*

## Abstract

While gravitational lensing can be used to trace the distribution of matter on cosmic and Galactic scales, it can also probe smaller scales, through an effect called microlensing, in order to detect exoplanets. There is remarkable synergy between requirements for dark energy probes by cosmic shear measurements and planet hunting by microlensing. Euclid therefore offers a unique scientific programme linking cosmology and exoplanets. It will use gravity as the tool to explore the full range of planet masses not accessible by any other means. Euclid is a 1.2m telescope proposed to the ESA Cosmic Vision programme and can be exploited for an exoplanet hunt using two microlensing programmes. A 3 month microlensing exoplanet hunt will efficiently detect planets down to the mass of Mars at the snow line, free floating terrestrial or gaseous planets and habitable super Earths. A 12+ month survey would give a census on habitable Earth planets around Sun-like stars. This is the perfect complement to the statistics that will be provided by the KEPLER satellite, and the combination of these missions will provide a full census of extrasolar planets from hot, warm, habitable, frozen to free floating.

## 17.1   Introduction

In the last fifteen years, astronomers have found over 400 exoplanets, including some in systems that resemble our very own solar system (Gaudi *et al.* 2008). These discoveries have already challenged and revolutionised our theories of planet formation and dynamical evolution. Several different methods have been used to discover exoplanets, including radial velocity, stellar transits, and gravitational microlensing. Exoplanet detection via gravitational microlensing is a relatively new method (Mao & Paczynski 1991; Gould & Loeb 1992; Wambsganss 1997) and is based on Einstein's theory of general relativity. So far 9 exoplanets have been published with this method. While this number is relatively modest compared with those discovered using the radial velocity method, microlensing probes a part of the parameter space (host separation vs. planet mass) not accessible in the medium term to other methods (see Figure 17.1).

   The mass distribution of microlensing exoplanets has already revealed that cold super-Earths (at or beyond the snow line and with a mass of around 5 to $15M_{\oplus}$) appear to be common (Beaulieu *et al.* 2006; Gould *et al.* 2006, 2007; Kubas *et al.* 2008). Microlensing is currently capable of detecting cool planets of super-Earth mass from the ground and, with a network of wide-field telescopes strategically located around the world, could detect planets with mass as low as the Earth. Old, free-floating planets can also be detected; a significant population of such planets are expected to be ejected during the formation of planetary systems (Jurić & Tremaine 2008). Microlensing is roughly uniformly sensitive to planets orbiting all types of stars, as well as white dwarfs, neutron stars, and black holes, while other method are most sensitive to FGK dwarfs and are now extending to M dwarfs. It is therefore an independent and complementary detection method for aiding a comprehensive understanding of the planet formation process. Ground-based microlensing mostly probes exoplanets outside the snow line, where the favoured core accretion theory of planet formation predicts a larger number of low-mass exoplanets (Ida & Lin 2005). The statistics provided by microlensing will enable a critical test of the core accretion model.

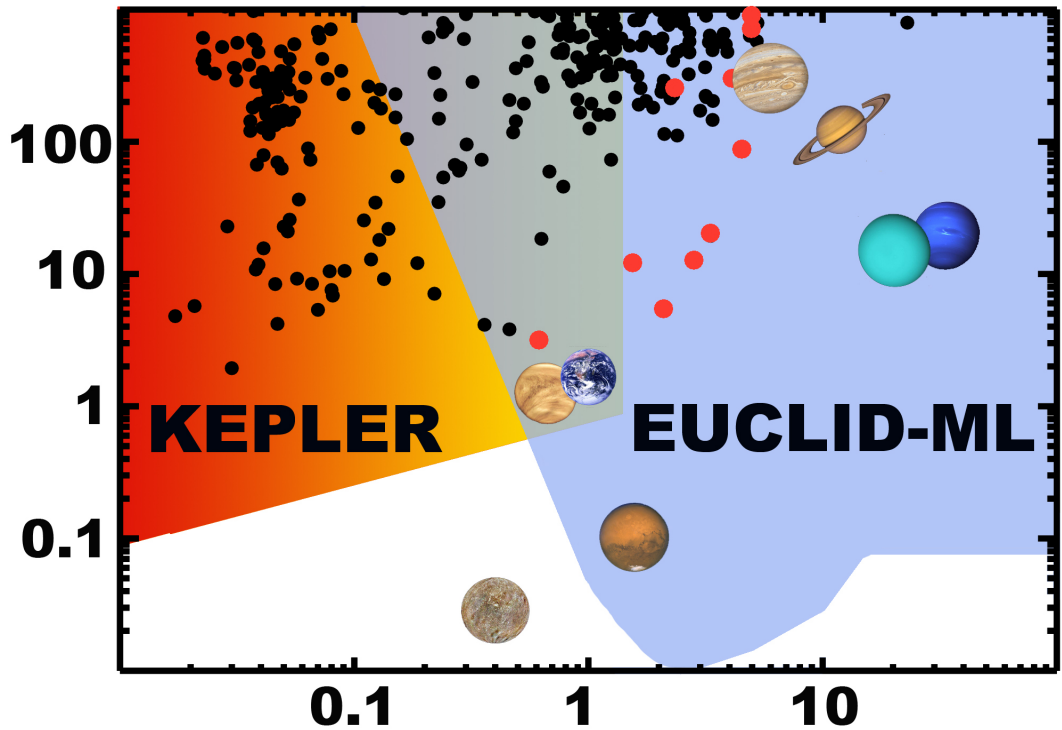

Figure 17.1: Semi major axis as a function of mass for all exoplanets discovered as of September 2009 (microlensing planets are plotted as red dots) and the planets from our solar system. We also plot the sensitivity of KEPLER and of space based microlensing observations.

Exoplanets probed by microlensing are much further away than those probed with other methods. They provide an interesting comparison sample with nearby exoplanets, and allow us to study the extrasolar population throughout the Galaxy. In particular, the host stars with exoplanets appear to have higher metallicity e.g. (Fischer & Valenti 2005). Since the metallicity is on average higher as one goes towards the Galactic centre, the abundance of exoplanets may well be somewhat higher in microlensing surveys.

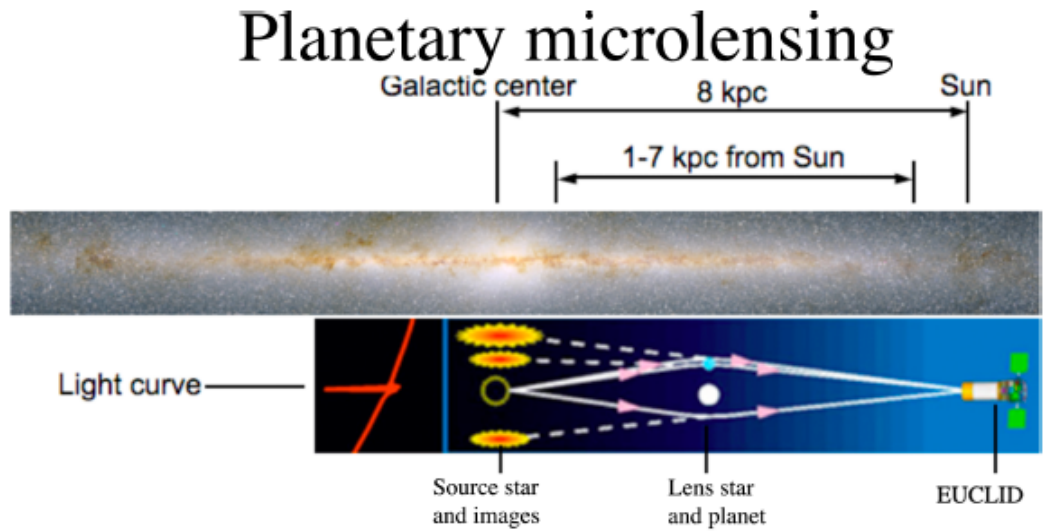

Figure 17.2: Euclid will monitor main sequence stars in the galactic bulge searching for those magnified by gravitational lensing due to foreground star-planet systems in the disk or in the bulge.

## 17.2 Basic microlensing principles

The physical basis of microlensing is the gravitational deflection of light rays by a massive body. A distant source star is temporarily magnified by the gravitational potential of an intervening star (the lens) passing near the line of sight, with an impact parameter smaller than the Einstein ring radius $R_E$, a quantity which depends on the mass of the lens, and the geometry of the alignment. For a source star in the Bulge, with a 0.3 $M_\odot$ lens, $R_E \sim 2$ AU, the angular Einstein ring radius is ∼1 mas, and the time to transit $R_E$ is typically 20-30 days, but can be in the range 5-100 days. The lensing magnification is determined by the degree of alignment of the lens and source stars (see Figure 17.2). The closer the alignment the higher the magnification.

A planetary companion to the lens star will induce a perturbation to the microlensing light curve with a duration that scales with the square root of the planet's mass, lasting typically a few hours (for an Earth) to a few days (for a Jupiter). Hence, planets can be discovered by dense photometric sampling of ongoing microlensing events (Mao & Paczynski 1991; Gould & Loeb 1992). The limiting mass for the microlensing method occurs when the planetary Einstein radius becomes smaller than the projected radius of the source star (Bennett & Rhie 1996). The ∼ 5.5$M_\oplus$ planet detected by Beaulieu *et al.* (2006) is near this limit for a giant source star, but most microlensing events have G or K-dwarf source stars with radii that are at least 10 times smaller than this. High angular enough resolution to resolve dwarf sources of the galactic bulge (≤ 0.5 arcsec) will open the sensitivity below a few Earth masses (Fig. 17.3).

The inverse problem, finding the properties of the lensing system (planet/star mass ratio, star-planet projected separation) from an observed light curve, is a complex non-linear one within a wide parameter space. In general, model distributions for the spatial mass density of the Milky Way, the velocities of potential lens and source stars, and a mass function of the lens stars are required in order to derive probability distributions for the masses of the planet and the lens star, their distance, as well as the orbital radius and period of the planet by means of Bayesian analysis. With complementary high angular resolution observations, currently done either by HST or with adaptive optics, it is possible to get additional constraints to the parameters of the system, and determine masses to 10 % by directly constraining the light coming from the lens and measuring the lens and source relative proper motion (Bennett *et al.* 2006, 2007b; Dong *et al.* 2009). A space-based microlensing survey can provide the planet mass, projected separation from the host, host star mass and its distance from the observer for most events using this method.

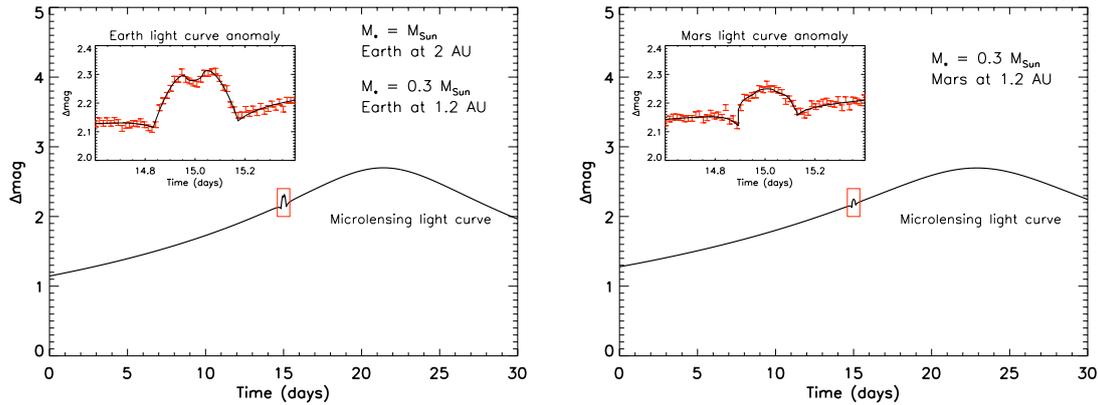

Figure 17.3: We illustrate the detection capability of the microlensing technique in the very low-mass exoplanet regime. The source star and the lens (a foreground star hosting a planet) are both located in the Galactic bulge. The sampling interval is twenty minutes, and the photometric precision is one percent. *Left Panel*: The planetary signal on the expected from an Earth-mass planet at 2 AU around a solar star, or from an Earth at 1.2 AU but orbiting an $0.3\,M_\odot$ M-dwarf star. *Right Panel*: A planet of the mass of Mars ($0.1M_\oplus$) at 1.2 AU can also be detected around such a low-mass host star. Both are typical examples of planets to be detected by Euclid Microlensing Survey.

Different papers have presented the future strategies in the near, medium and long term, with the ultimate goal of achieving a full census of Earth-like planets with either a dedicated space mission or advocating for synergy between dark energy probes and microlensing. There is a general consensus in the microlensing community about these milestones. White papers have been submitted such as: the ESA-EPRAT (Beaulieu *et al.* 2008, ExoPlanetary Roadmap Advisory Team), the ExoPTF (Bennett *et al.* 2007a; Gould *et al.* 2007, US ExoPlanet Task Force), the exoplanet forum (Gaudi *et al.* 2009b), the JDEM request for information and Astro2010 PSF (Bennett *et al.* 2009; Gaudi *et al.* 2009a).

We summarise the road map with the following milestones :

**The short term (5 years) : automatic follow up with existing networks**
A) An optimised planetary microlens follow-up network, including feedback from fully-automated real-time modelling.
B) The first census of the cold planet population, involving planets of Neptune to super-Earth (few $M_\oplus$ to 20 $M_\oplus$) with host star separations around 2 AU.
C) Under highly favourable conditions, sensitivity to planets close to Earth mass with host separations around 2 AU.

**The medium-term (5-10 years) : wide-field telescope networks**
A) Complete census of the cold planet population down to $\sim 10M_\oplus$ with host separations above 1.5 AU.
B) The first census of the free-floating planets population.

C) Sensitivity to planets close to Earth mass with host separations around 2 AU.

**The longer-term (10+ years) : a space-based microlensing survey**
A) A complete census of planets down to Earth mass with separations exceeding 0.5 AU.
B) Complementary coverage to Kepler of the planet discovery space.
C) Potential sensitivity to planets down to 0.1 $M_\oplus$, including all Solar System analogues except for Mercury.
D) Complete lens solutions for most planet events, allowing direct measurements of the planet and host masses, projected separation and distance from the observer.

## 17.3 A program on board Euclid to hunt for planets

**Space based microlensing observations**
The ideal satellite is a 1m class space telescope with a focal plane of 0.5 square degree or more in the visible or in the near infrared. The Microlensing Planet Finder or MPF is an example of such a mission (which was proposed to NASA's Discovery program, and endorsed by the ExoPTF, see Bennett *et al.* 2007a). Despite the fact that the designs were completely independent, **there is a remarkable similarity between the requirements for missions aimed at probing dark energy via cosmic shear and a microlensing planet hunting mission** (Beaulieu *et al.* 2008; Bennett *et al.* 2007a, 2009). Microlensing benefits from the strong requirement from cosmic shear on the imaging channel, and does not add any constraint to the design of Euclid.

**Observing strategy**
We will monitor 2 square degree of the area with highest optical depth to microlensing from the galactic Bulge with a sampling rate once every twenty minutes. Observations will be conducted in the optical and NIR channel.

**Angular resolution is the key to extend sensitivity below few earth masses**
Microlensing relies upon the high density of source and lens stars towards the Galactic bulge to generate the stellar alignments that are needed to generate microlensing events, but this high star density also means that the bulge main sequence source stars are not generally resolved in ground-based images. This means that the precise photometry needed to detect planets of $\leq 1 M_\oplus$ is not possible from the ground unless the magnification due to the stellar lens is moderately high. This, in turn, implies that ground-based microlensing is only sensitive to terrestrial planets located close to the Einstein ring (at ~2-3 AU). The full sensitivity to terrestrial planets in all orbits from $0.5 AU$ to free floating comes only from a space-based survey (17.1). In figure 17.3 we give examples of simulated detections of an Earth and a Mars-mass planet with Euclid.

**Microlensing from space yields precise star and planet parameters**
The high angular resolution and stable point-spread-functions available from space enable a space-based microlensing survey to detect most of the planetary host stars. When combined with the microlensing light curve data, this allows a precise determination of the planet and star properties for most events (Bennett et al. 2007b).

**Probing a parameter space out of reach of any other technique**
The Exoplanet Task Force (ExoPTF) recently released a report (Lunine *et al.* 2008) that evaluated all of the current and proposed methods to find and study exoplanets, and they expressed strong support for space-based microlensing. Their finding regarding space-based microlensing states that: *Space-based microlensing is the optimal approach to providing a true statistical census of planetary systems in the Galaxy, over a range of likely semi-major axes, and can likely be conducted with a Discovery-class mission.* **A program on board the European M class mission Euclid will provide a census of extrasolar planets that is complete (in a statistical sense) down to $0.1 M_\oplus$ at orbital separations $\geq 0.5$ AU** , and when combined with the results of the Kepler mission a space-based microlensing survey will give a comprehensive picture of all types of extrasolar planets with masses down to well below an Earth mass. This fundamental exoplanet census data is needed to gain a comprehensive understanding of processes of planet formation and migration, and this understanding of planet formation is an important ingredient for the understanding of the requirements for habitable planets and the development of life on extrasolar planets.

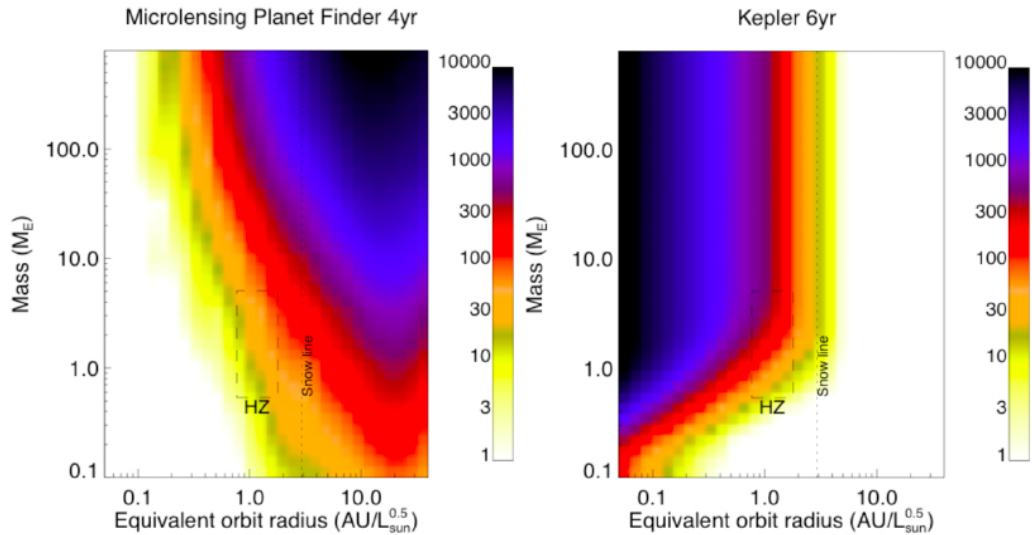

Figure 17.4:  Microlensing planet search with a 1m class telescope and 0.5 square degree with 36 months vs. Kepler exoplanet depth of survey estimates from the Exoplanet Task Force report (Lunine *et al.* 2008). The shading indicates the number of stars that are effectively searched for exoplanets as a function of mass and orbital radius, scaled by stellar luminosity so that the HZs of all types of stars are at ∼1 AU. A three months program on board Euclid would yield numbers 12 times smaller.

A subset of the science goals can be accomplished with an enhanced ground-based microlensing program (Gaudi et al. 2009), which would be sensitive to Earth-mass planets in the vicinity of the snow-line. But such a survey would have its sensitivity to Earth-like planets limited to a narrow range of semi-major axes, so it would not provide the complete picture of the frequency of exoplanets down to $0.1 M_\oplus$ that a space-based microlensing survey would provide. Furthermore, a ground-based survey would not be able to detect the planetary host stars for most of the events, and so it will not provide the systematic data on the variation of exoplanet properties as a function of host star type that a space-based survey will provide.

**Duration of the program**

One of the remarkable feature of such microlensing program is its linear sensitivity to allocated time and area of the focal plane. **The minimal time allocation of three months will already give important statistics on planets at the snow line, down to the mass** **of mars and of free floating planets. Habitable super Earth will also be probed.** Expected sensitivity can be extrapolated from Figure 17.4. Longer observing time (12 months of galactic bulge observing) would lead to good sensitivity to habitable Earth mass planets.

## 17.4 Conclusions

There is a remarkable synergy between requirements for dark energy probes by cosmic shear measurements and planet hunting by microlensing. Employing weak and strong gravitational lensing to trace and detect the distribution of matter on cosmic and Galactic scales, but as well as to the very small scales of exoplanets is a unique meeting point from cosmology to exoplanets. It will use gravity as the tool to explore the full range of masses not accessible by any other means. A modest allocation of telescope time on Euclid will already efficiently probe for planets down to the mass of Mars at the snow line, for free floating terrestrial or gaseous planets and habitable super Earth. This is the perfect complement to the statistics that will be provided by the KEPLER satellite, and these missions combined will provide a full census of extrasolar planets.

# The Milky Way and the Local Group

*Authors: Mark Cropper (Mullard Space Science Laboratory, University College London), Eva Grebel (ARI, University of Heidelberg), Serena Viti (Department of Physics & Astronomy, University College London).*

## Abstract

Euclid will make a huge impact on Milky Way science as a direct result of the Wide Survey. The scale of this survey at the visible and near-infrared imager sensitivity, spatial resolution and the extension to the infrared will transform studies of how our Galaxy was formed and will constrain the cosmological scenarios in which this process could have occurred. Euclid will also enable whole new areas of Local Group science. Beyond this, a Galactic Plane survey in an extended mission would be the reference dataset for star formation and Galactic structure studies, over the entire HR diagram and would very substantially increase the science return from Euclid, making it a fundamental astronomical resource. This survey would be technically feasible, and would be impacted minimally by the degradations expected in the instrumentation by the end of the nominal mission.

## 18.1 Introduction

Galactic astronomy and indeed Galactic archaeology are now experiencing a golden age. Without any special effort, Euclid is already primed to make a major contribution to Galactic astronomy. At high Galactic latitudes, it will provide a vast point source catalogue of stars in the Milky Way and nearby galaxies, enabling structural and stellar population studies with unprecedented depth, wavelength coverage, and spatial resolution.

The history of galaxy formation and mass assembly and the underlying physical mechanisms are key questions of modern astrophysics. In order to explore them, we can either turn to the high-redshift Universe to analyse the integrated light of distant massive galaxies, or we can observe our immediate neighbourhood to analyse the resolved stars in nearby galaxies. Our own Milky Way is the prime target for these studies as it allows us to explore the full range of stellar masses and ages, along with detailed information about chemistry and kinematics. Because of their proximity, our Galaxy and its neighbours are a vital and unique laboratory to test cosmological models of galaxy evolution.

Euclid's suite of imaging and spectroscopic instrumentation is highly suited to Galactic studies. The scale of the impact Euclid will have in this field can be extrapolated from the Sloan Digital Sky Survey (SDSS), in which, arguably, many of the most interesting results have been in Milky Way and Local Group science. Particular highlights include halo streams and new dwarf galaxies, with implications for cosmological structure formation at the lower mass extreme, and for how the Milky Way has been assembled. The Euclid Wide Survey covers 20,000 square degrees, twice that of the SDSS. It also reaches into the infrared to faint magnitudes, and a factor 10 deeper in the visible. It may also be possible to include a dedicated Galactic Plane Survey in an extended Euclid mission. It is clear that Euclid will be a discovery machine on an unprecedented scale for the Galaxy, and absolutely fundamental to more detailed and comprehensive studies with ALMA, JWST and E-ELT.

We specifically addressed the imaging capability of Euclid here, but it should be noted that the visible and near-infrared imagers will be complemented by the power of the Euclid near infrared spectrograph (NIS), which will obtain spectra of every object to $H_{ab} = 19.5$ in the continuum, in regions which are not too crowded.

## 18.2 Milky Way Science from the Wide Survey

By returning a vast catalogue of stars in the Milky Way, the Euclid surveys will enable a treasure of legacy science based on structural and stellar population studies with unprecedented depth, wavelength coverage, and spatial resolution. Euclid will massively augment the Gaia survey of our Milky Way, taking it several magnitudes deeper. Especially when combined with other deep surveys by ground-based surveys (SDSS, PanSTARRS, SkyMapper, DES), Euclid will provide 4-D information on the positions and velocities of hundreds of millions of medium and high latitude stars, highly constraining the integrals of motion when locked into the Gaia reference frame. Moreover, the 4 optical/IR colours will provide photometric distances (a 5th dimension) allowing us to trace large-scale Galactic streams and structure to much greater distances than Gaia will be able to, and accessing a wide and more representative range of the HR diagram (rather than relying on giant tracers). Below $V \sim 20$, the spatial sampling of Euclid will provide complementary information to the Gaia astrometry, photometry and spectroscopy, adding infrared colours, and providing infrared spectra for every single Gaia star it observes. This will go much of the way in breaking the age-metallicity degeneracy, which is critical for the chemical enrichment history of our Galaxy.

Euclid will permit us to derive Galactic structure and scale heights from the faintest, yet most numerous tracers – G and M main-sequence stars and brown dwarfs. Stellar overdensities and substructure will be mapped in unprecedented depth and detail, complemented by Gaia parallaxes and phase space information for the brighter stellar sources.

## 18.3 The Local Neighbourhood

The full sky coverage of Euclid will also yield the most detailed, most sensitive survey of structure and substructure in nearby galaxies, of their outer boundaries, of possible tidal disruption and streams, and of intergalactic or intragroup stars, again constraining merger and accretion histories. Euclid will provide the most complete census of extremely low-surface-brightness, low-mass dwarf galaxies in the Local Group and beyond, much less affected by extinction than any of the ongoing surveys, thus aiding in constraining the cosmological missing satellite problem. Euclid will make a substantial leap in the definition of both the faintest-end of the luminosity function, and the radial distribution of dwarfs, putting stringent constraints on the nature of dark matter, the epoch of reionisation, and the properties of star formation in small galaxies.

Euclid's imagers will also map the entire Magellanic Cloud/Stream system which is at relatively high Galactic latitude, allowing many of the studies discussed below for the Galactic plane but in the context of their different metallicities and star formation histories. In addition it will be possible to trace induced star formation as a result of the interaction with the Milky Way, and to place limits on the Galactic potential.

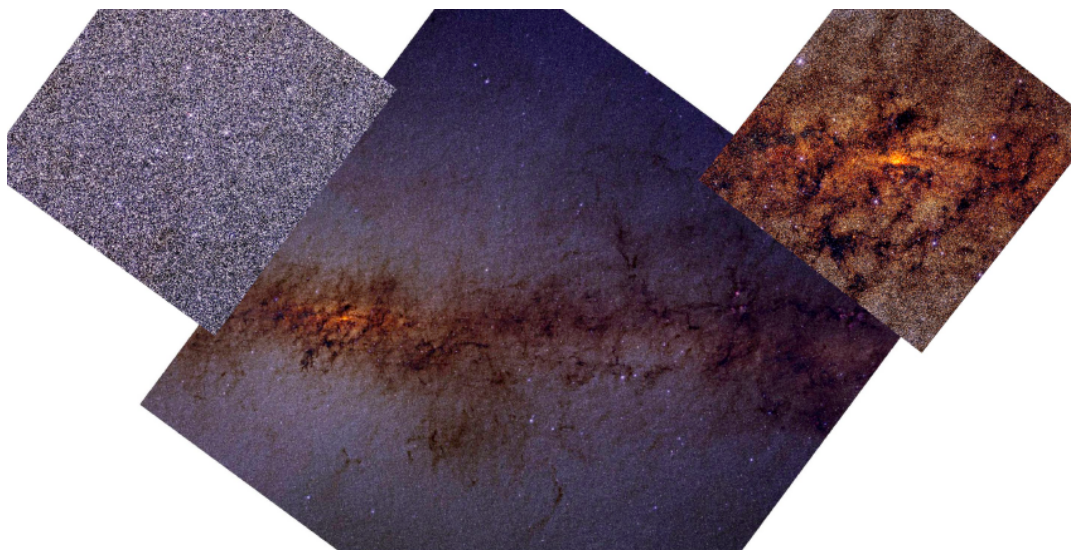

Figure 18.1: The Galactic plane in the infra-red from 2MASS data – the inset above right shows details around the Galactic centre, and above left shows starfields in regions of lower obscuration (from http://pegasus.astro.umass.edu/gc_images/). A Euclid Galactic Plane Survey (E-GPS) survey would reach $\sim 9$ magnitudes deeper, include a visible band and the 50 times finer spatial sampling in the infrared and 400 times finer sampling in the visible will militate against confusion in all but the most crowded regions.

Generally, Euclid's sensitivity will open up a new discovery space for rare stellar and other low-temperature objects. For nearby galaxies with resolved stellar populations, Euclid will permit us to obtain a complete census of (TP-)AGB and C/M stars, which in turn allow us to derive spatially resolved star formation histories for periods of up to 8 or 9 Gyr. Euclid will detect stars far from their parent galaxies, important for tracing those galaxy potentials.

## 18.4 A Euclid Galactic Plane Survey (E-GPS)

The Euclid high Galactic latitude surveys will generate large samples of disk stars, as we have discussed above. If an extended mission could include surveys at lower Galactic latitudes, the harvest of Galactic science would be exceptionally bountiful, given Euclid's fine spatial sampling and red/near-IR capabilities. Before Euclid, the state-of-the-art ground-based large-scale surveys will reach $J, H, K_s \sim 20$ (see Figure 18.1), with only narrower (10–20 square degrees) surveys reaching down to the Euclid near-infrared imager $Y, J, H \sim 24$ limits. It is in the near-IR that the increase in scientific capability is most pronounced. Such a survey is not prohibited by the current spacecraft operational envelope, and because these observations would be less sensitive to the deleterious effects of radiation damage to the detectors, there would be minimal performance impact even after the end of the nominal mission. Were the funds to be available to pursue a mission extension, a Galactic Plane Survey (E-GPS) should be feasible unless Euclid had by then suffered significant other deterioration.

### 18.4.1 Galactic Structure

E-GPS would probe the Galactic plane with many of the science goals of the currently-planned infrared surveys, but surpassing them by orders of magnitude. Such a survey would be the basis of understanding the bulge, bar and thin and thick disk structure as a function of age, metallicity and position. It will inform among others the chemodynamical evolution of the Galaxy; the relative importance of mergers and disk instabilities in bulge formation; the location and longevity of spiral arms; the structure and characteristics of the central bar and its connection to the spiral arms; the origin and nature of the thin and thick disk; and the disk heating and cooling mechanisms – particularly the relative importance of minor mergers, giant molecular clouds and globular clusters in the disk heating.

The Euclid imagers, and particularly the near infra-red imager, will trace disk structure and recent star formation through luminous main sequence and giant stars, to much greater distances or through more obscuration than existing or planned Galactic plane surveys, tracing the spiral arms and allowing the star formation efficiency to be determined as a function of Galactocentric distance. Among intermediate mass stars, the AGB phase, and planetary nebulae are extremely luminous in the infrared, and can be used as further tracers. Euclid is sufficiently sensitive that G-M main sequence stars will be a major resource, especially given the Euclid spatial resolution.

In short, E-GPS will address the entire field of Galactic science, from star formation processes, to dust formation, the recycling of elements in the interstellar medium, the end points of stellar evolution and their role in galactic evolution, to identifying stellar clusters and their remnant streams and so on. In conjunction with and building on Gaia data, E-GPS will determine stellar cluster distances, streaming motions, the stellar initial mass function and cluster luminosity functions to exceptionally faint limits. E-GPS will also provide significant samples of rare evolutionary stages and exotic astrophysically important systems, in addition to a vast sample of every other spectral type. Because the Gaia data will be more limited in the Galactic plane through obscuration, the E-GPS resource would be a key tool in determining the 3-D structure of the Milky Way disk.

The E-GPS will also complement the Herschel survey of the Galactic Plane in the far IR and will provide essential preparatory science for ALMA. Depending on the sequencing of observations within the E-GPS it may be possible to include the time dimension by returning to the same field at later times. This would allow the identification of huge numbers of variable stars, important as distance indicators and hence critical to our understanding of Milky Way dynamics and structure.

### 18.4.2 Young Stellar Objects

Young Stellar Objects (YSOs), in particular massive ones, shape the evolution and fate of galaxies by dominating their radiation budget and determining the ionization and metallicity of the intergalactic medium. Star formation and evolution ultimately drive the evolution of the Universe from its primordial composition to the present-day chemical diversity. Most YSO will still be embedded in their parental molecular clouds, hence observations in the far optical/near-IR are essential in order to determine their nature and distribution in the galaxy. The high sensitivity and resolution of Euclid will allow the identification of a statistically important number of YSOs. The survey will determine the number and spatial extent of YSOs and to measure their ages and masses. It will provide essential information for the characterization of the initial mass function; the star formation rate, history and efficiency; the formation of clusters and sequential star formation processes.

### 18.4.3 Brown Dwarfs

Brown dwarfs, and in general very low mass stars (VLMS), are the most numerous objects in the galaxy but they are also some of the most difficult objects to detect, owing to their faintness. They emit the bulk of their radiation in the near infrared ($1-3\mu$m). The astrophysical importance of identifying a large number of brown dwarfs and very low mass stars comes from the fields of cosmology (are some of the VLMS primordial) and Galactic dynamics (how much mass is stored in faint VLMS), from the field of star formation (does the initial mass function vary among star formation regions, and is there a lower mass limit below which no 'stars' form) as well as from an area of great current interest, that of extraSolar planets (what distinguishes a brown dwarf from a M dwarf, and an extrasolar planet from a brown dwarf). By determining their fundamental properties, such as effective temperature and metallicity, the parameters of very low mass stars can be derived.

### 18.4.4 Late Stages of Stellar Evolution

The Euclid imagers will be extraordinarily sensitive to late phases of stellar evolution, including cool white dwarfs, which are critical laboratories for fundamental physics (such as limits on the change of $G$ and the fine structure constant. In addition, the SDSS experience has shown that the survey will identify large numbers of exotic objects, such as accreting white dwarfs, double white dwarf systems (which are the brightest sources of gravitational radiation), X-ray binaries and even close-by isolated neutron stars. Such systems are important laboratories for the study of physical laws and processes in extreme conditions and are the observational driver for exciting and ground-breaking astrophysics.

## 18.5 Summary

It is clear that within its nominal mission, and without any special measures, Euclid will revolutionise local Universe and Milky Way studies, addressing formation histories, mergers and interactions, induced star formation, measuring galactic potentials, and exploring the nature of the thick disk and halo. The Solar neighbourhood will be characterised with complete samples to considerable distances, even for cool faint objects. If it proves possible to extend the Euclid survey to lower Galactic latitudes in a Galactic plane survey, Euclid will address a huge range of Galactic science, producing fundamental datasets of the disk and centre of the Galaxy, accumulating vast samples of every population of the Galaxy, even of rare objects. This resource will be used to address key questions in galactic evolution: chemical enrichment, feedback processes, star formation, the role of dynamical and other processes in driving disk structure, the nature and origin of the Milky Way bulge and so on. These surveys will identify objects and regions for more detailed followup by JWST, ALMA and SKA on a staggering scale, and the Euclid datasets will be an absolutely fundamental resource for a wide swathe of astronomical research for many years to come. Euclid's fine spatial resolution, high sensitivity and reach into the infra-red, coupled with infrared spectroscopy is the key: nothing of this scale, depth, and spatial sampling is anywhere near being matched from any other space- or ground-based facility, existing or planned.

# Part V

# SIMULATIONS

# Image Simulations

*Authors: Massimo Meneghetti (INAF - Osservatorio Astronomico di Bologna), Benjamin Dobke (JPL - Pasadena), Jason Rhodes (JPL - Pasadena), Lauro Moscardini (Dipartimento di Astronomia - Università di Bologna), Stéphane Paulin-Henriksson (CEA - Paris), Olivier Boulade (CEA - Paris), Jérôme Amiaux (CEA - Paris), Thomas Kitching (University of Edinburgh), Sarah Bridle (UCL London), Lisa Voigt (UCL London), Peter Melchior (Institut für Theoretische Astrophysik - Zentrum für Astronomie Heidelberg), Julian Merten (INAF - Osservatorio Astronomico di Bologna), Andrea Grazian (INAF - Osservatorio Astronomico di Monteporzio), Richard Massey (University of Edinburgh), David Johnston (Case Western).*

**Abstract**

The image simulation working group has developed tools for producing mock observations adopting the Euclid optical design. Among the many possible applications, these tools were used to estimate the number counts of galaxies in future Euclid observations. We produced images assuming different levels of background and changing the effective exposure time, in order to take into account possible changes of the instrument design. Our simulations show that an effective number density of galaxies between 30 and 40 arcmin$^{-2}$ is easily achievable with the current design of Euclid. The median redshift of the detected sources is around unity. These results allow the conclusion that Euclid will match all the requirements needed to reach a sub-percent accuracy on the estimate of cosmological parameters like the equation of state of dark energy.

## 19.1   Introduction

The precision with which cosmological parameters like the equation of state of dark energy can be constrained using weak lensing principally depends on the number of galaxies for which a shape measurement is achievable and on the redshift distribution of the sources. The work done by the image simulation working group aims at quantifying how many galaxies per square arcminute will be detectable in Euclid observations and usable for the weak lensing analysis. We tackle this issue by using state-of-the-art simulation pipelines that allows one to produce mock observations with specific instrumental set-ups, including all the relevant sources of noise. The images are subsequently analysed to compile catalogues of galaxies. By applying proper selection cuts to the catalogues, we estimate the effective number density of galaxies useful for lensing.

Reaching an accuracy of percent level precision on the measurement of dark energy requires an effective number density of, on average, 30–40 galaxies per square arcminute over the entire Euclid survey. We show here that the current Euclid mission and survey design will meet that goal.

## 19.2 Image simulation pipelines

The simulations are made using two pipelines, independently developed by two groups, one in Italy (`SkyLens`) and one in the US (`simage`). A detailed description of their implementation can be found in Meneghetti *et al.* (2008) and in Massey *et al.* (2004) (see also Dobke et al. in prep). Both the pipelines use the Hubble Ultra-Deep Field (HUDF; Beckwith *et al.* 2006) as a basis set to generate the galaxy population to be placed in the simulations. About 10,000 galaxies are decomposed into shapelets (Refregier 2003; Melchior *et al.* 2007; Massey & Refregier 2005) in the four $B, V, i$, and $z$ bands. The shapelets formalism allows to easily manipulate the galaxy shapes. For example, operations like convolutions (or de-convolutions), rotations, mirroring, shearing, etc. are straightforward. Moreover, the shapelets provide an analytical description of the galaxies. Thus, they can be "painted" on a virtual sky at any desired resolution, with the only limitation of the native resolution limit of the HUDF images. Using the HUDF galaxies as a reference basis set, ensures the coverage of a broad range of galaxy morphologies and sizes. Moreover, the HUDF is the deepest observation available to date. Briefly, the steps required to simulate an observation of a patch of the sky are: 1) they generate a population of galaxies using the luminosity and the redshift distribution of the galaxies in the HUDF; 2) they prepare the virtual observation, receiving from the user the pointing instructions, the exposure time, and the filter $F(\lambda)$ to be used. The pointing coordinates are used to calculate the level of the background, i.e. the surface brightness of the sky; 3) they assemble the virtual telescope. This implies that the user provides some a set of input parameters, like the effective diameter of the telescope, the field of view, the CCD specifications (gain, read-out-noise, pixel scale) and the additional information necessary to construct the total throughput function, defined as

$$T(\lambda) = C(\lambda)M(\lambda)R(\lambda)F(\lambda) . \tag{19.1}$$

In the previous formula, $C(\lambda)$ is the quantum efficiency of the CCD, $M(\lambda)$ is the mirror reflectivity, and $R(\lambda)$ is the transmission curve of the lenses in the optical system; 4) they calculate the fluxes in the band of the virtual observation. Then, using the shapelets coefficients of the galaxy decompositions, the surface brightness of the sources is calculated at each position on the sky and converted into a number of *Analogue to digital units* (ADUs) on the CCD pixels. Finally, noise is added according to the sky brightness and to the Read-Out-Noise (RON) of the CCD.

The two pipelines follow very similar approaches for modeling the galaxies in the images, although significant differences are present in the methods for creating the galaxy populations. While `SkyLens` uses the magnitude and redshift distributions of the HUDF, `simage` calibrates the number counts per magnitude bin such to match other existing surveys like COSMOS (Leauthaud *et al.* 2007). The pipelines also use different approaches when modeling the SEDs of the sources. Given these differences, the two pipelines have been extensively tested against each other. In particular, they have been validated by performing simulations of existing HST fields. The agreement between the outputs of two pipelines and the real catalogues is remarkable in terms of both limiting magnitudes and galaxy number counts. Both the pipeline produce galaxy number densities which differ from the observed ones by less than 10%. We conclude that we can safely combine the results of the two pipelines for the analyses shown in the next sections.

## 19.3 Instrumental set-up

The simulations used in this paper are based on the parameters listed in Tab. 19.1. Each Euclid observation will consist of four 450s-long dithered exposures of the same field. The total integration time will then be $t_{\text{exp}} = 1800$s. In the present study, we work with images obtained by combining four exposures with different noise patterns, and we do not include and we do not use dithering to increase the resolution of the final stacked image.

Table 19.1: Key parameters used in the simulations.

| parameter | value |
|---|---|
| Diameter | 1.19m |
| Obscuration (diameter) | 0.37m |
| Pixel scale | 0.1" |
| refl. of each mirror | 0.98 |
| number of mirrors | 5 |
| refl. of the dichroic | 0.98 |
| filter transmission | 0.9 |
| detector Q.E. (550nm) | 0.9 |
| detector Q.E. (700nm) | 0.9 |
| detector Q.E. (850nm) | 0.7 |
| detector Q.E. (900nm) | 0.5 |
| detector Q.E. (920nm) | 0.4 |
| R.O.N | 5 $e^-$ |
| dark current | 0.001 $e^-/\text{pixel/s}^{-1}$ |
| n. of. exposures | 4 |
| tot. exp. time | 1800s |

The key parameters listed above provide the reference configuration of the instrument. However, in order to take into account possible changes of their values, our simulations do not stick on a single set of parameters. Instead, we consider a range of possible values for the parameters listed above. For example, we allow for possible variations of the mirror and dichroic reflectivity.

Changes in the instrumental set-up are incorporated in the simulations by scaling by a proper factor the ADUs in each CCD pixel computed for the reference set-up. The pixel counts are proportional to the total throughput, being

$$\text{ADU}(\vec{x}) \propto t_{\exp} \int T(\lambda) S(\vec{x}, \lambda) \frac{\mathrm{d}\lambda}{\lambda} , \tag{19.2}$$

where $S(\vec{x}, \lambda)$ is the surface brightness (sources and sky) at the position $\vec{x}$ on the CCD array. The above equation shows that we can account for changes in the throughput by defining the *effective exposure time*, $t_{\text{eff}}$. This is defined as $t_{\text{eff}} = t_{\exp} f$, where the factor $f$ contains all the parameter variations with respect to the reference design. In order to explore several solutions, we allow for changes in $t_{\text{eff}}$ by a factor $f$ between 0.5 and 1.6.

The images are convolved with a realistic system PSF model of FWHM=0.180 arcsec, which takes into account the optical PSF, the jitter, and the detector effects.

## 19.4   Sky background

The principal background contributor in the wavelength range for Euclid will be the zodiacal light. A model for the zodiacal light near the North-Ecliptic-Pole (NEP) has been derived by

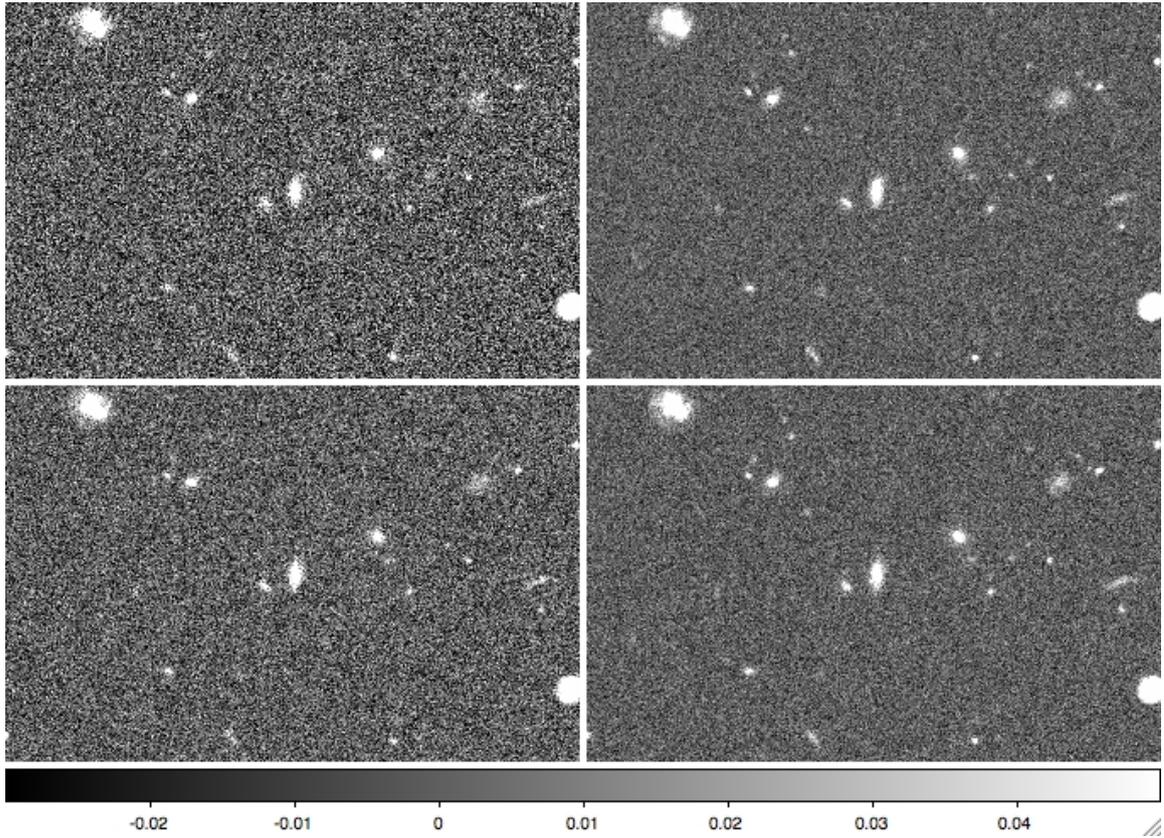

Figure 19.1: Simulated observation of a patch of the sky, corresponding to a region of 34" × 22", assuming the instrumental set-up of EUCLID. The images in the upper panels refer to simulations with effective exposure times of 900s (left) and 2880s (right) assuming an "average" sky background level of 22.342 $\mathrm{mag_{AB}\ arcsec^{-2}}$. The images in the bottom panels show two simulations with effective exposure time 1800s and sky background magnitudes of 21.75 $\mathrm{mag_{AB}\ arcsec^{-2}}$ and 22.95 $\mathrm{mag_{AB}\ arcsec^{-2}}$.

Leinert *et al.* (1998) and later revised by Aldering *et al.* (2004). This model has been used to derive "low", "average", and "high" background levels for Euclid observations. In terms of AB magnitudes, the model provides the surface brightnesses in the wavelength range [550 ÷ 920]nm of 22.95, 22.35, and 21.75 for the "low", "average", and "high" background cases, respectively. Thus, we carry out simulated observations assuming a continuous range of sky surface brightness values between 21.75 and 22.95 $\mathrm{mag_{AB}\ arcsec^{-2}}$.

## 19.5 Analysis of the images

Some examples of our simulations are shown in Fig. 19.1. The upper panels show a patch of the sky, which corresponds to a region of 34" × 22", as it might be seen by Euclid with $t_{\mathrm{eff}} = 900$s (left panel) and $t_{\mathrm{eff}} = 2880$s (right panel) assuming the "average" sky background. Instead, the bottom left and right panels show the same patch observed with $t_{\mathrm{eff}} = 1800$s and assuming the "high" and the "low" sky background levels, respectively. The colour bar on the bottom allows to convert the gray levels into ADUs $\mathrm{s^{-1}}$. Obviously the galaxy counts and the source signal-to-noises differ significantly in the limiting cases shown here. We now consider several selection criteria to show how the galaxy counts depend on both the effective exposure time and the sky surface brightness.

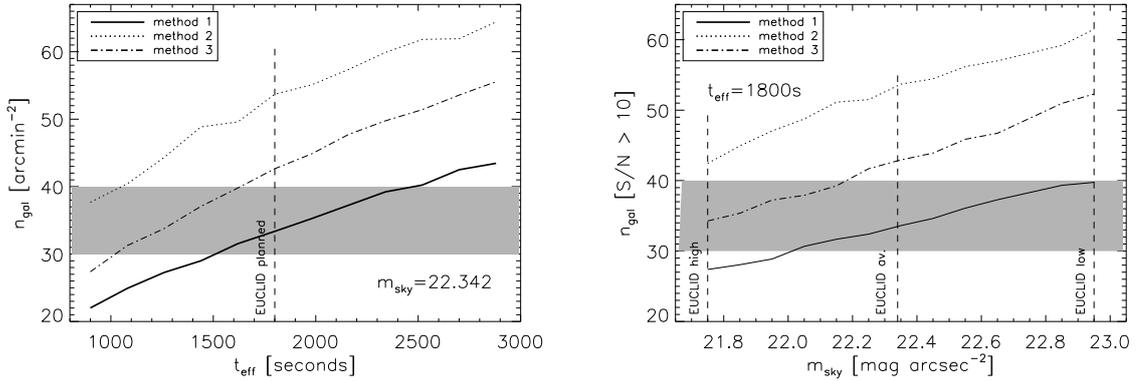

Figure 19.2: Left panel: Effective number density of galaxies in Euclid images as a function of effective exposure time, assuming the "average" sky background level. The vertical dashed line indicates the planned effective exposure time for this mission. Right panel: Effective number density of galaxies in Euclid images as a function of sky brightness, assuming nominal effective exposure time of 1800s (co-adding four short exposures of 450s each). The vertical dashed lines indicate the "low", "average", and "high" sky background levels defined in Sect. 19.4. In both panels, the shaded region denotes the number density interval which matches the scientific requirements for Euclid. We use different line-styles for the three selection methods discussed in this paper. The solid, dotted, and dash-dotted line indicate the results obtained with methods 1, 2, and 3, respectively.

**Method 1: Selection by signal-to-noise:** Firstly, among the galaxies detected in the images, we count those passing some cuts based on their signal-to-noise ratio (SNR) and size. These are motivated by the necessity of measuring the shape of the sources, given that the main goal is to perform shear measurements. Thus galaxies need to be well resolved. Note that, for this selection method, we do not calibrate the SNR limit on the effective capability of measuring the galaxy shape with a given accuracy. We pick a limit of $SNR_{min} > 10$, to enable good shape measurements with all the selected objects. We further only retain those galaxies whose $FWHM > 1.25 \times FWHM_{PSF}$.

**Method 2: Selection by shear measurability** *Lens*fit is a template fitting shape measurement method that makes use of Bayesian shape estimation techniques to measure the shear. The method is described in Miller *et al.* (2007) and Kitching *et al.* (2008) and has been applied to the STEP (Heymans *et al.* 2006; Massey *et al.* 2007) and GREAT08 simulations (Bridle *et al.* 2008). The Bayesian shape estimation procedure is optimised so that in the noisy low signal-to-noise regime any bias caused by noise should be removed. For lensfit we make no cuts at the catalogue level, every object in the SExtractor catalogues is analysed, and a weight and identification given.

For these simulations we performed two analyses, the first used the stacked images created by stacking the pixels in each individual exposure. For the second analysis we simultaneously estimated the galaxy shapes from each exposure and stacked the ellipticity probability distributions for each galaxy from each exposure. To calculate the effective number density we summed the lensfit weights of each galaxy for which we take the mean of the lensing sensitivities (as described in Miller *et al.* (2007) and Kitching *et al.* (2008)).

**Method 3: Selection by shear uncertainty**  An additional shear analysis is done with the Im2shape software. Running it on the simulations to measure the galaxy ellipticities, we calculate the total number of galaxies with a shear uncertainty smaller than 0.1 (ignoring shape noise), to obtain the effective number of galaxies.

### 19.5.1  Detections vs effective exposure time

In the left panel of Fig. 19.2 we show how the galaxy number counts change as a function of the effective exposure time for Euclid-like observations. Considering the selection method 1, the expected number density of galaxies with high SNR is $\sim 33$ arcmin$^{-2}$ for an effective exposure time of 1800s. We recall that, assuming a survey area of 20,000 sq. degrees, the effective galaxy number density required for achieving the Euclid objectives, is between 30-40 arcmin$^{-2}$. Such requirement is satisfied for $t_{eff} > 1500$s. The method 1 is the most conservative at estimating the effective galaxy number counts, since we intentionally adopt a large SNR limit for the detections. However, in many cases the ellipticity measurements are possible down to much lower SNRs ($\gtrsim 5$). This is shown by the analyses using the selection methods 2 and 3, whose results are given by the dotted and dash-dotted lines in Fig. 19.2. For example, using the *lens*fit pipeline to select those galaxies with achievable ellipticity measurements, we find $\sim 54$ arcmin$^{-2}$. This number decreases to $\sim 43$ arcmin$^{-2}$ when imposing a cut based on the ellipticity error. However, in a realistic situation we should consider that a number of detections will be missed due to pixel defects or CCD gaps, which may decrease the effective galaxy number counts. Decreasing the effective exposure time down to 1000s, i.e. in the most pessimistic case for Euclid, the expected number counts drop to $\sim 22$, 28, and 37 arcmin$^{-2}$ for the methods 1, 2, and 3, respectively. Increasing the effective exposure time to 2000s, 2500s, and 2900s would improve the number counts by $\sim 10\%$, $\sim 20\%$ and $\sim 30\%$.

### 19.5.2  Detections vs sky background level

The dependence of the effective galaxy number density in the Euclid images on the level of the sky background is shown in the right panel of Fig. 19.2. The analysis reveals that for sky brightnesses in the range $m_{sky} = 21.75$–$22.95$ mag$_{AB}$ arcsec$^{-2}$ the effective galaxy number density varies between $\sim 28$ arcmin$^{-2}$ and $\sim 40$ arcmin$^{-2}$ using the selection method 1 assuming the nominal effective exposure time, $t_{eff} = 1800$s. Thus, we expect variations of the order of $\sim 15$–$20\%$ in the effective number counts depending on the pointing directions. Similar trends are found with the method 2 and 3, for which the galaxy number counts always remain well above the threshold of 30 arcmin$^{-2}$. Therefore, the current reference design of Euclid matches the requirements on the effective galaxy number density for essentially any plausible level of sky background.

The zodiacal background can be minimised using a suitable observing strategy. In particular, the scanning strategy of Euclid will enable scans in ecliptic latitude, rolling around the Sun direction with the telescope pointing at right angle to the Sun. The whole sky outside a stripe of galactic latitude $|b| < 30$ degrees will be observed. The histogram in the left panel of Fig. 19.3 shows the distribution of sky levels obtained by dividing the sky into 3072 pixels of 13.4 sq. degrees, assuming each line of sight is observed at 90 degrees from the Sun and discarding the pixels at $|b| < 30$ degrees. In the right panel of Fig. 19.3 we show the distribution of $n_{gal}$ expected for the same survey strategy. The results are shown for the method 1, which, we remind again, is a very conservative method to define $n_{gal}$. In this case, we find that $\lesssim 28\%$ of the survey area will have $27 < n_{gal} < 30$ arcmin$^{-2}$ and $\gtrsim 38\%$ of the survey area will have $35 < n_{gal} < 40$ arcmin$^{-2}$ if observed with Euclid with an effective exposure time of 1800s.

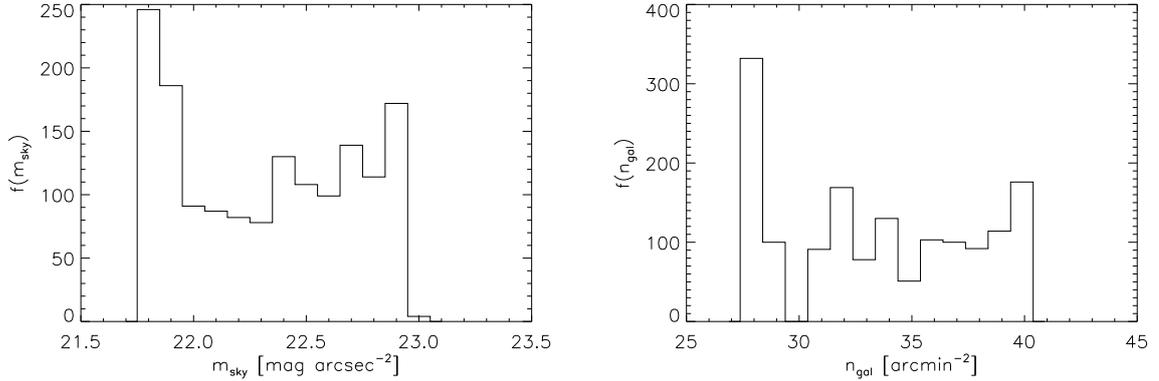

Figure 19.3: Left panel: distribution of sky magnitudes in the area of the sky at galactic latitudes $|b| > 30$ degrees, assuming to observe at 90 degrees from the direction to the Sun. Right panel: the magnitude distribution on the right is here converted into a distribution of effective galaxy number densities by means of the image simulations, using the selection method 1.

### 19.5.3  Redshift distribution

The last but not least important ingredient to establish the accuracy of on the measurements of the cosmological parameters is given by the redshift distribution of the sources. For each of the galaxies used in the simulations, we know the redshift, which is taken from the photometric redshift catalogue by Coe *et al.* (2006). Thus, we can use this information to determine the expected redshift distribution of the sources in the Euclid images. Assuming the reference design for the instrument and the "average" sky background level, the redshift distribution of all galaxies detected by SExtractor in the simulated images is given by the red dashed histogram in Fig. 19.4. If the selection method 1 is used to discard low SNR galaxies, the redshift distribution changes as shown by the solid histogram. In these two limiting cases, the median redshifts of the sources are 0.99 and 1.17, respectively, thus well matching the requirement $z_m \gtrsim 0.8$. Of course, this conclusion relies on the strong assumption that the redshift distribution of galaxies in the HUDF (shown by the dot-dashed blue histogram) can be extended to the whole sky.

## 19.6  Conclusions

The simulations discussed above show that the current optical design of the Euclid imaging channel ensures an average effective number density of galaxies detected at high significance $\gtrsim 33$–35 arcmin$^{-2}$. The effective number density of galaxies which will be available for the lensing analysis, i.e. for which it will be possible to obtain a reasonably accurate shear estimate, is likely to be significantly larger. Assuming the redshift distribution of the galaxies in the HUDF, we find that the median redshift in Euclid observations will be $z_m \sim 1$. Given these results, we conclude that the requirements needed for reaching the desired level of accuracy will be matched with the current Euclid design.

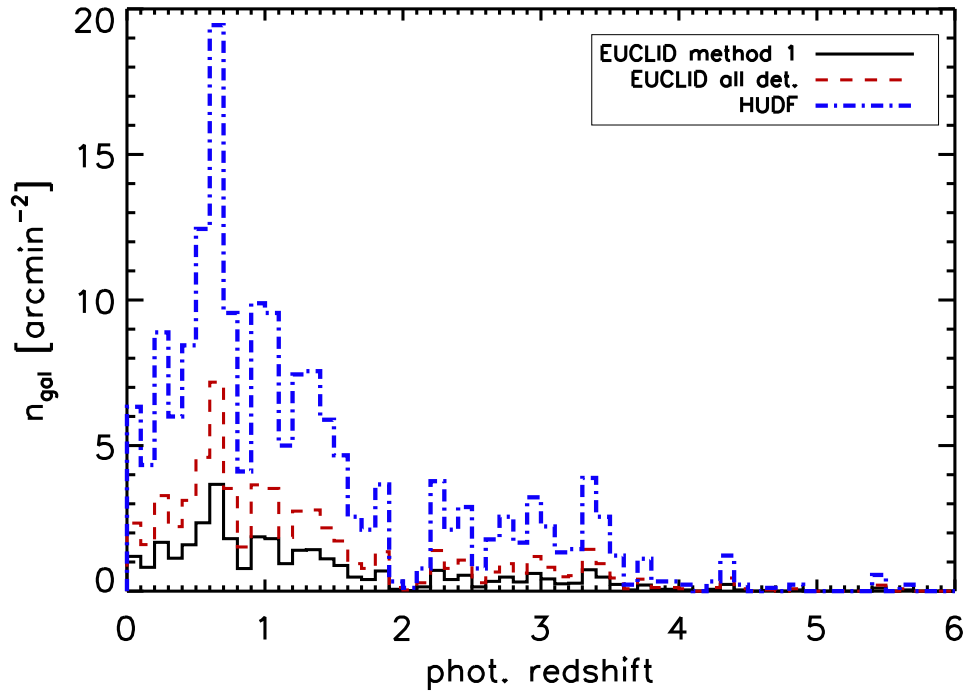

Figure 19.4: Galaxy number densities per redshift bins as detected in the Euclid simulated images. The solid and the dashed lines show the detections imposing and without imposing the cuts on the SNR and size outlines in Sect. 19.5 (Method 1). The dash-dotted line shows the input redshift distribution used in the simulations, which is based on the photometric redshift catalogue by Coe *et al.* (2006).

# Photometric redshifts

*Authors: Rongmon Bordoloi (ETH Zurich), Manda Banerji (UCL, IoA Cambridge), Filipe Abdalla (UCL), Adam Amara (ETH Zurich), Ofer Lahav (UCL), Simon Lilly (ETH Zurich).*

## Abstract

In this chapter we summarise the work done by the EIC Photo-z Working Group during the Euclid definition phase. The work involved contributions from teams in Barcelona, Rome, UCL and Zurich. We focus on two studies, presented in Abdalla *et al.* (2008a) and Bordoloi *et al.* (2009), that use different approaches to the photo-z measurements: a template fitting method ZEBRA (Feldmann *et al.* 2006) (ETH Zurich) and a Neural network method AnnZ (Collister & Lahav 2004) (UCL). Through extensive comparisons, we find that these two methods give similar performance, which in turn allows us to form robust conclusions. In this chapter, we address several of the issues associated with high performance photo-z measurements. These include the need for space-based Near Infrared imaging and our requirements for ground-based counterparts. We find that in combination with ground-based photometry (e.g. PanStarrs-2 or DES), the baseline Euclid mission can reach our requirements of sigma(z)$\leq 0.05(1 + z)$ after a simple cleaning methods have been applied. We find that the cleaning, which uses only information that is available in the photometry, is also able to reduce our catastrophic failure rate to below 0.25%. We also show that there are a number of approaches for reaching the demanding requirement that the mean redshift of galaxies in a redshift bin be known to a precision of $0.002(1 + z)$. Specifically, we show: (i) a case where the mean is measured directly using spectroscopic redshifts of a representative subsample; and (ii) a method for measuring the mean using the likelihoods coming from the photo-z estimation codes. Finally, we also summarise the effect of an imprecise correction for Galactic extinction and the effects of contamination by over-lapping objects in photo-z determination. The overall conclusion is that the acquisition of photometrically estimated redshifts with the precision required for the Euclid weak lensing science goals will be possible.

## 20.1 Introduction

Application of weak lensing for cosmology requires at least a statistical knowledge of the distances, i.e. redshifts, of large numbers of individual galaxies. At $I_{AB} < 24.5$, there are about 2.5 billion galaxies in the Euclid $2\pi$ sr survey area and so, realistically, reliance must be made on photometrically estimated redshifts (hereafter *photo-z*). In weak-lensing tomography, redshift information is required at two conceptually distinct steps. First, galaxies must be assigned to individual redshift bins. The shear signal is then extracted from the cross-correlation of the shape measurements of individual galaxies in different redshift bins, to exclude potential problems such as intrinsic alignments between physically associated galaxies. Second, the statistical $N(z)$ (i.e. effectively $\langle z \rangle$) of the galaxies in a given bin is required to then map the results onto cosmological distance and thereby extract the cosmological information. It is, of course, possible to do a similar correlation analysis with unbinned data (Castro *et al.* 2005; Kitching *et al.* 2008a), but for our purposes this distinction is unimportant.

The required accuracy of the individual photo-z for the first bin-construction task is set by the need to exclude overlaps in the $N(z)$ of individual bins (or the probability distribution for individual galaxies) and thereby remove physically close pairs (King & Schneider 2002). This typically sets a requirement on the precision of individual photo-z of about $\sigma_z = 0.05(1 + z)$ (Bridle & King 2007). This step must be done with photo-z because of the large number of galaxies involved (2.5 billion).

The required accuracy in the statistical redshift distribution $N(z)$ and the mean redshift $\langle z \rangle$ for each of these bins is set by the required accuracy of the cosmological parameters. The Euclid scientific goals require a precision of order $0.002(1 + z)$ in $\langle z \rangle$ (Amara & Réfrégier 2007; Ma *et al.* 2006; Kitching *et al.* 2008b). There are two approaches to constructing $N(z)$ and $\langle z \rangle$. The first involves the "direct" measurement of large numbers of spectroscopic redshifts for a representative set of objects in each bin (Abdalla *et al.* 2008a). The other relies on continued use of the photo-z themselves to characterise the bin, which relaxes the requirement on the spectroscopic redshifts which are used only to calibrate the photo-z scheme (Bordoloi *et al.* 2009).

These two requirements are both demanding and represent one of the observational challenges that lie along the path to enabling precision cosmology with weak lensing. Fortunately, there are some mitigating features of weak lensing analysis. For instance, the analysis is robust (aside from shot-noise statistics) to the exclusion of individual galaxies, provided only that the exclusion is unrelated to their shapes. One is free therefore to reject galaxies that are likely to have poor photo-z provided that they can be recognised *a priori*, i.e. from the photometric data alone.

This chapter summarises the photo-z work done within the EIC photo-z working group. These results appear in Abdalla *et al.* (2008a) and Bordoloi *et al.* (2009). Specifically, we discuss here the following topics:

- The choice of near-infrared filter configuration for Euclid.

- The required depth of ground-based complements.

- The photo-z performance on individual objects, emphasising the *a priori* identification and rejection of catastrophic failures.

- The number and sampling requirements on spectroscopic redshifts for the "direct" construction of $N(z)$.

- The construction of $N(z)$ using the photo-z alone, focussing on approaches that yield the least biased estimate of $N(z)$ and $\langle z \rangle$ for the ensemble.

- The systematic biases that can enter into the photo-z from an incorrect assumption about the level of foreground Galactic reddening, and how the reddening can be estimated from the photometric data themselves.

- The effects of the photometric superposition of two galaxies at different redshifts, leading to a mixed spectral energy distribution that may perturb the photo-z.

Our approach is to try to isolate these problems and to explore each in turn with the aim of providing an *existence proof* that gives a plausible route to achieve the very high photo-z performance required for *Euclid*.

## 20.2 Generation of photo-z from mock photometric catalogues

We have utilised the publicly available "neural network" photo-z package ANNz (Collister & Lahav 2004), and the publicly available template fitting package ZEBRA (Feldmann *et al.* 2006). It is now widely recognised, and demonstrated by direct comparison early in our work, that these two approaches to photo-z estimation give broadly similar results (Abdalla *et al.* 2008b).

These photo-z codes have been applied to mock photometric catalogues. The mock catalogues used at UCL are described in Abdalla *et al.* (2008a). The GOODS-N spectroscopic sample (Cowie *et al.* 2004; Wirth et al. 2004) was used. This is a flux limited sample with $R < 24.5$ and redshifts $z < 4$. The broadband photometry available for this sample in multiple wavebands, was used to generate a series of galaxy templates using a method similar to Budavári et. al (1999) but assuming a prior set of templates consisting of the observed Coleman *et al.* (1980) templates and the starburst galaxies of Kinney et al. (1996). Any reddening is removed from the photometry before template construction and the new templates used to calculate a best-fit SED value and reddening for each object assuming the reddening law of Calzetti (1997) as well as a correction for the IGM absorption according to the prescription of Madau (1995). Gaussian noise is added to the fluxes before calculating magnitudes and errors for the noisy sample in order to produce the final catalogue.

At ETH, the COSMOS mock catalogues were produced by Kitzbichler & White (2007). There are a total of 24 mock catalogues each representing 2 deg$^2$ and containing approximately 600,000 objects, the majority of which are at $z \leq 1$. To produce the photometric catalogue for a customised set of filters, SED templates were identified, from amongst 10,000 available, that well matched the quoted photometry for each galaxy in the COSMOS mock catalogue at its known redshift. These templates include a range of internal reddening, which ranges from $0 < A_v < 2$ magnitudes. These templates were then used to compute photometry for the galaxies in all other passbands of interest, adding Gaussian noise. Intergalactic absorption in each template is compensated for using the Madau law (Madau 1995). All 24 mocks were combined and a random sub-sample used to mimic a survey over a large area in the sky. Each of the simulations contains at least 100,000 objects. To match the proposed Euclid experiment objects with $I_{AB} \leq 24.5$ were considered. Table 20.1 provides the assumed $5\sigma$ AB magnitude limits in each of the different filters for different surveys that are described in this work. For this study the input photometric catalogues were constructed using exactly the same set of approximately ten thousand templates as were subsequently used in the ZEBRA photo-z code. This may strike some readers as being somewhat circular. However, this approach allows us to eliminate the choice of templates as a variable, or uncertainty, in our analysis. This is motivated by the exceptional performance that has already been achieved with the same templates coupled with the exquisite observational data in COSMOS. This strongly suggests to us that the choice of templates is unlikely to be the limiting factor with the degraded photometry that we can realistically expect to have over the whole sky within the timescale of a decade or so.

## 20.3 The Near Infra-Red Filter Configuration

In Abdalla *et al.* (2008a) we investigated the effect of adding deep NIR photometry to the visible bands we can expect from upcoming ground based wide field imaging surveys, namely DES, PanSTARRS and LSST. We find that there is a dramatic improvement in the redshift measurements when NIR photometry is added (see Figure 20.1). Further to this study we have varied the total exposure and find that we are able to meet our requirements of $\sigma(z) < 0.05(1+z)$, after cleaning, for the depths shown in Table 20.1.

We have also studied the effect of choosing different combinations of infrared filters (see

| Band | Survey-A | Survey-B | Survey-C |
|------|----------|----------|----------|
| g | 24.66 | 25.53 | 26.10 |
| r | 24.11 | 24.96 | 25.80 |
| i | 24.00 | 24.80 | 25.60 |
| z | 22.98 | 23.54 | 24.10 |
| y | 21.52 | 22.01 | 22.50 |
|   | Euclid | NIR |   |
| Y | 24.00 | 24.00 | 24.00 |
| J | 24.00 | 24.00 | 24.00 |
| H/K | 24.00 | 24.00 | 24.00 |

Table 20.1: The assumed filter sensitivities for different survey configurations considered. The values quoted here are $5\sigma$ errors in AB magnitude. Survey-A corresponds roughly to an optimistic PanSTARRS-1 like survey, Survey B is designed to correspond roughly to a DES/PanSTARRS-2 like survey and Survey C corresponds to depths achievable with LSST/PanSTARRS-4.

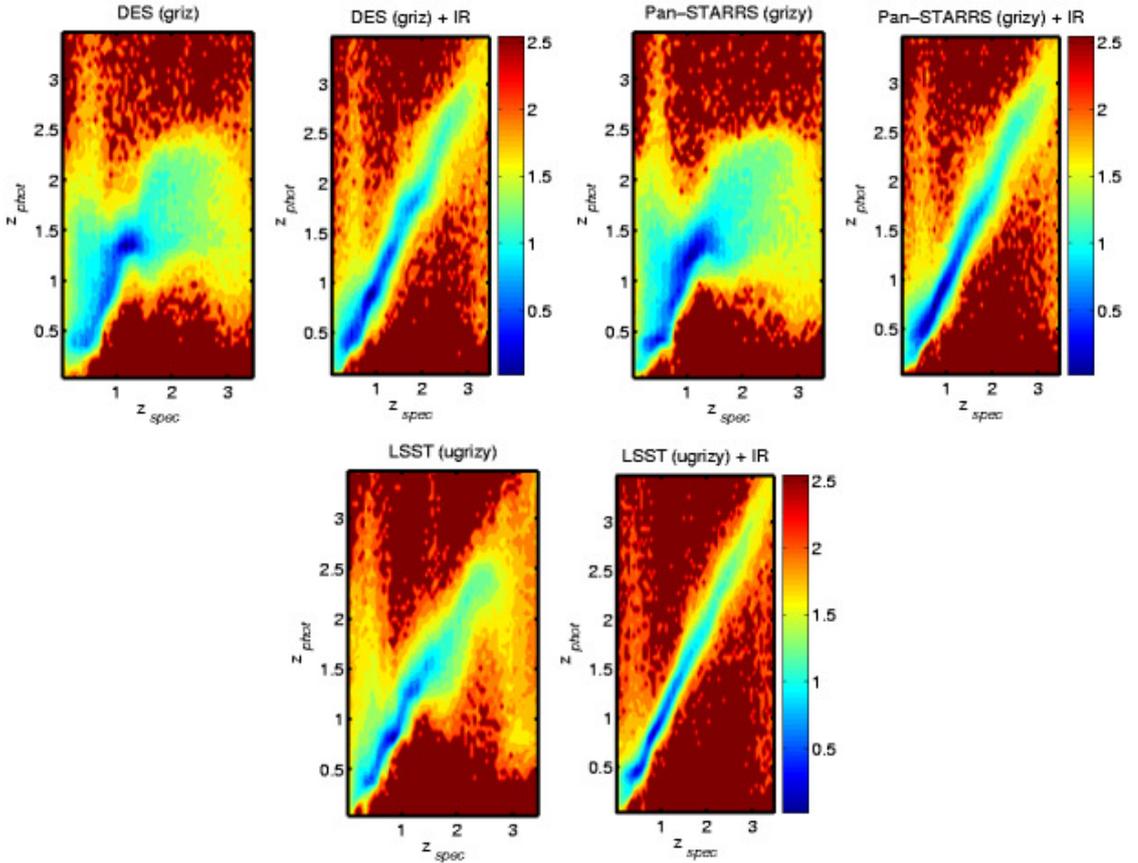

Figure 20.1: The expected photo-z performance for DES, PanSTARRS-4 and LSST. We show here the impact of combining these surveys with deep NIR coming from Euclid. We see that adding the NIR bands drastically improves the performance. Details of these calculations can be found in Abdalla *et al.* (2008a).

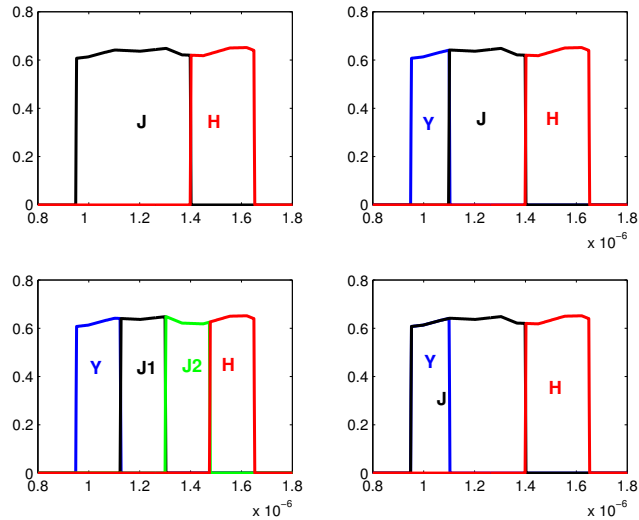

Figure 20.2: Four different NIR filter configurations considered for Euclid. These photo-z performance for each of these was investigated. For a fixed total exposure time we find that comparable results are achieved, with the configuration in the top right giving marginally better performance.

Figure 20.2), trading off the number of filters with the exposure time in each filter so as to preserve the total exposure time, and using the Euclid exposure time calculator [1] (Fontana *et al.* 2006), see chapter 21. It is found that the combination shown in the top right of Figure 20.2 produces marginally better results and this three-filter combination has been adopted as the default NIR configuration for Euclid. However, the trade-off is rather flat, and the photo-z performance is determined primarily by the total available exposure time for the near-infrared photometry, integrated across the available filters.

## 20.4 Ground-based photometry

### 20.4.1 Effect of depth of ground based photometry

As noted above, the basic performance of different photo-z codes is similar. The main challenge in reaching the Euclid requirements is in "cleaning" the photo-z of catastrophic failures, i.e. objects for which the photo-z is widely discrepant from the real redshift. This must of course be done *a priori*, i.e. without knowledge of the actual redshift from the photometry alone, and inevitably results in the loss also of objects with good photo-z. Increased depth in the photometry leads to a reduction in the percentage of these outliers and a reduction in the the associated "wastage" of good photo-z, as well as an increased precision of individual photo-z.

In Bordoloi *et al.* (2009) we examined the photo-z performance using the photometric catalogues degraded to simulate different surveys as listed in Table 20.1. In Figure 20.3 we show the r.m.s. spread ($\sigma_z(z)$) and the bias ($\Delta_z(z)$) for the different survey configurations. The blue curves in all the panels give the initial performance of the photo-z code, without attempting to remove outliers. As expected, increased depth in the optical ground-based photometry increases the reliability of the photo-z estimates. However, we always need to clean the photo-z catalogues to meet the requirement of $\sigma_z(z)/(1 + z) \leq 0.05$, especially at the lower redshifts $z \sim 0.5$ where many the galaxies in fact lie. The green curves show the effect of removing outliers, recognised purely photometrically, while the red curve shows the effect of modifying the individual $L(z)$ (discussed further below). All these procedures are described in detail in Bordoloi *et al.* (2009). After the *a priori* removal of doubtful photo-z, the errors in $\sigma_z(z)$ and mean bias $\Delta_z(z)$ are dramatically reduced, as shown by the green lines in Figure 20.3. This is summarised in Table 20.2. The large improvement in $\sigma_z(z)$ and $\Delta_z(z)$ comes from rejection of catastrophic failures rather than a tightening of the *good* photo-z. Furthermore, as the depth of the photometry increases, it is found that fewer objects need to be rejected to improve the photo-z estimates. In the case of survey-A, we find that 23% must be rejected to get below $\sigma_z(z) \leq 0.05(1 + z)$, for survey-B it is 12% and for survey-C, only 9%. The trade off between beneficial cleaning and the wasteful loss of objects determines the efficiency of the cleaning process. After the above cleaning has been performed, the fraction of $5\sigma$ outliers (catastrophic failures) is reduced below 1% in all the three cases Table 20.3.

---
[1] http://lbc.oa-roma.inaf.it/cgi-bin/ETC_DUNE.pl

| $\langle \frac{\sigma_z(z)}{1+z} \rangle$ for different surveys in the range $0.3 \leq z \leq 3.0$ | | | |
|---|---|---|---|
| Survey | Before Cleaning | After Cleaning | After Cleaning + Correction |
| Survey-A | 0.1703 | 0.0884 | 0.0675 |
| Survey-B | 0.1164 | 0.0640 | 0.0497 |
| Survey-C | 0.0876 | 0.0492 | 0.0398 |

Table 20.2: The $\langle \frac{\sigma_z(z)}{1+z} \rangle$ for the three surveys studied. After cleaning and correction has been performed survey-B just about reaches $\sigma_z(z)/(1 + z) \sim 0.05$ Euclid requirements.

| Percentage of $5\sigma$ outliers ($f_{cat}$) in | | |
|---|---|---|
| Survey | Before Cleaning | After Cleaning |
| Survey-A | 1.18 | 0.2300 |
| Survey-B | 0.8820 | 0.2138 |
| Survey-C | 0.8221 | 0.1776 |

Table 20.3: The percentage of $5\sigma$ outliers in various surveys studied. $f_{cat}$ reduces significantly once cleaning of the catalogue is performed, which identifies most of the outliers effectively.

## 20.5 Characterization of N(z)

### 20.5.1 Direct Sampling approach

One approach to determining $N(z)$ and $\langle z \rangle$ is to obtain spectroscopic redshifts for a subset of the objects in a given photo-z bin. As only a finite number of galaxies are available to determine this distribution, there will be uncertainties associated with its estimate, which will propagate into uncertainties in cosmological parameter estimates using weak lensing (Amara & Réfrégier 2007; Kitching *et al.* 2008b).

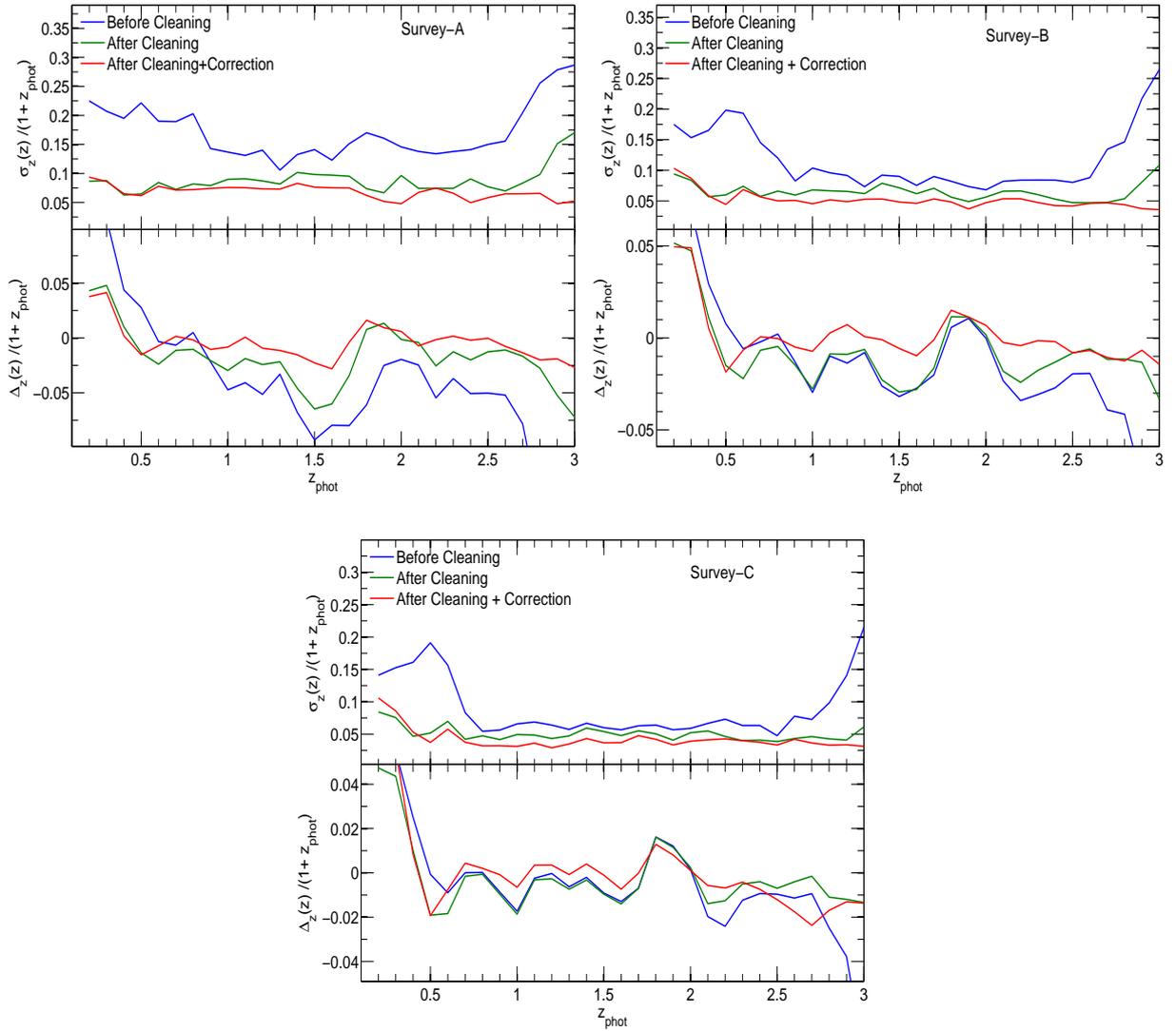

Figure 20.3: The overall performance of several surveys investigated whose depths are as quoted in Table 20.1. The blue lines give the performance without cleaning and the green lines after cleaning and the red line gives performance after cleaning and applying correction. For survey-A 23% for survey-B 13% and for survey-C 9% rejections were made after cleaning.

| z-range | $\mu_2$ | $\mu_4$ | $N_s$ from mean | $N_s$ from variance |
|---------|---------|---------|-----------------|---------------------|
| 0-0.561 | 0.013 | 0.0006 | 3300 | 7 |
| 0.561-0.741 | 0.02 | 0.001 | 5000 | 9 |
| 0.741-0.931 | 0.036 | 0.0035 | 8800 | 30 |
| 0.931-1.181 | 0.06 | 0.008 | 15000 | 70 |
| 1.181-3.531 | 0.14 | 0.07 | 35000 | 800 |

Table 20.4:  The second (variance) and fourth (kurtosis) moments about the mean of the redshift distributions in each of the redshift bins and resulting requirements on the size of the spectroscopic training set, $N_s$.

The two other photometric redshift statistics required for a Euclid weak lensing survey, are calibration of the mean of the redshift probability distribution, $p_i(z)$ to be less than $0.002(1 + z)$, and the calibration of variance of $p_i(z)$, to be less than $0.008$.

If we define the mean of the distribution as $\overline{z}$ and the $k^{th}$ moment of the distribution, $\mu_k$, as only a finite number of galaxies with spectroscopic redshifts, $N_s$ are available to determine this distribution, assuming Poisson statistics, the standard deviation on the mean redshift of each bin is given by:

$$rms(\overline{z}) = \frac{\sqrt{\mu_2}}{\sqrt{N_s}} \tag{20.1}$$

and similarly the standard deviation of the variance is given by, in the limit of large $N_s$ this becomes:

$$rms(\mu_2) = \frac{\sqrt{\mu_4 - \mu_2^2}}{\sqrt{N_{spec}}} \tag{20.2}$$

One can marginalise over both these uncertainties. The dependence on the number of spectroscopic redshifts is really an indirect expression of the scatter in $\overline{z}$ and in $\mu_2$. Using the simulations from the previous section the number of spectroscopic training set galaxies required in each redshift bin can be calculated. Estimates of $N_s$ from equations 20.1 and 20.2 are given in Table 20.4 assuming $rms(\overline{z}) < 0.002$ and $rms(\mu_2) < 0.008$. As the requirements on the spectroscopic training set is roughly the same for any optical survey with Euclid NIR photometry, an average value for $N_s$ is quoted in Table 20.4 across all the surveys.

The requirements for the calibration of the mean are clearly more demanding than those for calibrating the variance and it can be seen that a total of $\sim 10^5$ spectroscopic training set galaxies are required for calibration of a weak lensing survey such as Euclid assuming five redshift bins (Abdalla et al. 2008a).

Such direct spectroscopic determination of $N(z)$, and it's moments, requires impressively complete spectroscopic redshift determination on very large numbers of faint galaxies. A further complication is the need to have a large number of independent survey fields in order to limit the effects of cosmic variance in the $N(z)$ within a particular survey field. This aspect is studied in Bordoloi et al. (2009), by requiring that the effects of large scale structure in the galaxy distribution, also known as cosmic variance, are at most equal to the Poisson noise (considered above) in determining the error in $\langle z \rangle$. We approach this by determining, from the COSMOS mocks described above, how many galaxies can be observed within a given spectrograph field of view before the variation in $N(z)$ from field to field becomes dominated by the effects of large scale structure in the Universe. This then translates to a maximum on the "sampling rate" in each spectrograph field of view, and a minimum number of independent survey fields. Both will depend on the spectrograph field of view. For typical spectrographs on 8-m class telescopes, such as VIMOS on the ESO VLT, this maximum sampling rate is rather small (e.g. 2%), with an implied inefficiency in telescope utilization. The minimum number of independent fields that will be required in order to attain an uncertainty in the mean redshift of this particular redshift bin $\langle z \rangle$ that is equivalent to the Poisson variance from $10^4$ galaxies is shown in Figure 20.4.

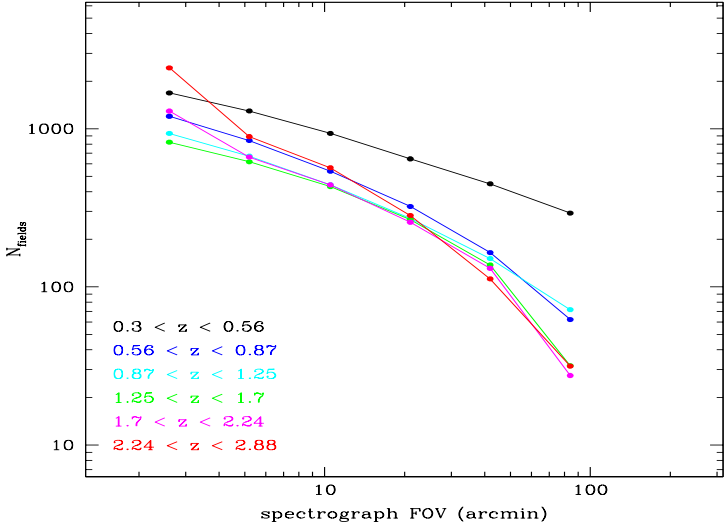

Figure 20.4: The minimum number of independent fields required (as a function of individual survey field size) in order to ensure that the uncertainty in the mean redshift of a given redshift bin is not dominated by large scale structure within the survey fields (i.e. by cosmic variance), and is equal to the Poisson variance from $10^4$ galaxies. Typically, the number of required fields is set by the lower redshifts where the effects of large scale structure are strongest.

### 20.5.2  Estimation of N(z) from photo-z themselves

Generally a single redshift estimator from the photo-z code (i.e. the maximum likelihood photo-z) is used to construct the photo-z bins. Not surprisingly, using these same redshifts, the $\Delta_{\langle z \rangle}$ requirement cannot be reached, as clearly shown in Figure 20.3, because the maximum likelihood redshifts cannot, by construction, then trace the wings of the N(z) that lie outside of the nominal bins, or sample the remaining catastrophic failures which produce distant outliers in the $N(z)$. Therefore a more sophisticated approach is required. We find (Bordoloi *et al.* 2009) that we can characterise $N(z)$ as the sum of the individual $L(z)$ likelihood functions for the galaxies in each redshift bin. These individual normalised $L(z)$ are output for each galaxy by the photo-z code.

The straight sum of the original likelihood functions is already able to characterise the redshift distribution well, as seen in Figure 20.5, which shows for survey-C the summed $L(z)$ traces (visually) both the the catastrophic failures and the wings of the redshift bins. However, we find that this approach must be further modified so as to characterise the $N(z)$ of the bins to the required precision. We do this by modifying the individual $L(z)$ using an approach described in detail in Bordoloi *et al.* (2009). In outline, this approach is based on a fundamental statistical requirement on the $L(z)$. Defining $P_i$ to be the cumulative position within an individual likelihood function $L_i(z)$ of the real "spectroscopic" redshift, $z_i$, then the distribution of $P_i$ across *any* subset of galaxies should be flat across $0 < P < 1$. We can therefore use a limited number of spectroscopic redshifts, in principle for any subset of galaxies, to determine an empirical correction to $L(P)$, that we then apply to all galaxies. This simple ad hoc approach works to bring us within the required precision. In Figure-20.6 the bias on the mean of the $N(z)$ is given for different redshift bins. The shaded region gives the Euclid requirement of $|\Delta_{\langle z \rangle}/(1+z)| \leq 0.002$ on the mean redshift of the redshift bins. The black dots are for survey-C, which easily reaches the Euclid requirements. The red open boxes are for survey-B and it just meets the Euclid requirement. The blue stars are for survey-A which do not meet the specifications as given by the shaded region. From this analysis we conclude that for a Euclid like survey, using a survey-B like ground based complement(similar to PanStarrs-2 or DES) we can characterise the $N(z)$ of the tomographic bins to a precision of $|\Delta_{\langle z \rangle}/(1+z)| \leq 0.002$ using around 800–1000 random spectroscopic redshifts per redshift bin.

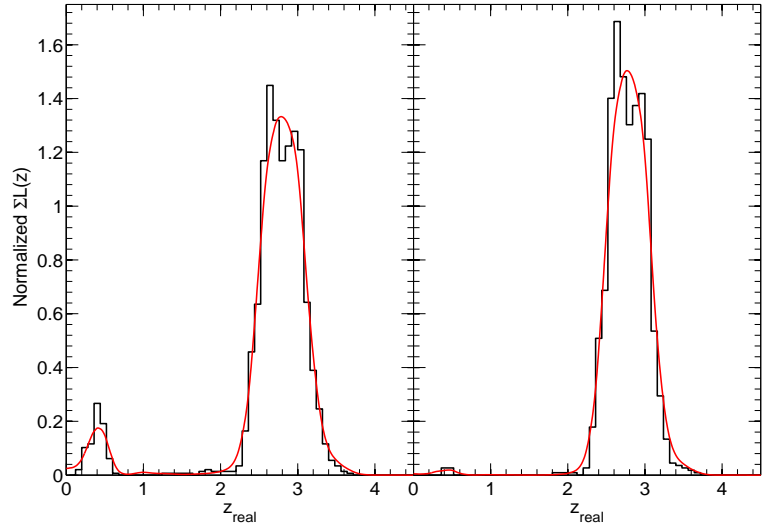

Figure 20.5: $N(z)$ constructed from the $\sum L(z)$ function before and after cleaning. Here the normalised histogram gives the real redshift distribution in the bin and line is the $N(z)$ constructed from the $\sum L(z)$ function. The left panel gives the redshift bin before cleaning and right panel gives after cleaning. The constructed $N(z)$ clearly traces the catastrophic failures.

## 20.6 Internal Calibration of Galactic Foreground Extinction

Extragalactic photometry is routinely corrected for the effects of foreground Galactic extinction using reddening maps and an assumed extinction curve. In practical terms, the effect that we should therefore worry about is an error or uncertainty in the $A_V$, i.e. a $\Delta A_v$, which may be positive or negative. This will cause galaxies to be either too red, or too blue, in the photometric input catalogue and will likely lead to systematic biasses in their photo-z, the sign of which may change with redshift. However, the photometry of the galaxies themselves can, in principle, be used to determine this error in the Galactic reddening, leading to an internal correction of the problem.

This is studied in Bordoloi *et al.* (2009) using a mock catalogue containing $10^4$ objects down to $I_{AB} \sim 24.5$, mimicing a roughly 0.1 deg$^2$ region of the Euclid survey. We consider photometry with the accuracy expected from both survey-A and survey-C, and then perturb these catalogues by applying a standard mean reddening law (Cardelli *et al.* 1989) with a relatively large $\Delta A_v \sim 0.155$, in both the positive and negative directions. We then compare the photo-z for the galaxies with and without these $\Delta A_V$ offsets. With this quite large assumed error, the effects on the photo-z are much larger than can be tolerated. Furthermore, the effect varies with redshift, making it very difficult to isolate this effect from cosmological effects.

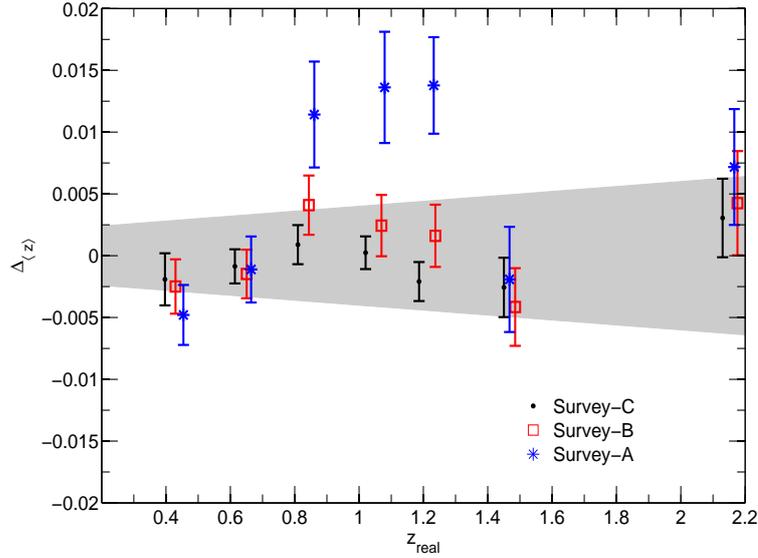

Figure 20.6: The bias in the mean of the tomographic bins estimates from the normalised $\sum L(z)$ functions for all surveys investigated. For survey-C, with cleaning for catastrophic failures and after applying correction gives $|\Delta_{(z)}/(1+z)| \le 0.002$. Here the shaded region gives the Euclid requirements. We have introduced a small offset in x-axis values of survey-B and survey-C for legibility.

However, we find that the photometry of the galaxies can be used to determine $\Delta A_v$ using a simple chi-squared minimization scheme. This works better, as expected, for galaxies with relatively high S/N photometry, and if the redshifts of the galaxies are known. To estimate the effect of photometric noise in estimating $\Delta A_v$ in this internal way, we consider objects in magnitude bins in $I_{AB}$ (see Figure-20.7). We find that it is worth using only galaxies with relatively high S/N photometry, i.e. with the adopted survey parameters, down to $I_{AB} \sim 22$. Below this level, the estimate degrades appreciably. The addition of spectroscopic redshift reduces the error bar significantly, but the method is still practicable down to the same magnitude limit and yields an error on $\Delta A_V$ of order 0.01 (with known redshifts) or 0.02 (without). This is sufficient to bring the effects of uncertain Galactic reddening within the precision requirements for Euclid.

## 20.7 Impact on photo-z from blended objects at different redshifts

Sometimes multiple galaxies will overlap on the sky, and the photometry will be a composite of the two spectral energy distributions. Even with spectroscopy, such objects have composite spectra rendering even their spectroscopic redshift estimation non-trivial. In this final section we explore the effects of this blending on photo-z estimates. We simulate many such blended objects by constructing composite spectral energy distributions constructed from galaxies at different redshifts, different colours and a wide range of relative brightnesses, from dominance of one through to dominance of the other. For definiteness we look at the photo-z behaviour for a survey-C like depth.

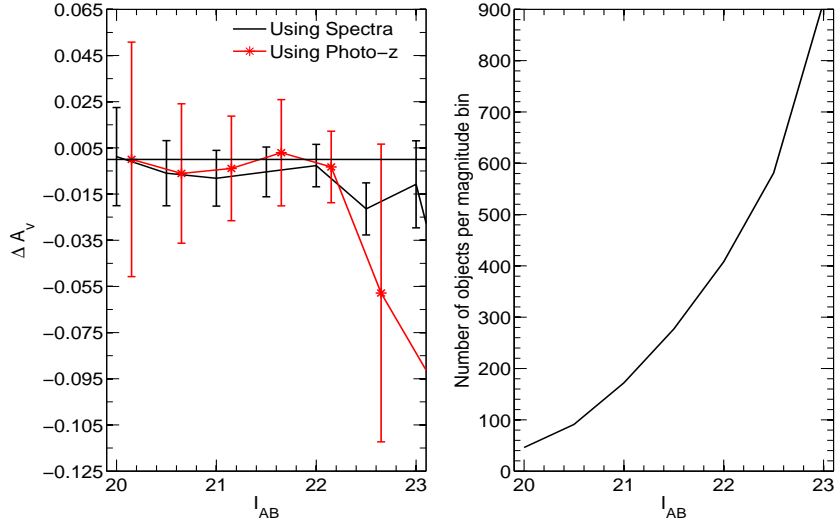

Figure 20.7: The left panel gives estimates of $\Delta A_v$ for different magnitude bins (in the configuration of Survey-C). The red line is obtained using internally computed photo-z, without knowledge of the redshifts of the galaxies, and the blue line is using spectroscopic information. Here we have introduced a small offset in x- axis on the red line for legibility. The right panel is the number of objects per magnitude bin. With spectra, at $I_{AB} \sim 22$ magnitude bin around 350 spectra are sufficient to estimate $\Delta A_v$ accurately.

Our general conclusion from this analysis is that the effect on the photo-z of blended objects is complex and varies in a rather unpredictable way from pair-to-pair. The photo-z are trustworthy only if the second component is at least two magnitudes fainter than the primary (in a waveband close to the middle of the spectral range of the photometry). At smaller magnitude differences the photo-z can be corrupted in a way that is not always recognizable. Sometimes the best fit photo-z varies smoothly between the two real redshifts, sometimes it jumps between the two, and sometimes there is an additional local maximum in between. The conclusion is that these blended objects should be recognised morphologically from the images, and excluded from the analysis. This is unlikely to be a problem, since the shape measurement of these objects would anyway be problematical.

## 20.8 Conclusions

In this chapter we have investigated a number of issues that could potentially impact the photo-z performance of deep all-sky surveys such as Euclid. In each case, we have found that the most basic approaches will not meet the demanding specifications. However, simple and logically motivated developments to these exiting techniques, yield a performance that meets the Euclid specification. Specifically, the following conclusions can be drawn from this photometric redshift analysis:

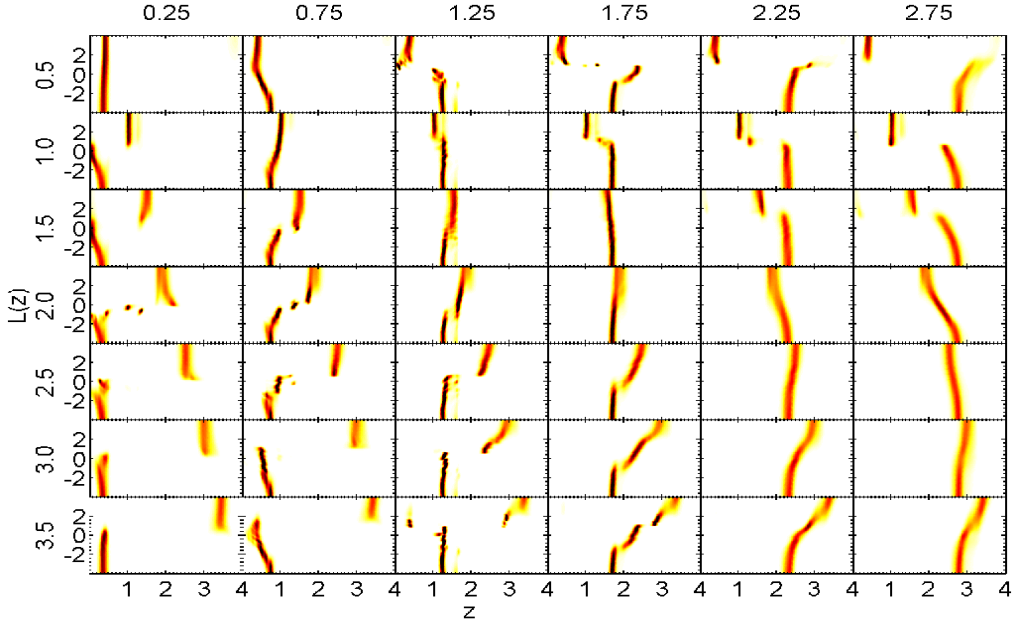

Figure 20.8: $L(z)$ functions of 42 pairs of blended objects. Here when the low redshift object is brighter than the high redshift one and the likelihood function traces low redshift function. The degeneracy of the likelihood functions at $\Delta m \sim 0$ means that the redshift estimation gets completely unreliable at that region. Here the objects are taken for survey-C like depth.

- We have studied the redshift performance of a number of ground based surveys with and without the space based Near-IR that Euclid is able to deliver and find that Near-IR photometry is crucial for securing reliable redshifts. The example in Table (20.2) shows the functioning point that sets our requirements.

- For a fixed integration time, we find that the photometric redshift estimate is not very sensitive to the details of near infra-red filter configuration. Optimising this configuration points to a three filter IR survey complemented by a multi band optical survey from the ground.

- Photo-z accuracy of $\sigma_z(z)/(1+z) \sim 0.05$ down to $I_{AB} \leq 24.5$ is achieved by combination of ground-based photometry with depth of Survey-B (similar to PanStarrs-2 or DES) and the deep all-sky NIR survey from Euclid itself. This requires the implementation of an *a priori* rejection scheme (i.e. based on the photometry alone, without knowledge of the actual redshifts of any galaxies) that rejects 13% of the galaxies and reduces the fraction of $5\sigma$ outliers to below $f_{cat} < 0.25\%$. There is a trade off between the rejection of outliers and the loss of *innocent* galaxies with usable photo-z. Deeper photometry improves both the statistical accuracy of the photo-z and reduces the wastage in eliminating the catastrophic failures, and the combination of Survey-C (similar to PanStaars-4) with Euclid near-infrared photometry achieves $\sigma_z(z)/(1+z) \sim 0.04$ after 9% rejection.

- If spectroscopic redshifts are to be used for the "direct" measurement of $N(z)$, then $\sim 10^5$ redshifts are required, with high completeness and reliability to meet the required precision on the mean redshift. Furthermore, these spectroscopic redshifts must be obtained from a large number of sparsely sampled fields or else the error on $\langle z \rangle$ will be dominated by the effects of large scale structure in the spectroscopic fields.

- Many of the difficulties of the direct spectroscopic determination of $N(z)$ can be circumvented by using the photo-z themselves. The $\sum L(z)$ function characterises both the wings of the $N(z)$ and the remaining catastrophic failures. However, to reach the required performance on the mean of the redshifts $|\Delta_{\langle z \rangle}| \leq 0.002(1 + z)$ with the Survey-B, or deeper Survey-C, combination (together with Euclid infrared photometry), an adaptive modification scheme on the individual $L(z)$ is implemented. This is based on the spectroscopic measurement of redshifts for a rather small number of galaxies ($\leq$1000) with relaxed requirements on statistical completeness (and no dependence on Large Scale Structure in the spectroscopic survey fields).

- Uncertainties in foreground Galactic reddening can have a serious effect in perturbing the photo-z. However, such errors in $A_v$ can be identified internally from the photometric data of galaxies, either with or without spectroscopic redshifts. This procedure is optimum for galaxies with relatively high S/N photometry $I_{AB} \leq 22$. The required number of galaxies suggests that a reddening map on the scales of 0.1 deg$^2$ can be internally constructed from the data on galaxies without known redshifts, or from a few hundred galaxies with spectroscopic redshifts.

- The photo-z of composite objects in which the photometry is a mixture of two objects at different redshifts behaves in a complex and unpredictable way. The photo-z of the composite SED is a good representation of the redshift of the brighter object as long as the magnitude difference is large, i.e. $\Delta I_{AB} > 2$. When the galaxies are closer in brightness, $\Delta I_{AB} < 2$, there is a wide-range of irratic behaviour. Our conclusion is that composite objects with $\Delta I_{AB} < 2$ should be recognised morphologically from imaging data and removed from the photo-z analysis.

As a consequence of this and other work undertaken by the photometric redshift working group, the current imaging component design for Euclid consists of a broad optical $RIZ$ filter as well as three near infra-red $YJH$ bands. The general conclusion of our study is that while reaching the photo-z performance required for weak-lensing surveys such as Euclid will not be trivial, implementation of new techniques, coupled with internal calibration of (e.g. foreground reddening from the photometric data itself), will allow this performance to be attained. If more reliance can be placed in this way on the photo-z themselves, then this will lead to a major simplification of the otherwise challenging requirement for spectroscopic calibration of large scale photometric surveys.

# An Exposure Time Calculator for Euclid

*Authors: Andrea Grazian (IAF-OAR), Stefano Gallozzi (IAF-OAR), Adriano Fontana (IAF-OAR).*

**Abstract**

We describe here the Exposure Time Calculator that has been developed for Euclid. It allows one to compute the predicted performances of Euclid under a variety of instrumental and observing conditions. The ETC is a Euclid-tailored version of one developed for the Large Binocular Camera at LBT, and is available at http://lbc.oa-roma.inaf.it/cgi-bin/calculateETC_DUNE.pl

## 21.1   Introduction

The variety of science applications and instrumental configurations developed during the design of the Euclid mission are so large that it is necessary to have a dedicated tool that can estimate the Euclid predicted performances. To accomplish this task, a dedicated Euclid Imaging Exposure Time Calculator (E-ETC in the following) has been developed. The E-ETC is a modified version of the one already developed and operational at the Large Binocular Camera of LBT.

The basic functions provided are typical of ETC's, with some improvement over standard ones. While typical ETC's allows the user to set the desired signal-to-noise (S/N) or an exposure time, the E-ETC allows to define two out of the three actual free parameters (for a given configuration): S/N, exposure time and magnitude limit. This flexibility allows the user to explore different combinations of parameters, and is very useful when designing a space mission like Euclid, where several constraints complicate the planned observations.

In addition to this an implementation, specific to Euclid, is the possibility of changing several instrumental parameters. For instance, the user can create on-the-fly new filters of arbitrary bandwidth and efficiency, or chose among several instrumental configurations for electronics.

In this section, we shall briefly discuss its use.

## 21.2   Layout overview

The front page of the ETC is organized in three panels shown in Fig.21.1. The main upper panel summarises the input parameters, which should be necessary filled by the common user. In this section the user can change the instrumental parameters, including the selected detector (optical/IR).

Figure 21.1: Overall layout of the front page of the E-ETC.

The second (left) panel operates with the Total Exposure Time, the Signal to Noise ratio and the Magnitude of a given object. In this panel the user can set the global parameters of the observations.

In case of observations resulting from the sum of different exposures, the details of the individual exposures can change the overall results. The details of the single exposure can be set in the third (right) panel, where the number of exposures, the background and the magnitude of saturation can be set.

In the following, we shall describe the assumed formula that sets the results for the Total Exposure Time panel.

## 21.3   Using the E-ETC

The key formula used in the ETC is described in the following, and allows to compute the S/N when the total exposure time $t$ and the total magnitude of an object $M$ are given. This is the reference formula for the ETC: the first and the third cases of this panel are derived by inverting this formula. To enhance the S/N it is possible to modify the total exposure time $t$ and/or on the number of exposures $n$

$$S/N = \frac{F}{\sqrt{(F + B * A)/g + n * A * (ron/g)^2 + \text{Dark}}}. \tag{21.1}$$

$F$ is the flux source (in ADUs), $B$ is the background, $n$ is the number of individual exposures and $A$ is the area of the collecting mirror.

A full description of the mathematics involved can be found here in `http://lbc.oa-roma.inaf.it/ETC/ETC_help/index.html`

In output, the ETC provides the basic number that describe the observations. The fundamental numbers (S/N, magnitude limits, exposure time and details of individual exposures) are shown in the first page, so that the user can easily iterate with few clicks. A more detailed input can be found in a separate page, where a simulated image - with the specified morphology and noise - is shown and can also be downloaded. (Fig.21.3)

It is also possible to see the final efficiency as a function of wavelength, both for filters only and including the detectors (Fig.21.3).

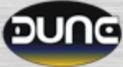

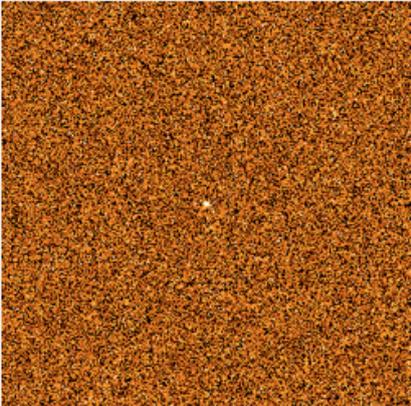

Figure 21.2: An example output page from the E-ETC.

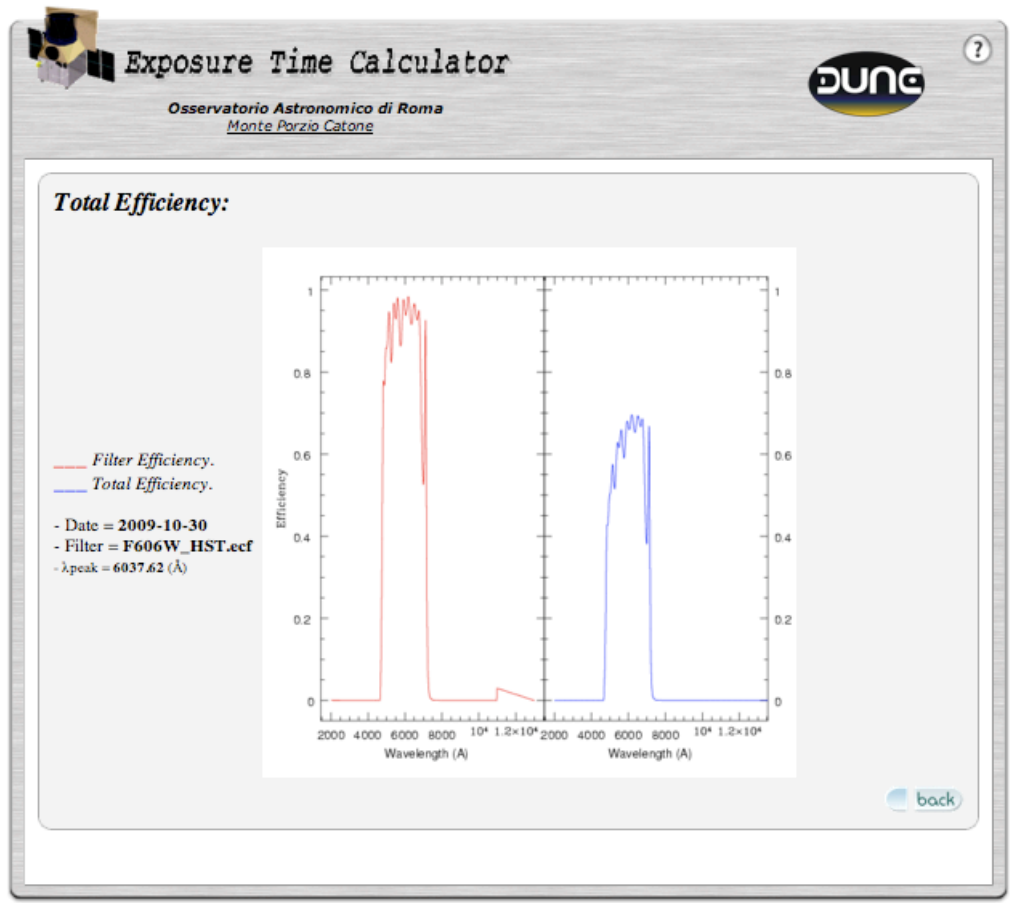

Figure 21.3: An example output page showing the efficiency as a function of wavelength.

# Cosmological Simulations

*Authors: Romain Teyssier (CEA Saclay, Univ. of Zurich), Alexandre Réfrégier (CEA Saclay, France), Adam Amara (ETH Zurich), Marcella Carollo (ETH Zurich), Alan Heavens (Uni. of Edinburgh).*

**Abstract**

In this section, we describe the work done by the EIC regarding cosmological simulations activities. N-body and gas dynamics simulations are great tools to compute theoretical expectations from the standard ΛCDM scenario, as well as for non-standard models. Given the required accuracy of the Euclid mission, these simulations are quite computationally demanding, and have to involve gas and galaxy formation physics. We outline recent progresses made by the EIC in running and analysing these challenging simulations.

## 22.1 Introduction

We have substantial evidence that dark matter is a collisionless, self-gravitating fluid. Its dynamical evolution is therefore described by the well understood Vlassov-Poisson equations, for which N-body models are known to capture the main bulk properties quite accurately. The challenge here is the dynamical range. Large-scale structure surveys such as Euclid will set very tight requirements on our N-body simulations. First, we need to cover a very large volume, in order to reproduce the same statistical sample of the forthcoming survey. In order to match the observational Universe, we need a box size of 6000 Mpc. Second, we need to describe the matter distribution down to very small scales, in order to achieve the angular resolution of those surveys (a few arcseconds). This requires a minimal dark halo mass of about $10^{11}$ Solar masses. Since we need at least $\sim 100$ particles to capture the formation of a dark matter halo, these requirements translate into a total number of particles in excess of 4 trillion. Obviously, we need the largest computers in the world in conjunction with highly optimised software to achieve this. Moreover, in order to reach the level of accuracy set by the low signal-to-noise galaxies that will be analysed in the the Euclid mission, numerical errors should controlled within a level lower than 0.1%, a rather strong requirement on existing N-body codes. Last but not least, at this level of precision, gas dynamics and galaxy formation physics, both relatively poorly known, should be also included. We now report the work done by our consortium in the last 3 years.

## 22.2 A 70 billion particle N-body simulation

We have performed in 2007 the largest N-body simulation so far: it features 70 billion particles (Teyssier *et al.* 2009). This "grand challenge" simulation required deploying the RAMSES code (Teyssier 2002) on a totally new hardware in CCRT, the CEA supercomputing centre. This rather unique effort was performed in collaboration between the Horizon project and computer scientists of CEA and BULL. Many technical bottlenecks have been solved and the simulation was completed in 2 months time, for a total of 6 million CPU hours (1000 years). We have analysed this 70 billion particle simulations in the context of the Euclid experiment. We have used for that purpose a rather advanced Ray-Tracing strategy to compute a high-resolution Healpix map (Górski *et al.* 2005) of the weak-lensing convergence over the whole sky. We have used this map to test various map-making algorithms that will be applied within the EIC consortium to detect point sources and to reconstruct the matter density field. We have shown that one of the main challenges in the Euclid All-Sky survey will be dealing optimally with both the large-scale, quasi-Gaussian signal and the small-scale, highly non-linear and non-Gaussian fluctuations.

## 22.3 A different strategy: many small simulations

In a completely different spirit, Teyssier & Pires have simulated a large number of small simulations, in order to explore various cosmological parameters. This strategy is interesting because a reliable statistical sample of weak-lensing maps can be generated without having to rely on world-class supercomputers. N-body simulations are only needed to reliably estimate the highly non-Gaussian likelihood functions. This series of simulations was performed on the Grid5000 system, in collaboration with a group of computer scientists from ENS Lyon. This was part of the LEGO project, funded by the French ANR in 2005-2009, and the feasibility of the approach was demonstrated in Caniou *et al.* (2006). This large number of small simulations was used to study various aspects of Euclid data processing and cosmological parameters estimation (Pires *et al.* 2009).

Using yet another approach Kiessling, Lynn, Taylor, Heavens & Peacock have run smaller-scale, fast simulations with halo replacement by NFW profiles (Lynn & Peacock, in prep), which provides greatly increased dynamic range at a very moderate cost. The method has been calibrated once and for all using one expensive, ultra-high-resolution simulation. This allows large numbers of reasonably accurate simulations to be run very fast, an essential requirement to provide good estimates of covariance properties of lensing statistics (Kiessling et al., in preparation).

## 22.4 Introducing the baryons

The other important work related to simulation activities is related to the effect of baryons on the predicted power spectrum. This has been investigated first by White (2004) and Zhan & Knox (2004). These authors showed that baryons decouple dynamically from dark matter at various scales within the halo virial radius. The hot halo gas, first, is known to be spatially extended, following the so-called "beta model". This density profile differs significantly from the dark matter profile (Navarro *et al.* 1997) and presents a core in the halo centre, while dark matter is cuspy. This translates into a deficit of power on intermediate scales (around a few arcminutes). The cold gas and the stars, finally, are more concentrated than dark matter on small scales. This translates into an excess of power at arcsecond scales. These early theoretical calculations have been confirmed by detailed cosmological simulations including galaxy formation physics (Jing *et al.* 2006; Rudd *et al.* 2008). Note however that these numerical simulations, like most present day galaxy formation simulations, suffer from the so-called over-cooling problem: too much baryonic mass ends up in the very centre of the galaxy, so that the effect on the matter power spectrum is probably over-estimated by these simulations. It is therefore of paramount importance be able to understand these effects, parameterised them in a sensible way and fit the baryonic parameters on the high-quality data of the future high-redshift surveys. Recently, we have proposed a model to incorporate baryonic effects onto the matter power spectrum (Guillet et al, submitted). We have shown that a simple model based on two components (the halo and a compact stellar disc in the centre) can account for the total matter distribution with great accuracy. The stellar disc has two effects on the matter distribution: first, this central mass concentration causes the dark halo to adiabatically contract, so that the dark matter density profile is denser and steeper than the original distribution; second, in the very centre, the stellar disc dominates the total mass and therefore contributes directly to the weak-lensing signal. These effects can be captured using rather simple analytical models, such as the "halo model" (Seljak 2000), but the model's parameters have to be calibrated on detailed galaxy formation simulations and/or direct observations of galaxy groups. The model predictions have to go beyond the simple 2-points correlations function of the weak-lensing signal and incorporate the effect of baryons on the bi-spectrum and higher order statistics.

## 22.5 Disc intrinsinc alignement

The other important effect related to baryons physics is the orientation of stellar discs with respect to the surrounding large-scale matter distribution. The main hypothesis of the weak-lensing analysis is that background galaxies are randomly oriented, so that systematic alignment between these galaxies is the signature of gravitational lensing. Any intrinsic alignment of galactic discs with the underlying dark matter distribution can be a very serious systematic effect for future high-redshift surveys. On intermediate scale, such alignment is predicted by N-body simulations of dark matter halos (Knebe *et al.* 2008). Surprisingly, however, observational results indicate that the orientations of galaxies are correlated up to scales of order 100 times larger than the virial radii of their host haloes. Mandelbaum *et al.* (2008), Okumura *et al.* (2009) and Faltenbacher *et al.* (2008) detect an alignment of galaxy shapes for the luminous red galaxies (LRGs) in the SDSS survey out to the largest probed scales (roughly 80 Mpc). Various effects related to gas physics can explain discs alignments on large scales. Ram pressure stripping of gaseous discs on the hot gas in filaments could be responsible for a systematic alignment of star-forming discs perpendicular to the satellite-central vector. These subtle effects have been explored in a detailed galaxy formation simulation by our group (Hahn et al., submitted). Analytic models based on the halo model have been also developed that also include prescriptions for small-scale satellite-central alignments (Schneider & Bridle 2009), and that can be tested using this type of simulation.

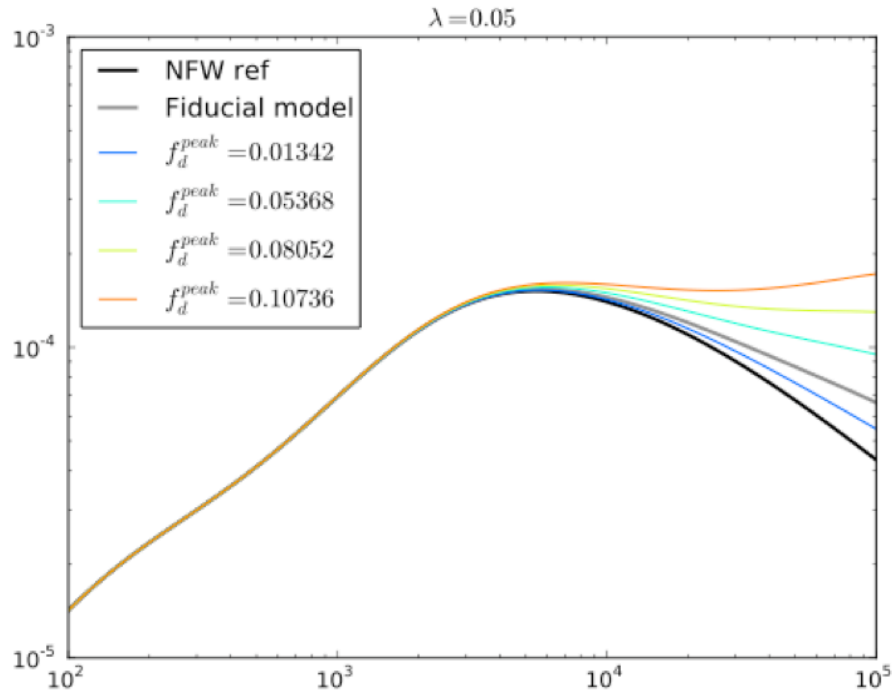

Figure 22.1: Effect of baryons on the total matter power spectrum for various central galaxy mass. We see a boost at small scale, from 1% around l=5000 to 50% at l=10⁵ (from Guillet et al. 2009).

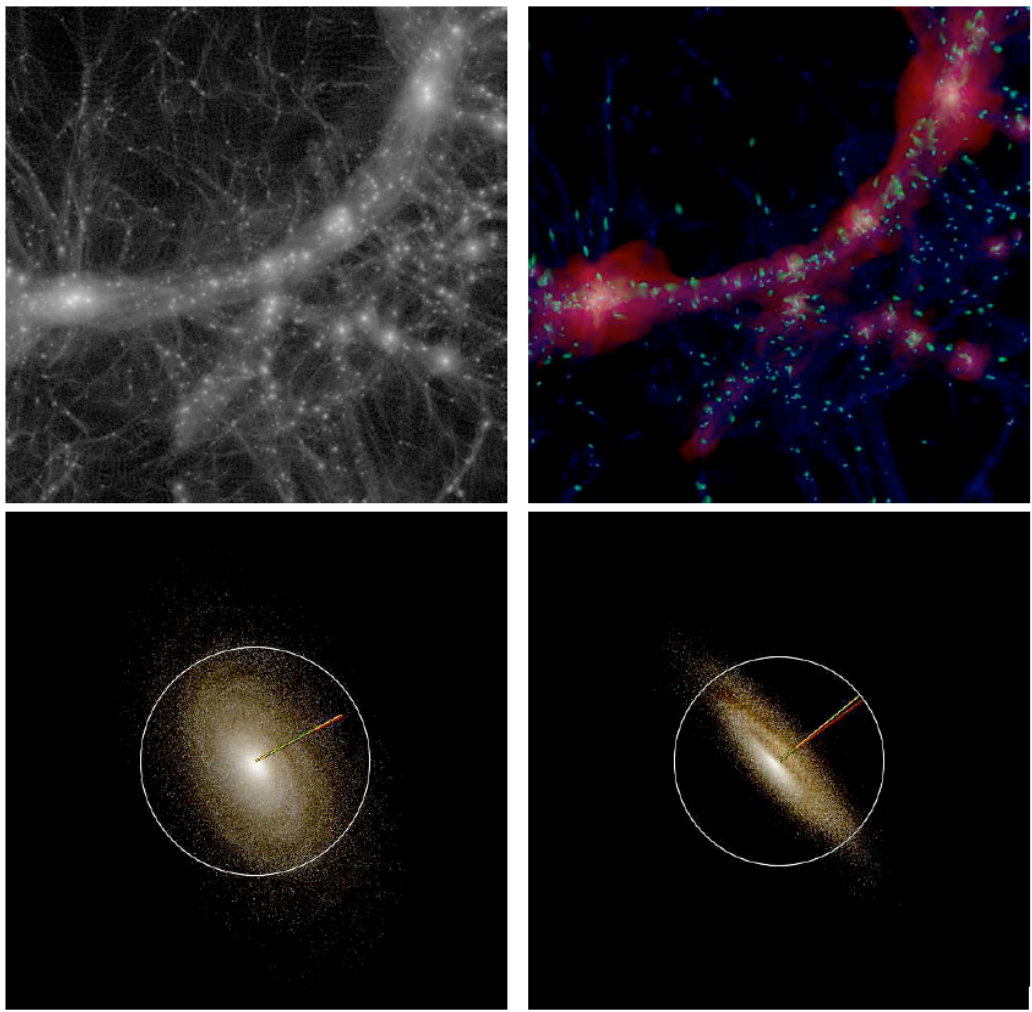

Figure 22.2: Cosmological simulation of a large filament (upper left: dark matter and upper right: gas distributions) from which we studied galactic disc alignment properties (see 2 examples at the bottom). From Hahn et al. (2009).

**Part VI**

# APPENDIX: Letters of Support from Ground-Based Projects

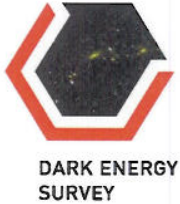

**DARK ENERGY
SURVEY**

September 1, 2009

Dear Sir/Madam,

I am writing to confirm my support for collaboration between the Dark Energy Survey (DES; www.darkenergysurvey.org) and the *Euclid* satellite mission. The details of this collaboration will be defined during the Definition Phase of the Cosmic Visions (CV) programme, namely January 2010 to September 2011.

I present a brief overview of DES here. It is an optical ground-based multi-color survey covering 5000 deg$^2$ of the South Galactic Gap region of the sky and it will be undertaken using "DECam", a new 3 deg$^2$ field-of-view camera, which is being built by the DES consortium. The camera is on schedule for deployment on the NOAO Blanco 4-meter telescope in Chile in 2011, with DES observations beginning in late 2011. DES has been allocated 525 nights on the Blanco telescope spread over 5 years of operations, and we expect to reach a depth of 24$^{th}$ magnitude in four filters (g,r,i,z,). We further expect that this will deliver photometric redshifts (photo-z's) to an accuracy of $\sigma_z \sim 0.12$ (up to $z \sim 1$) and provide a catalogue of 3x10$^8$ galaxies for both weak lensing and large-scale structure measurements. We will also make observation with a "DES Y band" filter to a depth of $\sim 22^{nd}$ magnitude. Finally, DES will spend part of its photometric time on undertaking a search for high redshift supernovae.

Discussions with the *Euclid* Survey Science Team (ESST) and the scientists working on both DES and *Euclid*, have led us to the conclusion that there is significant synergy between these two projects. DES could provide the needed ground-based optical data, and lower redshift photo-z's, for *Euclid* in the Southern Hemisphere. Likewise, the *Euclid* near-infrared data, when combined with DES colors, would significantly improve the photo-z's of galaxies beyond $z \sim 1$. Finally, the *Euclid* shape measurements would improve upon the DES measurements and would help DES scientists in their weak lensing measurements (e.g. assess systematic uncertainties).

The purpose of this letter is to confirm that the DES team plans to undertake a detailed discussion with the *Euclid* team and to continue the preliminary studies, which have been underway for some time, in order to determine how we would collaborate on joint *Euclid*-DES analyses and any exchange of data. These discussions and preliminary studies will include:

i) An understanding of when the DES data products become publicly available and the need for *Euclid* to access any of these data products prior to their release to the public. This could be the case for the latter years of DES operations (2015-2017) and/or include data from observation beyond what is presently planned. I note that the raw DES data will be released to the public one year after it has been obtained and the reduced images will be released on a similar time scale.

ii) An estimation of what DES data is required for the *Euclid* science requirements and whether *Euclid* would require the re-measurement of certain object parameters, e.g., re-measuring the brightness of DES objects in common apertures with *Euclid* objects. Likewise, we would need to assess how to join the datasets for estimating accurate photometric redshifts;

iii) The details of how to share *Euclid* and DES data between scientists in the two projects and develop procedures for any such data-sharing agreement;

iv) A plan for any additional resources within both DES and *Euclid* to facilitate this data sharing and potentially collaborating on joint science analyses. At present, this is not an exclusive relationship and DES and *Euclid* are free to explore additional collaborations with other surveys and projects.

In summary, I provide my full support for continued discussions between the DES and *Euclid* teams, and expect them to develop a plan for collaboration during the CV Definition Phase (if *Euclid* is successful). I see several mutual benefits in a closer relationship between these two surveys. Please feel free to contact me if you have any further questions or comments

Yours sincerely,

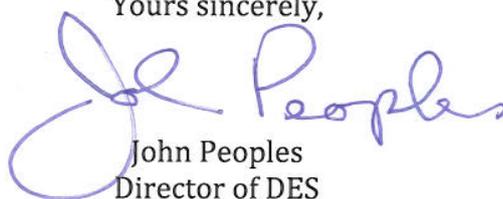

John Peoples
Director of DES

cc: J. Frieman
O. Lahav

# University of Hawai'i at Mānoa

**Institute for Astronomy**
2680 Woodlawn Drive · Honolulu, Hawai'i 96822
Phone: 808-956-8580 · Fax: 808-956-9590 · Email: `kaiser@ifa.hawaii.edu`

Alexandre Refregier
Euclid Project
CEA, Saclay

Dear Alexandre,

I am writing as Principal Investigator of the Pan-STARRS project to confirm my support for a collaboration between Pan-STARRS and the Euclid satellite mission. The details of this collaboration will be defined during the Definition Phase of the Cosmic Visions (CV) programme, namely January 2010 to September 2011.

Pan-STARRS is a wide-field optical and near-IR imager using 1.8m telescopes with 1.4 billion pixel detectors providing well sampled imaging of 7 square degree field of view. The project is funded under a cooperative agreement with the US Air Force to construct the telescopes, which, on completion become the property of the University of Hawaii. The project has completed the PS1 telescope, which is sited at Haleakala Observatory on Maui and which has embarked on a $\sim 3.5$ year mission to carry out a set of surveys. Operations for this mission are being funded by an institutional consortium. The project is in the process of construction of the second telescope PS2, which is planned to be sited temporarily on Maui next to PS1. At the completion of the PS1 mission, both telescopes will move to Mauna Kea where they will join two more telescopes that will be housed in a new enclosure that will replace the UH 88-inch telescope.

The full PS4 system is planned to be operated for a 10 year survey mission, during which it will carry out a combination of 3-pi, medium-deep, ultra-deep and possibly other surveys, observing in the g,r,i,z, and y bands as well as a broad w-band filter optimised for detection of asteroids.

PS4 will reach $\sim 24$th magnitude (5-sigma, point source) detection limit iin the $r$-band in the nominal $\sim 40$s exposure, with commensurate depths in the other bands, and will provide stacked images reaching to $\sim 26th$ magnitude over the 3-pi survey region and deeper still in the medium and ultra-deep survey regions. The project has developed an image processing pipeline that is now being used to process PS1 data and is in the process of developing the relational database system that will allow access too the catalogs, including catalogs of hundreds of millions of galaxies that will be detected with sufficient signal to noise for high precision cosmology with weak-lesing and large-scale structure.

Conversations with the Euclid Survey Science Team (ESST) have led us to the conclusion that there is significant synergy between these two projects. PS could provide the needed ground-based optical data, for Euclid. Likewise, the Euclid near-IR data, when combined with

PS colors, would significantly improve the photometric redshifts for galaxies beyond $z \sim 1$. Finally, the Euclid shape measurements would improve upon the PS measurements and would help PS scientists in their weak lensing measurements.

The purpose of this letter is to confirm that, should the Euclid project be successful in passing the next selection cuts, the PS project plans to undertake a detailed discussion with the Euclid team regarding how we would collaborate on joint Euclid-PS science and any exchange of data. These discussions will include:

- Discussions on the dates for public data release of PS data, and potential for prior access for Euclid scientists for the purpose of evaluation of data quality and refining data analysis pipelines. We note that the PS1 consortium is committed to allowing release of PS1 data one year after completion of the mission, but that the project does not currently have the resources necessary to allow dissemination of these data to the global community.

- Discussions on the quality of the data, with input from real data from PS1, PS2 etc., and assessment of the scientific performance that will realistically be achievable for the key cosmological parameter determination.

- Discussions on data-sharing and collaboration between the scientists in the two projects (and potentially other ground-based projects that may be entering into similar agreements with Euclid).

- Planning for obtaining additional resources that may be required to facilitate this collaboration.

I am very excited by the potential for collaboration with Euclid to substantially enhance the scientific return from both projects and look forward to developing this relationship further.

Yours sincerely,

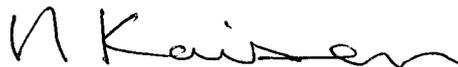

Nick Kaiser

Pan-STARRS Principal Investigator